\newif\ifabstract
\abstracttrue
\newif\iffull
\ifabstract \fullfalse \else \fulltrue \fi

\documentclass[11pt]{article}
\usepackage{amsfonts}
\usepackage{amssymb}
\usepackage{amstext}
\usepackage{amsmath}
\usepackage{xspace}
\usepackage{ntheorem}
\usepackage{graphicx}
\usepackage{url}
\usepackage{graphics}
\usepackage{colordvi}
\usepackage{hyperref}
\usepackage{subfigure}
\usepackage{xcolor}
\usepackage{cleveref}
\usepackage{textalpha}
\usepackage{aliascnt}

\advance \oddsidemargin by -0.8in
\newcommand{\myparskip}{3pt}
\parskip \myparskip

\newcommand{\vdeg}{\overline{\operatorname{deg}}}
\newcommand{\vw}{\overline{w}}
\newcommand{\vDelta}{\overline{\Delta}}

\newcommand{\td}{\tilde d}

\newcommand{\attime}[1][\tau]{^{(#1)}}

\newcommand{\shortpath}{\mbox{\sf{short-path}}}
\newcommand{\embedors}{\mbox{\sf{EmbedOrSeparate}}\xspace}
\newcommand{\embedorscatter}{\mbox{\sf{EmbedOrScatter}}\xspace}
\newcommand{\HSS}{Hierarchical Support Structure\xspace}
\newcommand{\pruning}{\mbox{\sf{NiceRouterPruning}}\xspace}

\newcommand{\DS}{\mathsf{DS}}

\newcommand{\hk}{\hat k}

\newcommand{\APSP}{\mbox{\sf{APSP}}\xspace}

\newcommand{\ceil}[1]{\ensuremath{\left\lceil#1\right\rceil}}
\newcommand{\floor}[1]{\ensuremath{\left\lfloor#1\right\rfloor}}

\newcommand{\set}[1]{\left\{ #1 \right\}}

\newcommand{\tset}{{\mathcal T}}
\newcommand{\iset}{{\mathcal{I}}}
\newcommand{\pset}{{\mathcal{P}}}
\newcommand{\qset}{{\mathcal{Q}}}

\newcommand{\aset}{{\mathcal{A}}}
\newcommand{\cset}{{\mathcal{C}}}

\newcommand{\wset}{{\mathcal{W}}}

\newcommand{\rset}{{\mathcal{R}}}

\newcommand{\hset}{{\mathcal{H}}}
\newcommand{\gset}{{\mathcal{G}}}
\newcommand{\sset}{{\mathcal{S}}}

\newcommand{\uset}{{\mathcal{U}}}
\newcommand{\dset}{{\mathcal{D}}}

\newcommand{\be}{\begin{enumerate}}
\newcommand{\ee}{\end{enumerate}}
\newcommand{\bd}{\begin{description}}
\newcommand{\ed}{\end{description}}
\newcommand{\bi}{\begin{itemize}}
\newcommand{\ei}{\end{itemize}}

\newtheorem{question}{Question}

\newtheorem{theorem}{Theorem}[section]

\newaliascnt{lemma}{theorem}
\newtheorem{lemma}[lemma]{Lemma}
\aliascntresetthe{lemma}

\newaliascnt{observation}{theorem}
\newtheorem{observation}[observation]{Observation}
\aliascntresetthe{observation}

\newaliascnt{corollary}{theorem}
\newtheorem{corollary}[corollary]{Corollary}
\aliascntresetthe{corollary}

\newaliascnt{claim}{theorem}
\newtheorem{claim}[claim]{Claim}
\aliascntresetthe{claim}

\newaliascnt{definition}{theorem}
\newtheorem{definition}[definition]{Definition}
\aliascntresetthe{definition}

\crefname{theorem}{theorem}{theorems}
\Crefname{theorem}{Theorem}{Theorems}

\crefname{lemma}{lemma}{lemmas}
\Crefname{lemma}{Lemma}{Lemmas}

\crefname{observation}{observation}{observations}
\Crefname{observation}{Observation}{Observations}

\crefname{corollary}{corollary}{corollaries}
\Crefname{corollary}{Corollary}{Corollaries}

\crefname{claim}{claim}{claims}
\Crefname{claim}{Claim}{Claims}

\newenvironment{proof}{\par \smallskip{\bf Proof:}}{\hfill\stopproof}
\def\stopproof{\square}
\def\square{\vbox{\hrule height.2pt\hbox{\vrule width.2pt height5pt \kern5pt
\vrule width.2pt} \hrule height.2pt}}

\newenvironment{proofof}[1]{\noindent{\bf Proof of #1.}}%
        {\hfill\stopproof}

\newcommand{\edel}{E^{\operatorname{del}}}

\newcommand{\tdel}{T^{\operatorname{del}}}

\newenvironment{prog}[1]{
\begin{minipage}{5.8 in}
\begin{center}
{\sc #1}
\end{center}
}
{
\end{minipage}
}
\usepackage{framed}
\usepackage[framemethod=tikz]{mdframed}
\newenvironment{wrapper}[1]
{
	\begin{center}
		\begin{minipage}{\linewidth}
			\begin{mdframed}[hidealllines=true, backgroundcolor=gray!20, leftmargin=0cm,innerleftmargin=0.5cm,innerrightmargin=0.5cm,innertopmargin=0.5cm,innerbottommargin=0.5cm,roundcorner=10pt]
				#1}
			{\end{mdframed}
		\end{minipage}
	\end{center}
} 

\renewcommand{\phi}{\varphi}
\newcommand{\eps}{\epsilon}

\newcommand{\half}{\ensuremath{\frac{1}{2}}}

\newcommand{\poly}{\operatorname{poly}}
\newcommand{\dist}{\mbox{\sf dist}}
\newcommand{\diam}{\mbox{\sf diam}}
\newcommand{\reals}{{\mathbb R}}

\newcommand{\expect}[2][]{\text{\bf E}_{#1}\left [#2\right]}

\newenvironment{properties}[2][0]
{
\begin{enumerate} \setcounter{enumi}{#1}}{\end{enumerate}}

\newcommand{\mynote}[2][red]{\textcolor{red}{\sc\bf{[#2]}}}
\newcommand{\mpnote}[2][blue]{\textcolor{blue}{\sc\bf{[MP: #2]}}}

\newcommand{\vol}{\operatorname{Vol}}

\newcommand{\EST}{\mbox{\sf{ES-Tree}}\xspace}

\newcommand{\hn}{\hat n}

\begin{document}
\begin{titlepage}
	\title{Fully Dynamic Algorithms for Graph Spanners via Low-Diameter Router Decomposition\footnote{Preliminary version appeared in SODA 2025}}
	\author{Julia Chuzhoy\thanks{Toyota Technological Institute at Chicago. Email: {\tt cjulia@ttic.edu}. Supported in part by NSF grant CCF-2402283 and NSF HDR TRIPODS award 2216899.}\and Merav Parter \thanks{Weizmann Institute of Science, Israel. Email: {\tt merav.parter@weizmann.ac.il}. Supported partially by the European Research Council (ERC) under the European Union’s Horizon 2020 research and innovation programme, grant agreement No. 949083.}}
	\maketitle
	
\pagenumbering{gobble}	
\thispagestyle{empty}

\begin{abstract}

A $t$-spanner of an undirected $n$-vertex graph $G$ is a sparse subgraph $H$ of $G$ that preserves all pairwise distances between its vertices to within multiplicative factor $t$, also called the \emph{stretch}. Spanners play an important role in the design of efficient algorithms for distance-based graph optimization problems, as they allow one to sparsify the graph, while approximately preserving all distances. It is well known that any $n$-vertex graph admits a $(2k-1)$-spanner with $O(n^{1+1/k})$ edges, and that this stretch-size tradeoff is optimal assuming the Erd{\" o}s Girth Conjecture. In this paper we investigate the problem of efficiently maintaining spanners in the fully dynamic setting with an adaptive adversary. Despite a long and intensive line of research, this problem is still poorly understood: for example, no algorithm achieving a sublogarithmic stretch, with a sublinear in $n$ update time, and a strongly subquadratic in $n$ bound on the size of the spanner is currently known in this setting.
One of our main results is a \emph{deterministic}  (and therefore, adaptive-adversary) algorithm, that, for any $512 \leq k \leq (\log n)^{1/49}$ and $1/k\leq \delta \leq 1/400$, maintains a spanner $H$ of a fully dynamic graph with stretch $\poly(k)\cdot 2^{O(1/\delta^6)}$ and size $|E(H)|\leq O(n^{1+O(1/k)})$, with worst-case update time  $n^{O(\delta)}$ and recourse  $n^{O(1/k)}$. 

Our algorithm relies on a new technical tool that we develop, and that we believe to be of independent interest, called \emph{low-diameter router decomposition}. Specifically, we design a deterministic algorithm that maintains a decomposition of a fully dynamic graph into edge-disjoint clusters with bounded vertex overlap, where each cluster $C$ is guaranteed to be a bounded-diameter router, meaning that any reasonable multicommodity demand over the vertices of $C$ can be routed along short paths and with low congestion inside $C$. 
 A similar graph decomposition notion was  introduced by 
  [Haeupler et al., STOC 2022] and recently strengthened by [Haeupler et al., FOCS 2024]; the latter result was already used to obtain fast algorithms for multicommodity flows [Haeupler et al., STOC 2024] and dynamic distance oracles [Haeupler et al., FOCS 2024].
However, in contrast to these and other prior works, the decomposition that our algorithm maintains is guaranteed to be \emph{proper}, in the sense that the routing paths between the pairs of vertices of each cluster $C$ are contained inside $C$ (rather than in the entire  graph $G$). 
 Additionally, our algorithm maintains,  for each cluster $C$ of the decomposition, a subgraph $C'\subseteq C$ of a prescribed density, that also has strong routing properties.

We show additional applications of our low-diameter router decomposition, by obtaining new deterministic dynamic algorithms for  fault-tolerant spanners and low-congestion spanners. Several of these applications crucially rely on the fact that our router decomposition is proper.
\end{abstract}

\end{titlepage}

\tableofcontents

\newpage

\pagenumbering{arabic}

\section{Introduction}

A spanner of a graph $G$ is a sparse subgraph of $G$, that approximately preserves all pairwise distances between its vertices. 
The notion of spanners naturally falls within the broader class of graph-theoretic objects called \emph{graph sparsifiers}. A sparsifier of a given graph $G$ is a sparse subgraph of $G$, that approximately preserves some central properties of $G$, such as, for example, cut values, flows, or, in this case, distances. 
A natural and powerful paradigm for designing fast algorithms for graph optimization problems is to first compute a sparsifier $H$ of the input graph $G$, that preserves properties of $G$ relevant to the problem, and then solve the problem on the much sparser graph $H$. It is then not surprising that various types of sparsifiers increasingly play a central role in the design of fast algorithms for various combinatorial optimization problems, in static, dynamic, and other settings. Since their introduction by Peleg and Schaffer \cite{PelegS89}, spanners have been studied extensively in various settings (see, e.g. \cite{PelegU87,ElkinP01,ElkinZ04,Baswana06,BaswanaS07,Baswana08,DerbelGPV08,BaswanaKS12,GhaffariK18,DoryFKL21}), and have been used in a wide range of applications: from distance oracles \cite{TZ}, shortest path computation \cite{BaswanaK06} and routing schemes \cite{PelegU:89-routing} to spectral sparsification \cite{kapralov2012spectral}. 

Formally, given an $n$-vertex graph $G$ with non-negative edge lengths, a $k$-spanner for $G$ is a subgraph $H \subseteq G$, such that,  for every pair $u,v$ of vertices of $G$, $\dist_H(u,v)\leq k \cdot \dist_G(u,v)$. In addition to the stretch parameter $k$, a central measure of the quality of the spanner $H$ is its size $|E(H)|$. It is well known that any $n$-vertex graph admits a $(2k-1)$-spanner with $O(n^{1+1/k})$ edges, and this tradeoff is tight under the Erd{\" o}s Girth Conjecture \cite{Erd}. In their influential paper, Baswana and Sen \cite{BaswanaS07} presented a randomized algorithm for computing a $(2k-1)$-spanner of a static $n$-vertex and $m$-edge graph $G$, that contains $O(k\cdot n^{1+1/k})$ edges, in expected time $O(km)$.
Roditty, Thorup and Zwick \cite{RodittyTZ05} derandomized this construction, while achieving similar parameters.

In this paper we focus on the problem of efficiently maintaining a spanner of a fully dynamic graph $G$, that is, a graph undergoing an online sequence of edge insertions and deletions. As usual in the area of dynamic algorithms, we distinguish between the \emph{oblivious-adversary} and the \emph{adaptive-adversary} settings.
In the case of oblivious adversary, it is assumed that the sequence of updates to the input graph $G$ is fixed in advance, and may not depend on the algorithm's behavior and random choices. 
In contrast, in the \emph{adaptive-adversary} setting, each update to the graph may depend on the algorithm's past behavior and inner state arbitrarily. 
It is often the case that the oblivious adversary assumption considerably limits the applicability of a dynamic algorithm. As an example, one common use of such algorithms is in speeding up algorithms for static problems, which typically requires that the dynamic algorithm can withstand an adaptive adversary. 
A large number of breakthrough result in recent years, that obtained fast algorithms for central graph optimization problems in the static setting, crucially rely on dynamic algorithms that can withstand an adaptive adversary. 
%
With this motivation in mind, we focus in this paper on the adaptive-adversary setting.
We note that deterministic algorithms are always guaranteed to work against an adaptive adversary, and as such they are especially desirable. For brevity, we will refer to algorithms that can withstand oblivious or adaptive adversaries as \emph{oblivious-update} or \emph{adaptive-update} algorithms, respectively. 

\subsection*{Dynamic Spanners} 
Dynamic algorithms for graph spanners have been studied extensively, since the pioneering work of Ausiello et al. \cite{AusielloDFIR07}. This  line of research can be partitioned into three main eras. 
%
%

\textbf{Era 1: Optimal Spanners with Low Amortized Update Time.}
Earlier work studied the problem of maintaining spanners of fully dynamic graphs with near-optimal 
tradeoff between the stretch and the size of the spanner, with the focus on  optimizing the \emph{amortized} update time, that is, the {\bf average} amount of time that the algorithm spends in order to process each update to the input graph $G$. 
 Ausiello, Fraciosa and Italiano \cite{AusielloDFIR07} provided deterministic dynamic algorithms for spanners with stretch $3$ and $5$, with amortized update time proportional to the maximum vertex degree in the graph. 
Elkin \cite{Elkin07} presented a randomized oblivious-update algorithm for maintaining $(2k-1)$-spanners for arbitrary values of $k$, with amortized update time $O(mn^{-1/k})$, where $m$ is the number of edges in the initial graph; note that, for sufficiently dense graphs, this amortized update time may be superlinear in $n$.
This line of work culminated with the result of Baswana, Khurana, and Sarkar \cite{BaswanaKS12}, that achieved expected amortized update time  $\min\set{O(k^2\log^2 n),2^{O(k)}}$ against an oblivious adversary with spanner size $O(kn^{1+1/k}\log n)$.  Forster and Goranci \cite{ForsterG19} later improved the size and the update time by factor $k$. The main drawbacks of these algorithms  are that (i) their worst-case update time may be as high as $\Omega(n)$; and (ii) except for the work of \cite{AusielloDFIR07}, all results cited above work in the oblivious-adversary setting only.

\textbf{Era 2: Low Worst-Case Update Time against an Oblivious Adversary.}
Bodwin and Krinninger \cite{BodwinK16} designed the first algorithms for maintaining spanners of dynamic graphs with worst-case update time that is sublinear in $n$, for the special case of stretch values $3$ and $5$. Specifically, they design randomized oblivious-update  algorithms that maintain $3$-spanners with $\widetilde{O}(n^{3/2})$ edges and worst-case update time  $\widetilde{O}(n^{3/4})$, and $5$-spanners with $\widetilde{O}(n^{4/3})$ edges and worst-case update time $\widetilde{O}(n^{5/9})$.  Later, Bernstein, Forster and Henzinger \cite{BernsteinFH19} significantly narrowed the gap between amortized and worst-case update time guarantees by ``deamortizing'' the result of Baswana et al. \cite{BaswanaKS12}. Their algorithms are based on a  general technique for randomized data structures, that converts amortized update time guarantees to worst-case ones. This approach yields a randomized oblivious-update algorithm for dynamic $k$-spanners with $k=O(1)$, with spanner size $O(nk^{1+1/k})$ and worst-case expected update time $2^{O(k)}\log^3 n$, nearly matching the amortized update time bound of \cite{BaswanaKS12}. 

A key limitation of the algorithm of \cite{BernsteinFH19}, as well as all of the above-mentioned previous dynamic algorithms for stretch values $k>5$, is that they are randomized, and more crucially, they only work against an \emph{oblivious adversary}. As mentioned above, this significantly limits the applicability of such algorithms, and addressing this drawback became the next major goal in this area.

%
%
%

\textbf{Era 3: The Quest for Adaptive-Update Algorithms.} Bernstein et al. \cite{sparsifiers} provided the first dynamic algorithms for maintaining spanners that can withstand an adaptive adversary for stretch values $k>5$. Their algorithm is randomized, and with high probability maintains a spanner with stretch $O(\poly\log n)$ and size $\widetilde{O}(n)$, with $O(\poly\log n)$ \emph{amortized} update time. Their algorithm can  also be adapted to obtain worst-case update time at most $n^{o(1)}$, at the cost of increasing the stretch bound to  $n^{o(1)}$.  Their approach relies on expander decomposition, and as such is unlikely to lead to the construction of spanners with sublogarithmc stretch, as was also noted in \cite{BhattacharyaSS22}. 
Lastly, the recent work of \cite{ChenKLPGS22,brand2023deterministic} provided a deterministic algorithm for maintaining an $n^{o(1)}$-spanner of a fully dynamic graph, whose size is $n^{1+o(1)}$, and worst-case update time bound is at least as high as $n^{o(1)}$ times the maximum vertex degree in the graph. Their algorithm has several other important properties, such as, for example, it has a low recourse, and it maintains a low-congestion embedding of $G$ into the sparsifier, which we discuss below. This algorithm, in turn, serves as one of the main subroutines used in their breakthrough almost-linear time algorithm for min-cost flow, and it was also used in a recent work of \cite{kyng2024dynamic} for obtaining low-congestion vertex sparsifiers and algorithms for dynamic All-Pairs Shortest Paths.
We provide a  summary of known results for fully-dynamic spanners in Table \ref{tab:priorwork}. To summarize, to the best of our knowledge, the following fundamental question still remains open: 

\begin{wrapper}\vspace{-3pt}
	\begin{question}\label{qestion: main}
		Is there an adaptive-update algorithm for maintaining a spanner of an $n$-vertex dynamic graph with  any stretch value $k<o(\log n)$, whose size is at most $n^{2-\eps}$ for any constant $\eps$, and worst-case (or even amortized) update time is $o(n)$?
	\end{question}\vspace{-7pt}
\end{wrapper}

\textbf{Dynamic Spanners with Small Recourse.} An additional desired property of dynamic algorithms that maintain spanners (as well as other types of sparsifiers) is to ensure that the resulting spanner changes slowly, as the input graph $G$ undergoes updates.
The reason is that, as mentioned already, a commonly used paradigm is to apply algorithms for various distance-based problems to the spanner rather than the original graph, in order to obtain faster running time guarantees. If, however, following an update to the input graph $G$, the resulting spanner $H$ may undergo a large number of changes, then it is unlikely that such an algorithm may provide low worst-case update time guarantees. This naturally leads to the notion of a \emph{recourse}: the largest number of edge insertions and deletions that the spanner $H$ may undergo after each update to the input graph $G$. This measure naturally plays an important role in dynamic algorithms, see e.g. \cite{Censor-HillelHK16,BernsteinC18,AssadiOSS19,ChechikZ19}. Bhattacharya, Saranurak and Sukprasert \cite{BhattacharyaSS22} presented a deterministic (and therefore, adaptive-update) algorithm for maintaining $(2k-1)$ spanners with near-optimal size, $O(\log n)$ amortized recourse and $\poly(n)$ worst-case update time. 
The algorithm of  \cite{ChenKLPGS22,brand2023deterministic} mentioned above also has recourse at most $n^{o(1)}$.

Our first main result answers Question \ref{qestion: main} in the affirmative, by providing an algorithm for maintaining dynamic spanners that also has a low recourse:


\begin{wrapper}\vspace{-3pt}
\begin{theorem}[Dynamic Deterministic Spanners, Informal]\label{thm:dynamic-spanners}
There is a deterministic algorithm, whose input is an $n$-vertex simple undirected graph $G$ with integral poly-bounded edge lengths, a  stretch parameter $512 \leq k \leq (\log n)^{1/49}$, and an additional parameter $1/k \leq \delta \leq 1/400$. Graph $G$ initially has no edges, and it undergoes an online sequence of edge insertions and deletions. The algorithm maintains a spanner $H$ of $G$ with stretch $k^{O(1)} \cdot 2^{O(1/\delta^6)}$ and $|E(H)|\leq O(n^{1+O(1/k)})$. The worst-case update time of the algorithm is $n^{O(\delta)}$, and its recourse is $n^{O(1/k)}$. The algorithm can also be extended to maintain fault-tolerant spanners, connectivity certificates, and low-congestion spanners.
\end{theorem}\vspace{-5pt}
\end{wrapper}

At the heart of our approach is a new algorithmic tool that we believe is of independent interest, that we call a  \emph{low-diameter router decomposition}, or just a \emph{router decomposition} for brevity. Given an $n$-vertex graph $G$ and a parameter $k$, a router decomposition of $G$ consists of a collection $\cset$ of edge-disjoint \emph{router} subgraphs of $G$, that we call clusters, each of which has a small diameter, such that $\sum_{C\in \cset}|V(C)|\leq n^{1+O(1/k)}$ holds, and the vast majority of the edges of $G$ lie in some cluster of $\cset$. 
The notion of routers was introduced in a very recent work of  \cite{HHT2024,HaeuplerH0RS24},  as an object that is closely related to length-constrained expanders of \cite{haeupler2022hop,haeupler2022cut}. Routers are also closely related to well-connected graphs introduced in \cite{chuzhoy2023distanced}; we discuss this connection in more detail in Section \ref{subsec: routers and wellconnected}. Intuitively, a graph $G$ is a router, if any ``reasonable'' multicommodity demand can be routed in $G$ via short paths (the length is typically sublogarithmic in $n$), and with low congestion. One example of a reasonable demand is where the total demand on every vertex is bounded by its degree. 
%
%

Before we continue, we need to define the notion of \emph{demands} and their \emph{routing}. A \emph{demand} $\dset$ in a graph $G$ consists of a collection $\Pi$ of pairs of vertices of $G$, and, for every pair $(a,b)\in \Pi$, a demand value $D(a,b)>0$. For a vertex $v\in V(G)$, the \emph{demand on $v$} is the sum of demands $D(u,v)$ over all pairs $(u,v)\in \Pi$ containing $v$. We say that the demand $\dset$ is \emph{unit}, if the demand on every vertex $v$ is bounded by its degree, and we say that it is $\Delta$-bounded, for some value $\Delta>0$, if the demand on every vertex $v$ is at most $\Delta$. Lastly, we say that demand $\dset$ is $h$-\emph{length} if, for every pair $(a,b)\in \Pi$ of vertices, $\dist_G(u,v)\leq h$.
The \emph{routing} of a demand $\dset$ is a flow $f$, where, for every pair $(a,b)\in \Pi$, $D(a,b)$ flow units are sent from $a$ to $b$. The \emph{congestion} of the routing is the maximum amount of flow through any edge of $G$. We say that the flow is over paths of length at most $d$, if every flow-path $P$ with $f(P)>0$ has length at most $d$. We say that an algorithm for a static problem has \emph{almost linear} running time, if its running time is $m^{1+o(1)}$, where $m$ is the number of edges in the input graph, and we say that its running time is \emph{near linear}, if it is bounded by $\tilde O(m)$; throughout, $\tilde O$ is used to hide $\poly\log n$ factors.
We now  provide an overview of recent work on expander decompositions, routers and  related graph notions. 
%

%
%

\begin{table*}[t]
	\begin{center}
		\begin{tabular}{|llllc|}
			\hline
			\textbf{Stretch} & \textbf{Spanner Size} & \textbf{Update Time \hfill [W/A]} & \textbf{Deterministic?} &  \textbf{Citation} \\
			\hline
			$(2k-1)$ &    $O\left(k \cdot n^{1+1/k} \right)$ & $O(m\cdot n^{-1/k})$ \hfill [A] & Oblivious & \cite{Elkin07} \\
			$(2k-1)$ &    $O\left(kn^{1+1/k} \log n\right)$ & $\min\{O(k^2\log^2 n),2^{O(k)}\}$ \hfill [A] & Oblivious & \cite{BaswanaKS12}\\
						$(2k-1)$ &    $O\left(n^{1+1/k} \log n\right)$ & $O(k\log^2 n)$ \hfill [A] & Oblivious & \cite{ForsterG19}\\
			$(2k-1)$ &    $O \left(k \cdot n^{1+1/k} \right)$  & $2^{O(k)}\cdot \log^3 n$\hfill [W] &  Oblivious & \cite{BernsteinFH19} \\
			$\poly\log n$ &    $\widetilde{O}(n)$  & $\poly\log n$ \hfill [A] &  Adaptive & \cite{sparsifiers} \\
			$n^{o(1)}$ &    $O \left(n^{1+o(1)} \right)$  & $n^{o(1)}$\hfill  [W] &  Adaptive & \cite{sparsifiers} \\
			$n^{o(1)}$ &    $O \left(n^{1+o(1)} \right)$  & $n^{o(1)}$ \hfill [W] &  Deterministic & \cite{brand2023deterministic} \\
			$\poly(k)\cdot 2^{O(1/\delta^6)}$ & $O \left(n^{1+O(1/k)} \right)$ & $n^{O(\delta)}$\hfill [W] & Deterministic  &\textbf{(this paper)}\\
			\hline
		\end{tabular}
		\caption{\label{tab:priorwork} Summary of known algorithms for fully dynamic spanners. Amortized update time is marked by [A] and worst-case update time by [W]. In the penultimate column we mark wherther the algorithm is deterministic, and, if it is randomized, whether it works against an adaptive or an oblivious adversary. Our result holds for any $512\leq k \leq (\log n)^{1/49}$ and $1/k\leq \delta \leq1/400$ where $1/\delta$ is an integer2.}
	\end{center}
	\vspace{-15pt}
\end{table*}

\subsection*{Expanders, Routers and Graph Decompositions.} 
%
Expanders are a fundamental notion in both graph theory and algorithms. Thanks to a number of powerful algorithmic techniques that were developed for expanders over the years, they now play a central role in the design of efficient algorithms for a wide variety of graph problems, in static, dynamic, and other settings. One of the central properties that make the expanders useful in many applications is that, as shown in the seminal work of Leighton and Rao \cite{LR}, they have strong routing properties, in the sense that any unit demand can be routed across an expander with polylogarithmic congestion, via paths of polylogarithmic length. 
%
An almost-linear time algorithm for computing such routing was shown in \cite{GhaffariKS17,ghaffari2018new}.

Another central tool in this area is \emph{expander decomposition},
that decomposes the input graph $G$ into vertex-disjoint subgraphs called clusters, such that each cluster is a strong enough expander, and only a small number of edges of $G$ connect vertices that lie in different clusters. Expander decompositions are often used in order to reduce a given optimization problem in general graphs to that on expanders, which, due to the many useful properties of expanders, is often more tractable. 
Some examples of algorithmic results relying on expander decomposition include graph sketching and sparsification \cite{andoni2016sketching,jambulapati2018efficient,chu2020graph}, dynamic algorithms for minimum spanning forest 
\cite{nanongkai2017dynamic,dynamic-spanning-forest}, and recent breakthrough almost-linear time algorithms for Min-Cost Flow in directed graphs \cite{ChenKLPGS22,brand2023deterministic}, to name just a few.


In their seminal work, Spielman and Teng  \cite{ST04} provided the first almost-linear time algorithm that computes a decomposition of a graph into clusters, with somewhat weaker guarantees than the expander decomposition:  their decomposition only guarantees that every cluster is contained inside some expander. In other words, if we are given, for every cluster $C$ in the decomposition $\cset$,  a unit demand $\dset_C$, then all these demands can be routed in $G$ simultaneously via short paths and with low congestion, but the routing of the demand $\dset_C$ for a cluster $C$ may not be confined to $C$, and may use edges and vertices that lie outside of $C$. While their decomposition algorithm found many applications, the use of this weakened notion of expander decomposition makes some of the resulting algorithms quite complex. Moreover, in some applications, this weakened notion of expander decomposition is not sufficient. The work of Saranurak and Wang \cite{saranurak2019expander}  overcomes this shortcoming, by providing a near-linear time randomized algorithm for computing a proper (or strong) expander decomposition. This so-called ``user-friendly''  decomposition led to simplifying the analysis of existing algorithms, e.g., \cite{ST04,KelnerLOS14,CohenKPPRSV17,jambulapati2018efficient}, and to new applications that crucially require the stronger guarantee of the proper decomposition, that include algorithms for dynamic minimum spanning forest \cite{nanongkai2017dynamic2,Wulff-Nilsen17,NanongkaiSW17} and short-cycle decomposition \cite{chu2020graph}. Deterministic almost-linear time algorithms for computing proper expander decompositions were later developed by \cite{gao2019deterministic,chuzhoy2020deterministic,chuzhoy2023distanced}.

While expanders provide a valuable and powerful tool for the design of algorithms, they have some shortcomings. One such central shortcoming is that the diameter of an $n$-vertex expander may be as large as logarithmic in $n$, and moreover, given a unit demand $\dset$, one can only guarantee its routing with low congestion via paths of polylogarithmic length, even if all pairs of vertices with non-negative demand are very close to each other.
One implication of this shortcoming is that, when relying on expanders for distance-based problems, such as, for example, dynamic All-Pairs Shortest Paths (APSP), distance oracles, and graph spanners, it seems inevitable that one has to settle for super-logarithmic approximation guarantees. 
To address this challenge, and in order to generalize the successful paradigm of using expanders in distance-based problems to shorter distances,
Ghaffari, Haeupler and R{\"{a}}cke \cite{haeupler2022hop} introduced the notion \emph{length-constrained (LC) expanders}. We provide here a \emph{routing characterization} of length-constrained expanders, that 
was shown by \cite{HHT2024} to be equivalent, up to constant factors, to the original definition of \cite{haeupler2022hop}. Under this definition,
 a graph $G$ is an $(h,s)$-length $\phi$-expander, if any $h$-length unit demand $\dset$ can routed in $G$ via paths of length $O(h \cdot s)$, with congestion at most $\widetilde{O}(1/\phi)$. 
 A somewhat similar notion of well-connected graphs was introduced in \cite{chuzhoy2023distanced}, with the same motivation. They also designed a number of algorithmic tools for well-connected graphs, that may be thought of as analogues of similar tools for expanders, and that  were recently used in \cite{ChuzhoyZ23} to obtain the first adaptive-update algorithm for dynamic APSP with sublogaritmic approximation and amortized update time $n^{o(1)}$. We provide a formal definition of well-connected graphs and discuss their relationship with LC-expanders in Section \ref{subsec: routers and wellconnected}.

Due to the success of the expander decompositions as a powerful tool in algorithm design, it is natural to ask whether one can also decompose a given graph into LC-expanders. Ghaffari, Haeupler and R{\"{a}}cke \cite{haeupler2022hop} introduced the notion of LC-expander decomposition, that, in turn, relies on the notion of a \emph{moving cut}. Given a graph $G$ with lengths $\ell(e)\geq 0$ on its edges $e\in E(G)$ and integers $h$ and $s$, an $(hs)$-length moving cut $L$ assigns an integral \emph{length increase} value $L(e)\in \set{1,\ldots,(hs)}$ to each edge $e\in E(G)$. A graph that is obtained from $G$ by setting the length of every edge $e$ to $\ell(e)+L(e)$ is denoted by $G-L$. 
 The \emph{size} of the $(hs)$-length moving cut $L$ is $|L|=\frac 1 {hs}\cdot \sum_{e\in E(G)}L(e)$. 
An $(h,s)$-length $\phi$-expander decomposition is simply an $(hs)$-length moving cut $L$, such that graph $G-L$ is an $(h,s)$-length $\phi$-expander. 


The work of \cite{haeupler2022hop} provided a proof that there exists an $(h,s)$-length $\phi$-expander decomposition of any $n$-vertex $m$-edge graph,  for $s=O(\log n)$, such that the resulting $(hs)$-length moving cut has size $\tilde O(\phi m)$. Additionally, they provide an algorithm for computing a weaker notion of an $(h,s)$-length $\phi$-expander decomposition. 

 A very recent work by Haeupler, Hershkowitz and Tan \cite{HHT2024} provided an algorithm that, given an $n$-vertex $m$-edge graph $G$, with parameters $0<\eps<1$, $h$ and $\phi$,  computes an $(h,s)$-length $\phi$-expander decomposition of $G$ for $s=\exp(\poly(1/\epsilon))$, where the size of the resulting moving cut is at most $n^{\epsilon}\cdot m \cdot \phi$, in time $O(m \cdot n^{\poly(\epsilon)}\cdot \poly(h))$. Their algorithm was recently extended to the dynamic setting by \cite{HLT2024}.
A key notion underlying these results is that of \emph{routers}. As mentioned already, a graph $G$ is a router if any unit demand can be routed in $G$ via short paths with low congestion. Equivalently, a router can be thought of as an $(h,s)$-length $\phi$-expander, whose diameter is bounded by $h$. 
In order certify that a graph $H$ is an $(hs)$-length $\phi$-expander, \cite{HHT2024} use a variation of the notion of \emph{neighborhood covers}. In this variation, a neighborhood cover is a collection $\set{\cset_1,\ldots,\cset_r}$ of clusterings, where, for all $1\le i\leq r$, clustering $\cset_i$ is a collection of subgraphs of $H$ called clusters, of radius at most $h$ each. Additionally, every pair of clusters in $\cset_i$ must be at distance at least $hs$ from each other. Lastly, it is required that every vertex of $H$ belongs to a relatively small number of clusters in $\bigcup_i\cset_i$, and, for every vertex $v$ of $G$, its $h$-neighborhood $B_{H}(v,h)$ must be contained in some cluster  in $\bigcup_i\cset_i$.
As shown in \cite{HHT2024}, in order to certify that a graph $H$ is an $(h,s)$-length $\phi$-expander, it is sufficient to compute, for every cluster $C\in \bigcup_i\cset_i$, a router graph $R_C$ with $V(R_C)=V(C)$, together with an embedding of all such resulting routers $R_C$ into $G$ via short paths, that cause low congestion. Their algorithm then computes a moving cut $L$ for graph $G$, and certifies that the resulting graph $G'=G-L$ is an $(h,s)$-length $\phi$-expander by computing the neighborhood cover $\set{\cset_1,\ldots,\cset_r}$ of $G'$, and then embedding, for every cluster $C\in \bigcup_i\cset_i$, the corresponding router $R_C$ into $G'$. 
Therefore, the algorithms of \cite{HHT2024,HLT2024} can also be thought of as computing and dynamically maintaining a weaker notion of router decomposition of the input graph $G$, similar to the weaker notion of expander decomposition of \cite{ST04}. Specifically, every cluster $C$ in the neighborhood cover is provided with a certificate router graph $R_C$, with $V(C)=V(R_C)$, and an embedding of $R_C$ into $G$ via short paths. However, this embedding may use edges and vertices that lie outside of $C$.
Our main technical result is an algorithm for maintaining a \emph{proper} router decomposition of a dynamic  graph:

\begin{wrapper}\vspace{-4pt}
\begin{theorem}[Dynamic Router Decomposition, Informal]\label{thm:dynamic-routers}
There is a deterministic algorithm, whose input is an $n$-vertex undirected graph $G$, a stretch parameter $512 \leq k \leq (\log n)^{1/49}$, and two additional parameters $1/k \leq \delta \leq 1/400$, and $\Delta \geq n^{20/k}$. Graph $G$ initially has no edges, and undergoes an online sequence of edge insertions and deletions. The algorithm maintains  a collection $\cset$ of edge-disjoint subgraphs (clusters) of $G$, and, for every cluster $C\in \cset$, a subgraph $C'\subseteq C$ with $V(C')=V(C)$, such that the following hold at all times: (i) for every cluster $C\in \cset$, any unit demand $\dset_C$ for $C$ can be routed inside $C$ via paths of length at most $d\leq \poly(k) \cdot 2^{O(1/\delta^6)}$, with congestion at most  $\eta=n^{O(1/k)}$; (ii) $\sum_{C\in \cset}|V(C)|\leq n^{1+O(1/k)}$; (iii) for every cluster $C\in \cset$, any $\Delta$-restricted demand on $V(C)$ can be routed in $C'$ via paths of length at most $d$, with congestion at most $\eta$; (iv) $\sum_{C \in \mathcal{C}}|E(C')|\leq \Delta\cdot n^{1+O(1/k)}$; and (v) the total number of edges $e\in E(G)$ that do not lie in any cluster of $\cset$ is  at most $\Delta\cdot n^{1+O(1/k)}$. The worst-case update time of the algorithm is $n^{O(\delta)}$. 
\end{theorem}\vspace{-5pt}
\end{wrapper}

In a sense, our decomposition result can be thought of as strengthening the results of \cite{HHT2024,HLT2024} to a proper router decomposition, in the same manner as the result of \cite{saranurak2019expander}  strengthened the fast algorithm  of \cite{ST04} for weak expander decomposition to obtain a proper expander decomposition. While this may appear as a technicality, just as with expander decompositions, a proper router decomposition provides a cleaner structure that is easier to use in applications. In fact some of the applications of our router decomposition described below crucially rely on the fact that the decomposition is proper. Additionally, for every cluster $C$ in the decomposition, we maintain a sparsified subgraph $C'\subseteq C$ of a prescribed average density that remains a sufficiently strong router. These subgraphs are useful for computing various sparsifiers of $G$. 
We note that, unlike the decompositions of \cite{HHT2024,HLT2024}, our decomposition does not provide an LC-expander decomposition of $G$. Specifically, if we denote by $\edel$ the set of edges that do not lie in any cluster, then our decomposition does not guarantee that $G\setminus \edel$ is a length-constrained expander. This is due to the fact that the clusters in our decomposition do not necessarily define a neighborhood cover in $G\setminus \edel$.

We obtain the result of Theorem \ref{thm:dynamic-spanners} from
Theorem \ref{thm:dynamic-routers}  as follows: we use the parameter $\Delta=n^{O(1/k)}$; the spanner $H$ is obtained by taking the union of the subgraphs $C'$ for all $C\in \cset$, together with all edges of $G$ that do not lie in the clusters of $\cset$. We show that $H$ can be maintained with worst-case update time $n^{O(\delta)}$ and recourse $n^{O(1/k)}$.
Our results also extend to low-congestion and fault-tolerant spanners, that we discuss next. 
\subsection*{Low-Congestion and Fault-Tolerant Spanners.} 
The following general recipe, pioneered by \cite{sparsifiers}, can be used in order to compute spanners via expander decomposition and routing. Initially, the spanner $H$ contains $n$ vertices and no edges. Next, we compute an expander decomposition $\mathcal{C}$ of $G$, and, for every cluster $C\in \cset$, we compute a sparse subgraph $C'\subseteq C$ that is an expander. 
For every cluster $C\in \cset$ of the decomposition, we then add the edges of $C'$ to $H$, and we consider the edges of $C$ as ``settled''. The algorithm then continues recursively with edges that are not settled yet: namely, edges whose endpoints lie in different clusters of $\cset$. It is easy to verify that the resulting spanner $H$ has polylogarithmic stretch and that it is sparse. 
The recent breakthrough almost-linear algorithm of  \cite{ChenKLPGS22,brand2023deterministic}  for Min Cost Flow exploits the fact that the  spanner $H$ obtained via this approach has additional useful properties.  Specifically, they show that one can embed all edges of $G$ into $H$, such that all the embedding paths are short, and cause low vertex-congestion. This is in contrast to the standard notion of spanners, that only guarantees that, for each edge $(u,v)$ of $G$, there is a short $u$-$v$ path in $H$, though the resulting paths may cause arbitrarily high congestion.

In this work, we formalize this notion by introducing \emph{low-congestion spanners}, and show an algorithm that dynamically maintains such a spanner, by exploiting our result for router decomposition. Formally, let $G$ be a graph, and let $H$ be a subgraph of $G$. We say that $H$ is a $(d,\eta)$-low congestion spanner for $G$, if there is an embedding of the edges of $G$ into $H$, such that all embedding paths have length at most $d$, and cause edge-congestion at most $\eta$.
We use our router decomposition to show that for any stretch parameter $512\leq k\leq  (\log n)^{1/49}$ and degree threshold $\Delta$, any $n$-vertex graph $G$ admits a $(d,\eta)$-low congestion spanner with $\widetilde{O}(\Delta\cdot  n^{1+O(1/k)})$ edges for $d=\poly(k)$ and $\eta=O\left(n^{O(1/k)}\cdot\max\set{ \frac{\Delta_{\max}(G)}{\Delta},1}\right )$, where $\Delta_{\max}(G)$ is the maximum vertex degree in $G$. In fact we show that the graph $H$ that is maintained by the algorithm from \Cref{thm:dynamic-spanners}, and which, in turn, relies on the router decomposition from \Cref{thm:dynamic-routers}, has this property.
We believe that low-congestion spanner is an interesting  object that may find other applications in the future. 

%
%

\textbf{Fault-Tolerant Spanners.} An additional useful property satisfied by expander-based spanners is their resilience to edge faults. As many applications of spanners arise in the context of distributed networks, which are inherently prone to failures of edges and vertices, it is especially desirable to obtain robust spanners, that maintain their functionality in the presence of such faults. \emph{Fault-tolerant (FT) spanners} are specifically designed to provide such  guarantees: given a graph $G$ and parameters $f$ and $t$, a subgraph $H\subseteq G$ is
an $f$-FT $t$-spanner if, for every collection $F\subseteq E(G)$ of at most $t$ failed edges, $H\setminus F$ is a $t$-spanner for $G \setminus F$. FT-spanners were first introduced in the context of geometric graphs by Levcopoulos et al. \cite{levcopoulos1998efficient}, and were later considered for general graphs by Chechik et al. \cite{ChechikLPR:10}.
 Recently, Parter \cite{Parter22} presented a near-linear time randomized algorithm for constructing an $f$-FT $(2k-1)$-spanner with $\widetilde{O}(f^{1-1/k}\cdot n^{1+1/k})$ edges, which is also resilient to up to $f$ vertex faults. 
 Bodwin, Dinitz and Robelle \cite{BodwinDR22} provided an existential result for $f$-FT $(2k-1)$-spanners: for odd values of $k$, they show that there exist spanners of size $O(k^2f^{1/2-1/(2k)}n^{1+1/k}+kfn)$, and for even $k$, the size is $O(k^2f^{1/2}n^{1+1/k}+kfn)$.  These bounds almost match the  $\Omega\left(f^{1/2-1/(2k)}\cdot n^{1+1/k}\right )$ size lower bound that is based on the Erd{\" o}s Girth Conjecture \cite{bodwin2018optimal}.

Notice that, if we use the notion of $f$-FT spanners, that can only tolerate at most $f$ edge failures, the spanner size must be at least $fn$, even when allowing a large stretch, since it is possible that all of the edge faults are incident to a single vertex. It is possible, however, that one can obtain sparse spanners that can withstand a significantly higher number of edge faults, as long as these faults are distributed evenly in the graph, and are not concentrated around a small number of vertices. With this motivation in mind, a very recent work of Bodwin, Haeupler and Parter \cite{BodwinHP24} introduced a stronger notion of fault-tolerant spanners, called \emph{Faulty-Degree (FD) spanners}. Specifically, suppose we are given an $n$-vertex graph $G$, and parameters $t$ and $f$. An $f$-FD $t$-spanner for $G$ is a graph $H\subseteq G$, with the following property: if $F$ is any subset of edges of $G$, such that every vertex of $G$ is incident to at most $f$ edges of $F$, then $H\setminus F$ is a $t$-spanner for $G\setminus F$. Notice that, for $f=1$, the set $F$ of edges can be any matching, whose size may be as high as linear in $n$. Bodwin et al. \cite{BodwinHP24} provided two approaches for computing sparse $f$-FD spanners. The first approach uses the LC-expander decomposition, and obtains $f$-FD spanners with stretch $k^{O(k)}$ and size $\widetilde{O}(fn^{1+1/k})$ via an algorithm with a high polynomial running time. Combining this approach with the recent time-efficient algorithm for LC-expander decomposition and routing of \cite{HHT2024,HLT2024} yields an algorithm with running time $O(m\cdot n^{O(1/k)})$, albeit with a slightly higher stretch $2^{O(\poly(k))}$.
 Their second approach,  based on a greedy algorithm, provides an $f$-FD spanner with stretch $(2k-1)$ and size $\widetilde{O}\left (f^{1-1/k} \cdot n^{1+1/k}\cdot k^{O(k)}\right )$, which is nearly optimal for fixed  values of $k$ under the Erd{\" o}s Girth Conjecture. The main drawback of the latter approach is the large polynomial running time of their algorithm.
 In their paper, \cite{BodwinHP24} noted that improving the $k^{O(k)}$ stretch bound obtained via the LC-expanders is closely related to the question of the resilience of LC-expanders to bounded degree faults, which was left open therein (see Theorem 1.14 in \cite{BodwinHP24}). In this work, we build on our results for router decomposition 
to obtain $f$-FD spanners with stretch $\poly(k)$, that contain $\tilde{O}(fn^{1+O(1/k)})$ edges. This allows us to address the open problem of \cite{BodwinHP24}, only on router graphs (which can viewed as LC-expanders with small diameter), and improve the stretch bound from $k^{O(k)}$ to $\poly(k)$. 
We also obtain an algorithm for maintainning such a spanner for a fully dynamic graph $G$: given a parameter $\delta\geq 1/k$, the stretch of the spanner is bounded by $\poly(k)\cdot 2^{O(1/\delta^6)}$, and the worst-case update time of the algorithm is $n^{O(\delta)}$; the algorithm also guarantees that the recourse is bounded by $n^{O(1/k)}$. In particular, this yields an algorithm for constructing a spanner for a static $m$-edge graph with stretch $\poly(k)\cdot 2^{O(1/\delta^6)}$ in time  $m^{1+O(\delta)}$, which outperforms the (at least quadratic) runtime of the greedy-based approach of \cite{BodwinHP24}. 

%

\textbf{Connectivity Certificates.}
Finally, a graph structure closely related to FT-spanners is the $f$-edge connectivity certificates, introduced by \cite{NagamochiI92}. For an integer $f\geq 1$, we say that a graph $G$ is $f$-edge connected, if, for any subset $F$ of $f-1$ edges of $G$, graph $G\setminus F$ is connected. An $f$-edge connectivity certificate for $G$ is a subgraph $H\subseteq G$, that has the following property: $H$ is $f$-edge connected if and only if $G$ is. Connectivity certificates play a key role in fast algorithms for minimum cuts; see, e.g. \cite{Matula93,CheriyanKT93,KargerM97,GhaffariK13,DagaHNS19,ForsterNYSY20,LiNPSY21,GoranciHNSTW23}. 
 Nagamochi and Ibaraki \cite{NagamochiI92} presented a linear-time static algorithm for computing $f$-edge connectivity certificates of $n$-vertex graphs $G$, with $O(f n)$ edges. 
In the dynamic setting, Thorup and Karger \cite{Thorup00swat} presented a deterministic algorithm for maintaining an $f$-edge connectivity certificate of a dynamic $n$-vertex graph, of size $O(fn)$, with $\widetilde{O}(f^2)$ amortized update time. Combining the latter with the deterministic algorithm  of \cite{chuzhoy2020deterministic} for dynamic minimum spanning forest  yields $O(f^2 \cdot n^{o(1)})$ worst-case update time. The dependency of the update time  on $f$ can be avoided using randomness: the recent work of Assadi and Shah \cite{AssadiS23} implies a \emph{randomized} algorithm for maintaining an $f$-connectivity certificate with $O(f \cdot n\log n)$  edges in a fully dynamic $n$-vertex graph, with $O(\poly\log n)$ worst-case update time. To the best of our knowledge, there is no deterministic dynamic algorithm for maintaining sparse $f$-connectivity certificate with worst-case update time bound that is independent of $f$. We show that our algorithm for maintaining  a router decomposition can also be used to maintain  an $f$-connectivity certificate with $f \cdot n^{1+o(1)}$ edges (and therefore of almost optimal size) for a fully dynamic graph, with worst-case update time and recourse bounded by $n^{o(1)}$. In fact, we maintain an even stronger version of an $f$-connectivity certificate recently introduced in \cite{BodwinHP24}, called $f$-FD connectivity certificate, that can withstand the failure of any edge subset $F$, as long as every vertex of $G$ is incident to at most $f$ edges of $F$.

\subsection{Our Results and Techniques}
\label{subsec: intro: results and techniques}

Our main technical result is an algorithm that maintains a low-diameter router decomposition of a fully dynamic graph with low worst-case update time. Before we state the result, we need to define the notions of demands and routers; formal definitions can be found in Sections \ref{subsec: demands} and \ref{subsec: routers}.
Consider a graph $G=(V,E)$. A \emph{weighting} of the vertices of $G$ is an assignment $w(v)\geq 0$ of a non-negative weight to every vertex $v\in V$. We also denote such a weighting by an $n$-dimensional vector $\vw\in \reals^n$. We sometimes use two special vertex weightings: in the first weighting, denoted by $\vdeg_G$, the weight of every vertex $v\in V$ is its degree in $G$. In the second weighting, we are given some value $\Delta>0$, and the weight of every vertex is set to be $\Delta$; we denote such a weighting by $\vDelta$. 
A \emph{demand} $\dset$ in $G$ consists of a collection $\Pi$ of pairs of vertices, and, for every pair $(a,b)\in \Pi$, a value $D(a,b)>0$, that we refer to as the \emph{demand between $a$ and $b$}. For a vertex weighting $\vw$, we say that a demand $\dset=\left(\Pi,\set{D(a,b)}_{(a,b)\in \Pi}\right )$ is
	\emph{$\vw$-restricted}, if, for every vertex $v\in V$, the total demand between all pairs in $\Pi$ that contain $v$ is at most $w(v)$.
 A \emph{routing} of the demand $\dset$ is a flow $f$, in which, for every pair $(a,b)\in \Pi$, $D(a,b)$ flow units are sent from $a$ to $b$. The \emph{congestion} of the routing is the maximum amount of flow through any edge of $G$. We say that the routing is via \emph{flow-paths of length at most $d$}, if every flow-path with non-zero flow value contains at most $d$ edges.
	
A central notion that we use in our paper is that of a \emph{router}, that was first defined in \cite{HHT2024} as an object that is closely related to length-constrained expanders. Given a graph $G$ with vertex weighting $\vw$, and parameters $\eta$ and $d$, we say that $G$ is a $(\vw,d,\eta)$-router\footnote{in Section \ref{subsec: routers} we provide a slightly more general definition of routers, that includes a set $S$ of supported vertices, where the router is only required to route demands defined over the vertices of $S$. But all our results hold for the case where $S=V(G)$, so the definition above is sufficient in order to state our results.}, if every $\vw$-restricted demand $\dset$ can be routed with congestion at most $\eta$ in $G$, via paths of length at most $d$. 


Our main technical result, summarized in the following theorem, is an algorithm that maintains a router decomposition of a fully dynamic graph, with small worst-case update time. The algorithm also maintains a subgraph $H$ of $G$ of a prescribed density that has additional useful properties. In particular, as we show later, graph $H$ is a (low-congestion and fault-tolerant) spanner of $G$.

\begin{theorem}\label{thm: router decomposition main}
	There is a deterministic algorithm, whose input is a graph $G$ with $n$ vertices that initially has no edges, which undergoes an online sequence of at most $n^2$ edge deletions and insertions, such that $G$ remains a simple graph throughout the update sequence. Additionally, the algorithm is given parameters $512\leq k\leq (\log n)^{1/49}$, $\frac 1 k\leq \delta<\frac{1}{400}$  and $\Delta\geq n^{20/k}$, such that $k,\Delta$ and $1/\delta$ are integers. The algorithm maintains the following:
	
	\begin{itemize}
		\item a collection $\cset$ of edge-disjoint subgraphs of $G$ called \emph{clusters} with $\sum_{C\in \cset}|V(C)|\leq n^{1+O(1/k)}$, such that every cluster $C\in \cset$ is a $(\vdeg_C, d,\eta)$-router for $d=k^{19}\cdot 2^{O(1/\delta^6)}$ and $\eta=n^{O(1/k)}$. Moreover, if we denote by  $\edel=E(G)\setminus \left(\bigcup_{C\in \cset}E(C)\right )$, then, at all times, $|\edel|\leq n^{1+O(1/k)}\cdot \Delta$ holds;
		
		\item for every cluster $C\in \cset$, a subgraph $C'\subseteq C$  with $V(C')=V(C)$, that is a $(\vDelta,d,\eta)$-router, with $\sum_{C\in \cset}|E(C')|\leq n^{1+O(1/k)}\cdot \Delta$; and
		
		\item a graph $H\subseteq G$ with $|E(H)|\leq  n^{1+O(1/k)}\cdot \Delta$, that contains all edges of $\edel\cup\left(\bigcup_{C\in \cset}E(C')\right )$.
	\end{itemize}
	
The worst-case update time of the algorithm is $n^{O(\delta)}$ per operation, and the total number of edge insertions and deletions that graph $H$ undergoes after each update to $G$ is at most $n^{O(1/k)}$.
\end{theorem}

We note that, if we set $\Delta=\ceil{n^{20/k}}$, then it is guaranteed that $|E(H)|\leq n^{1+O(1/k)}$ holds, and moreover, it is easy to verify that $H$ is a $d$-spanner for $G$. In Section \ref{sec:bd-fault} we show that, in addition to  being a $d$-spanner, $H$ has many other useful properties; for example, it is also a low-congestion spanner. Moreover, given a fault parameter $f$, by setting the bound $\Delta$ appropriately, we can guarantee that $H$ is an $f$-FD spanner. Lastly, graph $H$ can also be used as an $f$-connectivity certificate for $G$. Interestingly, the worst-case update time to maintain these  structures is independent of the  parameter $f$. 


We now provide an overview of the techniques that we employ in the proof of \Cref{thm: router decomposition main}. Using a standard deamortization technique (see \cite{nanongkai2017dynamic2,jin2022fully}, and \Cref{subsec: online batch}), it is sufficient to provide an algorithm with guarantees similar to those in \Cref{thm: router decomposition main} in the \emph{online-batch dynamic model}. In this model, the initial graph $G$ may contain an arbitrary number of edges, and it undergoes $k$ batches $\pi_1,\ldots,\pi_k$ of updates, where, for all $1\leq i\leq k$, the $i$th batch $\pi_i$ of updates is a collection of edge insertions and deletions for graph $G$. In order to obtain the desired worst-case update time and recourse bound from \Cref{thm: router decomposition main}, it is enough to design an algorithm in the online-batch model, whose initialization time is $m^{1+O(\delta)}$; the time required to process each batch $\pi_i$ is bounded by $|\pi_i|\cdot n^{O(\delta)}$; and the total number of edge insertions and deletions that graph $H$ undergoes following each batch $\pi_i$ of updates is bounded by $|\pi_i|\cdot n^{O(1/k)}$. We develop several technical tools in order to solve this problem, that we describe next.

\noindent{\bf First tool: nice router $W_k^{N,\Delta}$ and its pruning.}
Our first tool is an explicit construction of a nice well-structured router graph, and an algorithm for its pruning. This construction is very similar to that of \cite{HLT2024}. However we modify their construction slightly, and we modify their pruning algorithm significantly, in order to obtain a better distance-time and distance-size tradeoffs.
Suppose we are given parameters $N,\Delta$ and $k$. The corresponding graph $W_k^{N,\Delta}$ (that, for simplicity, we will denote by $W_k$), consists of $N^k$ vertices, that are partitioned into a set $V^c$ of $N^{k-1}$ \emph{center} vertices, and a set $V^{\ell}$ of remaining vertices, called \emph{leaf} vertices. The set of edges of $W_k$ is partitioned into $k$ subsets $E_1,\ldots,E_k$, where, for $1\leq i\leq k$, we refer to the edges of $E_i$ as \emph{level-$i$ edges}. For all $1\leq i\leq k$, the subgraph of $W_k$ induced by the edges of $E_i$ can be thought of as consisting of a  union of $N^{k-1}$ disjoint star graphs (called \emph{level-$i$ stars}), each of which has $N-1$ leaves, after we replace every edge of every star by a collection of $\Delta$ parallel edges (that we refer to as a \emph{bundle}; we also refer to the corresponding star edge as a \emph{superedge}). The center vertices of all stars  lie in $V^c$, and their leaves  lie in $V^{\ell}$. 

The construction of the graph $W_k$ is inductive. In order to obtain the graph $W_1$, we start with a star $S$ that has $N-1$ leaves. We then replace every edge of $S$ with $\Delta$ parallel edges, obtaining the graph $W_1$. The center of the star $S$ becomes the unique center vertex of $W_1$, and the edges of $W_1$ are \emph{level-1 edges}.
Assume now that, for an integer $1<  i\leq k$, we have defined the graph $W_{i-1}$. In order to obtain a graph $W_i$, we start by constructing $N$ disjoint copies of $C_1,\ldots,C_N$ of the graph $W_{i-1}$; we refer to these copies as \emph{level-$(i-1)$ clusters}. The set of the center vertices of $W_i$ is the union of the sets of center vertices of the graphs $C_1,\ldots,C_N$, and the remaining vertices of $W_i$ are leaf vertices. We connect the clusters $C_1,\ldots,C_N$ to each other by level-$i$ edges, as follows. We construct $N^{k-1}$ disjoint stars $S_1,\ldots,S_{N^{k-1}}$ (level-$i$ stars), so that each star $S_j$ has $N-1$ leaves; each vertex of $S_j$ lies in a distinct cluster of $\set{C_1,\ldots,C_N}$; and the center of the star $S_j$ is a center vertex of $W_k$. For each such star $S_j$, we then replace each of its edges with $\Delta$ parallel edges, which are referred to as level-$i$ edges. 

As mentioned already, this construction is very similar to that used in \cite{HLT2024}; the main difference is that they used cliques instead of stars, and in their construction $\Delta=1$. 
Next, we define the notion of a \emph{properly pruned subgraph} of $W_k$. Suppose we are given a subgraph $W\subseteq W_k$, and $k$ collections $U_1,\ldots,U_k$ of vertices of $W$, with $U_k\subseteq U_{k-1}\subseteq\cdots\subseteq U_1$. We say that $W$ is a \emph{properly pruned subgraph} of $W_k$ with respect to the sets $U_1,\ldots,U_k$ of its vertices, if $V(W)=U_1$, and the graph obeys the following additional rules. First, for each level $1\leq i\leq k$, for each level-$i$ superedge $(u,v)$, where $v$ is a leaf vertex, if $v$ remains in set $U_i$, then at least $\Delta/2$ parallel edges $(u,v)$ must remain in $W$. Second, for each level $1\leq i\leq k$, and each level-$i$ star $S$, if the center vertex $v$ of $S$ remains in $U_i$, then a large fraction of the leaves of $S$ must remain in $U_i$. Lastly, for all $1\leq i<k$, for every level-$i$ cluster $C$, if any vertex of $C$ remains in $U_1$, then a significant fraction of vertices of $C$ must lie in $U_{i+1}$. 
We prove that, if $W$ is a properly pruned subgraph of $W_k$, then it is a strong router: namely, any $\Delta$-restricted demand can be routed in $W$ via paths of length $O(k^2)$, with congestion $k^{O(k)}$. The routing algorithm uses the star structure of the graph $W_k$ in a natural manner. As mentioned already, our algorithm for pruning the graph $W_k$ is more involved than that of \cite{HLT2024}; this is in order to obtain stronger routing properties in $W$. For example, in the construction of \cite{HLT2024}, the bound on the lengths of the routing paths was at least exponential in $k$ (see Theorem 5.1 in \cite{HLT2024}).

Finally, we provide an algorithm for pruning the graph $W_k$ in the online-batch dynamic model. The algorithm is given $k$ batches $\pi_1,\ldots,\pi_k$ of edge deletions from $W_k$, and maintains a properly pruned decremental subgraph $W$ of $W_k$ throughout the update sequence. The algorithm ensures that the processing time of each batch $\pi_i$ is bounded by $O(|\pi_i|\cdot k^{O(k)})$, and the number of edges and vertices deleted from $W$ after each batch $\pi_i$ is small compared to $|\pi_i|$. The pruning algorithm is quite straightforward: whenever any of the above rules are not obeyed, we perform vertex deletions in order to ensure that the current graph $W$ is compliant with the rules.

One additional advantage of our construction is that it is easy to maintain a sparsified subgraph $W'$ of $W_k^{N,\Delta}$ of any prescribed density $\Delta'<\Delta$ that remains a strong router, even as the graph undergoes edge deletions and pruning: for every superedge $(u,v)$, we simply include in $W'$ arbitrary $\Delta'$ parallel copies of $(u,v)$. This, in turn, allows us to maintain the sparsified routers $C'\subseteq C$ for the clusters $C$ in the router decomposition $\cset$ that our algorithm maintains. 

\textbf{Second tool: embedding a nice router.}
Our second tool is an algorithm that can be used in order to embed a nice router $W_k$ into a given graph. Specifically, suppose we are given a graph $C$, and parameters $k$ and $d=k^{O(1)}$, such that, for some vertex $v\in V(C)$, the number of vertices of $C$ that lie within distance $d$ from $v$ is significantly larger than $|V(C)|^{1-1/k}$. Assume further that $|E(C)|\approx |V(C)|\cdot \Delta\cdot n^{\Theta(1/k)}$.
Our algorithm either successfully embeds a graph $W_k^{N,\Delta}$ into $C$ via short paths that cause small congestion, where $N^k$ is quite close to $|V(C)|$; or computes a subset $E'$ of $O(n\Delta)$ edges, so that $C\setminus E'$ becomes \emph{scattered}: that is, for every vertex $u\in V(C)$, the ball of radius $d$ around $u$ in $C\setminus E'$ contains fewer than $|V(C)|^{1-1/k}$ vertices. More specifically, our algorithm uses an additional parameter $\delta$ to control the tradeoff between the lengths of the embedding paths and the running time. In the first case, the algorithm computes a relatively small collection $F$ of edges of $W_k^{N,\Delta}$, and an embedding $\pset$ of $W_k^{N,\Delta}\setminus F$ into $C$ via paths of length $\poly(k)\cdot 2^{O(1/\delta^6)}$, with congestion at most $n^{O(1/k)}$. The running time of the algorithm is $m^{1+O(\delta)}$. If the algorithm terminates with an embedding of $W_k^{N,\Delta}\setminus F$ into $C$, then we can employ our pruning algorithm, in order to compute a large enough properly pruned subgraph $W$ of $W_k^{N,\Delta}$ that does not contain any edges of $F$, together with its embedding $\pset'$ via paths whose length and congestion are bounded as before. Moreover, we can compute a large enough subgraph $\hat C\subseteq C$ that contains all embedding paths in $\pset'$, such that every vertex of $\hat C$ participates in a large number of such paths. As we show later, this is sufficient in order to prove that $\hat C$ is a $(\vdeg_{\hat C}, d, \eta)$-router.

The algorithm for computing the embedding of $W_k^{N,\Delta}\setminus F$ (or a subset $E'$ of edges whose removal makes $C$ scattered), employs the algorithmic techniques that were developed in \cite{chuzhoy2023distanced} for \emph{well-connected graphs}. Specifically, we start by selecting a large subset $T$ of vertices of $C$ that are close to each other, and then use an algorithm from \cite{chuzhoy2023distanced} in order to attempt to embed a well-connected graph, whose vertex set is a large subset of $T$, into $C$, via short paths that cause a low congestion. 
If the algorithm fails to do so, then it computes a small collection $E''$ of edges, so that, in $C\setminus E''$, many pairs of vertices of $T$ become far from each other. In the latter case, we either compute a new large subset $T$ of vertices of $C$ that remain close to each other and restart the same algorithm; or correctly establish that the graph becomes scattered and terminate the algorithm. Using simple accounting, we show that the number of such iterations in the algorithm is not too large. If we successfully embed a large well-connected graph $R$ into $C$, we employ the algorithm of \cite{chuzhoy2023distanced} for \APSP in well-connected graphs, together with the standard Even-Shiloach trees,  in an attempt to embed the edges of $W_k^{N,\Delta}$ into $C$ one by one. If we fail to embed a large fraction of the edges of $W_k^{N,\Delta}$, then we again obtain a small subset $E''$ of edges, such that, in $C\setminus E''$, many pairs of vertices that were originally close to each other, become far from each other. All such sets $E''$ of edges removed from $C$ over the course of the algorithm are added to a set $E'$, and, since the number of iterations in the algorithm is relatively small, we can ensure that $|E'|$ remains sufficiently small as well.

\textbf{Third tool: router witness.}
Suppose we are given a graph $W$, that is a properly pruned subgraph of the graph $W_k^{N,\Delta}$. Assume further that we are given some graph $C$, and an embedding of $W$ into $C$ via paths of length $d^*$ that cause congestion at most $\eta^*$. Assume further that all vertex degrees in $C$ are at most $\alpha\cdot \Delta$, and every vertex $v\in V(C)$ lies on at least $\Delta/\beta$ of the embedding paths. We prove that, in such a case, $C$ must be a $(\vdeg_C,\tilde d,\tilde\eta)$-router, for $\tilde d=d^*\cdot \poly(k)$, and $\tilde \eta=\alpha\cdot\beta\cdot d^*\cdot \eta^*\cdot k^{O(k)}$. Therefore, we can view the graph $W$ and its embedding $\pset$ into $C$ with the above properties as a \emph{witness} that $C$ is a $(\vdeg_C,\tilde d,\tilde\eta)$-router. In our construction of the router decomposition, we employ this type of witnesses to ensure that all clusters in the decomposition remain strong routers.

Assume now that we are given a target parameter $\Delta^*\ll\Delta$, a cluster $C$, and a router witness $(W,\pset)$ as described above. Assume further that our goal is to compute a \emph{sparsified router} $C'\subseteq C$ for $C$, so that $V(C')=V(C)$, $|E(C')|\leq \Delta^*\cdot |V(C)|$, and $C'$ is a $(\Delta^*,\tilde d,\tilde \eta')$-router, where $\tilde d$ remains the same as before, and $\tilde \eta'$ is only slightly higher than $\tilde \eta$. We show that such a cluster $C'$ can be constructed as follows.
We let $\Delta'<\Delta^*$ be a parameter that is sufficiently close to $\Delta^*$.
 First, for every superedge $(a,b)$ of $W$, we select a subset $E'(a,b)$ of $\Delta'$ parallel edges $(a,b)$, and we let $\qset'\subseteq \pset$ be the set of all embedding paths of all edges in sets $E'(a,b)$, for all superedges $(a,b)$ of $W$. Next, we construct a collection $\qset''\subseteq \pset$ of paths as follows: every vertex $v\in V(C)$ selects a subset of $\Delta'$ paths of $\pset$ that contain $v$, and adds them to $\qset''$; but we require that every leaf vertex of $W$ serves as an endpoint of at most $\gamma\cdot \Delta'$ such paths, for $\gamma=n^{O(1/k)}$. We then let $C'\subseteq C$ be the graph that contains all edges and vertices participating in the paths of $\qset'\cup \qset''$. We prove that such a graph $C'$ must be a $(\Delta^*,\tilde d,\tilde \eta')$-router, by using the routing properties of properly pruned subgraphs of $W_k^{N,\Delta}$ in a straightforward manner. We also show that $|E(C')|\leq |V(C')|\cdot \Delta^*$.
 
 \textbf{Fourth tool: decremental low-diameter clustering.}
 Our last tool is a simple algorithm for decremental low-diameter clustering. Suppose we are given a graph $G$ and a parameter $k<\log n$. We say that a subgraph $C\subseteq G$ is a \emph{settled cluster}, if either $|V(C)|\leq n^{1/k}$, or there is a vertex $v\in V(C)$, such that the ball of radius $\poly(k)$ around $v$ contains at least $|V(C)|^{1-1/k}$ vertices. If a subgraph $C\subseteq G$ is not a settled cluster, then we call it \emph{unsettled}.
 
 In the decremental low-diameter clustering problem, the goal is to maintain a decomposition $\cset$ of $G$ into edge-disjoint clusters under edge deletions, so that each cluster in $\cset$ is settled, and $\sum_{C\in\cset}|V(C)|\leq n^{1+O(1/k)}$ holds (for conciseness, we refer to the latter property as a \emph{low overlap}).  We note that the settled clusters in the decomposition that we maintain are not guaranteed to have a low diameter; each such cluster either has relatively few vertices, or it has a single vertex with a sufficiently large neighborhood. 

Specifically, the algorithm is required to first compute an initial decomposition $\cset$ of $G$ into settled clusters with low overlap.
 The set $\cset$ of clusters is then partitioned by the adversary into a set $\cset^I$ of \emph{inactive} and a set $\cset^A$ of \emph{active} clusters. At the beginning, all clusters are active. Once a cluster becomes inactive, it may not undergo any further updates.
 Then the algorithm proceeds in iterations. In every iteration, the adversary may (i) move some clusters from $\cset^A$ to $\cset^I$; and (ii) delete some edges and vertices from clusters $C\in \cset^A$, after which each such cluster must become unsettled. After that the algorithm is required to ``fix'' the clusters in $\cset^A$ via \emph{cluster-splitting} updates: given a cluster $C\in \cset$ and a subgraph $C'\subseteq C$, add $C'$ to $\cset$ and to the set $\cset^A$ of active clusters, delete all edges of $C'$ from $C$, and delete any resulting isolated vertices from $C$. The algorithm must perform such cluster splitting operations, until all clusters in $\cset^A$ become settled.
  We provide a simple algorithm for the decremental low-diameter clustering problem that is based on the standard ball-growing technique.
 
 \textbf{Combining all tools together.}
 We are now ready to describe the initialization algorithm for 
\Cref{thm: router decomposition main}.
Assume first that all vertex degrees in the input graph $G$ are close to $\Delta\cdot n^{\Theta(1/k)}$, for some $\Delta\gg \Delta^*$.
We use the algorithm for  the low-diameter clustering problem, to compute an initial collection $\cset$ of edge-disjoint clusters with low overlap, so that all clusters in $\cset$ are settled. We set $\cset^I=\emptyset$, $\cset^A=\cset$, and then iterate, as long as $\cset^A\neq \emptyset$. In every iteration, we consider every cluster $C\in \cset^A$ one by one. For each such cluster $C$, we attempt to embed a graph $W_k^{N_C,\Delta}\setminus F$ into $C$, where $F$ is a relatively small set of fake edges, and $N_C$ is chosen so that $\frac{|V(C)|}{n^{\Theta(1/k)}}\leq N_C^k\leq |V(C)|$, via short paths that cause low congestiton. If we fail to do so, then we obtain a relatively small set $E_C$ of edges, such that graph $C\setminus E_C$ is scattered. We delete the edges of $E_C$ from $C$ (and add them to the set $\edel$ of deleted edges that we maintain), and notify the algorithm for the low-diameter clustering problem regarding these edge deletions; we show that $C$ now becomes unsettled. If, however, we manage to embed a graph $W_k^{N_C,\Delta}\setminus F$ into $C$ as above, then we move $C$ to the set $\cset^I$ of inactive clusters. Using the pruning algorithm, we then compute a large subgraph $\hat C\subseteq C$, together with a router witness for $\hat C$, so that $\hat C$ is $(\vdeg_{\hat C},\tilde d,\tilde\eta)$-router. We also construct a subgraph $\hat C'\subseteq \hat C$, with $V(\hat C')=V(\hat C)$, $|E(\hat C')|\leq \Delta^*\cdot |V(\hat C)|$, such that $\hat C'$ is a $(\vDelta^*,\tilde d,\tilde \eta)$-router. We then add $\hat C$ to the router decomposition that we are constructing.

Once the algorithm terminates, $\cset^A=\emptyset$ holds, and so all clusters are inactive. For each such inactive cluster $C$, we have constructed a large enough router $\hat C\subseteq C$. If the number of edges lying outside of the routers that we have constructed so far is too high, then we repeat the algorithm with these remaining edges. However, since our algorithm ensures that a large enough fraction of edges of each cluster $C$ are included in the corresponding router $\hat C$, we can show that the number of times that we need to restart the algorithm is relatively small.

Our algorithm for processing each batch $\pi_i$ of updates to graph $G$ proceeds as follows. Assume first that all updates in $\pi_i$ are edge deletions. Then we simply use the pruning algorithm for the nice routers that we described above, inside each cluster $C$ of the decomposition, in order to maintain the corresponding router witness.

Lastly, so far we have assumed that all vertex degrees in the graph $G$ are close to each other, and that the update batches $\pi_i$ only contain edge deletions. In order to handle a general graph $G$, we partition its edges among at most $k$ subgraphs $G_1,G_2,\ldots,$ where, for each $j$, $|E(G_j)|\leq \Delta^*\cdot n^{1+j/k}$, and all vertex degrees in $G_j$ are at least $\Delta^*\cdot n^{(j-2)/k}$. For each such subgraph $G_j$, we further perform \emph{splitting} of its high-degree vertices, to ensure that all vertex degrees become close to $\Delta^*\cdot n^{(j-2)/k}$, and the number of vertices in the resulting graph $G_j'$ remains at most $2n$. For all $j>2$, we initialize the above algorithm on graph $G'_j$. In fact, the initialization proceeds from higher to lower indices $j$, and edges that are left over from the initialization algorithm for graph $G'_j$ (that is, edges that do not lie in any of the routers), become the edges of graph $G_{j-1}$. The set $\edel$ of edges is then set to contain all edges of $E(G_1)\cup E(G_2)$. Consider now a batch $\pi_i$ of updates. We compute the smallest integer $j$, so that $\Delta^*\cdot n^{1+(j-2)/k}\geq |\pi_i|$. For indices $j'>j$, for each edge $e$ of $G_{j'}'$ that is  deleted in $\pi_i$, we delete $e$ from $G_{j'}'$, and update the router decomposition of $G_{j'}'$ following these edge deletions; this may result in an additional collection $E_{j'}$ of edges to be deleted from $G_{j'}$, where $|E_{j'}|$ is roughly bounded by the number of edges of $G_{j'}$ that are deleted by $\pi_i$. Next, we construct a set $E^*$ of edges, that contains (i) all edge insertions from $\pi_i$; (ii) all edges lying in graphs $G_1,\ldots,G_j$; and (iii) all edges in the sets $E_{j'}$ that were deleted from graphs $G_{j'}$ for $j'>j$ while the batch $\pi_i$ was processed. We let the new graph $G_j$ contain all edges of $E^*$, and we initialize the algorithm for computing the router decomposition on the corresponding vertex-split graph $G'_j$ from scratch. As before, all leftover edges that are not included in the resulting router decomposition are added to the graph $G_{j-1}$, and we recursively initialize the algorithm for computing the router decomposition on the corresponding graph $G'_{j-1}$, and so on.

\paragraph{Organization.}
We start with Preliminaries in Section \ref{prelims}. We formally define router decompositions, and state our main technical result: namely, an algorithm that maintains a router decomposition of a graph that undergoes a small number of batches of updates, in Section \ref{sec:well-conn-decomp}. We also show that the proof of \Cref{thm: router decomposition main} follow from this main technical result. Then in sections Sections \ref{sec: nice router}--\ref{sec: clustering} we develop our main technical tools for maintaining a router decomposition. The first tool is a simple construction of a nice router, and an algorithm for its pruning, that are presented in Section \ref{sec: nice router}. The second tool is an algorithm for embedding a nice router into a given graph, and is presented in Section \ref{sec: embed-well-connected}. 
The third tool is a router witness, that is presented in \Cref{sec: router witness}. Finally, the last tool is a decremental algorithm for low-diameter decomposition, that appears in \Cref{sec: clustering}. 
In \Cref{sec: tieup}, we combine all these tools to complete the proof of the main technical theorem. Lastly, in Section \ref{sec:bd-fault}, we provide an algorithm for maintaining (low congestion, fault tolerant) spanners for bounded-degree faults in a fully dynamic graph, and connectivity certificates. 

\section{Preliminaries}
\label{prelims}
All logarithms in this paper are to the base of $2$. 
Throughout the paper, we use the $\tilde O(\cdot)$ notation to hide multiplicative factors that are polynomial in $\log m$ and $\log n$, where $m$ and $n$ are the number of edges and vertices, respectively, in the initial input graph.
All graphs in this paper are undirected. By default, we allow parallel edges but we do not allow self loops. Graphs in which parallel edges are not allowed are explicitly referred to as \emph{simple} graphs.

We follow standard graph-theoretic notation.   Given a graph $G=(V,E)$ and two disjoint subsets $A,B$ of its vertices, we denote by $E_G(A,B)$ the set of all edges with one endpoint in $A$ and another in $B$, and by $E_G(A)$ the set of all edges with both endpoints in $A$. We also denote by $\delta_G(A)$ the set of all edges with exactly one endpoint in $A$. For a vertex $v\in V(G)$, we denote by $\delta_G(v)$ the set of all edges incident to $v$ in $G$, and by $\deg_G(v)$ the degree of $v$ in $G$.
We also denote by $\vdeg_G$ the $n$-dimensional vector of all vertex degrees in $G$.
We may omit the subscript $G$ when clear from context. Given a subset $S\subseteq V$ of vertices of $G$, we denote by $G[S]$ the subgraph of $G$ induced by $S$.

If $G$ is a graph, and $\pset$ is a collection of paths in $G$, we say that the paths in $\pset$ cause \emph{congestion $\eta$}, if every edge $e\in E(G)$ participates in at most $\eta$ paths in $\pset$, and some edge participates in exactly $\eta$ such paths.


Assume now that we are given a graph $G$  with capacities $c(e)> 0$ on edges $e\in E(G)$. 
A \emph{flow} in $G$ is an assignment of non-negative flow values $f(P)$ to paths $P$ in $G$. If $P$ is a path whose corresponding flow value $f(P)$ is non-zero, then we say that $P$ is a \emph{flow-path}. A flow $f$ is typically defined by providing a list of flow-paths $P$ together with their corresponding flow value $f(P)$. For an edge $e\in E(G)$, the flow $f(e)$ through  $e$ is the total sum of flow values $f(P)$ over all paths $P$ containing $e$. We say that flow $f$ causes {\em congestion} $\eta$ if $\max_{e\in E(G)}\set{\frac{f(e)}{c(e)}}=\eta$.
Notice that it is possible that $\eta<1$. If $f(e)\leq c(e)$ holds for every edge $e\in E(G)$, we may sometimes say that $f$ is a \emph{valid flow}, or that it \emph{respects edge capacities}. When the edge capacities $c(e)$ are not explicitly given, we assume that they are unit.

\paragraph{Distances and Balls.}
Suppose we are given a graph $G$ with lengths $\ell(e)> 0$ on its edges $e\in E(G)$. For a path $P$ in $G$, we denote its length by $\ell_G(P)=\sum_{e\in E(P)}\ell(e)$. For a pair of vertices $u,v\in V(G)$, we denote by $\dist_G(u,v)$ the \emph{distance} between $u$ and $v$ in $G$: the smallest length $\ell_G(P)$ of any path $P$ connecting $u$ to $v$ in $G$.  For a pair $S,T$ of subsets of vertices of $G$, we define the distance between $S$ and $T$ to be $\dist_G(S,T)=\min_{s\in S,t\in T}\set{\dist_G(s,t)}$.
For a vertex $v\in V(G)$, and a subset $S\subseteq V(G)$ of vertices, we also define the distance between $v$ and $S$ as $\dist_G(v,S)=\min_{u\in S}\set{\dist_G(v,u)}$.
The \emph{diameter} of the graph $G$, denoted by $\diam(G)$, is the maximum distance between any pair of vertices in $G$.

Consider now some vertex $v\in V(G)$, and a distance parameter $D\geq 0$. The \emph{ball of radius $D$ around $v$} is defines as: $B_G(v,D)=\set{u\in V(G)\mid \dist_G(u,v)\leq D}$.
Similarly, for a subset $S\subseteq V(G)$ of vertices, we let the ball of radius $D$ around $S$ be $B_G(S,D)=\set{u\in V(G)\mid \dist_G(u,S)\leq D}$.
We will sometimes omit the subscript $G$ when clear from context.

\paragraph{Embeddings of Graphs.}
Let $G$, $X$ be two graphs with $|V(X)|\leq |V(G)|$. An \emph{embedding} of $X$ into $G$ consists of a mapping $g:V(X)\rightarrow V(G)$, such that the vertices of $X$ are mapped to {\bf distinct} vertices of $G$, and a collection $\pset=\set{P(e)\mid e\in E(X)}$ of paths in graph $G$, such that, for every edge $e=(x,y)\in E(X)$, path $P(e)$ connects $g(x)$ to $g(y)$. The \emph{congestion} of the embedding is the maximum, over all edges $e'\in E(G)$, of the number of paths in $\pset$ containing $e'$. Given an embedding of $X$ into $G$ as above, we will often identify vertices $v\in V(X)$ with the corresponding vertices $g(v)\in V(G)$ to which they are mapped, and so the embedding will only be specified by the collection $\pset$ of embedding paths. Whenever we say that we are given an embedding of a graph $X$ into a graph $G$, this implicitly means that $|V(X)|\leq |V(G)|$.

\subsection{Vertex Weighting, Conductance and Expanders}

Given an $n$-vertex graph $G=(V,E)$,
a \emph{vertex weighting} $w: V \mapsto \mathbb{Z}_{\geq 0}$ is an assignment of non-negative integral weight $w(v)$ to every vertex $v\in V(G)$. We denote by $\vw$ the corresponding $n$-dimensional vector of all vertex weights.
We may sometimes consider two special vertex weightings: weighting $\vdeg_G$, where the weight of every vertex $v$ is $\deg_G(v)$; and a weighting $\vDelta$, where $\Delta>0$ is an real number, and $\vDelta$ is an $n$-dimensional vector with all entries equal to $\Delta$, so the weight of every vertex is $\Delta$.

Consider now a graph $G=(V,E)$ with a vertex weighting $\vw$. For a   subset $S \subseteq V$ of vertices, the weight of $S$ is $w(S)=\sum_{v \in S} w(v)$. 
When $\vw=\vdeg_G$, we may denote $w(S)$ by $\vol_G(S)$, and refer to it as the \emph{volume of $S$ in $G$}.   
If $S,V\setminus S\neq \emptyset$, we sometimes refer to $(S,V\setminus S)$ as a \emph{cut}. 
Given a cut $(S,V\setminus S)$ with $w(S),w(V\setminus S)>0$, the \emph{conductance} of the cut with respect to vertex weighting $\vw$ is:
$$\Phi_{G,\vw}(S)=\frac{|\delta_G(S)|}{\min\{w(S), w(V \setminus S)\}}~.$$

For simplicity, if $w(S)$ or $w(V\setminus S)$ is $0$, we define $\Phi_{G,\vw}(S)=0$. 

The \emph{conductance} of a graph $G$ with respect to vertex weighting $\vw$ is denoted by
$\Phi(G,\vw)$, and it is defined to be the minimum conductance  $\Phi_{G,\vw}(S)$ of any cut $(S,V\setminus S)$.

If no vertex-weighting is explicitly given for a graph $G=(V,E)$, we use the default weighting $\vw=\vdeg_G$. In such a case, the conductance of a cut $(S,V\setminus S)$ is defined with respect to the standard volume $\vol_G(\cdot)$ notion defined above.

\begin{definition}[$\phi$-expander]\label{def:expander}
	Let $G=(V,E)$ be a graph, let $\vw$ be a weighting of its vertices, and let $\phi \geq 0$ be a parameter. We say that $G$ is a $\phi$-expander with respect to $\vw$, if $\Phi(G,\vw) \geq \phi$.  If $G$ is $\phi$-expander with respect to vertex weighting $\vw = \vdeg_G$, then we may simply say that $G$ is a $\phi$-expander.
\end{definition}

\subsection{Demands and their Routing}
\label{subsec: demands}

\begin{definition}[$\vw$-restricted demand]\label{def: restricted demand}
	Let $G=(V,E)$ be a graph, and let $\vw$ be a weighting of its vertices. A \emph{demand} in $G$ consists of a collection $\Pi$ of pairs of vertices, and, for every pair $(a,b)\in \Pi$, a value $D(a,b)>0$, that we refer to as the \emph{demand between $a$ and $b$}. We say that a demand $\dset=\left(\Pi,\set{D(a,b)}_{(a,b)\in \Pi}\right )$ is
	\emph{$\vw$-restricted}, if, for every vertex $v\in V$, the total demand between all pairs in $\Pi$ that contain $v$ is at most $w(v)$.
	When $\vw=\vDelta$ for an integer $\Delta$, we may sometimes refer to a $\vw$-resticted demand as  a \emph{$\Delta$}-restricted demand. 

	 The \emph{support} of the demand $\dset=\left(\Pi,\set{D(a,b)}_{(a,b)\in \Pi}\right )$ is the set of all vertices participating in the pairs in $\Pi$. We say that the demand is between two pairs $A,B\subseteq V$  of subsets of vertices of $G$, if $\Pi\subseteq A\times B$.
\end{definition}

\begin{definition}[Routing of a demand]\label{def: routing}
	Let $G=(V,E)$ be a graph,  and let $\dset=\left(\Pi,\set{D(a,b)}_{(a,b)\in \Pi}\right )$  be a demand in $G$. For every pair $(a,b)\in \Pi$, let $\pset_{a,b}$ denote the collection of all $a$-$b$ paths in $G$. A \emph{fractional routing}, or just a routing, of the demand $\dset$, is a flow $f$, that assigns non-zero values $f(P)>0$ only to flow-paths $P\in \bigcup_{(a,b)\in \Pi}\pset_{a,b}$, such that, for all $(a,b)\in \Pi$, $\sum_{P\in \pset_{a,b}}f(P)=D(a,b)$. The routing is defined by only specifying values $f(P)$ that are non-zero. The \emph{congestion} of the routing is the congestion caused by the flow $f$ in $G$. We say that the routing is \emph{over paths of length at most $d$}, if every flow-path $P$ with $f(P)>0$ has length at most $d$. If, for every path $P$, $f(P)\in\set{0,1}$, then we say that the routing is \emph{integral}. In this case, we may represent the routing as a collection of paths $\qset=\set{P\mid f(P)=1}$
\end{definition}

\subsection{Routers}
\label{subsec: routers}

A central object that we use is routers. Routers were introduced in \cite{HHT2024} as an object that is closely related to bounded-hop expanders. As we show below, routers are also closely related to the well-connected graphs, that were introduced in \cite{chuzhoy2023distanced}. The definition below slightly generalizes the definition of \cite{HHT2024}.

\begin{definition}[$(\vw,d,\eta)$-Router.]\label{def: router}
	Let $G=(V,E)$ be a graph, let $\vw$ be a weighting of its vertices, and let $S\subseteq V$ be a subset of vertices of $G$. Additionally,  let $d,\eta \geq 1$ be parameters. We say that $G$ is a \emph{$(\vw,d,\eta)$-router} for the set $S$ of supported vertices, if every $\vw$-restricted demand $\dset$ defined over the vertices of $S$ can be routed with congestion at most $\eta$ in $G$, via paths of length at most $d$. 
\end{definition}

We note that the inclusion of the set $S$ of supported vertices in the above definition may seem redundant, since we could simply set the weight of each such vertex to $0$, without the need to include the vertices of $S$ in the definition. However, supported vertices will be  useful when we consider special vertex weightings, such as $\vDelta$ for $\Delta \in \reals^{>0}$, and $\vdeg_G$.

\subsection{Well-Connected Graphs, their Relationship to Routers, and Related Algorithmic Tools }

\label{subsec: routers and wellconnected}

We employ well-connected graphs, that were introduced in  \cite{chuzhoy2023distanced}. We start with a formal definition of such graphs, which is identical to that from \cite{chuzhoy2023distanced}.

\begin{definition}[Well-Connected Graph] \label{def: well-connected}
	Given an $n$-vertex graph $G$,  a set $S$ of its vertices called \emph{supported vertices}, and parameters $\eta,d>0$, we say that graph $G$ is $(d,\eta)$-well-connected with respect to $S$, if, for every pair $A,B\subseteq S$ of disjoint equal-cardinality subsets of supported vertices, there is a collection $\pset$ of paths in  $G$, that connect every vertex of $A$ to a distinct vertex of $B$, such that the length of each path in $\pset$ is at most $d$, and every edge of $G$ participates in at most $\eta$ paths in $\pset$.
\end{definition}	

For intuition, it may be convenient to think of $d=2^{\poly(1/\eps)}$, $\eta=n^{O(\eps)}$, and $|S|\geq |V(G)|-n^{1-\eps}$, for some precision parameter $0<\eps<1$.

\subsubsection{The Relationship between Routers and Well-Connected Graphs}

The notion of routers is closely related to the notion of well-connected graphs.  It is immediate to see that, if $G$ is a $(\vw,d,\eta)$-router with respect to any set $S$ of supported vertices, for any vertex weighting $\vw$ where $w(v)\geq 1$ for all $v\in V(G)\setminus S$, then $G$ is $(d,O(\eta\cdot \log n))$-well-connected with respect to $S$ (the $O(\log n)$ factor in the congestion is lost when converting a fractional flow into integral one via standard randomized rounding). 

For the connection in the opposite direction, at first glance, it appears that the notion of well-connected graphs is weaker, for two reasons. The first reason is that, from the definition, well-connected graphs only provide routings between pairs $A,B$ of vertex subsets, and it appears that one cannot control the specific pairs that are being routed. The second reason is that they only guarantee the routing of $|A|=|B|$ flow units from $A$ to $B$, which roughly corresponds to routing with vertex weighting $\vDelta$ where $\Delta=1$. We first address the former issue, and show, in the following claim, that in fact a $(d,\eta)$-well-connected graph is a sufficiently strong router.
We do not use this claim directly, but we feel that it is useful in that it illuminates the relationship between these two objects. The proof of the claim is deferred to Section \ref{subsec: appx: proof well connected to router1} of Appendix.

\begin{claim}\label{claim: well connected to routers}
	Let $G$ be an $n$-vertex graph, and assume that it is an $(d,\eta)$-well-connected graph with respect to a set $S$ of supported vertices. Then for all $1\leq k\leq \log n$, $G$ is a $(1,d',\eta')$-router with respect to $S$, for $d'=512k\cdot d$ and $\eta'= \left(2^{17}\cdot kdn^{1/k}\log^2n\right )\cdot \eta$.
\end{claim}

There are three approaches to address the second issue that is raised above -- namely, that the routing guarantees provided by well-connected graphs correspond to vertex weighting $w(v)=1$ for all vertices $v$. The first approach has to do with the manner in which well-connected graphs are used. Most if not all applications of such graphs known today  construct a well-connected graph $W$ that is embedded into a given input graph $G$ via the \emph{distanced-matching game}, that can be thought of as the analogue of the cut-matching game for constructing a well-connected graph. From the construction, graph $W$ has a rather low degree, so it can be thought of as providing routing for the weighting $\vdeg_W$ of its vertices. Moreover, in the typical application of well-connected graphs as described above, the embedding of $W$ into $G$ is usually then used as a certificate that $G$ itself is a well-connected graph, and is thus a router with respect to vertex weighting $w(v)=1$ for all vertices $v$, from \Cref{claim: well connected to routers}. In order to obtain a certificate of routability for a general weighting $\vw$ of the vertices of $G$, we can construct a larger graph $W$, containing $\sum_{v\in V(G)}w(v)$ vertices, and then embed it into $G$ via the distanced-matching game, where, for every vertex $v\in V(G)$, we embed $w(v)$ vertices of $W$ into $G$. Such an embedding would then serve as a certificate that $G$ is a router for the vertex weighting $\vw$.
The second approach is that, given any graph $H$ with arbitrary integral vertex weighting $\vw$, we can obtain a new graph $H'$ via the following simple transformaiton: for every vertex $v\in V(H)$, add a new collection $T(v)$ of $w(v)$ vertices to the graph, connecting them to $v$ with new edges. Let $S=\bigcup_{v\in V(H)}T(v)$. Then if graph $H'$ is well-connected with respect to the set $S$ of supported vertices, the original graph $H$ is a router with respect to vertex weighting $\vw$.
Lastly, the third approach is to generalize the definition of well-connected graphs directly to arbitrary vertex weighting. In view of the first two approaches, it is not clear to us if such a generalization provides an additional value.

\paragraph{Algorithmic tools for well-connected graphs.}
The work of \cite{chuzhoy2023distanced} provides a fast algorithm, that, given a graph $G$ and a set $T$ of its vertices called terminals, either computes a well-connected graph $X$ that is defined over a large subset of $T$, together with its embedding into $G$, or returns two large sets $T_1,T_2\subseteq T$ of terminals, together with a relatively small set $E'$ of edges, such that $\dist_{G\setminus E'}(T_1,T_2)$ is large. Another algorithm that  \cite{chuzhoy2023distanced} provides, and that we will exploit, is for decremental \APSP in well-connected graphs. Given a graph $X$ that undergoes an online sequence of edge deletions, and a set $S$ of its vertices, such that $X$ is well-connected with respect to $S$, the algorithm maintains a large subset $S'\subseteq S$ of vertices, and supports \shortpath\ queries between vertices of $S'$: given a pair $x,y\in V(S')$ of such vertices, it returns a path $P$ connecting $x$ to $y$ in $X$, such that the length of $P$ is small, and the time required to respond to the query is $O(|E(P)|)$. However, the algorithm for decremental \APSP in well-connected graphs requires one additional input, called a \emph{\HSS}, for graph $X$. Intuitively, \HSS is a hierarchy of well-connected graphs that are embedded into each other, with graph $X$ being the topmost graph in the hierarchy. The algorithm for embedding a well-connected graph into a given graph $G$, that we mentioned above, in case it produces a well-connected graph $X$ and its embedding into $G$, also returns the required \HSS for graph $X$, that can then be used by the algorithm for the \APSP problem on the well-connected graph $X$. Therefore, the specifics of the definition of the \HSS are not important for us: the algorithm for embedding a well-connected graph will produce exactly the kind of \HSS that the algorithm for \APSP needs to use. 
But for completeness, we provide the definition of the \HSS from \cite{chuzhoy2023distanced} here.

\paragraph{\HSS.}
The \HSS uses two main parameters: the base parameter $N>0$, and a level parameter $j>0$. We also assume that we are given a precision parameter $0<\eps<1$. The notion of Hierarchical Support Structure is defined inductively, using the level parameter $j$. If $X$ is a graph containing $N$ vertices, then a level-1 \HSS for $X$ simply consists of a set $S(X)$ of vertices of $X$, with $|V(X)\setminus S(X)|\leq N^{1-\eps^4}$. Assume now that we are given a graph $X$ containing exactly $N^j$ vertices. A level-$j$ Hierarchical Support Structure for $X$ consists of a collection $\hset=\set{X_1,\ldots,X_r}$ of $r=N-\ceil{2N^{1-\eps^4}}$ graphs, such that for all $1\leq i\leq r$, $V(X_i)\subseteq V(X)$; $|V(X_i)|=N^{j-1}$; and $|E(X_i)|\leq N^{j-1+32\eps^2}$. We also require that $V(X_1),\ldots,V(X_r)$ are all mutually disjoint. 
Additionally, it must contain, for all $1\leq i\leq r$, a level-$(j-1)$ Hierarchical Support Structure for $X_i$, which in turn must define the set $S(X_i)$ of supported vertices for graph $X_i$. We require that each such graph $X_i$ is $(\td_{j-1},\eta_{j-1})$-well-connected with respect to $S(X_i)$, where $\tilde d_{j-1}=2^{\tilde c(j-1)/\eps^4}$ for some constant $\tilde c$, and $\eta_{j-1}=N^{6+256(j-1)\eps^2}$. Lastly, the Hierarchical Support Structure for graph $X$ must contain an embedding of graph $X'=\bigcup_{i=1}^rX_i$ into $X$, via path of length at most $2^{O(1/\eps^4)}$, that cause congestion at most $N^{O(\eps^2)}$. 
We then set $S(X)=\bigcup_{i=1}^rS(X_i)$, and we view $S(X)$ as the set of supported vertices for graph $X$, that is defined by the \HSS.

\paragraph{Embedding of a well-connected graph.}
We will use the following theorem from \cite{chuzhoy2023distanced}.

\begin{theorem}[Corollary 5.3 from \cite{chuzhoy2023distanced}]\label{cor: HSS witness}
	There is a deterministic algorithm, whose input consists of an $n$-vertex graph $G$, a set $T$ of $q$ vertices of $G$ called terminals, and parameters $\frac{2}{(\log q)^{1/12}}< \eps<\frac{1}{400}$, $d>1$ and $\eta>1$, such that $1/\eps$ is an integer. The algorithm computes one of the following:

	\begin{itemize}
		\item either a pair $T_1,T_2\subseteq T$ of disjoint subsets of terminals, and a set $E'$ of edges of $G$, such that:
		
		\begin{itemize}
			\item $|T_1|=|T_2|$ and $|T_1|\geq \frac{q^{1-4\eps^3}}{4}$;
			\item $|E'|\leq \frac{d\cdot |T_1|}{\eta}$; and
			\item for every pair $t\in T_1,t'\in T_2$ of terminals, $\dist_{G\setminus E'}(t,t')>d$;
		\end{itemize} 
		
		\item or a graph $X$ with $V(X)\subseteq T$, $|V(X)|=N^{1/\eps}$, where $N=\floor{q^{\eps}}$,  and maximum vertex degree at most   $q^{32\eps^3}$, together with an embedding $\pset$ of $X$ into $G$ via paths of length at most $d$ that cause congestion at most $\eta\cdot q^{32\eps^3}$, and a level-$(1/\eps)$ \HSS for $X$, such that $X$ is $(\td,\eta')$-well-connected with respect to the set $S(X)$ of vertices defined by the support structure, where $\eta'=N^{6+256\eps}$, and $\td=2^{\tilde c/ \eps^5}$, where $\tilde c $ is the constant used in the definition of the \HSS.
	\end{itemize}
	The running time of the algorithm is  $O\left (q^{1+O(\eps)}+|E(G)|\cdot q^{O(\eps^3)}\cdot(\eta+d\log n)\right )$.
\end{theorem}

\paragraph{\APSP in Well-Connected Graphs.}

Lastly, we need an algorithm for decremental \APSP in well-connected graphs from \cite{chuzhoy2023distanced}. Assume that we are given a graph $X$ that is obtained from \Cref{cor: HSS witness}. In other words, we are given a level-$(1/\eps)$ \HSS for $X$, together with a large set $S(X)$ of vertices of $X$, so that $X$ is well-connected with respect to $S(X)$. We then assume that graph $X$ undergoes a sequence of edge deletions. As edges are deleted from $X$, the well-connectedness property may no longer hold, and the \HSS may be partially destroyed. Therefore, we only require that the algorithm maintains a large enough subset $S'(X)\subseteq S(X)$ of supported vertices, and that it can respond to \shortpath\ queries between pairs of vertices in $S'(X)$: given a pair $x,y$ of such vertices, the algorithm needs to return a path of length at most $2^{O(1/\eps^6)}$ in the current graph $X$ connecting them. We also require that the set $S'(X)$ is \emph{decremental}, so vertices can leave this set but they may not join it. The algorithm is summarized in the following theorem.

\begin{theorem}[Theorem 2.3 in \cite{chuzhoy2023distanced}]\label{thm: APSP in HSS full}
	There is a deterministic algorithm, whose input consists of:
	
	\begin{itemize}
		\item a parameter $0<\eps<1/400$, so that $1/\eps$ is an integer;
		\item an integral parameter $N$ that is sufficiently large, so that $\frac{N^{\eps^4}}{\log N}\geq 2^{128/\eps^6}$ holds;
		\item  a graph $X$ with $|V(X)|=N^{1/\eps}$; and
		\item a level-$(1/\eps)$ \HSS for  $X$, such that $X$ is $(\td,\eta')$-well-connected with respect to the set $S(X)$ of vertices defined by the \HSS, where $\td$ and $\eta'$ are parameters from \Cref{cor: HSS witness}.  
	\end{itemize} 
	Further, we assume that graph $X$ undergoes an online sequence of at most $\Lambda=|V(X)|^{1-10\eps}$ edge deletions. The algorithm maintains a set $S'(X)\subseteq S(X)$ of vertices of $X$, such that, at the beginning of the algorithm, $S'(X)=S(X)$, and over the course of the algorithm, vertices can leave $S'(X)$ but they may not join it. The algorithm ensures that $|S'(X)|\geq \frac{|V(X)|}{2^{4/\eps}}$ holds at all times, and it supports short-path queries between vertices of $S'(X)$: given a pair $x,y\in S'(X)$ of vertices, return a path $P$ connecting $x$ to $y$ in the current graph $X$, whose length is at most $2^{O(1/\eps^6)}$, in time $O(|E(P)|)$. The total update time of the algorithm is $O( |E(X)|^{1+O(\eps)})$. 
\end{theorem}

\subsection{Length-Constrained Expanders}

Length-constrained expanders were initially introduced in \cite{haeupler2022hop} as a generalization of the standard notion of expanders, that allows routing via paths of sublogarithmic length; this was also the main motivation behind the well-connected graphs introduced in \cite{chuzhoy2023distanced}.
We do not use length-constrained expanders directly (except that we use routers that can be viewed as bounded-diameter length-constrained expaners). We provide a short discussion here for completeness.

Consider any graph $G$, and a demand $\dset=\left(\Pi,\set{D(a,b)}_{(a,b)\in \Pi}\right )$ for $G$. We say that the demand is \emph{$h$-length} if, for every pair $(a,b)\in \Pi$ of vertices, $\dist_G(a,b)\leq h$.  The following description of length-constrained expanders follows the summary in \cite{haeupler2022cut}. The formal definition of $(h,s)$-length $\phi$-expanders 
relies on the notion of so-called ``moving cuts'', and can be found in \cite{haeupler2022hop}. 
Instead, as in prior work on graph spanners (e.g., \cite{HHT21} and \cite{BodwinHP24}), we use the more convenient \emph{routing characterization} of length-constrained expanders. Informally, up to a constant factor in the length parameter, and a poly-logarithmic factor in the congestion parameter,  a graph $G$ is an $(h,s)$-hop $\phi$-expander for a vertex weighting $\vw$, if any $h$-length $\vw$-bounded demand $\dset$ can be routed in $G$ via paths of length at most $(s \cdot h)$, with congestion roughly $\frac{1}{\phi}$. The precise characterization of length-constrained expanders in terms of routing appears in the following theorem.

\begin{theorem}
[Theorem 5.17 of \cite{HHT2024}]\label{thm:flow character} Let $G$ be a graph, let $\vw$ be a weighting of its vertices, and let $h \geq 1$, $0<\phi < 1$, and $s \geq 1$ be parameters. Then:
\begin{enumerate}
\item if $G$ is an $(h,s)$-length $\phi$-expander for vertex weighting $\vw$, then every $h$-length $\vw$-restricted demand $\dset$ can be routed in $G$ via paths of length at most $(s \cdot h)$, with congestion $O\left (\frac{\log(n)}{\phi}\right )$; and

\item if $G$ is not an $(h,s)$-length $\phi$-expander for vertex weighting $\vw$, then there is some $h$-length $\vw$-bounded demand $\dset$ that cannot be routed in $G$ via paths of length at most  $(\frac{s}{2}\cdot h)$ with congestion at most $\frac{1}{2\phi}$.
\end{enumerate}
\end{theorem}

The next claim summarized a  bidrectional connection between routers and length-restricted expanders that was shown by
\cite{haeupler2022hop,HHT2024}, and also follows from  \Cref{thm:flow character}. 

\begin{claim}[Immediate from Lemma 5.24 in \cite{HHT2024}]\label{cl:router-to-bounded-hop} 
	If $G$ is an $(h,s)$-length $\phi$-expander for vertex weighting $\vw$ with $\diam(G)\leq h$, then $G$ is a $\left (\vw,hs,O\left (\frac{\log n}{\phi}\right ) \right )$-router for the set $S=V(G)$ of supported vertices. Conversely, if an $G$ is a graph of diameter at most $h$ that is not an $(h,s)$-length $\phi$-expander for vertex weighting $\vw$, then $G$ is not a $\left (\vw,\frac{hs}{2},O\left(\frac{1}{2\phi}\right )\right )$-router.
\end{claim}

To summarize, one can intuitively think of a $(\vdeg,d,\eta)$-router as being roughly equivalent to a bounded-diameter length-constrained $\phi$-expander, where $\phi=\tilde \Theta(1/\eta)$, and the bound on the diameter is roughly $d$.

\subsection{Even-Shiloach Trees}
Suppose we are given a graph $G=(V,E)$ with integral lengths $\ell(e)\geq 1$ on its edges $e\in E$, a source $s$, and a distance bound $d\geq 1$. The Even-Shiloach Tree (\EST) algorithm of~\cite{EvenS,Dinitz,HenzingerKing}  maintains, for every vertex $v\in V$ with $\dist_G(s,v)\leq d$, the distance $\dist_G(s,v)$, under the deletion of edges from $G$. It also maintains a shortest-path tree $\tau$ of $G$ rooted at  $s$, that includes all vertices $v$ with $\dist_G(s,v)\leq d$. We denote the corresponding data structure by $\EST(G,s,d)$. 
The total update time of the algorithm  is $O(m\cdot d\cdot \log n)$, where $m$ is the initial number of edges in $G$ and $n=|V|$.

\subsection{The Online-Batch Dynamic Model and Deamortization}
\label{subsec: online batch}

Our algorithm uses the by now standard online-batch dynamic model, and a reduction from this model to the standard dynamic model with bounded worst-case update time, that was formalized in \cite{jin2022fully}. Much of the exposition in this subsection is borrowed from Section 10.1 of \cite{jin2022fully}

We assume that we are given two main parameters: batch number $\sigma$ and sensitivity parameter $w$. We let $\DS$ be a data stucture that we would like to maintain for an input graph $G$ undergoing updates. A dynamic algorithm for maintaining $\DS$ in the online-batch dynamic model consists of $\sigma+1$ phases $\Phi_0,\ldots,\Phi_{\sigma}$, where $\Phi_0$ is the \emph{preprocessing}, or the \emph{initialization} phase, and $\Phi_1,\ldots,\Phi_{\sigma}$ are \emph{regular} phases. For all $1\leq i\leq \sigma$, at the beginning of Phase $\Phi_i$, the algorithm is given the $i$th batch $\Pi_i$ of updates  to graph $G$; in our case, all updates are insertions and deletions of edges to and from $G$. The number of such updates in each batch $\Pi_i$ must be bounded by $w$. 
The algorithm is required to maintain the data structure $\DS$, so that,  at the end of the $0$th phase, the data structure $\DS$ that it obtains correctly corresponds to the initial graph $G\attime[0]$. Additionally, for all $1\leq i\leq \sigma$, the data structure $\DS$ that the algorithm obtains at the end of Phase $\Phi_i$ must correctly correspond to the graph $G\attime[i]$, that is obtained from the initial graph $G\attime[0]$, after the sequence $\Pi_1\circ\cdots\circ\Pi_i$ of updates is applied to it. The algorithm has amortized time $t$, if, for all $1\leq i\leq \sigma$, the running time of the algorithm during Phase $\Phi_i$ is bounded by $t\cdot |\Pi_i|$. Here, $t$ is allowed to be some function of the graph parameters, such as, for example, $n^{\eps}$, where $n=|V(G)|$, and $\eps$ a given parameter.

The following lemma was formally proved in \cite{jin2022fully}, and is implied by the previous work of \cite{nanongkai2017dynamic2}. We provide a slightly simplified version of the lemma that is sufficient for us.

\begin{lemma}[Lemma 10.1 in \cite{jin2022fully}; see also Section 5 in  \cite{nanongkai2017dynamic2}]\label{lem: deamortization} Let $G$ be a graph undergoing batch updates, and let $\sigma>0$ and $w\geq 2\cdot 6^{\sigma}$ be parameters.
Assume that there is an algorithm in the batch update model with batch number $\sigma$ and sensitivity parameter $w$, that maintains some data structure $\DS$ for $G$, whose preprocessing time is $T_0$, and amortized update time is $T'$, where both $T_0$ and $T'$ are functions that map some graph measure (e.g. $|E(G\attime[0])|$ or $|V(G)|$) to non-negative numbers. 
	
Then there is a dynamic algorithm, with preprocessing time $O\left (2^{O(\sigma)}\cdot T_0\right)$, and worst case update time $O\left(4^{\sigma}\cdot \left(\frac{T_0}{w}+w^{1/\sigma}\cdot T'\right )\right )$, that maintains a set of $O(2^{O(\sigma)})$ instances of the data structure $\DS$, such that, after each update, the algorithm indicates one of the maintained instances of the data structure as the ``current'' one. The current instance of data structure $\DS$ satisfies the following conditions:
	
\begin{itemize}
	\item it is up-to-date with respect to the current graph; and
	\item the online-batch update algorithm underwent at most $\sigma$ batches of updates, with each update batch size at most $w$.
\end{itemize}
\end{lemma}

\subsection{Vertex-Split Graphs}
\label{subsec: vertex split}

Some of our subroutines work best when all degrees in the given graph are close to each other. In order to extend such subroutines to general graphs, we will employ \emph{vertex-split graphs}.

Consider a graph $G=(V,E)$. A \emph{vertex split operation} in $G$ is defined as follows. The input to the vertex split operation is a vertex $v\in V$, and a subset  $E_1\subseteq \delta_G(v)$ of its incident edges. In order to execute the split operation, we insert a new vertex $v'$ into $G$, that we call a \emph{copy} of $v$. For every edge $e=(u,v)\in E_1$, we then delete $e$ from $G$, and insert the edge $(u,v')$, that we call a \emph{copy of $e$} into $G$. We will not distinguish between the original edges of $G$ and their copies in the resulting graph. We are now ready to define vertex-split graphs.

\begin{definition}\label{def: vertex split graph}
	Let $G$ and $G'$ be two graphs. We say that $G'$ is a \emph{vertex-split graph for $G$}, if $G'$ can be obtained from $G$ via a sequence of vertex split operations.
\end{definition}

Next, we provide a simple claim that allows us to produce a vertex-split graph for a given graph $G$, in which no vertex degree is too high.

\begin{claim}\label{claim: uniformize degree}
There is a deterministic algorithm, that, given a graph $G$ with  average vertex degree $\Delta$, computes a vertex-split graph $G'$ for $G$, such that the degrees of all vertices in $G'$ are at most $2\Delta$, and $|V(G')|\leq 2|V(G)|$. The running time of the algorithm is $O(|E(G)|)$.
\end{claim}
\begin{proof}
	Let $S\subseteq V(G)$ be the set of all vertices $v$ with $\deg_G(S)>2\Delta$. 
		Our algorithm processes every vertex $v\in S$ one by one. In order to process a vertex $v\in S$, we perform iterations, as long as $\deg_G(v)>2\Delta$. In every iteration, we select an arbitrary subset  $E_1\subseteq \delta_G(v)$ of $\Delta$ edges, and then perform a vertex splitting of $v$ using $E_1$. Notice that the splitting operation can be executed in time $O(|E_1|)$. We then continue to the next iteration. It is easy to verify that the number of new vertices inserted into $G$ while processing $v$ is at most $\frac{\deg_G(v)}{\Delta}$. 
	
	Once all vertices in $S$ are processed, we obtain the final graph $G'$, that is a vertex-split graph for $G$. Clearly, all vertex degrees in $G'$ are bounded by $2\Delta$. Since  $\sum_{v\in S}\deg_G(v)\leq \Delta \cdot |V(G)|$, the total number of new vertices inserted into $G'$ is bounded by $|V(G)|$, and so $|V(G')|\leq 2|V(G)|$. Lastly, it is easy to verify that the running time of the algorithm is $O(|E(G)|)$.
\end{proof}

Next, we obtain the following immediate observation.

\begin{observation}\label{obs: split graph router}
	Let  $G$ is a graph, and let $G'$ be a vertex-split graph for $G$. Assume that $G'$ is a $(\vdeg_{G'}, d, \eta)$-router, for some parameters $ d$ and $ \eta$, with respect to the set $V(G')$ of supported vertices. Then $G$ is a $(\vdeg_G, d,\eta)$-router with respect to the set $V(G)$ of supported vertices.
\end{observation}

\begin{proof}
	For every vertex $v\in V(G)$, let $S(v)$ be the collection of its copies in $G'$. Recall that $\sum_{v'\in S(v)}\deg_{G'}(v')=\deg_G(v)$.
	
	Let $\dset=\left(\Pi,\set{D(a,b)}_{(a,b)\in \Pi}\right )$ be any $\vdeg_G$-restricted demand in $G$. It is enough to show that it can be routed in $G$ via paths of length at most $d$, with congestion at most $\eta$.
	
	We use the demand $\dset$, in order to define a demand $\dset'=\left(\Pi',\set{D'(a',b')}_{(a',b')\in \Pi'}\right )$ in $G'$, that is $\vdeg_{G'}$-restricted, so that a routing of $\dset'$ in $G'$ via paths of length at most $d$ and congestion at most $\eta$ defines a routing of $\dset$ in $G$ via paths of length at most $d$ and congestion at most $\eta$. 
	
	In order to define the demand $\dset'$, we consider every pair $(a,b)\in \Pi$ one by one. Let $\Pi'(a,b)=S(a)\times S(b)$. For every pair $(a',b')\in \Pi'(a,b)$, we define a demand $D'(a',b')=D(a,b)\cdot \frac{\deg_{G'}(a')\cdot \deg_{G'}(b')}{\deg_G(a)\cdot \deg_G(b)}$.
	Notice that: 
	
	\[\sum_{(a',b')\in \Pi'(a,b)}\deg_{G'}(a')\cdot\deg_{G'}(b')=\left(\sum_{a'\in S(a)}\deg_{G'}(a')\right ) \cdot \left(\sum_{b'\in S(b)}\deg_{G'}(b')\right ) =\deg_G(a)\cdot \deg_G(b). 
	\]
	
	Therefore: 
	
	\begin{equation}\label{eq: route all flow}
	\sum_{(a',b')\in \Pi'(a,b)}D'(a',b')=D(a,b). 
	\end{equation}
	
	Moreover, for every vertex $a'\in S(a)$: 

\begin{equation}\label{eq: load on a}	
\sum_{b'\in S(b)}D'(a',b')=D(a,b)\cdot \frac{\deg_G(a')}{\deg_G(a)\cdot \deg_G(b)}\cdot \sum_{b'\in S(b)}\deg_{G'}(b')=
	D(a,b)\cdot \frac{\deg_{G'}(a')}{\deg_G(a)}.
\end{equation}

Similarly, for every vertex $b'\in S(b)$, 	$\sum_{a'\in S(a)}D'(a',b')=
D(a,b)\cdot \frac{\deg_{G'}(b')}{\deg_G(b)}$.

We are now ready to define the demand $\dset'=\left(\Pi',\set{D'(a',b')}_{(a',b')\in \Pi'}\right )$ for $G'$: we set $\Pi'=\bigcup_{(a,b)\in \Pi}\Pi'(a,b)$, and we set the demand $D'(a',b')$ for every pair $(a',b')\in \Pi'$ as above.

Consider now any vertex $v'\in V(G')$. Assume w.l.o.g. that in every pair $(a',b')\in \Pi'$ that contains $v'$, $v'=a'$ holds. Then, from Inequality \ref{eq: load on a}:

\[\sum_{(v',u')\in \Pi'}D'(v',u')=\sum_{\stackrel{(a,b)\in \Pi:}{v'\in S(a)}}D(a,b)\cdot \frac{\deg_{G'}(v')}{\deg_G(a)}\leq \deg_{G'}(v'),\]

since the demand $\dset$ is $\vdeg_G$-restricted. Therefore, demand $\dset'$ is $\vdeg_{G'}$-restricted, and so there is a routing $f'$ of $\dset'$ in $G'$ via paths of length at most $d$, with congestion at most $\eta$. Since every edge of $G'$ corresponds to a distinct edge of $G$, routing $f'$ immediately defines a routing $f$ of $\dset$ via paths of length at most $d$, with congestion at most $\eta$.
\end{proof}

We will also use the following observation, whose proof is immediate.

\begin{observation}\label{obs: split graph router fixed delta}
	Let  $G$ is a graph, and let $G'$ be a vertex-split graph for $G$. Assume that $G'$ is a $(\vDelta, d, \eta)$-router, for some parameters $ d$, $ \eta$ and $\Delta$, with respect to the set $V(G')$ of supported vertices. Then $G$ is a $(\vDelta, d,\eta)$-router with respect to the set $V(G)$ of supported vertices.
\end{observation}

\subsection{Chernoff Bound}

We use the following standard version of Chernoff Bound (see e.g., \cite{dubhashi2009concentration}).

\begin{lemma}[Chernoff Bound]
	\label{lem: Chernoff}
	Let $X_1,\ldots,X_n$ be independent random variables taking values in $\set{0,1}$. Let $X=\sum_{1\le i\le n}X_i$, and let $\mu=\expect{X}$. Then for any $t>2e\mu$,
	\[\Pr\Big[X>t\Big]\le 2^{-t}.\]
\end{lemma}

\section{Low-Diameter Router Decomposition}
\label{sec:well-conn-decomp}

In this section we formally define one of our main combinatorial objects -- a Low-Diameter Router Decomposition, that, for conciseness, we refer to as ``Router Decomposition''. We then provide the statement of our main technical result, namely, an algorithm that maintains such a decomposition in the online-batch dynamic model, and show that the proof of \Cref{thm: router decomposition main} follows from it. We complete the proof of the main technical result in Sections \ref{sec: nice router}--\ref{sec: tieup}.
We start with a formal definition of a router decomposition.

\begin{definition}[Router Decomposition of a Graph]\label{def: router decomposition}
Let $G$ be a simple graph, and let $\Delta^*,\tilde d,\tilde \eta$ and $\rho$ be non-negative parametes. A \emph{router decomosition} of $G$ with parameters $\Delta^*,\tilde d,\tilde \eta$  and $\rho$ consists of the following ingredients:

	\begin{itemize}
		\item a collection $\cset$ of subgraphs of $G$ called \emph{clusters} with $\sum_{C\in \cset}|V(C)|\leq \rho \cdot |V(G)|$, such that every cluster $C\in \cset$ is a $(\vdeg_C,\tilde d,\tilde \eta)$-router with respect to the set $V(C)$ of supported vertices, and every edge of $G$ lies in at most one cluster $C\in \cset$;  and
		
		\item for every cluster $C\in \cset$, a subgraph $C'\subseteq C$ with $V(C')=V(C)$, such that $C'$ is a $(\vDelta^*, \tilde d,\tilde \eta)$-router for the set $V(C)$ of supported vertices, and  $\sum_{C\in \cset}|E(C')|\leq \rho\cdot \Delta^* \cdot |V(G)|$.
	\end{itemize}

We denote by $\edel=E(G)\setminus \left(\bigcup_{C\in \cset}E(C)\right )$ the set of all edges of $G$ that do not lie in any cluster of $\cset$.
\end{definition}

We note that generally we will ensure that $|\edel|$ is sufficiently small in a router decomposition.
The following theorem summarizes our main technical result, and a significant part of the paper is devoted to proving it.

\begin{theorem}\label{thm: one level router decomposition}
	There is a deterministic algorithm, that receives as input a parameter $n$ that is greater than a large enough constant, and additional parameters $512\leq k\leq (\log n)^{1/49}$, $\frac{1}k\leq \delta<\frac{1}{400}$,  $\Delta^*$, and $\Delta$, such that $n^{18/k}\leq \Delta^*\leq \frac{\Delta}{n^{8/k^2}}\leq n$ holds,  and $k,\Delta,\Delta^*$ and $1/\delta$ are integers.
	The algorithm is also given as input a simple graph $G$ with $|V(G)|\leq n$, such that, initially, all vertex degrees in $G$ are at most $\Delta\cdot n^{16/k}$.
	Lastly, the algorithm receives $k$ online batches $\pi_1,\ldots,\pi_k$ of updates to graph $G$, where for $1\leq i\leq k$, the $i$th batch $\pi_i$ is a collection of at most $\Delta\cdot n$ edges that are deleted from $G$.

	The algorithm maintains a router decomposition $\left(\cset,\set{C'}_{C\in \cset}\right )$ of the graph $G$, for parameters $\Delta^*$, $\tilde d= k^{19}\cdot 2^{O(1/\delta^6)}$, 
	$\tilde \eta= n^{17/k}$, and $\rho=n^{20/k}$.
	After the initialization, $|\edel|\leq |V(G)|\cdot \Delta \cdot n^{2/k^2}$ holds, and, for all $1\leq i\leq k$, at most $|\pi_i|\cdot n^{13/k^2}$ edges may be added to $\edel$ while processing the batch $\pi_i$.
	For all $1\leq i\leq k$, after the $i$th batch of updates is received, the clusters $C\in \cset$ may only be updated by removing some of their edges (which are then added to $\edel$) and some of their vertices. The corresponding graphs $C'\subseteq C$ may be updated by removing or inserting edges into them, and by removing vertices from them. Additionally, clusters of $\cset$ that become empty are removed from $\cset$. These are the only updates to the router decomposition.
	The initialization time of the algorithm is $O\left((n\Delta)^{1+O(\delta)} \right )$, and for all $1\leq i\leq k$, the time required to process the $i$th update batch $\pi_i$ is bounded by  $O\left(|\pi_i|\cdot n^{O(1/k^2)}\right )$.  The total number of edges inserted into or deleted from all subgraphs in $\set{C'\mid C\in \cset}$ while batch $\pi_i$ is processed is at most $O\left(|\pi_i|\cdot n^{O(1/k^2)}\right )$.
\end{theorem}

We will prove \Cref{thm: one level router decomposition} in sections 
 \ref{sec: nice router}--\ref{sec: tieup}. In the remainder of this section, we complete the proof of \Cref{thm: router decomposition main} using it. We start with the following simple corollary of \Cref{thm: one level router decomposition}; its main difference from the theorem is that we only require that the total number of edges in $G$ is bounded by $\Delta\cdot n^{1+15/k}$, while the degrees of individual edges may be higher. Other than this and slight changes to the input parameters, the corollary is essentially identical to \Cref{thm: one level router decomposition}. The corollary follows  from \Cref{thm: one level router decomposition} in a straightforward manner by employing vertex-split graphs. We defer its proof to Section \ref{sec:appx:split one level} of Appendix.

\begin{corollary}\label{cor: one level router decomposition}
	There is a deterministic algorithm, that receives as input a simple $n$-vertex graph $G$, where $n$ is greater than a large enough constant, and additional parameters $512\leq k\leq (\log n)^{1/49}$, $\frac{1}k\leq \delta<\frac{1}{400}$,  $\Delta^*$, and $\Delta\leq n$, such that $n^{19/k}\leq \Delta^*\leq \frac{\Delta}{n^{9/k^2}}$ holds, $|E(G)|\leq \Delta\cdot n^{1+15/k}$,  and $k,\Delta^*$ and $1/\delta$ are integers.
	Lastly, the algorithm receives $k$ online batches $\pi_1,\ldots,\pi_k$ of updates to graph $G$, where for $1\leq i\leq k$, the $i$th batch $\pi_i$ is a collection of at most $\Delta\cdot n$ edges that are deleted from $G$.

	The algorithm maintains a router decomposition $\left(\cset,\set{C'}_{C\in \cset}\right )$ of the graph $G$, for parameters $\Delta^*$, $\tilde d= k^{19}\cdot 2^{O(1/\delta^6)}$, 
	$\tilde \eta= n^{19/k}$, and $\rho=n^{21/k}$.
	After the initialization step, $|\edel|\leq n^{1+4/k^2}\cdot \Delta$ holds, and, for all $1\leq i\leq k$, at most $|\pi_i|\cdot n^{15/k^2}$ edges may be added to $\edel$ while processing the batch $\pi_i$.
	For all $1\leq i\leq k$, after the $i$th batch of updates is received, the clusters $C\in \cset$ may only be updated by removing some of their edges (which are then added to $\edel$) and some of their vertices. The corresponding graphs $C'\subseteq C$ may be updated by removing or inserting edges into them, and by removing vertices from them. Additionally, clusters of $\cset$ that become empty are removed from $\cset$. These are the only updates to the router decomposition.
	The initialization time of the algorithm is $O\left((n\Delta)^{1+O(\delta)} \right )$, and for all $1\leq i\leq k$, the time required to process the $i$th update batch $\pi_i$ is bounded by  $O\left(|\pi_i|\cdot n^{O(1/k^2)}\right )$.  The total number of edges inserted into or deleted from all subgraphs in $\set{C'\mid C\in \cset}$ while batch $\pi_i$ is processed is at most $O\left(|\pi_i|\cdot n^{O(1/k^2)}\right )$.
\end{corollary}

Next, we prove a corollary of \Cref{cor: one level router decomposition}, where we remove the bound on the number of edges in $G$. Other than that, the corollary is very similar to \Cref{cor: one level router decomposition}, and can be thought of as an analogue of our main technical result -- \Cref{thm: router decomposition main}, for the online-batch dynamic model.

\begin{corollary}\label{cor: full batch dynamic}
	There is a deterministic algorithm, that receives as input a simple $n$-vertex graph $G$, where $n$ is greater than a large enough constant, and additional parameters $512\leq k\leq (\log n)^{1/49}$, $\frac{1}k\leq \delta<\frac{1}{400}$  and  $\Delta^*\geq n^{19/k}$, such that $k,\Delta^*$ and $1/\delta$ are integers.
	Lastly, the algorithm receives $k$ online batches $\pi_1,\ldots,\pi_k$ of updates to graph $G$, where for $1\leq i\leq k$, the $i$th batch $\pi_i$ is a collection of edge insertions and deletions for $G$; graph $G$ is guaranteed to remain simple throughout the update sequence.

	The algorithm maintains a router decomposition $\left(\cset,\set{C'}_{C\in \cset}\right )$ of the graph $G$, for parameters $\Delta^*$, $\tilde d= k^{19}\cdot 2^{O(1/\delta^6)}$, 
	$\tilde \eta= n^{19/k}$, and $\rho=n^{O(1/k)}$, such that $|\edel|\leq n^{1+O(1/k)}\cdot \Delta^*$ holds. The algorithm also maintains a graph $H$ with $V(H)=V(G)$ and $E(H)=\edel\cup\left(\bigcup_{C\in \cset}E(C')\right )$, with $|E(H)|\leq n^{1+O(1/k)}\cdot \Delta^*$.
	The initialization time of the algorithm is $O\left(m^{1+O(\delta)} \right )$, where $m$ is the initial number of edges and vertices in $G$. For all $1\leq i\leq k$, the time required to process the $i$th update batch $\pi_i$ is bounded by  $O\left(|\pi_i|\cdot n^{O(\delta)}\right )$. The number of edge insertions and deletions that graph $H$ undergoes while batch $\pi_i$ is processed is bounded by  $O\left(|\pi_i|\cdot n^{O(1/k)}\right )$.
\end{corollary}

\begin{proof}
	Let $z$ be the smallest integer, so that $\Delta^*\cdot n^{z/k}\geq 2n$. Clearly, $z\leq k$ must hold.
	For $1\leq j\leq z+1$, let $\Delta_j=\Delta^*\cdot n^{j/k}$. Notice that all vertex degrees in $G$ are bounded by $\Delta_z$ at all times, and $\Delta_1= \Delta^*\cdot n^{1/k}\geq n^{9/k^2}\cdot \Delta^*$.
	 Throughout the algorithm, we maintain subgraphs $G_1,\ldots,G_z$ of $G$, where $V(G_1)=\cdots=V(G_z)=V(G)$, and every edge of $G$ lies in exactly one of the graphs $G_1,\ldots,G_z$. 	 We will ensure that, for all $1\leq j\leq z$, for all $1\leq i\leq k$, just before the $i$th batch $\pi_i$ of updates arrives, $|E(G_j)|\leq i\cdot \Delta_{j}\cdot n$ holds.
	  For all $3\leq j\leq z$, we also maintain a router decomposition $\left(\cset_j,\set{C'}_{C\in \cset_j}\right )$ of graph $G_j$ with parameters $\Delta^*$, $\tilde d= k^{19}\cdot 2^{O(1/\delta^6)}$, 
	 $\tilde \eta= n^{19/k}$, and $\rho'=n^{21/k}$, such that $\bigcup_{C\in \cset_j}E(C)=E(G_j)$. We refer to the graph $G_j$ and its router decomposition as \emph{level-$j$ data structure} (for levels $1$ and $2$, the correponding data structure only contains the graph $G_j$).

 We will then let $\cset=\bigcup_{j=3}^z\cset_j$, obtaining a router decomposition $\left(\cset,\set{C'}_{C\in \cset}\right )$ of graph $G$ with parameters $\Delta^*$, $\tilde d= k^{19}\cdot 2^{O(1/\delta^6)}$, 
 $\tilde \eta= n^{19/k}$, and $\rho=k\cdot \rho'\leq n^{O(1/k)}$. Notice that the edges of $\edel=E(G)\setminus \left(\bigcup_{C\in \cset}E(C)\right)$ are precisely the edges of $E(G_1)\cup E(G_2)$, and so $|\edel|\leq |E(G_1)|+|E(G_2)|\leq 2nk\Delta_2\leq n^{1+O(1/k)}\cdot \Delta^*$ holds throughout the algorithm.
 
 The main subroutine that we use is the initialization of the data structures from levels $1,\ldots,z$. Consider some level $3\leq j\leq z$, and assume that we are given a graph $G'_j$ that contains at most $kn\Delta_j$ edges. In order to initialize the level-$j$ data structure, we initialize the algorithm from \Cref{cor: one level router decomposition} on graph $G'_j$, with parameters $k,\delta$ and $\Delta^*$ remaining unchanged, and parameter $\Delta$ replaced with 
 $\Delta_{j-2}$. Since $j\geq 3$ and $\Delta_{j-2}=\Delta^*\cdot n^{(j-2)/k}\geq \Delta^*\cdot n^{1/k}$, it is easy to verify that  $\Delta^*\leq \frac{\Delta}{n^{9/k^2}}$, as required, and, from the definition of $z$, $\Delta_{j-2}\leq \Delta_{z-2}<n$.
 Moreover, $|E(G'_j)|\leq kn\Delta_j\leq kn^{1+2/k}\cdot \Delta_{j-2}\leq n^{1+15/k}\cdot \Delta$, as required.
 	The algorithm from  \Cref{cor: one level router decomposition}  then computes a router decomposition $\left(\cset,\set{C'}_{C\in \cset}\right )$ of the graph $G'_j$, for parameters $\Delta^*$, $\tilde d$, $\tilde \eta$, and $\rho'=n^{21/k}$. We then define the graph $G_j$ to contain the same vertex set as $G'_j$, and all edges that lie in the clusters of $\cset$. Let $\tilde E_j$ denote the resulting set $\edel=E(G_j')\setminus E(G_j)$ of edges. Then we are guaranted that:
 	
\[|\tilde E_j|\leq n^{1+4/k^2}\cdot \Delta=n^{1+4/k^2}\cdot \Delta_{j-2}\leq n^{1+4/k^2+(j-2)/k}\cdot \Delta^*\leq n^{1+(j-1)/k}\cdot \Delta^*=n\cdot \Delta_{j-1}.\]

We then construct a graph $G'_{j-1}$, with $V(G'_{j-1})=V(G'_j)$, and $E(G'_{j-1})=\tilde E_j$. If $j=3$, then we set $G_{j-1}=G'_{j-1}$, and terminate the initialization algorithm. Otherwise, we call the initialization algorithm for level $(j-1)$ with the graph $G'_{j-1}$. Notice that the time required to initialize the level-$j$ data structure (excluding the time spent on recursive calls to initialize data structures from lower levels) is bounded by $O\left((kn\Delta_j)^{1+O(\delta)}\right )\leq O\left((n\Delta_j)^{1+O(\delta)}\right )$.
We are now ready to complete the proof of \Cref{cor: full batch dynamic}. We start by describing the initialization algorithm, followed by the algorithm for processing the batches $\pi_i$ of edge deletions.

\paragraph{Initialization.}
Consider the initial graph $G$, and let $m=|E(G)|+|V(G)|$. Let $z'\geq 1$ be the smallest integer, for which $m\leq n\cdot \Delta_{z'}$ holds. It is easy to verify that $z'\leq z$ must hold, and moreover, $n\cdot \Delta_{z'}\leq mn^{1/k}\leq m^{1+1/k}$. 
For all for all $z'<j\leq z$, the graph $G_j$ is initialized to contain $n$ vertices and no edges.
If $z'<3$, then we let $G_{z'}=G$. Otherwise, we apply the initialization algorithm described above to graph $G'_{z'}=G$ in order to initialize level $z'$. The algorithm then recursively initializes levels $z'-1,\ldots,1$. Notice that $|E(G'_{z'})|\leq n\Delta_{z'}$, and our initialization algorithm ensures that, for all $1\leq j\leq z'$, $|E(G'_j)|\leq n\cdot \Delta_j$ holds.
We then let $\cset=\bigcup_{j=3}^z\cset_j$, obtaining a router decomposition $\left(\cset,\set{C'}_{C\in \cset}\right )$ of graph $G$ with parameters $\Delta^*$, $\tilde d$, 
$\tilde \eta$, and $\rho$ as required, where  $\edel=E(G)\setminus \left(\bigcup_{C\in \cset}E(C)\right)=E(G_1)\cup E(G_2)$, and so $|\edel|\leq  n^{1+O(1/k)}\cdot\Delta^*$ holds.
From our discussion, it is easy to verify that the total time required for the initialization algorithm is bounded by:

\[\sum_{j=1}^{z'}O\left((n\Delta_j)^{1+O(\delta)}\right )\leq O\left((n\Delta_{z'})^{1+O(\delta)}\right )\leq O\left(m^{1+O(\delta)}\right )\,\]

since $\delta\geq 1/k$. We also initialize the graph $H$ with $V(H)=V(G)$ and $E(H)=\edel\cup\left(\bigcup_{C\in \cset}E(C')\right )$. It is easy to verify that graph $H$ can be initialized without increasing the asymptotic running time of the algorithm. Additionally, since, for all $3\leq j\leq z'$, $\sum_{C\in \cset_j}|E(C')|\leq n^{1+21/k}\cdot \Delta^*$, while $|\edel|\leq n^{1+O(1/k)}\Delta^*$, we get that $|E(H)|\leq  n^{1+O(1/k)}\cdot \Delta^*$ holds.

\paragraph{Processing update batch $\pi_i.$}
We now provide an algorithm for processing the $i$th batch $\pi_i$ of updates to graph $G$. Assume first that $|\pi_i|\geq n^{2-4/k}$. Let $G\attime[i]$ be the graph obtained from $G$ after applying the first $i$ batches of updates to it. Since $G\attime[i]$ is guaranteed to be a simple graph, $|E(G\attime[i]|)\leq n^2\leq |\pi_i|\cdot n^{4/k}$.  In this case, we simply initialize all data structures from scratch, using the initialization algorithm described above. The time required to process the batch $\pi_i$ is then bounded by:

\[ O\left((|E(G\attime[i])|+n)^{1+O(\delta)}\right )\leq O\left(|\pi_i|\cdot n^{4/k}\right )^{1+O(\delta)}\leq O(|\pi_i|\cdot n^{O(\delta)}),\]

since $\delta\geq 1/k$.
In order to update the graph $H$, we initially delete all edges from it, and then add the edges obtained from initializing the graph $H$ as described above. The total number of edge deletions and insertions that graph $H$ undergoes is bounded by $O(n^2)\leq |\pi_i|\cdot n^{O(1/k)}$.

From now on we assume that $|\pi_i|<n^{2-4/k}$. We let $\hat z$ be the smallest integer, so that $|\pi_i|\leq n\cdot \Delta_{\hat z-2}$. Since $\Delta_z\geq 2n$, while $|\pi_i|<n^{2-4/k}$, it is easy to verify that $n\cdot \Delta_{z-3}>|\pi_i|$, and so $\hat z<z$.  Additionally, $|\pi_i|\geq n\cdot \Delta_{\hat z-3}$ must hold. Let $\hat z'=\max\set{\hat z,4}$. For each integer $\hat z'\leq j\leq z$, we let $\pi_i^j$ be the set of all edges of graph $G_j$ that are deleted from $G$ by the batch $\pi_i$ of edge deletions. From our discussion, $|\pi_i^j|\leq |\pi_i|\leq n\cdot \Delta_{j-2}$ holds. We then provide $\pi_i^j$ as a batch of edge deletions to the algorithm that maintains a router decomposition of graph $G_j$. Recall that, as the result, the clusters $C\in \cset_j$ may be modified, with edges or vertices deleted from $C$. We denote by $E_i^j$ the set of all edges that were deleted from the clusters of $\cset_j$ during the update. Recall that the algorithm from \Cref{cor: one level router decomposition} ensures that $|E_i^j|\leq |\pi_i^j|\cdot n^{15/k^2}$. Additionally, for every cluster $C\in \cset_j$, the corresponding graph $C'$ may be updated by removing or inserting edges into it, and possibly removing vertices from it. The total number of edges inserted into or deleted from all subgraphs in $\set{C'\mid C\in \cset_j}$ is bounded by $O\left (|\pi_i^j|\cdot n^{O(1/k^2)}\right )$. We update the graph $H$ accordingly, so that at most $O\left (|\pi_i^j|\cdot n^{O(1/k^2)}\right )$ edges are deleted from or inserted into $H$. The time that is required in order to process batch $\pi_i^j$ of deletions from graph $G_j$ is at most $O\left (|\pi_i^j|\cdot n^{O(1/k^2)}\right )\leq O\left (|\pi_i^j|\cdot n^{O(\delta)}\right )$. We also delete the edges of $E_i^j$ from graph $G_j$, so that $E(G_j)=\bigcup_{C\in \cset_j}E(C)$ continues to hold.

To summarize, the time required to update data structures for all levels $\hat z'\leq j\leq z$ is bounded by:

\[\sum_{j=\hat z'}^zO\left (|\pi_i^j|\cdot n^{O(\delta)}\right )\leq O\left (|\pi_i|\cdot n^{O(\delta)}\right ).\]

Using similar reasoning, the total number of edge insertions and deletions that graph $H$ undergoes due to updating the data structures from levels $\hat z'\leq j\leq z$ is at most $O\left (|\pi_i|\cdot n^{O(1/k)}\right )$.

Assume now that $\hat z\geq 4$, so that $\hat z=\hat z'$ holds.
We construct a set $E^*$ of edges, that consists of all edges in graphs $G_1,G_2,\ldots,G_{\hat z-1}$ (excluding the edges that are deleted in batch $\pi_i$); all edges that are inserted into $G$ via the batch $\pi_i$; and all edges in sets $E_i^{\hat z},\ldots,E_i^z$ that we have computed (the edges that were deleted from graphs $G_{\hat z},\ldots,G_z$).

Recall that, for all $\hat z\leq j\leq z$, $|E_i^j|\leq |\pi_i^j|\cdot n^{15/k^2}$, and so $\sum_{j=\hat z}^z|E_i^j|\leq |\pi_i|\cdot n^{15/k^2}$. Recall also that $|\pi_i|\leq n\cdot \Delta_{\hat z-2}$. Lastly, from our invariants, for all $1\leq j<\hat z$, $|E(G_j)|\leq i\cdot \Delta_j\cdot n$
holds. Therefore, we get that:

\[|E^*|\leq \sum_{j=1}^{\hat z-1}i\cdot \Delta_j\cdot n+2|\pi_i|\cdot n^{15/k^2}\leq i\cdot \Delta_{\hat z-1}\cdot n+2i\cdot \Delta_{\hat z-2}\cdot n+2n^{1+18/k^2}\cdot \Delta_{\hat z-2}\leq (i+1)\cdot \Delta_{\hat z-1}\cdot n.\]

We let $G'_{\hat z-1}$ be the graph whose vertex set is $V(G)$, and edge set is $E^*$. We then initialize the level-$(\hat z-1)$ data structure from scratch, using the initialization algorithm described above, which also recursively initializes data structures from levels $1,\ldots,\hat z-2$. The running time required to initialize the data structures from levels $1,\ldots,\hat z-1$ is bounded by:

\[O\left((n\cdot \Delta_{\hat z-1})^{1+O(\delta)}\right )\leq O\left(|\pi_i|\cdot n^{O(\delta+1/k)}\right )\leq O\left(|\pi_i|\cdot n^{O(\delta)}\right ),\] 

since $|\pi_i|\geq n\cdot \Delta_{\hat z-3}\geq n^{1-2/k}\cdot \Delta_{\hat z-1}$.

Lastly, we delete from $H$ all edges that originally lied in graphs $C'$ for $C\in \bigcup_{j=3}^{\hat z-1}\cset_j$ and in the original graphs $G_1$ and $G_2$, and insert edges that lie in the new graphs $C'$ for  $C\in \bigcup_{j=3}^{\hat z-1}\cset_j$, and in the new graphs $G_1$ and $G_2$. Recall that, for all $3\leq j\leq z$, $\sum_{C\in \cset_j}|E(C')|\leq n^{1+O(1/k)}\cdot \Delta^*$ holds. Since we have assumed that $\hat z\geq 4$, we get that $|\pi_i|\geq n^{1-1/k}\cdot \Delta^*$, and so $\sum_{C\in \cset_j}|E(C')|\leq n^{1+O(1/k)}\cdot \Delta^*\leq |\pi_i|\cdot n^{O(1/k)}$ holds. Therefore, the total number of updates the graph $H$ undergoes while processing batch $\pi_i$ is bounded by $|\pi_i|\cdot n^{O(1/k)}$.

Finally, if $\hat z<4$, then $|\pi_i|\leq n\cdot \Delta_1$. In this case, let $\hat E$ be the collection of edges that includes all edges inserted into $G$ via the batch $\pi_i$ of updates, and the edges of $\bigcup_{j=\hat z'}^zE_i^j$.
Since $\sum_{j=\hat z'}^z|E_i^j|\leq |\pi_i|\cdot n^{15/k^2}$, we get that $|\hat E|\leq 2|\pi_i|\cdot n^{15/k^2}\leq n\cdot \Delta_2$.
 We simply insert the edges of $\hat E$ into $G_2$ and into $H$, and we delete from them all edges that were deleted by $\pi_i$. This completes the algorithm for processing the batch $\pi_i$ of updates. The total time required to process the batch $\pi_i$ is bounded $O\left (|\pi_i|\cdot n^{O(\delta)}\right )$, and the number of edge insertions and deletions that $H$ undergoes is bounded by $O\left (|\pi_i|\cdot n^{O(1/k)}\right )$.
\end{proof}

We are now ready to complete the proof of \Cref{thm: router decomposition main}. Let $G$ be an initial graph with $n$ vertices and no edges,  that undergoes an online sequence of at most $n^2$ edge deletions and insertions, such that $G$ remains a simple graph throughout the update sequence. We can assume w.l.o.g. that $n$ is greater than a sufficiently large constant, since, if it is not the case, we can simply increase it, and add some dummy vertices to $G$. We also assume that we are given parameters $512\leq k\leq (\log n)^{1/49}$, $\frac 1 k\leq \delta<\frac{1}{400}$  and $\Delta^*\geq n^{19/k}$, such that $k,\Delta^*$ and $1/\delta$ are integers (if $n$ was initially too small, then $\Delta^*$ may need to be increased to accommodate the increase in $n$). 

Notice that the algorithm from \Cref{cor: full batch dynamic} allows us to maintain the collection $\cset$ of edge-disjoint clusters of $G$, the subgraphs $C'\subseteq C$ for all clusters $C\in \cset$, and the graph $H$ with all required properties, in the online-batch dynamic model, with batch number $k$ and sensitivity parameter $w=n^2$. The initialization time of the algorithm is bounded by $O(n^{2+O(\delta)})$, and, for all $1\leq i\leq k$, the time required to process the $i$th batch $\pi_i$ of updates is bounded by $O(|\pi_i|)\cdot n^{O(\delta)}$. We denote the data structure maintained by the algorithm from \Cref{cor: full batch dynamic} by $\DS$.

From \Cref{lem: deamortization}, we then obtain an algorithm that maintains the collection $\cset$ of clusters, and, for every cluster $C\in \cset$, a subgraph $C'\subseteq C$ with all required properties, whose worst-case update time is $O\left(2^{O(k)}\cdot \left(n^{O(\delta+1/k)}\right )\right )\leq O\left(n^{O(\delta)}\right )$ (since $2^{O(k)}\leq n^{O(1/k)}$ as $k \leq (\log n)^{1/49}$, and $\delta\geq 1/k$). 

The only subtlety is in maintaining the graph $H$, so that, after each update to graph $G$, graph $H$ undergoes at most $n^{O(1/k)}$ updates. The reason is that the algorithm from \Cref{lem: deamortization} can be thought of as maintaining, at all times, $3^k$ different instances of data structure $\DS$, each of which is associated with a different graph $H$. At any given time, the algorithm indicates which of the data structures is up to date. However, following a single update to $G$, the algorithm may decide to switch the designated current data structure, which may lead to a large number of changes in the graph $H$. In order to overcome this difficulty, we will simply include in our final sparsifier $H^*$ the edges of all graphs $H$ that are maintained by all currently maintained copies of data structure $\DS$.

We now provide more details of the algorithm from  \Cref{lem: deamortization}, and the algorithm for maintaining the final graph $H^*$.
It is convenient to think of the algorithm from  \Cref{lem: deamortization}  as consisting of $k+1$ levels, which are associated with a hierarchical partition of the time line into intervals. Let $T$ be the time horizon of the algorithm. We first partition $T$ into consecutive intervals of length $w$ each, and denote the resulting collection of intervals by $\iset_0$; we refer to these intervals as \emph{level-0 intervals}. We can think of the level-$0$ algorithm as having $3^{k+1}$ copies of data structure $\DS$ from \Cref{cor: full batch dynamic}, that are partitioned into 3 groups $\gset_1,\gset_2,\gset_3$. Each group $\gset_r$ contains $3^{k}$ identical copies of data structure $\DS$. For each interval $I\in \iset_0$, every data structure $\DS$ is in one of the following three states: (i) preparation: given some initial graph $G'$ at the beginning of interval $I$, the data structure needs to initialize the algorithm from \Cref{cor: full batch dynamic} for graph $G'$, during the first half of interval $I$; or (ii) active: the data structure will be used, during time interval $I$, by data structures from levels $1,\ldots,k$; or (iii) rollback: we can think of this part as ``undoing'' all computation that has been done during the preparation step. If data structure $\DS$ is active during interval $I$, then it is in preparation state during the preceding interval, and in the rollback state during the following interval. Moreover, all data structures in a single group $\gset_r$ are always in the same state, and exactly one group is active during each interval $I\in \iset_0$. Recall that the time to initialize the data structure $\DS$ is bounded by $n^{2+O(\delta)}$, and during this time the initial graph $H_{\DS}$, that contains all edges of $\edel$ and $\bigcup_{C\in \cset}E(C')$ is constructed. We can think of the construction of the graph $H_{\DS}$ as being stretched over the course of the first half of a single level-$0$ interval when the data structure is in the preparation state, where we start with $H_{\DS}$ containing no edges, and then insert at most $n^{O(1/k)}$ edges into $H_{\DS}$ per time unit. Then during the rollback state, we will delete the edges from $H_{\DS}$, at the rate of $n^{O(1/k)}$ edge deletions per time unit, until no edges remain in $H_{\DS}$. We define the level-$0$ graph $H^0$ to contain the edges of all graphs $H_{\DS}$, for all level-$0$ data structures $\DS$, excluding the edges that were already deleted from $G$. Note that graph $H^0$ undergoes at most $n^{O(1/k)}\cdot 2^{O(k)}\leq n^{O(1/k)}$ updates per time unit.

Consider now some integer $1\leq j\leq k$, and some level-$(j-1)$ interval $I$. This interval is partitioned into $\ceil{n^{2/k}}$ level-$j$ intervals of identical length; we denote the length of each such interval by $L_j$. Moreover, there are $3^{k+1-j}$ identical copies of data structure $\DS$ that are active during time interval $I$. Each of these copies is obtained by applying the algorithm from  \Cref{cor: full batch dynamic} to some initial graph $G$, which then undergoes $j-1$ batches of updates. These copies are in turn partitioned into 3 groups $\gset^j_1,\gset^j_2,\gset^j_3$, each of which contains $3^{k-j}$ identical copes of the data structure. For each level-$j$ interval $I'\subseteq I$, each such data structure is in one of the following three states: (i) preparation: the data structure is processing the $j$th batch of updates, whose size is at most $4L_j\cdot n^{2/k}$, and the processing needs to complete by the middle of interval $I'$; (ii) active: the data structure is ready to be used by levels $j+1,\ldots k$; and (iii) rollback: we undo the changes that the data structure underwent during the preparation state. As before, if data structure $\DS$ is active during a level-$j$ interval $I$, then it is in preparation state in the preceding interval, and in rollback state during the subsequent interval. Consider now any copy $\DS$ of the data structure, and suppose it is in preparation state during the level-$j$ interval $I'$. Recall that the length of the interval is denoted by $L_j$, and the length of the update batch to the data structure is at most $4L_j\cdot n^{2/k}$. During this update sequence, the graph $H_{\DS}$ associated with the data structure undergoes at most $ L_j\cdot n^{O(1/k)}$ edge insertions and deletions. We think of these updates as being spread over the course of interval $I$, so graph $H_{\DS}$ undergoes at most $n^{O(1/k)}$ updates per time unit. During the following rollback stage, we undo the changes that graph $H_{\DS}$ underwent during the preparation stage, with the graph undergoing at most $n^{O(1/k)}$ updates per time unit. We now define the level-$j$ graph $H^j$ to be the union of all graphs $H_{\DS}$ for all data structures $\DS\in \gset^j_1\cup \dset^j_2\cup\dset^j_3$, excluding edges that have been deleted from $G$. As before, graph $H^j$ undergoes at most $n^{O(1/k)}\cdot 2^{O(k)}\leq n^{O(1/k)}$ updates per time unit.

We let $j^*\leq k-1$ be smallest integer, such that level-$j^*$ time intervals have length at most $n^{2/k}$ each. In this case, we think of level-$(j^*+1)$ time intervals as consisting of a single time slot each. For each level-$(j^*+1)$ time interval $I=\set{\tau}$, there is at least one data structure $\DS$, that is up to date with respect to the graph $G$ at time $\tau$. This is the data structure that the algorithm from \Cref{lem: deamortization} considers as current. The corresponding graph $H_{\DS}$ must then be contained in the graph $H^*=H^0\cup \cdots\cup H^{j^*+1}$ that our algorithm maintains. From our discussion, graph $H^*$ undergoes at most $n^{O(1/k)}$ edge insertions or deletions per time unit. Since the algorithm from \Cref{cor: full batch dynamic} ensures that each graph $H$ maintained by a copy of the data structure contains at most $n^{1+O(1/k)}\cdot \Delta^*$ edges, and since, at all times, our graph $H^*$ is contained in the union of at most $2^{O(k)}$ such graphs, we get that $|E(H^*)|\leq 2^{O(k)}\cdot n^{1+O(1/k)}\cdot \Delta^*\leq n^{1+O(1/k)}\cdot \Delta^*$ holds. 
This completes the proof of \Cref{thm: router decomposition main}.


\section{First Tool: a Nice Router and its Pruning}
\label{sec: nice router}

In this section we provide an explicit construction of  a router graph, and provide an algorithm for pruning this graph. We show that the graph remains a router even after pruning. As mentioned in the introduction, our construction is very similar to that of \cite{HLT2024} (see Section 5 in the full version). Our pruning algorithm is more involved, and ensures somewhat better routing parameters, in terms of the lengths of the routing paths, in the resulting pruned graph.

\subsection{Construction of a Nice Router Graph}
We assume that we are given three integral parameters:  $k\geq 256$, $N\geq k^{3k}$, and $\Delta\geq 4k^2$. We sometimes refer to $\Delta$ the \emph{target degree}. We now describe a construction of a corresponding router, that we denote by $W_k^{N,\Delta}$. To avoid clutter, in the remainder of this section, we assume that $N$ and $\Delta$ are fixed, and we do not include them in superscripts, so in this section we denote graph $W_k^{N,\Delta}$ by $W_k$.
The construction of graph $W_k$ is recursive. We start by constructing  a graph $W_1$, and then, for $1< i\leq k$, we construct graph $W_i$ by combining $N$ copies of graph $W_{i-1}$. Parameters $N$ and $\Delta$ remain unchanged for these graphs.

\paragraph{Graph $W_i$ -- a High Level View.}
Consider an integer $1\leq i\leq k$. We now provide a high-level description of graph $W_i=(V_i,E_i)$. The set $V_i$ contains exactly $N^i$ vertices, and it is partitioned into two subsets: set $V_i^c$ of $N^{i-1}$ vertices called \emph{center} vertices, and set $V_i^{\ell}=V_i\setminus V_i^c$ of remaining vertices, called \emph{leaf} vertices. The set $E(W_i)$ of edges of the graph is partitioned into $i$ subsets $E^i_1,\ldots,E^i_i$, where for $1\leq z\leq i$, we refer to the edges of $E^i_z$ as \emph{level-$z$ edges}. If we let $H_z=G_i[E^i_z]$ be the subgraph of $G_i$ induced by the set $E^i_z$ of level-$z$ edges, and denote by $\rset_z$ the collection of the connected components of $H_z$, then each such connected component $C\in \rset_z$ contains exactly $N$ vertices, out of which exactly one is a center vertex. Additionally, if we let $S$ be a star defined over the vertices of $C$, whose center is the unique  vertex of $V(C)\cap V_i^c$, then $C$ can be obtained from $S$ by replacing every edge of $S$ with $\Delta$ parallel copies (we refer to these copies as a \emph{bundle}; the edges of $S$ are referred to as \emph{superedges}). We denote by $\sset_z$ a collection of star graphs that contains, for every connected component $C\in \rset_z$, the corresponding star graph $S$ on $V(C)$ (without the parallel edges). Note that every edge in $W_i$ connects a leaf vertex to a center vertex.
We now describe the construction of the graph $W_i$ in more detail. The construction is recursive, from smaller to larger values of $i$.

\paragraph{Graph $W_1$.}
The set of vertices of the graph $W_1$ is $V_1=\set{v_1,\ldots,v_N}$, where vertex $v_1$ is the \emph{center} vertex, and vertices $v_2,\ldots,v_N$ are \emph{leaf vertices}.
In other words, $V_1^c=\set{v_1}$, and $V_1^{\ell}=\set{v_2,\ldots,v_N}$.
 Let $S$ be a star defined on the vertices of $V_1$, whose center is vertex $v_1$, and leaves are $v_2,\ldots,v_N$. Graph $W_1$ is obtained from $S$ by replacing every edge of $S$ with a collection of $\Delta$ parallel edges. All edges of $W_1$ belong to level $1$, so $E(W_1)=E_1^1$. The collection $\sset_1$ of level-$1$ stars contains a single star $S$.
 
\paragraph{Graph $W_i$ for $i>1$.}
We assume that we are given an integer $i>1$, and that we have already defined the graph $W_{i-1}$. We now describe the construction of graph $W_i$.

We start by creating $N$ copies of graph $W_{i-1}$, that we denote by $C_1,\ldots,C_N$. We refer to $C_1,\ldots,C_N$ as \emph{level-$(i-1)$ clusters}. We let the set $V_i$ of vertices of $W_i$ be $\bigcup_{j=1}^NV(C_j)$. For all $1\leq j\leq N$, if $v\in V(C_j)$ is a center vertex for $C_j$, then it becomes a center vertex for $W_i$, and otherwise it becomes a leaf vertex for $W_i$.
Recall that, for all $1\leq j\leq N$, $|V(C_j)|=N^{i-1}$, and the number of center vertices in $C_j$ is $N^{i-2}$. Therefore, we get that $|V_i|=N^i$ and the number of center vertices in $W_i$ is $N^{i-1}$, as required.

For all $1\leq i'<i$, the set $E^i_{i'}$ of level-$i'$ edges of $W_i$ is the union of the sets of level-$i'$ edges of $C_1,\ldots,C_N$. It now remains to define the set $E_i^i$ of level-$i$ edges for graph $W_i$.
In order to do so, we will construct a collection $\sset_i$ of $N^{i-1}$ disjoint stars, where each star $S\in \sset_i$ contains exactly one vertex from each of the clusters $C_1,\ldots,C_N$, and the center of the star lies in $V_i^c$. The set $E_i^i$ of level-$i$ edges is then  obtained by replacing each edge $e\in \bigcup_{S\in \sset_i}E(S)$ with  $\Delta$ parallel edges.

In order to construct the collection $\sset_i$ of stars, we first define, for each $1\leq j\leq N$, an ordering $v^j_1,\ldots,v^j_{N^{i-1}}$ of the vertices of $C_j$, such that, for all $1\leq z\leq N^{i-1}$, if we consider the vertices $ v^1_z,\ldots,v^N_z$ (that is, the vertices that appear in the $z$th position in the orderings of $V(C_1),\ldots,V(C_N)$), then exactly one of these vertices is a center vertex in $V^c_i$.
 
 Such orderings are easy to define: consider the collection $I=\set{1,\ldots,N^{i-1}}$ of indices, and partition it into $N$ disjoint subsets $I_1,\ldots,I_N$, each of which contains exactly $N^{i-2}$ contiguous indices.
  For each $1\leq j\leq N$, we order the vertices of $C_j$ arbitrarily, but we ensure that the center vertices of $C_j$ lie in the positions whose indices belong to $I_j$ in this ordering. This ensures that, for all $1\leq z\leq N^{i-1}$, exactly one vertex that lies in the $z$th position in the orderings of $C_1,\ldots,C_N$ is a center vertex.

 For all $1\leq z\leq N^{i-1}$, we let $S_z$ be the star that consists of the vertices $\set{v^1_z,\ldots,v^N_z}$, where the center of the star is the unique vertex $v^j_z$ that lies in $V_i^c$. Then for every edge $(x,y)$ of the star $S_z$, we create $\Delta$ parallel edges $(x,y)$, which are added to set $E^i_i$. These parallel edges are called a \emph{bundle}, and the edge of the star corresponding to them is called a \emph{superedge}.
 We denote by $\sset_i=\set{S_1,\ldots,S_{N^{i-1}}}$ the resulting collection of level-$i$ stars. This completes the construction of the set $E^i_i$ of level-$i$ edges of graph $W_i$. Lastly, we let $E(W_i)=\bigcup_{i'=1}^NE^i_{i'}$, completing the construction of the graph $W_i$. For all $1\leq i'<i$, the collection $\sset_{i'}$ of level-$i'$ stars is the union of the collections of level-$i'$ stars of the graphs $C_1,\ldots,C_N$.

Consider now the level-$k$ graph $W_k$. For all $1\leq i<k$, graph $W_k$ contains $N^{k-i}$ copies of the level-$i$ graph $W_i$. We let $\cset_i$ denote the partition of the vertices of $W_k$ defined by these copies of $W_i$. We refer to $\cset_i$ as \emph{level-$i$ clustering}, and we refer to each subgraph $C\in \cset_i$ as a \emph{level-$i$ cluster}. Notice that for all $1\leq i<k$, each level-$i$ cluster is contained in a single level-$(i+1)$ cluster, and moreover, each level-$(i+1)$ cluster contains the union of exactly $N$ level-$i$ clusters. There is a single level-$k$ cluster, containing all vertices of $W_k$. From our construction, it is easy to verify that, for all $1\leq i\leq k$, if we denote by $H_i=W_k[E^k_i]$ the subgraph of $W_k$ induced by the level-$i$ edges, then there is a collection $\sset_i$ of $N^{k-1}$ disjoint star graphs, where each star $S\in \sset_i$ has $N$ vertices, and the center of $S$ lies in $V_k^c$, such that the set $E^k_i$ of edges can be obtained from $E'_i=\bigcup_{S\in \sset_i}E(S)$, by replacing each edge $e\in E'_i$ with $\Delta$ parallel edges. We refer to the set $\sset_i$ of stars as \emph{level-$i$ stars}. Lastly, note that for all $1\leq i\leq k$, the number of level-$i$ superedges in $W_k$ is at most $N^k$, and $|E^k_i|\leq \Delta\cdot N^k$. Overall, $W_k$ contains at most $k\cdot N^k$ superedges, and $|E(W_k)|\leq k\cdot \Delta\cdot N^k$. Moreover, the degree of every center vertex in $W_k$ is exactly $(N-1)\cdot \Delta \cdot k$, and the degree of every leaf vertex is exactly $\Delta\cdot k$.

As mentioned already, the construction above is almost identical to that of \cite{HLT2024}. The main difference is that they used cliques instead of stars, and their construction used the parameter $\Delta=1$.
We do not directly prove that the graph $W_k$ is a router. Instead, we first define \emph{properly pruned} subgraphs of $W_k$, and we prove that each such graph is a router. This definition is somewhat more involved than that in \cite{HLT2024}, but it achieves better routing parameters in the resulting graph.

\subsection{Properly Pruned Subgraphs of $W_k$}
\label{subsec: properly pruned}

For all $1\leq i\leq k$, we now define the notion of a \emph{properly pruned subgraph} of $W_i$. We then show that any $\Delta$-restricted demand can be routed with low congestion on paths of length $O(k^2)$ in a properly pruned subgraph of $W_k$.

\begin{definition}\label{def: properly pruned}
	For $1\leq i\leq k$, a subgraph $W\subseteq W_i$ is a \emph{properly pruned} subgraph of $W_i$ with respect to  sets $U_1,U_2,\ldots,U_i\subseteq V(W_i)$ of its vertices, where $U_i\subseteq U_{i-1}\subseteq\cdots\subseteq U_1$, if the following hold:
	
	\begin{properties}{P}
		\item for every level $1\leq i'\leq i$, for every leaf vertex $v\in V^{\ell}_{i}$:\label{prop: leaf vertices}
		\begin{itemize}
			\item if $v\in U_{i'}$, then at least $\ceil{\Delta/2}$ level-$i'$ edges that are incident to $v$ in $W_i$ lie in $W$ (recall that all these edges must be parallel edges that correspond to the unique level-$i'$ superedge that is incident to $v$);
			\item otherwise, $W$ contains no level-$i'$ edges that are incident to $v$; 
		\end{itemize}
	\item for each level $1\leq i'\leq i$, for every center vertex $v\in V^{c}_{i}$, if we denote by $S\in \sset_{i'}$ the unique level-$i'$ star for which $v$ serves as a center, then:\label{prop: center vertices}
	
	\begin{itemize}
		\item if $v\in U_{i'}$, then at least $N\cdot \left(1-\frac{1}{k}\right )$ leaf vertices of $S$ lie in $U_{i'}$ (and we say that star $S$ \emph{survives} in this case); 
		\item otherwise, no leaf vertices of $S$ lie in $U_{i'}$ (and we say that star $S$ is \emph{destroyed}); 
	\end{itemize}

\item $V(W)=U_1$; \label{prop: U1 is all vertices} and
\item for every level $1\leq i'<i$, for every level-$i'$ cluster $C\in \cset_{i'}$ of $W_i$, either $V(C)\cap U_1=\emptyset$ (in which case we say that cluster $C$ is \emph{destroyed}), or at least $\frac{|V(C)|}{k^{3k}}$ vertices of $C$ lie in $U_{i'+1}$ (and we say that cluster $C$ \emph{survives}). \label{prop: cluster survival}
\end{properties}
\end{definition}

The following observation follows immediately from the definition of a properly pruned subgraph.

\begin{observation}\label{obs: hereditary pruning}
	Let $W$ be a subgraph of $W_i$, for some $1\leq i\leq k$, and let $U_1,\ldots,U_i$ be subsets of vertices of $W$ with $U_i\subseteq U_{i-1}\subseteq\cdots\subseteq U_1$.
	Assume that $W$ is a properly pruned subgraph of $W_i$ with respect to $U_1,\ldots,U_i$.
	 Consider a level $1\leq i'<i$, a level-$i'$ cluster $C$ of $W_i$ (which is a copy of $W_{i'}$), and let $W^C$ be the subgraph of $C$ induced by the set $V(W)\cap V(C)$ of vertices. 
	For $1\leq i''\leq i'$, let $U^C_{i''}=U_{i''}\cap V(C)$. 
	Then  $W^C$ is a properly pruned subgraph of $W_{i'}$, with respect to the sets $U_1^C,\ldots,U^{C}_{i'}$ of its vertices.
	\end{observation}

In the following subsection, we prove that a properly pruned subgraph of $W_k$ is a good router. After that we show an algorithm that maintains  a properly pruned subgraph of $W_k$, as it undergoes adversarial edge deletions.

\subsection{Routing in Properly Pruned Subgraphs of $W_k$}
\label{subsec: routing in properly pruned}

The main goal of this subsection is to prove the following theorem, that shows that a properly pruned subgraph of $W_k$ is a good router.

\begin{theorem}\label{thm: pruning to routing}
	Let $W$ be a properly pruned subgraph of $W_k$, with respect to sets $U_1,\ldots,U_k$ of its vertices, and let $\dset=\left (\Pi,\set{D(a,b)}_{(a,b)\in \Pi}\right )$ be a  $\frac{\Delta}{k^{4k}}$-restricted demand  for $W$. Then there is a fractional routing of $\dset$ in $W$ with no edge-congestion, such that every flow-path $P$ with $f(P)>0$ has length at most $20k^2$.	Moreover, there is an algorithm, that, given the graph $W$, the sets $U_1,\ldots,U_k$ of its vertices, and the demand $\dset$, computes such a routing in time  $O\left (k^2\cdot\Delta\cdot N\cdot (N^k+|\Pi|)\right )$.
\end{theorem}

The remainder of this subsection is dedicated to the proof of \Cref{thm: pruning to routing}. 
For all $1\leq i\leq k$, let ${r_i=\frac{\Delta}{32^i}}$. 
We start by proving a slightly weaker result: namely, that  for all $1\leq i\leq k$, if $W$ is a properly pruned subgraph of $W_i$ with respect to sets $U_1,\ldots,U_i$ of its vertices, and $\dset$ is any $r_i$-restricted demand that is defined over the vertices of $U_i$, then $\dset$ can be fractionally routed in $W_i$ over short paths with no edge-congestion. 

\begin{lemma}\label{lem: pruned to routing}
	Fix an integer $1\leq i\leq k$, let $W$ be a subgraph of $W_i$, and let $U_1,\ldots,U_i$ be subsets of vertices of $W$ with $U_i\subseteq U_{i-1}\subseteq\cdots\subseteq U_1$. Assume further that $W$ is a properly pruned subgraph of $W_i$ with respect to $U_1,\ldots,U_i$. Then for any  $r_i$-restricted demand $\dset=\left (\Pi,\set{D(a,b)}_{(a,b)\in \Pi}\right )$ with $\Pi\subseteq U_i\times U_i$, there is a routing of $\dset$ in $W$ with no edge-congestion over flow-paths of length at most $4i$. If the demand $\dset$ is integral, then so is the routing.	Moreover, there is an algorithm, that, given the graph $W$ and the sets $U_1,\ldots,U_i$ of its vertices, computes such a routing in time $O\left (i\cdot\Delta\cdot N\cdot (N^i+|\Pi|)\right )$.
\end{lemma}

Notice that the main difference between \Cref{lem: pruned to routing} and \Cref{thm: pruning to routing} is that in \Cref{lem: pruned to routing} all demands must be between the vertices of $U_i$, while in \Cref{thm: pruning to routing} the demands are between the vertices of $V(W)=U_1$. 

\begin{proof}
	The proof is by induction on $i$. For the induction base, we consider a subgraph $W$ of $W_1$. Recall that $W_1$ is a graph on $N$ vertices, that can be defined as follows: let $S$ be a star graph on $N$ vertices. Then $W_1$ is obtained from $S$ by replacing every edge of $S$ by a set of $\Delta$ parallel edges. From Properties \ref{prop: leaf vertices}--\ref{prop: U1 is all vertices}, $U_1=V(W)$ must hold. Additionally, if $W\neq \emptyset$, then there is a star graph $S'\subseteq S$, that contains the center vertex of $S$, such that $W$ can be obtained from $S'$ by replacing every edge of $S'$ with parallel edges, whose number is between $\ceil{\Delta/2}$ and $\Delta$.
	Consider now any $r_1$-restricted demand $\dset=\left (\Pi,\set{D(a,b)}_{(a,b)\in \Pi}\right )$, with $\Pi\subseteq U_1\times U_1$,	and recall that $r_1=\Delta/32$. It is immediately to see that demand $\dset$ can be routed in $W$, via paths of length at most $2$. Each such path simply connects a pair of leaves in the star, or it connects the center of the star to one of its leaves. It is also immediate to design an algorithm that computes such a routing in time $O(N^2\cdot \Delta)$. If the demand is integral, then so is the routing.

Consider now an integer $i>1$, and assume that the claim holds for $i-1$. Let $W$ be a subgraph of $W_i$, and let
$U_1,\ldots,U_i$ be subsets of its vertices with $U_i\subseteq U_{i-1}\subseteq\cdots\subseteq U_1$, such that $W$ is a properly pruned subgraph of $W_i$ with respect to $U_1,\ldots,U_i$.
 Recall that $W_i$ contains a disjoint union of $N$ copies of $W_{i-1}$, that we denote by $\cset=\set{C_1,\ldots,C_N}$, and refer to as \emph{clusters}. For each such cluster $C\in \cset$, let $W^C=W\cap C$, and, for all $1\leq i'\leq i$, let $U^C_{i'}=U_{i'}\cap V(C)$.

Let $\cset'\subseteq \cset$ be the set of clusters $C\in \cset$, whose corresponding graph $W^C$ is non-empty.
For every cluster $C\in \cset'$,
 we denote the sets of its leaf and center vertices  by $V^{\ell}(C)$ and $V^c(C)$, respectively.
From \Cref{obs: hereditary pruning}, graph 
 $W^C$ is a properly pruned subgraph of $W_{i-1}$, with respect to the sets $U_1^C,\ldots,U_{i-1}^C$ of its vertices. 

Consider now an $r_i$-restricted demand $\dset=\left (\Pi,\set{D(a,b)}_{(a,b)\in \Pi}\right )$ for $W$, where $\Pi\subseteq U_i\times U_i$. 
For every pair $(a,b)\in \Pi$, we will define a \emph{proxy pair} $(a',b')$ of vertices, and we will set a \emph{proxy demand} $D'(a',b')=D(a,b)$. We will ensure that $a$ and $a'$ lie in the same level-$i$ star, and that the same is true for $b$ and $b'$. We will also ensure there is some cluster $C\in \cset'$ with $a',b'\in U^C_{i-1}$. We say that $a'$ is a \emph{proxy vertex for $a$ with respect to pair $(a,b)$}, and we similarly say that $b'$ is a \emph{proxy vertex for $b$ with respect to pair $(a,b)$}. Additionally, we will define a flow $F_{(a,b)}$ in $W$, in which $a$ sends $D(a,b)$ flow units to $a'$, and $b$ sends $D(a,b)$ flow units to $b'$. The flow $F_{(a,b)}$ will only use level-$i$ edges. 

For every cluster $C\in \cset'$, we will therefore obtain a new demand $\dset_C=\set{\Pi_C, \set{D'(a',b')}_{(a',b')\in \Pi_C}}$, where $\Pi_C\subseteq U_{i-1}^C\times U_{i-1}^C$. The demand $\dset_C$ is obtained by combining all proxy demands $D'(a',b')$ where $a',b'\in V(C)$. We will ensure that this demand is $r_{i-1}$-restricted. We can then use the induction hypothesis to route, for each cluster $C\in \cset'$, the corresponding demand $\dset_C$ in $W^C$, obtaining a flow $f_C$. The final routing of $\dset$ is obtained by combining the flows $f_C$ for all clusters $C\in \cset'$ with the partial routings $F_{(a,b)}$ for $(a,b)\in \Pi$ that we have computed.

We now describe the algorithm more formally. Throughout the algorithm, for every vertex $v\in V(W)$, the \emph{load} on vertex $v$, denoted by $\lambda(v)$, is the total amount of the proxy demand $\sum_{u}D'(u,v)$ in which $v$ currently participates. Initially, the proxy demands are undefined, and so the load on every vertex is $0$.

We process the pairs $(a,b)\in \Pi$ one by one. Consider an iteration when the pair $(a,b)$ is processed. 
Let $S,S'\in \sset_i$ be the level-$i$ stars containing $a$ and $b$, respectively. Assume first that $S=S'$. In this case, we let $F_{(a,b)}$ be the flow that sends $D(a,b)$ flow units from $a$ to $b$ by employing the edges of $E_i\cap E(W)$ (the level-$i$ edges of $W_i$ that lie in $W$), that correspond to the edges of $S$ that lie on the unique $a$-$b$ path in $S$, completing the routing of $D(a,b)$ flow units from $a$ to $b$, via paths of length at most $2$.

Assume now that $S\neq S'$. Recall that every star in $\sset_i$ contains exactly one vertex from each of the graphs $C_1,\ldots,C_N\in \cset$. Moreover, from Property \ref{prop: center vertices}, at least $N\cdot\left(1-\frac{1}{k}\right )$ leaves of $S$, and at least $N\cdot\left(1-\frac{1}{k}\right )$ leaves of $S'$ lie in $U_i$. Therefore, there is a collection $\cset_{(a,b)}\subseteq \cset'$ of clusters, with $|\cset_{(a,b)}|\geq N\cdot\left(1-\frac{2}{k}\right )$, so that, for every cluster $C\in \cset_{(a,b)}$, vertex set $V(S)\cap U_i$ contains a leaf vertex of $C$, and so does $V(S')\cap U_i$. 
For a cluster $C\in \cset_{(a,b)}$, let $a^C$ be the unique vertex of $C$ that lies in $V(S)\cap U_i$, and let $b^C$ be the unique vertex of $C$ that lies in $V(S')\cap U_i$. We say that cluster $C\in \cset_{(a,b)}$ is \emph{admissible} for pair $(a,b)$ if $\lambda(a^C),\lambda(b^C)\le r_{i-1}-D(a,b)$. We will later show that at least one cluster $C\in \cset_{(a,b)}$ must be admissible for $(a,b)$. Let $C$ be any such cluster. Then we define the proxy pair  for $(a,b)$ as $(a',b')=(a^C,b^C)$, and we set $D'(a',b')=D(a,b)$. We also increase $\lambda(a'),\lambda(b')$ by $D(a,b)$. Finally, we construct the flow $F_{(a,b)}$, in which $a$ routes $D(a,b)$ flow units to $a'$ and $b$ routes $D(a,b)$ flow units to $b'$ in a straightforward manner: the flow from $a$ to $a'$ simply follows the unique $a$-$a'$ path in star $S$, and the flow from $b$ to $b'$ follows the unique $b$-$b'$ path in star $S'$. Therefore, all flow-paths in $F_{(a,b)}$ have length at most $2$. This completes the desription of the algorithm for constructing proxy demands. Let $F$ be the flow obtained by taking the union of all flows $F_{a,b}$ for all $(a,b)\in \Pi$. Then all flow-paths in $F$ have length at most $2$, and they only use level-$i$ edges. Since the demand $\dset$ is $r_i$-restricted, and $r_i\leq \Delta/4$, it is immediate to verify that flow $F$ causes no edge-congestion. It is also immediate to verify that every pair $(a,b)\in \Pi$ can be processed in time $O(N+\Delta)$.
Next, we show that, when a pair $(a,b)\in \Pi$ is processed, an admissible cluster $C\in \cset_{(a,b)}$ must exist.

\begin{observation}\label{obs: admissible cluster}
	For every pair $(a,b)\in \Pi$, when the pair is processed, an admissible cluster $C\in \cset_{(a,b)}$ must exist.
\end{observation}
\begin{proof}
 Assume otherwise. Let $A$ be the set of all leaf vertices $v\in V(S)\cap U_i$, such that $v$ lies in some cluster of $\cset_{(a,b)}$, and let $B$ be the set of all leaf vertices $u\in V(S')\cap U_{i'}$, such that $u$ lies in some cluster of $\cset_{(a,b)}$. Recall that $|A|=|B|=|\cset_{(a,b)}|\geq N\cdot \left(1-\frac{2}{k}\right )\geq \frac{N}{2}$. Let $A'\subseteq A$ be the set of vertices $v$ whose current load $\lambda(v)>r_{i-1}-D(a,b)$, and let $B'\subseteq B$ be defined similarly. Since no cluster of $\cset_{(a,b)}$ is admissible, it must be the case that either $|A'|\geq \frac{|A|}{2}-1> \frac{|A|}{4}$, or $|B'|> \frac{|B|}{4}$. Assume w.l.o.g. that it is the former. Then the total load on the vertices of $A'$ is:

\[\sum_{v\in A'}\lambda(v)> \frac{|A|}{4}\cdot \left(r_{i-1}-D(a,b)\right )\geq \frac{|\cset_{(a,b)}|}{4}\cdot \left(r_{i-1}-r_i\right )\geq \frac{N\cdot r_{i-1}}{16}, \]

since $D(a,b)\leq r_i$ and $r_{i-1}> 2r_i$.
However, the total demand of the vertices of $S$ in $\dset$ is bounded by $N\cdot r_i$. The vertices of $A'$ may only serve as proxy vertices of the vertices of $S$, and so the total load on the vertices of $A'$ is bounded by $N r_i< \frac{N\cdot r_{i-1}}{16}$, since $r_i=\frac{\Delta}{32^i}$, a contradiction.
\end{proof}

For each cluster $C\in \cset'$, we now obtain an $r_{i-1}$-restricted demand $\dset_C=\set{\Pi_C, \set{D'(a',b')}_{(a',b')\in \Pi_C}}$, that includes all proxy pairs $(a',b')$ with $a',b'\in V(C)\cap U_i$. Clearly, for each such proxy pair $(a',b')$, $a',b'\in U_{i-1}^C$. Using the induction hypothesis, we compute, for each such cluster $C\in \cset'$, a routing $F^C$ of the demand $\dset_C$ in graph $W^C$ via paths of length at most $4(i-1)$ with no congestion. 
The time required to compute the flow $F^C$ is bounded by $O\left ((i-1)\cdot\Delta\cdot N\cdot  (N^{i-1}+|\Pi_C|)\right )$. Since $\sum_{C\in \cset}|\Pi_C|\leq |\Pi|$, we get the total time requried to compute all such flows $F^C$ is bounded by $O\left ((i-1)\cdot\Delta\cdot N\cdot  (N^{i}+|\Pi|)\right )$.
Lastly, by combining the partial flow $F$ that we have computed with the flows $F^C$ for clusers $C\in \cset'$, we obtain a routing of the demand $\dset$ over flow-paths of length at most $4i$. The routing causes no congestion, since the flow $F$ only uses level-$i$ edges, while the flows $F_C$ for $C\in \cset'$ cannot use any level-$i$ edges. The total running time of the algorithm is bounded by:

\[O\left ((i-1)\cdot\Delta\cdot N\cdot (N^{i}+|\Pi|)\right )+O(|\Pi|\cdot (N+\Delta))\leq O\left (i\cdot\Delta\cdot N\cdot (N^{i}+|\Pi|)\right ).  \]

If the demand $\dset$ is integral, then it is easy to verify that the resulting routing is integral as well.
\end{proof}

The following lemma summarizes our second step in the proof of \Cref{thm: pruning to routing}. The lemma provides an algorithm for routing vertices of $U_1$ to vertices of $U_i$ in a properly pruned subgraph $W$ of $W_i$.

\begin{lemma}\label{lem: U1 to Ui}
Fix an index $1\leq i\leq k$, and let $W$ be a properly pruned subgraph of $W_i$ with respect to sets $U_1,\ldots,U_i$ of its vertices. Then there is an integral flow $f$ in $W$, in which every flow-path $P$ with $f(P)>0$ connects a vertex of $U_1$ to a vertex of $U_i$; every vertex of $U_1$ sends $\Delta$ flow units; every vertex of $U_i$ receives at most $\Delta\cdot k^{3k}$ flow units; the congestion of the flow is at most $32^{i}\cdot k^{3k}$; and the length of every flow-path is at most $(4i)^2$. Additionally, for every vertex $v\in U_1$, all flow originating from $v$ terminates at a single vertex of $U_i$.	Moreover, there is an algorithm, that, given $W$ and the sets $U_1,\ldots, U_i$ of its vertices, computes such a flow in time $O(i^2\cdot \Delta\cdot N^{i+1})$.
\end{lemma}

\begin{proof}
	The proof is by induction on $i$. The base case is when $i=1$. 
	Let $W$ be a properly pruned subgraph of $W_1$ with respect to $U_1$. We can then let $f$ be a trivial flow, in which every vertex $v\in U_1$ sends $\Delta$ flow units to itself, via a flow-path $P_v=(v)$.
	
Consider now an integer $i>1$, and assume that the claim holds for $i-1$. Let $W$ be any properly pruned subgraph of $W_i$ with respect to sets $U_1,\ldots,U_i$ of its vertices.  
Recall that $W_i$ contains a disjoint union of $N$ copies of $W_{i-1}$, that we denote by $\set{C_1,\ldots,C_N}$, and refer to as \emph{clusters}. 
Let $\cset\subseteq \set{C_1,\ldots,C_N}$ be the set of clusters that survived, and consider any such cluster $C\in \cset$.

Denote $W^C=W\cap C$, and, for all $1\leq i'\leq i$, we denote by $U^{C}_{i'}=U_{i'}\cap V(C)$. From \Cref{obs: hereditary pruning}, graph $W^C$ is a properly pruned subgraph of $W_{i-1}$ with respect to vertex sets $U_1^C,\ldots,U_{i-1}^C$, and, from Property \ref{prop: cluster survival} of properly pruned subgraphs, $|U_i^C|\geq \frac{V(C)}{k^{3k}}$.
By using the induction hypothesis, we compute, in time $O((i-1)^2\cdot \Delta\cdot N^i)$, a flow $f_C$ in graph $W^C$, where every vertex of $U_1^C$ sends $\Delta$ flow units, and every vertex of $U_{i-1}^C$ receives at most $\Delta\cdot k^{3k}$ flow units, so that the congestion of the flow is at most $32^{i-1}\cdot k^{3k}$, and the length of every flow-path is at most $(4(i-1))^2$. Additionally, for every vertex $v\in U^C_1$, all flow originating at $v$ terminates at a unique vertex of $U_{i-1}^C$.

In the remainder of the proof, we will compute another flow $\hat f_C$, whose purpose is to route the flow that the vertices in $U_{i-1}^C$ receive via $f_C$ to the vertices of $U_i^C$, inside the graph $W^C$. In order to do so, we will first define an $r_{i-1}$-restricted demand $\dset_C$ over the vertices of $U_{i-1}^C$, and then use \Cref{lem: pruned to routing} in order to route it in $W^C$. After that, by properly scaling the resulting flow, we will obtain the desired flow $\hat f_C$. Lastly, by composing the flows $f_C$ and $\hat f_C$, we obtain a flow $f^*_C$ from vertices of $U_1^C$ to vertices of $U_i^C$, where every vertex in $U^C_1$ sends $\Delta$ flow units and every vertex in $U_i^C$ receives at most $\Delta\cdot k^{3k}$ flow units. The final flow is obtained by taking the union of flows $f^*_C$  for all $C\in \cset$.

We now turn to define the demand $\dset_C=\set{\Pi^C,\set{D'(a,b)}_{(a,b)\in \Pi^C}}$. Recall that, from Property \ref{prop: cluster survival} of properly pruned subgraphs, $|U_i^C|\geq \frac{|V(C)|}{k^{3k}}\geq \frac{|U_1^C|}{k^{3k}}$. Using a simple greedy algorithm we can then compute a mapping $\sigma$, that maps every vertex $v\in U_1^C$, to some vertex $\sigma(v)\in U_i^C$, such that, for every vertex $u\in U_i^C$, at most $k^{3k}$ vertices of $U_1^C$ are mapped to $u$. Recall that every vertex $v\in U_1^C$ sends $\Delta$ flow units to some vertex $u'\in U_{i-1}^C$ via the flow $f_C$. We denote $\sigma'(v)=u'$. We are now ready to define the demand $\dset_C$. For every vertex $v\in U_1^C$, we add the pair $(\sigma'(v),\sigma(v))$ to $\Pi^C$, with demand $D'(\sigma'(v),\sigma(v))=\Delta$. It is easy to verify that $\Pi^C\subseteq U_{i-1}^C\times U_{i-1}^C$, and that the resulting demand $\dset_C$ is $(\Delta\cdot k^{3k})$-restricted. 

Let $\dset'_C$ be the demand that is identical to $\dset_C$, except that we scale all demands down by factor $32^{i-1}\cdot k^{3k}$. Then $\dset'_C$ is an $r_{i-1}$-restricted demand, defined over the set $U_{i-1}^C$ of vertices. We can now apply the algorithm from \Cref{lem: pruned to routing} to graph $W^C$ and demand $\dset'_C$, to compute a routing $f'_C$ of the demand $\dset'_C$ in $W^C$ with no edge-congestion, with flow-paths of length at most $4(i-1)$. The time required to compute this flow is bounded by: 

\[O\left ((i-1)\cdot\Delta\cdot N\cdot (N^{i-1}+|\Pi^C|)\right )\leq O\left ((i-1)\cdot\Delta\cdot N^i\right ),\]

since $|\Pi^C|\leq |U_1^C|\leq N^{i-1}$. By scaling the flow $f'_C$ up by factor $32^{i-1}\cdot k^{3k}$, we obtain a flow $\hat f^C$, that routes the demand $\dset_C$ via paths of length at most $4(i-1)$, with congestion at most $32^{i-1}\cdot k^{3k}$.

Next, we combine flow $f_C$ and $\hat f_C$, obtaining a flow $f^*_C$, in which every vertex $v\in U_1^C$ sends $\Delta$ flow units to vertex $\sigma(v)\in U_i^C$; the length of each flow-path is at most $(4(i-1))^2+4(i-1)\leq (4i)^2$, and the flow causes congestion at most $2\cdot 32^{i-1}\cdot k^{3k}\leq  32^i\cdot k^{3k}$.

The final routing is obtained by taking the union of the flows $f^*_C$ for all clusters $C\in \cset$.

The time required to compute the flows $\hat f_C$ for all $C\in \cset$ is bounded by $O\left ((i-1)\cdot\Delta\cdot N^{i+1}\right )$. The time required to compute the flows $f_C$ for all $C\in \cset$ by the induction hypothesis is bounded by $O((i-1)^2\cdot \Delta\cdot N^{i+1})$. Therefore, the total running time of the algorithm is bounded by $O(i^2\cdot \Delta\cdot N^{i+1})$.
\end{proof}

We are now ready to complete the proof of \Cref{thm: pruning to routing}. 
We assume that we are given a properly pruned subgraph $W$ of $W_k$, with respect to sets $U_1,\ldots,U_k$ of its vertices, together with a $\frac{\Delta}{k^{4k}}$-restricted demand demand $\dset=\left (\Pi,\set{D(a,b)}_{(a,b)\in \Pi}\right )$ for $W$.

We use the algorithm from \Cref{lem: U1 to Ui} to compute an integral flow $f$, in which every flow-path $P$ with $f(P)>0$ connects a vertex of $U_1$ to a vertex of $U_k$; every vertex of $U_1$ sends $\Delta$ flow units; every vertex of $U_k$ receives at most $\Delta\cdot k^{3k}$ flow units; the congestion of the flow is at most $32^k\cdot k^{3k}\leq \frac{k^{4k}}{4}$ (since $k\geq 256$); and the length of every flow-path is at most $(4k)^2$. Recall that we are guaranteed that, for every vertex $v\in U_1$, all flow originating from $v$ terminates at a single vertex of $U_k$, that we denote by $\sigma(v)$.	The running time of the algorithm so far is  $O(k^2\cdot \Delta\cdot N^{k+1})$.
We slightly modify the flow $f$ by scaling the flow on every flow-path down by factor $k^{4k}$, obtaining a flow in which every edge carries at most $\frac 1 4 $ flow units. Notice that now every vertex $u\in U_k$ receives at most $\frac{\Delta}{k^k}\leq \frac{\Delta}{2\cdot 32^k}$ flow units, and every vertex $v\in U_1$ sends $\frac{\Delta}{k^{4k}}$ flow units.

Next, we define a new demand $\dset'=\set{\Pi',\set{D'(a,b)}_{(a,b)\in \Pi'}}$, as follows. For every pair $(a,b)\in \Pi$, we add the pair $(\sigma(a),\sigma(b))$ to $\Pi'$, with demand $D'(\sigma(a),\sigma(b))=D(a,b)$. Clearly, $\Pi'\subseteq U_k\times U_k$. We claim that the resulting demand $\dset'$ is $\frac{r_k}{2}$-restricted, where $r_k=\frac{\Delta}{32^k}$. Indeed, consider any vertex $u\in U_k$, and let $A\subseteq U_1$ be the collection of all vertices $v$ with $\sigma(v)=u$. Then $|A|\leq k^{3k}$ must hold, and, since the original demand $\dset$ was $\frac{\Delta}{k^{4k}}$-restricted, we get that $\dset'$ must be $\frac{\Delta}{k^k}\leq \frac{\Delta}{2\cdot 32^k}=\frac{r_k}{2}$-restricted. We can now use the algorithm from \Cref{lem: pruned to routing} in order to compute a routing $f'$ of $\dset'$ in $W$ via flow-paths of length at most $4k$, such that every edge of $W$ carries at most $1/2$ flow units (in order to do this, we first scale all demands in $\dset'$ up by factor $2$ to obtain an $r_k$-restricted demand; compute its routing with no congestion via \Cref{lem: pruned to routing}, and then scale the flow down by factor $2$). The running time required to compute flow $f'$ is bounded by $O\left (k\cdot\Delta\cdot N\cdot (N^k+|\Pi|)\right )$.
Lastly, we combine the flows $f$ and $f'$, possibly reducing the flow $f$ on some of its flow-paths as needed, to obtain the final routing $f^*$ of the demand $\dset$ in $W$ via flow-paths of length at most $(4k)^2+4k\leq 20k^2$, with no edge congestion. The total running time of the algorithm is bounded by $O\left (k^2\cdot\Delta\cdot N\cdot (N^k+|\Pi|)\right )$.

\subsection{Efficient Pruning of Graph $W_k$}


In this subsection we provide an algorithm to efficiently prune the graph $W_k^{N,\Delta}$, as it undergoes an online sequence of edge deletions. 
In order to motivate the model that we use in this section, we note that our approach for proving \Cref{thm: one level router decomposition} uses the online-batch dynamic update model (see Section \ref{subsec: online batch}). Intuitively, during initialization, we will construct an initial clustering $\cset$ of the input graph $G$, and, for every cluster $C\in \cset$, we will compute an embedding of a graph $W_k^{N,\Delta}$, for suitably chosen parameters $N$ and $\Delta$, into $C$. In order to ensure that $C$ has strong routing properties, we need to ensure that every vertex of $C$ participates in the embeddings of many edges of $W_k^{N,\Delta}$. Recall that in the online-batch dynamic update model, the algorithm receives updates in $\sigma$ batches $\pi_1,\ldots,\pi_{\sigma}$. In our case, we will use $\sigma=k$, and the only type of updates we will need to process is the deletion of edges from the input graph. Consider now some cluster $C$, and let $E'$ be the set of edges deleted from $C$ in a single batch $\pi_i$. Then every edge $e\in E(W_k^{N,\Delta})$ whose embedding path contains an edge from $E'$ needs to be deleted from $W_k^{N,\Delta}$. In order to ensure that we obtain a properly pruned subgraph $W\subseteq W_k^{N,\Delta}$, we may delete some additional edges from $W$. As the result of these deletions, it is possible that, for some vertices $v\in V(C)$, the number of edges $ e\in E(W)$ whose embedding path contains $v$ becomes too small. When this happens, we need to delete $v$ from $C$, and we need to further delete all edges $e\in E(W)$, whose embedding path uses $v$, from $W$. This, in turn, may trigger another round of pruning, and so on. Therefore, even though our algorithm for maintaining a router decomposition works in the online-batch dynamic model, the input to the pruning algorithm does not arrive in batches. Instead, we partition the timeline into phases $\Phi_0,\ldots,\Phi_k$. Phase $\Phi_0$ is special and is only used during initialization, while phases $\Phi_1,\ldots,\Phi_k$ can be thought of as corresponding to the batches $\pi_1,\ldots,\pi_k$ of updates to the router decomposition algorithm. For all $1\leq \tau \leq k$, the pruning algorithm receives as input an online sequence $E'_{\tau}$ of edges that are deleted from $W_k^{N,\Delta}$ over the course of the phase $\Phi_{\tau}$. We emphasize that these edges do not arrive simultaneously, but rather their arrival occurs over the course of the phase $\Phi_{\tau}$.
Intuitively, the algorithm needs to maintain subsets $U_1,\ldots,U_k$ of vertices of $W_k^{N,\Delta}$ with $U_k\subseteq\cdots\subseteq\cdots U_1$, and a subgraph $W\subseteq W_k^{N,\Delta}$, such that, at all times, $W$ is a properly pruned subgraph of $W_k^{N,\Delta}$ with respect to $U_1,\ldots,U_k$. We require that, for all $0\leq \tau\leq k$,  the total number of vertices deleted from $U_k$ over the course of the phase is not much larger than $|E'_{\tau}|/\Delta$, and the total number of edges deleted from $W$ over the course of the phase is not much larger than $|E'_{\tau}|$. 
We now define a new problem, that we call $\pruning$, which abstracts our pruning algorithm.

\begin{definition}[The $\pruning$ Problem]
The input to the \pruning problem consists of parameters $k\geq 256$,  $N\geq k^{3k}$ and $\Delta\geq 4k^2$, and an online sequence $e_1,\ldots,e_r$ of edge deletions from the corresponding graph $W_k=W^{N,\Delta}_k$. We denote the resulting dynamic graph by $H$, so $H\attime[0]=W_k$, and, for $t>0$, $H\attime[t]=H\attime[0]\setminus\set{e_1,\ldots,e_t}$.

The algorithm for the problem is required to maintain subsets $U_1,\ldots,U_k$ of vertices of $H$ with $U_k\subseteq \cdots\subseteq U_1$, and a subgraph $W\subseteq H$ for which the following properties hold:

\begin{itemize}
	\item for all $1\leq i\leq k$, at the beginning of the algorithm, $U_i=V(W_k)$, and, as the algorithm progresses, vertices may be deleted from $U_i$, but no vertices may be inserted into $U_i$; 
	\item the only changes that graph $W$ undergoes is the deletion of edges and vertices; and
	\item at all times, $W$ must be a properly pruned subgraph of $W_k$ with respect to the current sets $U_1,\ldots,U_k$.
\end{itemize}

The algorithm's timeline is partitioned into $k+1$ phases $\Phi_0,\Phi_1,\ldots,\Phi_k$, and the algorithm is notified when a new phase starts. For all $1\leq \tau\leq k$, if we denote by $E'_{\tau}$ the sequence of edge deletions from $W_k$ that algorithm receives as input during the phase $\Phi_{\tau}$, then it is guaranteed that $|E'_{0}|\leq \frac{N^k\cdot \Delta}{2k^{4k}}$, and, for all $1\leq \tau\leq k$, $|E'_{\tau}|\leq \frac{N^k\cdot \Delta}{k^{8k}}$. We require that, for all $0\leq \tau\leq k$, the total number of vertices deleted from $U_k$ over the course of phase $\Phi_{\tau}$ is at most $\frac{|E'_{\tau}|}{\Delta}\cdot k^{4k}$; the total number of vertices deleted from $U_1$ over the course of pahse $\Phi_{\tau}$ is at most $\frac{2k^{7k+2}\cdot |E'_{\tau}|}{\Delta}$, and the total number of edges deleted from $W$ over the course of phase $\Phi_{\tau}$ is at most $|E'_{\tau}|\cdot k^{7k+3}$.
\end{definition}
 
The main result of the current subsection is the following theorem, that provides an efficient algorithm for the $\pruning$ problem.

\begin{theorem}[Algorithm for $\pruning$]\label{thm: pruning of Wk}
	There is a deterministic algorithm for the \newline $\pruning$ problem, whose initialization time is $O(k^2\cdot N^k)$, and the running time of each phase $\Phi_{\tau}$, for  $0\leq \tau\leq k$,  is bounded by $O\left(|E'_{\tau}|\cdot k^{O(k)}\right )$.
\end{theorem}

The remainder of this subsection is dedicated to the proof of \Cref{thm: pruning of Wk}. At the beginning of the algorithm, we set $U_1=U_2=\cdots=U_k=V(W_k)$ and $W=W_k$. 
We start by describing the data structures that the algorithm maintains, together with the algorithm for their initialization. After that we describe our algorithm for implementing a single phase.

\subsubsection{Data Structures and their Initialization}

For every superedge $\hat e=(u,v)$ and phase $0\leq \tau\leq k$, our algorithm maintains a counter $n_{\tau}(\hat e)$, that counts the number of copies of the superedge that have been deleted from $H$ during Phase $\Phi_{\tau}$. All such counters are initialized to $0$.

For every level $1\leq i\leq k$, for every level-$i$ star $S\in \sset_i$, and for every phase $0\leq \tau\leq k$, our algorithm maintains a counter $n'_{\tau}(S)$, that counts the number of leaf vertices of $S$ that have been deleted from $U_i$ over the course of Phase $\Phi_{\tau}$. All these counters are initialized to $0$.

Lastly, for every level $1<i\leq k$, for every level-$(i-1)$ cluster $C\in \cset_{i-1}$, our algorithm maintains a counter $n''(C)$, that counts the number of vertices of $C$ that have been deleted from $U_i$ over the course of the entire algorithm so far. All these counters are initialized to $0$.

It is easy to verify that all these data structures can be initialized in time $O(k^2\cdot N^k)$.

We now turn to describe our algorithm for implementing a single phase.

\subsubsection{Algorithm for a Single Phase}
\label{subsubsec: one phase alg}

Our algorithm will ensure that, for all $0\leq \tau\leq k$, at all times $t$ during Phase $\Phi_{\tau}$, the following invariants hold:

\begin{properties}{I}
		\item For every level $1\leq i\leq k$, for every leaf vertex $v$ that lies in $U_i$ at time $t$, the total number of level-$i$ edges that are incident to $v$ in $W_k$ and were deleted over the course of Phase $\Phi_{\tau}$ so far is at most $\frac{\Delta}{4k}$. Moreover, if $v\not\in U_i$, then no level-$i$ edges are incident to $v$ in the current graph $W$; \label{Inv: leaf vertices}

		\item For every level $1\leq i\leq k$, for every level-$i$ star $S\in \sset_i$, if the center vertex of $S$ currently lies in $U_i$, then the total number of leaf vertices of $S$ that have been deleted from $U_i$ by the algorithm over the course of the current phase so far is at most $\frac{N}{4k^2}$. Moreover, if the center vertex of $S$ does not lie in $U_i$ at time $t$, then no leaf vertex of $S$ lies in $U_i$ at time $t$; and \label{Inv: stars}

		\item For every level $1< i\leq k$, for every level-$(i-1)$ cluster $C\in \cset_{i-1}$ of $W_k$, if $V(C)\cap U_1\neq \emptyset$ currently holds, then $|V(C)\cap U_i|\geq \frac{\hat n(C)}{16k^2}$ must hold, where $\hat n(C)$ is the number of vertices of $C$ that lied in $U_i$ at the beginning of the current phase. \label{Inv: cluster survival}
\end{properties}

\begin{observation}\label{obs: valid pruned subgraph}
	Assume that the above invariants hold throughout the algorithm, and that $V(W)=U_1$ holds at all times. Then, at all times $t$, the current graph $W$ is a properly pruned subgraph of $W_k$ with respect to the current sets $U_1,\ldots,U_k$.
\end{observation}
\begin{proof}
Consider any time $t$ during the algorithm's execution. Let $1\leq i\leq k$ be any level, and let $v$ be a leaf vertex that lies in $U_i$ at time $t$. Let $\hat e=(u,v)$ be the unique level-$i$ superedge incident to $v$. Then, from Invariant \ref{Inv: leaf vertices}, the total number of copies of $\hat e$ deleted from $H$ by the adversary up to time $t$ is bounded by $\sum_{\tau=0}^k\frac{\Delta}{4k}< \frac{\Delta}{2}$. 
If $v\not\in U_i$ at time $t$, then Invariant \ref{Inv: leaf vertices} ensures that no level-$i$ edges are incident to $v$ in the current graph $W$.
Therefore, Property \ref{prop: leaf vertices} holds throughout the algorithm.

Consider now a level $1\leq i\leq k$, and a level-$i$ star $S\in \sset_i$, whose center vertex $v$ lies in $U_i$ at time $t$. Then, from Invariant \ref{Inv: stars}, the total number of leaf vertices of $S$ that have been deleted from $U_i$ by the algorithm so far is at most $\sum_{\tau=0}^k\frac{N}{4k^2}\leq \frac{N}{k}$. If the center vertex of $S$ does not lie in $U_i$ at time $t$, then Invariant \ref{Inv: stars} ensures that no leaf vertex of $S$ lies in $U_i$ at time $t$.
Therefore, Property \ref{prop: center vertices} holds throughout the algorithm.

Lastly, consider a level $1< i\leq k$, and a level-$(i-1)$ cluster $C\in \cset_{i-1}$ of $W_k$, for which $V(C)\cap U_1\neq \emptyset$ currently holds. Let $\tau$ be the index of the current phase at time $t$.
For all $1\leq \tau'\leq \tau$, let $\hat n_{\tau'}(C)$ be the cardinality of the set $V(C)\cap U_i$ at the beginning of Phase $\Phi_{\tau'}$. Then, 
from Invariant \ref{Inv: cluster survival}, 
for all $1\leq \tau'<\tau$, $\hat n_{\tau'+1}(C)\geq \frac{\hat n_{\tau'}(C)}{16k^2}$, and moreover, at time $t$, $|V(C)\cap U_i|\geq \frac{\hat n_{\tau}(C)}{16k^2}$. Therefore, at time $t$:

\[|V(C)\cap U_i|\geq \frac{\hat n_{\tau}(C)}{16k^2}\geq \frac{|V(C)|}{(16k^2)^{\tau}}\geq \frac{|V(C)|}{k^{3k}},  \]

since $k\geq 32$ and $\tau\leq k$. Therefore, Property \ref{prop: cluster survival} holds throughout the algorithm.

Since the algorithm explicitly ensures that $V(W)=U_1$ holds at all time, we conclude that, at all times $t$, the current graph $W$ must be a properly pruned subgraph of $W_k$ with respect to the current sets $U_1,\ldots,U_k$. 
\end{proof}

From now on, we fix an index $0\leq \tau\leq k$, and consider Phase $\Phi_{\tau}$. It is now enough for us to provide an algorithm for implementing Phase $\Phi_{\tau}$ that ensures that Invariants \ref{Inv: leaf vertices}--\ref{Inv: cluster survival}  hold over the course of the algorithm, and that $V(W)=U_1$ always holds. For convenience, we denote $\Phi_{\tau}$ by $\Phi$, and we denote counters $n_{\tau}(\hat e)$ for superedges $\hat e$ and $n'_{\tau}(S)$ for stars $S$, by $n(\hat e)$ and $n'(S)$, respectively. We also denote the set of edges that the adversary deletes from $H$ over the course of the phase by $E'$ (this set is not known to the algorithm beforehand). It is enough for us to ensure that the running time of the algorithm to implement the phase is bounded by $O(|E'|\cdot k^{O(k)})$; that the total number of vertices deleted from $U_1$ and from $U_k$ over the course of the phase is bounded by $\frac{|E'|\cdot 2k^{7k+2}}{\Delta}$ and $\frac{|E'|}{\Delta}\cdot k^{4k}$ respectively; and that the total number of edges deleted from $W$ over the course of the phase is bounded by $|E'|\cdot k^{7k+3}$.

\subsubsection*{The Algorithm Description}
 
 We now formally describe the algorithm for implementing Phase $\Phi$.
 Throughout the algorithm, we maintain a list $\Lambda$, that can be thought of as a "to do" list. This list contains entries of three types. An entry of type 1 is a pair $(i,v)$, where $1\leq i\leq k$ is a level, $v\in U_i$ is a vertex, and the entry indicates that $v$ should be deleted from $U_i$. The \emph{level} of entry $(i,v)$ is $i$. An entry of type $2$ is a pair $(i,S)$, where $1\leq i\leq k$ is a level, and $S\in \sset_i$ is a level-$i$ star. The entry indicates that star $S$ must be destroyed, either because its center vertex was just deleted from $U_i$, or because the number of leaf vertices of $S$ that were deleted from $U_i$ over the course of the current phase is at least $\frac{N}{4k^2}$. The \emph{level} of entry $(i,S)$ is $i$. Lastly, an entry of type $3$ is a pair $(i,C)$, where $1<i\leq k$ is a level, and $C\in \cset_{i-1}$ is a level-$(i-1)$ cluster. The entry indicates that cluster $C$ needs to be destroyed, because $|V(C)\cap U_i|$ fell below $\frac{\hat n(C)}{16k^2}$, where $\hat n(C)$ is the number of vertices of $C$ that lied in $U_i$ at the beginning of the current phase. The \emph{level} of entry $(i,C)$ is $i$.

We are now ready to describe the algorithm for processing the deletion of a single edge $e=(u,v)$ from graph $H$. If $e$ does not lie in  the current graph $W$, then no further processing is needed. Assume now that $e$ lies in $W$, and let $\hat e=(u,v)$ be the corresponding superedge. We increase the counter $n(\hat e)$ by $1$. If counter $n(\hat e)$ remains bounded by $\frac{\Delta}{4k}$, then no further updates are needed. From now on we assume that $n(\hat e)>\frac{\Delta}{4k}$ holds. We assume w.l.o.g. that $v$ is a leaf vertex, and $u$ is a center vertex. Let $1\leq i\leq k$ be the level to which superedge $(u,v)$ belongs. We initialize the list $\Lambda=\set{(i,v)}$, indicating that vertex $v$ must be deleted from set $U_i$. We say that entry $(i,v)$ was added to $\Lambda$ \emph{directly}.
Next, we iterate, as long as $\Lambda\neq\emptyset$. In every iteration, we select a single entry of $\Lambda$ to process. The entry that we choose to process in each iteration must be that of the smallest level in $\Lambda$, and among entries of the same level, it must be of the lowest type (that is, type-1 entries are processed before type-2 entries, which are processed before type-3 entries of a single level). We now describe the procedure for processing each entry.

\paragraph{Processing type-1 entries.}
Suppose the entry of $\Lambda$ that needs to be processed in the current iteration is of type $1$, and denote it by $(i,v)$, where $v$ is a vertex that currently lies in $U_i$. We start by deleting vertex $v$ from $U_{i}$, and removing entry $(i,v)$ from $\Lambda$. We also delete all level-$i$ edges that are incident to $v$ from $W$. If $i<k$, then we add type-1 entry $(i+1,v)$ to $\Lambda$, so that $v$ is also deleted from  subsequent layers. In this case, we say that entry $(i+1,v)$  was added to $\Lambda$ \emph{directly}.
Next, we consider the unique level-$i$ star $S$ that contains $v$. If $S$ is already marked as destroyed then no further  updates to it needed. Otherwise, if $v$ is the center of the star $S$, then we add a type-2 entry $(i,S)$ to $\Lambda$, to indicate that the star $S$ needs to be destroyed. Assume now that $v$ is a leaf of $S$. We increase the counter $n'(S)$ by $1$, and, if the counter becomes greater than $\frac{N}{4k^2}$, then we add a type-2 entry $(i,S)$ to $\Lambda$, indicating that star $S$ must be destroyed. Lastly, if $i>1$, then then we also consider the level-$(i-1)$ cluster $C$ of $W_k$ to which $v$ belongs. 
If no vertex of $C$ has been deleted from $U_{i}$ in the current phase yet, then we mark $C$ to indicate that one of its vertices was deleted from $U_{i}$ in the current phase, and we record that the number of vertices of $C$ that lied in $U_i$ at the beginning of the current phase is $\hat n(C)=|V(C)|-n''(C)$. If $C$ is already marked as destroyed, then no further processing of $C$ is needed. Otherwise, we increase $n''(C)$ by $1$, and, if $|V(C)|-n''(C)<\frac{\hat n(C)}{16k^2}$ holds, we add a type-3 entry $(i,C)$ to $\Lambda$, indicating that cluster $C$ must be destroyed.

\paragraph{Processing type-2 entries.}
Assume now that the entry of $\Lambda$ that needs to be processed in the current iteration is of type $2$, and denote it by $(i,S)$, where $S$ is a level-$i$ star. For every vertex $x\in V(S)$ that still lies in $U_{i}$, such that entry $(i,x)$ does not currently lie in $\Lambda$, we add the type-1 entry $(i,x)$ to $\Lambda$, and we say that this entry was added to $\Lambda$ \emph{indirectly}. We then delete entry $(i,S)$ from the list $\Lambda$, and we mark $S$ as being destroyed.

\paragraph{Processing type-3 entries.}
Finally, assume that the entry of $\Lambda$ that needs to be processed in the current iteration is of type $3$, and denote it by $(i,C)$, where $C$ is a level-$(i-1)$ cluster. For every vertex $x\in V(C)\cap U_1$, we delete $x$ from sets $U_1,\ldots,U_{i-1}$, and we delete from $W$ all edges of levels $1,\ldots,i-1$ that are incident to $x$ in the current graph $W$. Lastly, we add a type-1 entry $(i,x)$ to list $\Lambda$, and we say that the entry was added to the list \emph{indirectly}. We delete entry $(i,C)$ from $\Lambda$, and we mark $C$ as being destroyed.

The algorithm for processing a single edge deletion from $H$ terminates when $\Lambda$ becomes empty. 
Notice that, if some vertex $v$ was deleted from set $U_1$ during the update procedure, then $v$ was also deleted from $U_2,\ldots,U_k$, and it has become an isolated vertex in $W$. In such a case, we also delete $v$ from $V(W)$, to ensure that $V(W)=U_1$ always holds.

For all $1\leq i\leq k$, let $U^D_i$ be the set of vertices that were deleted from $U_i$ over the course of the update procedure, and let $M=\sum_{i=1}^k|U^D_i|$.  Then the running time of the procedure is $O(1)+O(M\cdot \Delta)$, and the total number of edges deleted from $W$ over the course of the procedure is at most $M\Delta$.
This is since, when a center vertex $u$ is deleted from $U_i$ by the algorithm, every leaf vertex that lies in the level-$i$ star $S$ for which $u$ serves as a center is also deleted from $U_i$, and so the deletion of the level-$i$ edges incident to $u$ in $H$ can be charged to the corresponding leaf vertices; the deletion of a leaf vertex $v$ from a set $U_i$ can trigger the deletion of at most $\Delta$ edges from $W$. It is immediate to verify that, at the end of the update procedure Invariants \ref{Inv: leaf vertices}--\ref{Inv: cluster survival} hold.

Consider now some level $1\leq i\leq k$, and a type-1 level-$i$ entry $(i,v)$ that ever lied in $\Lambda$ over the course of the update procedure. Each such entry was added to $\Lambda$ either directly or indirectly. The entry can only be added to $\Lambda$ indirectly during the processing of a level-$i$ type-2 entry $(i,S)$, if $v$ is a vertex of the star $S$ that is being destroyed, or during the processing of a level-$i$ type-3 entry $(i,C)$, where $C$ is a level-$(i-1)$ cluster being destroyed, and $v\in V(C)$. Entry $(i,v)$ may be added to $\Lambda$ directly only in one of the following two cases: either (i) while processing a level-$(i-1)$ type-1 entry $(i-1,v)$; or (ii) if edge $(u,v)$ was just deleted from $H$ by the adversary, $v$ is a leaf vertex, and the number of copies of superedge $(u,v)$ deleted from $H$ in the current phase became greater than $\frac{\Delta}{4k}$. 
We denote by $J_i$ the set of all leaf vertices $v$ that lie in $U_i$ at the beginning of the phase, so that at least $\frac{\Delta}{4k}$ level-$i$ edges incident to $v$ lie in $E'$. Clearly, $|J_i|\leq \frac{4k|E'|}{\Delta}$.

We denote by $R_i$ the set of all vertices $v$, such that type-1 entry $(i,v)$ ever lied in $\Lambda_i$ during the current phase, and it was added to $\Lambda_i$ directly. 
We also include in $R_i$ all vertices of $J_i$, even if, for some such vertex $v$, entry $(i,v)$ was first added to $\Lambda$ indirectly.
Similarly, we denote by $R'_i$ the set of all vertices $v\not\in J_i$, such that entry $(i,v)$ ever lied in $\Lambda_i$ during the current phase, and it was added to $\Lambda_i$ indirectly. Lastly, we denote by $R_i''$ all vertices $v$ that were deleted from $U_i$ during the current phase, but the type-1 entry $(i,v)$ never lied in $\Lambda$ during the current phase. The latter may only happen if, for some level $i'>i$, the level-$(i'-1)$ cluster $C$ containing $v$ was destroyed after the corresponding type-3 entry $(i',C)$ was added to $\Lambda$.

Notice that, when a type-3 entry $(i,C)$ is processed, every vertex of $C$ that currently lies in $U_1$ is deleted from sets $U_1,\ldots,U_{i-1}$. Since, as we showed in \Cref{obs: valid pruned subgraph}, at the beginning of the current phase Property \ref{prop: cluster survival} held, we get that $\hat n(C)\geq \frac{|V(C)|}{k^{3k}}$ at the beginning of the phase. By the time the algorithm starts processing entry $(i,C)$, at least $\frac{\hat n(C)}{2}\geq \frac{|V(C)|}{2k^{3k}}$ type-1 entries $(i,x)$ for vertices $x\in V(C)$ have been added to $\Lambda$, and so at least $\frac{|V(C)|}{2k^{3k}}$ vertices of $C$ lie in $R_i\cup R'_i$.
Intuitively, we can \emph{charge} these vertices for the deletion of the vertices of $C$ from sets $U_1,\ldots,U_{i-1}$, when cluster $C$ is destroyed. Let $\Pi$ denote the collection of all pairs $(i,x)$, where $x$ is a vertex that was deleted from $U_i$ over the course of the current phase.
Then, from the above discussion:

\[|\Pi| \leq 2k\cdot k^{3k}\cdot \sum_{i=1}^k(|R_i|+|R'_i|)\leq k^{3k+2}\sum_{i=1}^k(|R_i|+|R'_i|).\]

Additionally, from our discussion, the total number of edges deleted from $W$ over the course of the current phase is at most:

\[|E'|+\Delta\cdot \Pi\leq |E'|+\Delta \cdot k^{3k+2}\sum_{i=1}^k(|R_i|+|R'_i|),\]

and the total running time for processing the current phase is bounded by $O(|E'|)+O(\Delta\cdot |\Pi|)\leq O(|E'|)+k^{O(k)}\cdot\Delta\cdot \sum_{i=1}^k(|R_i|+|R'_i|)$.

Notice also that $R''_k=\emptyset$, and so the total number of vertices deleted from $U_k$ in the current phase is bounded by $|R_k|+|R'_k|$. 
Next, we will bound the cardinalities of the sets $R_i,R'_i$ of vertices for $1\leq i\le k$. This will allow us to bound the running time of  a phase, the number of edges deleted from $W$ over the course of the phase, and the total number of vertices deleted from $U_1$ and from $U_k$ in the current phase.

Before we proceed, consider an index $1\leq i\leq k$, and denote $\lambda_i=|R_i|$. Consider now some center vertex $v\in R_i$. Recall that the only way that entry $(i,v)$ may be added to $\Lambda$ directly when $v$ is a center vertex, is if the type-1 entry $(i-1,v)$ was processed by the algorithm. If $S$ is the unique level-$(i-1)$ star whose center is $v$, then star $S$ must have been destroyed by our algorithm, and so every leaf vertex $x$ of $S$ that lied in $U_{i-1}$ at the beginning of the current phase was deleted from $U_{i-1}$, and the corresponding type-1 entry $(i,x)$ was added to $\Lambda$ directly. Since, from Property \ref{prop: center vertices}, at least $N/2$ leaf vertices of $S$ lied in $U_{i-1}$ at the beginning of the phase, we get that, for each each center vertex $v\in R_i$, there are at least $N/2$ leaf vertices that lie in $R_i$, and these sets of leaf vertices that correspond to each center vertex are disjoint. In other words, the total number of center vertices in $R_i$ must be bounded by $2|R_i|/N$.

\subsubsection{Analysis}

\label{subsubsec: analyzing Ris}

In this subsection, we again consider the phase $\Phi$, and the set $E'=\set{e_1,\ldots,e_q}$ of edges that were deleted from $H$ during this phase. 
Our goal is to bound the cardinalities of the sets $R_i$ and $R'_i$ for all $1\leq i\leq k$. Recall that we have denoted $\lambda_i=|R_i|$. We start with the following claim.

\begin{claim}\label{claim: bound indirect}
For all $1\leq i\leq k$, $|R_i|+|R'_i|\leq 32k^2\lambda_i$.
\end{claim}
\begin{proof}
	Let $U'_i$ be the set $U_i$ at the beginning of the current phase.
In order to prove the claim, we assign, to every vertex $v\in U'_i$, a potential $\psi(v)$, that may change over the course of the phase.

 Recall that we denoted by $J_i$ the set of all leaf vertices $v\in U'_i$, so that at least $\frac{\Delta}{4k}$ level-$i$ edges incident to $v$ lie in $E'$.
For the sake of the analysis, we will view the set $R_i$ of vertices as fixed, so $R_i$ is the set of all vertices $v$, whose corresponding type-$1$ entry $(i,v)$ was ever added to $\Lambda$ directly during the current phase. This set also includes all vertices in $J_i$, even if, for some such vertex $v$, entry $(i,v)$ was first added to $\Lambda$ indirectly. We also define a dynamically evolving set $\hat R_i$, that, at any time $t$ during the current phase, contains all vertices $v\in U'_i\setminus R_i$, whose corresponding type-1 entry $(i,v)$ was added to $\Lambda$ indirectly, since the beginning of the phase, and until time $t$ (inclusive). We emphasize that set $\hat R_i$ does not contain vertices of $J_i$, even if, for some such vertex $v$, entry $(i,v)$ was first added to $\Lambda$ indirectly.
Therefore, at the beginning of the phase, $\hat R_i=\emptyset$, at the end of the phase, $\hat R_i=R'_i$, and over the course of the phase, vertices may join $\hat R_i$, but they may never leave this set.

Consider some time $t$ during the execution of the algorithm, and let $v\in U'_i$ be any vertex. If $v\not\in R_i\cup \hat R_i$, then we set the potential $\psi(v)=0$.

Assume now that $v\in R_i\cup \hat R_i$ currently holds. We now define values $\psi'(v)$ and $\psi''(v)$, and we will eventually set the potential $\psi(v)=\psi'(v)+\psi''(v)$. Let $S$ be the unique level-$i$ star containing $v$. If a type-2 entry $(i,S)$ has not been processed in the current phase yet (either because this entry was never added to $\Lambda$, or because it was added but not yet processed), then we set $\psi'(v)=8 k^2$ if $v$ is a leaf vertex, and we set $\psi'(v)=2N$ if it is a center vertex. Otherwise, if entry $(i,S)$ was processed in the current phase already, then we set $\psi'(v)=0$. 

If $i=1$, then we set $\psi''(v)=0$. Otherwise,
let $C$ be the unique level-$(i-1)$ cluster containing $v$. If type-$3$ entry $(i,C)$ was not processed in the current phase yet (either because it was never added to $\Lambda$, or because it was added but was not processed yet), then we set $\psi''(v)=8$ if $v\in J_i$ and $\psi''(v)=2$ if $v\not\in J_i$. Otherwise, if  type-$3$ entry $(i,C)$ was already processed, we set $\psi''(v)=1$. Lastly, we let $\psi(v)=\psi'(v)+\psi''(v)$ be the potential of $v$. 
Notice that for every vertex $v\in U'_i$, if $v$ is a leaf vertex then $\psi(v)\leq 8k^2+8$ always holds, and if $v$ is a center vertex, then $\psi(v)\leq 2N+2$ always holds.

Recall that vertices may be deleted from set $U_i$ when, for some level $i'>i$, some level-$(i'-1)$ cluster is destroyed. In this proof we ignore all such updates to $U_i$. They do not affect the potentials of the vertices, and they do not cause the additions of any new level-$i$ entries into $\Lambda$. Moreover, observe that, for every level-$i$ star $S$, following such an update, either all vertices of $S$ are deleted from $U_i$, or none of them are. Similarly, for every level-$(i-1)$ cluster $C$, following such an update, either all vertices of $C$ are deleted from $U_i$, or none of them are. Therefore, all such updates can be ignored in this proof.

Let $\Psi=\sum_{v\in U'_i}\psi(v)$ be the total potential at the system.
Note that, at the beginning of the phase, for every vertex $v\in R_i$, if $v$ is a leaf vertex, then its potential is $\psi(v)=8k^2+8\leq 9k^2$, and if it is a center vertex, then  $\psi(v)\leq 2N+8$. Moreover, if $v\not\in R_i$, then $\psi(v)=0$. Since the number of center vertices in $R_i$ is bounded by $\frac{2\lambda_i}{N}$ and $|R_i|=\lambda_i$, we get that $\Psi\leq 32k^2\lambda_i$. Throughout the phase, every vertex $v\in R_i\cup \hat R_i$ has potential $\psi(v)\geq 1$. In the remainder of the proof, we show that the potential $\Psi$ never increases over the course of the phase. Since, at the end of the phase, the potential of every vertex in $R_i\cup \hat R_i$ is at least $1$, it will then follow that $|R_i|+|\hat R_i|\leq 32 k^2\lambda_i$ always holds, and so in particular, $|R_i|+|R_i'|\leq 32k^2\lambda_i$.

Notice that the potential of the vertices of $U'_i$ may only change when a level-$i$ entry of $\Lambda$ is being processed. This is since the processing of entries whose level is $i'<i$ may only add new level-$i$ entries into $\Lambda$ that are of type $1$, and such entries $(i,v)$ are added to $\Lambda$ directly, so the corresponding vertices $v$ lie in $R_i$ already; the processing of entries of levels $i'>i$ may not affect the potentials of the vertices.
Consider now some time $t$ during the phase, when a level-$i$ entry is being processed. We show that the potential $\Psi$ may not grow as the result of processing the entry.

\paragraph{Type-1 entries.}
Assume first that the entry is of type $1$, and denote it by $(i,v)$. Then at time $t$, $v\in R_i\cup \hat R_i$ already holds. It is easy to verify that processing this entry does not change the potential of any vertex in $U'_i$.

\paragraph{Type-2 entries.}
Assume now that the processed entry is of type $2$, and denote it by $(i,S)$, where $S$ is the corresponding level-$i$ star. Let $X$ be the set of all leaf vertices of $S$ that currently lie in $R_i\cup \hat R_i$, and let $Y$ be the set of all leaf vertices of $S$ that lie in $U'_i\setminus(R_i\cup \hat R_i)$. Let $u$ be the center vertex of $S$.

Assume first that $u\in R_i\cup \hat R_i$. Then, since entry $(i,S)$ has not yet been processed, $\psi'(u)=2N$ currently holds, while $|Y|\leq N$. Once entry $(i,S)$ is processed, $\psi'(u)$ is set to $0$, while $\psi''(u)$ remains unchanged. For every vertex $v\in X$, $\psi(v)$ decreases, while for every vertex $v\in Y$, $\psi(v)$ grows from $0$ to at most $2$ (since $v$ may not lie in $J_i\subseteq R_i$ in this case). Overall, the potential of $u$ decreases by at least $2N$, while the potential of the vertices in $Y$ grows by at most $2|Y|\leq 2N$. The potential of the vertices in $X$ may only decrease.
Therefore, $\Psi$ does not increase.

Assume now that $u\not\in R_i\cup \hat R_i$. Then, for every vertex $v\in X$, $\psi'(v)= 8 k^2$ holds before entry $(i,S)$ is processed, and $\psi'(v)=0$ holds after it is processed. Therefore, the potential of the vertices in $X$ decreases by $8 k^2\cdot |X|$. 
For every vertex $v'\in Y\cup\set{u}$, $\psi(v')$ grows from $0$ to at most $2$ once entry $(i,S)$ is procesed. So the total potential of the vertices in $Y\cup \set{u}$ increases by at most $2(|Y|+1)$. Since entry $(i,S)$ was added to $\Lambda$, while the center vertex of $S$ does not lie in $R_i\cup \hat R_i$ and hence remains in $U'_i$, it must be the case that $|X|\geq \frac{N}{ 4k^2}\geq \frac{|Y|+1}{4k^2}$. Therefore, the potential of the vertices in $X$ decreases by at least $8 k^2\cdot |X|\geq 2(|Y|+1)$, and overall the potential $\Psi$ does not increase over the course of processing type-2 entries.

\paragraph{Type-3 entries.}
It now remains to consider the case where a type-3 entry $(i,C)$ is being processed, where $C$ is a level-$(i-1)$ cluster. Let $X'$ be the subset of vertices of $V(C)\cap U'_{i}$ that lie in $R_i\cup \hat R_i$ and let $Y'$ be the set containing the remaining vertices of $V(C)\cap U'_{i}$. Since entry $(i,C)$ was added to $\Lambda$,  $|Y'|\leq \frac{\hat n(C)}{16 k^2}$ holds, while $|X'|=\hat n(C)-|Y'|$. When entry $(i,C)$ is processed, for every vertex $v\in Y'$, entry $(i,v)$ is added to $\Lambda$ indirectly, and so $v$ is added to $\hat R_i$, and the  potential $\psi(v)$ may grow from $0$ to $8k^2+1$ if $v$ is a leaf vertex, and from $0$ to $2N+1$ if it is a center vertex.

Let $Y'_c\subseteq Y'$ be the set of all center vertices that lie in $Y'$.
Consider any vertex $v\in Y'_c$, and let $S$ be the unique level-$(i-1)$ star for which $v$ serves as a center. Since our algorithm processes the entries of $\Lambda$ from lower to higher levels, entry $(i-1,S)$ has not been added to $\Lambda$ yet. From Invariant \ref{Inv: stars}, and from the fact that $W$ is a properly pruned subgraph of $W_k$ at all times, at least $\frac{3N}{4}$ leaf vertices of $S$ currently lie in $U_{i-1}$; we denote the set of all these leaf vertices by $Z(v)$. Notice that $Z(v)\subseteq V(C)$ must hold. Consider now a vertex $u\in Z(v)$. From our discussion, when entry $(i,C)$ is processed, entry $(i-1,u)$ has not been added to $\Lambda$ yet. 
Since $u$ is deleted from $U_{i-1}$ when entry $(i,C)$ is processed, entry $(i-1,u)$ is never be added to $\Lambda$ over the course of the current phase. Therefore, the only way that $u$ may lie in $R_i$ is if $u\in J_i$. We partition the set $Z(v)$ of vertices into two subsets: $Z'(v)=Z(v)\cap J_i$ and $Z''(v)=Z(v)\setminus J_i$. From our discussion, $Z''(v)\cap R_i=\emptyset$, and so $Z''(v)\subseteq Y'$ must hold.
Since, for every vertex $v\in Y'_c$, $|Z(v)|\geq \frac{3N}{4}$ we get tha either (i) $\sum_{v\in Y'_c}|Z'(v)|\geq \frac{|Y'_c|\cdot N}{2}$, or (ii) $\sum_{v\in Y'_c}|Z''(v)|\geq \frac{|Y'_c|\cdot N}{4}$ must hold. We consider each of these two cases separately.

Assume first that $\sum_{v\in Y'_c}|Z''(v)|\geq \frac{|Y'_c|\cdot N}{4}$. 
It is easy to verify that the vertex sets $\set{Z''(v)}_{v\in Y'_c}$ are mutually disjoint, and, as observed already, for every vertex $v\in Y'_c$, $Z''(v)\subseteq Y'$ must hold. Therefore, in this case, $|Y'_c|\leq \frac{4|Y'|}{N}$.
 Altogether, the total potential $\psi(v)$ of all vertices $v\in Y'$ grows by at most: 
 
 \[
 \begin{split}
 (8k^2+1)\cdot |Y'|+(2N+1)\cdot |Y'_c|&\leq (8k^2+1)\cdot |Y'|+(2N+1)\cdot \frac{4|Y'|}{N}\\
 &\leq (8k^2+10)\cdot |Y'|\\
 &\leq (8k^2+10)\cdot \frac{\hat n(C)}{16k^2}. 
 \end{split}\]
 
 At the same time, after entry $(i,C)$ is processed, the potential $\psi(v')$ of every vertex $v'\in X'$ decreases by at least $1$. Since $|X'|=\hat n(C)-|Y'|\geq 0.99\hat n(C)$, the total potential $\Psi$ does not grow.

It remains to consider the case where $\sum_{v\in Y'_c}|Z'(v)|\geq \frac{|Y'_c|\cdot N}{2}$. As before, the vertex sets $\set{Z'(v)}_{v\in Y'_c}$ are mutually disjoint. As observed already, for every vertex $v\in Y'_c$, $Z'(v)\subseteq J_i$ must hold. Therefore, in this case, $|Y'_c|\leq \frac{2|V(C)\cap J_i|}{N}$.
Altogether, the total potential $\psi(v)$ of all vertices $v\in Y'$ grows by at most: 

\[
\begin{split}
(8k^2+1)\cdot |Y'|+(2N+1)\cdot |Y'_c|&\leq (8k^2+1)\cdot |Y'|+(2N+1)\cdot \frac{2|V(C)\cap J_i|}{N}\\
&\leq (8k^2+1)\cdot \frac{\hat n(C)}{16k^2}+5|V(C)\cap J_i|\\
&\leq 0.51\hat n(C)+5|V(C)\cap J_i|.
\end{split}\]

(We have used the fact that $k\geq 256$.)
At the same time, after entry $(i,C)$ is processed, the potential $\psi(v')$ of every vertex $v'\in X'$ decreases by at least $1$, while the potential of every vertex in $V(C)\cap J_i$ decreases by at least $7$. Therefore, the total decrease in the potential is at least:

\[7|V(C)\cap J_i|+(|X'|-|V(C)\cap J_i|)\geq 5|V(C)\cap J_i|+|X'|\ge 5|V(C)\cap J_i|+0.51 \hat n(C).
\]

(We have used the fact that $|X'|=\hat n(C)-|Y'|\geq 0.99\hat n(C)$). Altogether, in either case, the total potential does not grow when a type-3 entry of $\Lambda$ is processed. 

We conclude that, throughout the phase, $\Phi\leq 32k^2\lambda_i$ must hold. Since, throughout the phase, for every vertex $v\in R_i\cup \hat R_i$, $\psi(v)\geq 1$ holds, and since, at the end of the phase, $\hat R_i=R'_i$, we get that $|R_i|+|R'_i|\leq 32k^2\lambda_i$.

\end{proof}

We now obtain the following easy corollary.

\begin{corollary}\label{cor: all deleted from Ui}
For all $1\leq i\leq k$, $|R_i\cup R'_i|\leq k^{4i}\cdot \frac{q}{\Delta}$, where $q=|E'|$ is the number of edges deleted from $H$ in the current phase.
\end{corollary}
\begin{proof}
Consider the set 	$E'$ of edges deleted over the course of the algorithm, and let $\hat E$ contain all superedges $\hat e=(u,v)$, such that at least  $\frac{\Delta}{4k}$ copies of $\hat e$ lie in $E'$. Clearly, $|\hat E|\leq \frac{4qk}{\Delta}$. For all $1\leq i\leq k$, let $\hat E_i\subseteq \hat E$ contain the subset of level-$i$ superedges of $\hat E$. Recall that we denoted by $J_i$ be the set of all leaf vertices $v$ that lie in $U_i$ at the beginning of the phase and serve as endpoints of the superedges in $\hat E_i$. Clearly, $|J_i|\leq |\hat E|\leq \frac{4qk}{\Delta}$ for all $i$.

We prove the corollary by induction.
The induction base is when $i=1$. Notice that the only type-1 entries $(1,v)$ that are added to $\Lambda$ directly over the course of the phase are those corresponding to vertices $v\in J_1$, so $\lambda_1=|R_1|\leq \frac{4qk}{\Delta}$. From \Cref{claim: bound indirect} we then get that $|R_1|+|R'_1|\leq 32k^2\lambda_1\leq \frac{128k^3q}{\Delta}\le \frac{k^4q}{\Delta}$, since $k\geq 256$.

Consider now an integer $i>1$, and assume that the claim holds for $i-1$. Note that the only type-1 entries $(i,v)$ that are added to $\Lambda$ directly are those corresponding to vertices $v\in J_i$, or those whose corresponding level-$(i-1)$ entry $(i-1,v)$ was added to $\Lambda$ as well. Therefore, $\lambda_i=|R_i|\leq |J_i|+|R_{i-1}|+|R'_{i-1}|\leq \frac{4qk}{\Delta}+k^{4i-4}\cdot \frac{q}{\Delta}\leq 2k^{4i-4}\cdot \frac{q}{\Delta}$. From \Cref{claim: bound indirect}, we then get that
$|R_i|+|R'_i|\leq 32k^2\lambda_i\leq 64k^{4i-2}\cdot \frac{q}{\Delta}\leq k^{4i}\cdot \frac{q}{\Delta}$.
\end{proof}

We conclude that the total number of vertices deleted from $U_k$ over the course of the phase is bounded by $|R_k\cup R'_k|\leq k^{4k}\cdot\frac{q}{\Delta}$. From our discussion, the running time of the phase is bounded by:

\[O(q)+k^{O(k)}\cdot\Delta\cdot \sum_{i=1}^k(|R_i|+|R'_i|)\leq O\left(k^{O(k)}\cdot q\right ). \]

Recall that we denoted by $\Pi$ the collection of all pairs $(i,v)$, such that $v$ is a vertex that was deleted from $U_i$ in the current phase, and we have shown that $|\Pi| \leq k^{3k+2}\sum_{i=1}^k(|R_i|+|R'_i|)$.
We then get that the total number of vertices deleted from $U_1$ over the course of the current phase is bounded by:

\[|\Pi| \leq k^{3k+2}\sum_{i=1}^k(|R_i|+|R'_i|)\leq \frac{2k^{3k+2}\cdot k^{4k}\cdot q}{\Delta}\leq \frac{2k^{7k+2}\cdot q}{\Delta}.\]

Lastly, from our discussion, the total number of edges deleted from $W$ in this phase is bounded by: 

\[ q+\Delta \cdot k^{3k+2}\cdot \sum_{i=1}^k(|R_i|+|R'_i|)\leq q+2q\cdot k^{3k+2}\cdot k^{4k}\leq q\cdot k^{7k+3}. \]


\section{Second Tool: Embedding a Nice Router}
\label{sec: embed-well-connected}

One of the main notions that we use in this section is that of a scattered graph, that we define next.

\begin{definition}[Scattered Graph.]
Let $G$ be a graph, and let $d\geq 1$ and $0<\eps<1$ be parameters. We say that $G$ is \emph{$(d,\eps)$-scattered} if, for every vertex $v\in V(G)$, $|B_G(v,d)|\leq |V(G)|^{1-\eps}$.
\end{definition}

Next, we provide a simple algorithm, that, given a graph $G$ and parameters $d$ and $\eps$, either correctly certifies that $G$ is $(d,\eps)$-scattered, or computes a vertex $v\in V(G)$, such that $|B_G(v,4d/\eps)|\geq |V(G)|^{1-\eps}$. The proof follows standard techniques and is deferred to Section \ref{sec: appx: scattered or large ball} of Appendix.

\begin{claim}\label{claim: scattered or large ball}
	There is an algorithm, that, given an $n$-vertex graph $G$ and parameters $d\geq 1$ and $0<\eps<1$, either correctly certifies that $G$ is $(d,\eps)$-scattered, or computes a vertex $v\in V(G)$, such that $|B_G(v,4d/\eps)|\geq |V(G)|^{1-\eps}$. The running time of the algorithm is $O(m^{1+\eps})$, where $m=|E(G)|$.
\end{claim}

Next, we define a new problem, that we call \embedorscatter.
The input to the problem consists of an $n$-vertex graph $G$, and parameters $\Delta$ and $k$. Additionally, we are given a distance parameter $d$, and a congestion parameter $\eta$. The goal of the algorithm is to either (i) compute an embedding of graph $W^{N,\Delta}_k$ into $G$ via paths whose length is not much larger than $d$, that cause congestion at most $\eta$, where parameter $N$ is chosen so that $N^k\geq n^{1-2/k}$; or (ii) compute a set $E'$ of at most $2\Delta n$ edges of $G$, such that $G\setminus E'$ is $(d,1/k)$-scattered. In the former case, we also allow the algorithm to compute a small subset $F\subseteq E(W^{N,\Delta}_k)$ of edges, that we call \emph{fake edges}, for which the algorithm does not need to provide an embedding. In addition to the above input, the algorithm is given a parameter $0<\delta<1$, that will determine the tradeoff between the lengths of the embedding paths (which will be bounded by $d\cdot k\cdot  2^{O(1/\delta^6)})$, and the running time of the algorithm (that will depend on $m^{\delta}$, where $m=|E(G)|$).
We now provide a formal definition of the \embedorscatter problem.

\begin{definition}[\embedorscatter problem]\label{def: embedorscatter}
	The input to the \embedorscatter problem is a simple $n$-vertex graph $G$ with no isolated vertices, and parameters $16\leq \Delta\leq n$, $d\geq 1$,  $4\leq k<(\log n)^{1/3}$, and $\frac{2}{(\log n)^{1/24}}<\delta<\frac{1}{400}$, such that $\Delta$, $k$ and $1/\delta$ are integers. We require that $n$ is sufficiently large, so that $\sqrt{\log n}>100\log\log n$ holds.
	Let $N=\floor{n^{1/k-1/k^2}}$ and let $\eta=n^{4/k}\cdot d\cdot k^{ck}\cdot 2^{c/\delta^6}$ for a large enough constant $c$.
		The output of the problem is one of the following:
	\begin{itemize}
		\item either a collection $F\subseteq E(W_k^{N,\Delta})$ of at most $\frac{N^k\cdot \Delta}{(512k)^{4k}}$ edges of $W_k^{N,\Delta}$ (that we call \emph{fake edges}), together with an embedding of $W_k^{N,\Delta}\setminus F$ into $G$ via paths of length at most $d\cdot k\cdot 2^{O(1/\delta^6)}$ that cause congestion at most $\eta$; or
		
		\item a subset $E'$ of at most $2n\Delta$ edges, such that $G\setminus E'$ is $(d,1/k)$-scattered.
	\end{itemize}
\end{definition}

The main goal of the current section is to provide an algorithm for solving the $\embedorscatter$ problem, that is summarized in the following theorem.

\begin{theorem}\label{thm: embedorscatter}
	There is a deterministic algorithm for the \embedorscatter problem, whose running time on input $(G,\Delta,k,d,\eta,\delta)$ is at most $O\left(m^{1+O(\delta+1/k)}\cdot d^2 \cdot 2^{O(1/\delta^6)} \right )$, where $m=|E(G)|$.
\end{theorem}

The remainder of this section is dedicated to the proof of \Cref{thm: embedors}. We start by defining a slightly different problem, called \embedors, and then show that it is sufficient to obtain an efficient algorithm for \embedors in order to prove \Cref{thm: embedorscatter}. We then provide an efficient algorithm for \embedors.


\subsection{EmbedOrSeparate Problem}

In this subsection, we define the \embedors problem, and state our main result for this problem, namely an algorithm that solves the problem efficiently. We then show that this algorithm can be used in order to prove \Cref{thm: embedorscatter}.

 Intuitively, in the \embedors problem, we are given a graph $G$ and a subset $T$ of its vertices that we call terminals. We are also given parameters $\Delta,N,k $, together with a distance parameter $d$. We require that $|T|=N^k$, and every vertex in $T$ has degree at least $\Delta$ in $G$. The goal is to either (i) compute a large subset $\tdel\subseteq T$ of terminals together with a relatively small set $\edel$ of edges of $G$ such that, for every terminal $t\in\tdel$, $B_{G\setminus E'}(t,d)$ contains relatively few terminals; or (ii) to compute an embedding of $W^{N,\Delta}_k$ into $G$, via paths that are relatively short and cause a relatively small congestion, such that the vertices of $W^{N,\Delta}_k$ are mapped to the terminals in $T$. In the latter case, we allow the algorithm to select a small subset $F\subseteq E(W^{N,\Delta}_k)$ of edges, that we call \emph{fake edges}, that do not need to be embedded, so the algorithm only needs to embed the non-fake edges of $W^{N,\Delta}_k$.  
In addition to the parameters $N,\Delta$ and $k$, that are used in the definition of the graph $W^{N,\Delta}_k$, and the distance parameter $d$, the algorithm is given as input a congestion parameter $\eta$, specifying the desired bound on the congestion caused by the embedding paths, and an additional parameter $0<\delta<1$. The latter parameter  plays a role in ensuring that the resulting algorithm is efficient, and is used in a tradeoff between the running time of the algorithm and the lengths of the embedding paths.
We now define the \embedors problem formally.

\begin{definition}[\embedors problem]\label{def: embedors}
The input to the \embedors problem is a graph $G$, a subset $T\subseteq V(G)$ of its vertices that we call \emph{terminals}, and parameters $\Delta\geq 16$, $N,k,d, \eta\geq 1$ and $\frac{2}{(\log (|T|))^{1/12}}<\delta<\frac{1}{400}$, such that $N,k,\Delta$ and $1/\delta$ are integers. We require that $|T|= N^{k}$, that every vertex in $T$ has degree at least $\Delta$ in $G$. We also require that $\eta>N^4\cdot k\cdot d\cdot 2^{c/\delta^6}$ holds, for a large enough constant $c$, and that $|T|\cdot \Delta$ is sufficiently large, so that  $(|T|\cdot \Delta)^{\delta^5/2}\geq 2\log(|T|\cdot \Delta)$ holds.

The output of the problem is one of the following:
\begin{itemize}
	\item either a collection $F\subseteq E(W_k^{N,\Delta})$ of at most $\frac{N^k\cdot \Delta}{(512k)^{4k}}$ edges of $W_k^{N,\Delta}$ (that we call \emph{fake edges}), together with an embedding of $W_k^{N,\Delta}\setminus F$ into $G$ via paths of length at most $d\cdot 2^{O(1/\delta^6)}$ that cause congestion at most $\eta$, such that the vertices of $W_k^{N,\Delta}$ are mapped to the vertices of $T$; or
	
	\item a subset $\tdel\subseteq T$ of at least $\frac{N^{k-1}}{k\cdot (512k)^{4k}}$ terminals, and a subset $\edel\subseteq E(G)$ of at most $N^{k+1}\cdot \frac{k\cdot \Delta\cdot d\cdot 2^{O(1/\delta^6)}}{\eta}$ edges, such that, for all $t\in \tdel$, $B_{G\setminus \edel}(t,d)$ contains at most $\frac{N^k}{2}$ vertices of $T$.
\end{itemize}
\end{definition}

The following theorem summarizes our algorithm for the $\embedors$ problem.

\begin{theorem}\label{thm: embedors}
	There is an algorithm for the \embedors problem, whose running time on input $(G,T,\Delta,N,k,d,\eta,\delta)$ is at most $O\left(m^{1+O(\delta)}\cdot\eta^2\right )$, where $m=|E(G)|$.
\end{theorem}

We provide the proof of \Cref{thm: embedors} in \Cref{subsec: algorithm for embedorseparate}, after we complete the proof of \Cref{thm: embedorscatter} using it.

\subsection{Algorithm for \embedorscatter -- Proof of \Cref{thm: embedorscatter}}

We assume that we are given an 
$n$-vertex graph $G$ with no isolated vertices, and parameters  $16\leq \Delta\leq n$, $d\geq 1$,  $2\leq k<(\log n)^{1/3}$, and $\frac{2}{(\log n)^{1/24}}<\delta<\frac{1}{400}$, such that $\Delta,k$ and $1/\delta$ are integers.  We also assume that $n$ is sufficiently large, so that $\sqrt{\log n}>100\log\log n$ holds.
Let $N=\floor{n^{1/k-1/k^2}}$ and let $\eta=n^{4/k}\cdot d\cdot k^{ck}\cdot 2^{c/\delta^6}$ for a large enough constant $c$.


The algorithm consists of a number of phases. Throughout the algorithm, we maintain a set $E'$ of deleted edges, that is initialized to $E'=\emptyset$, and then we may gradually add edges to $E'$. We will ensure that $|E'|\leq 2n\Delta$ always holds.
We also maintain a graph $G'=G\setminus E'$.
Let $d'=8kd$. We now describe the $i$th phase, for $i\geq 1$.

\paragraph{Description of the $i$th phase.}
We start with a preprocessing step, where, for every vertex $v\in V(G')$, if $\deg_{G'}(v)<\Delta$, then we add all edges incident to $v$ in $G'$ to $E'$, and delete them from $G'$. We continue this process until, for every vertex $v$ of $G'$, if $v$ is not isolated in $G'$, then its degree is at least $\Delta$. If $e$ is an edge that is added to $E'$ during this step, then we say that $e$ is added to $E'$ \emph{directly}. Notice that the total number of edges that are aver added to $E'$ directly is bounded by $n\cdot\Delta$. The preprocessing step can be implemented in time $\tilde O(m)$: we start by computing the degree of every vertex in $G'$, and then maintain a heap $H$, containing all non-isolated vertices of $G'$, with the key being the current degree of the vertex in $G'$. In each iteration, we remove the smallest-degree vertex $v$ from $H$. If $\deg_{G'}(v)<\Delta$, then we delete all edges incident to $v$ from $E'$, and we update the degrees of its neighbors, reinserting them into $H$. We can charge the time required to update the neighbors of $v$ and to reinsert them into $H$ to the edges incident to $v$ that were delted from $E'$, so that the total running time of the algorithm is $\tilde O(m)$.

In our second step, we apply the algorithm from \Cref{claim: scattered or large ball} to graph $G'$, with parameter $\eps=1/k$, and parameter $d$ remaining unchaged. Recall that the running time of the algorithm is $O(m^{1+\eps})=O(m^{1+1/k})$. If the algorithm correctly certified that $G'$ is $(d,1/k)$-scattered, then we terminate the algorithm, and return the current set $E'$ of edges. Therefore, we assume from now on that the algorithm from \Cref{claim: scattered or large ball} returned a vertex $v\in V(G')$ with $|B_{G'}(v,4d/\eps)|\geq n^{1-1/k}$. 

Let $T'=B_{G'}(v,4d/\eps)$. Clearly, every vertex $x\in T'$ has $\deg_{G'}(x)\geq \Delta$, and moreover, for every pair $x,y\in T'$ of vertices, $\dist_{G'}(x,y)\leq 8d/\eps=d'$.
Recall that ${N=\floor{n^{1/k-1/k^2}}}$, and so $N^k\leq n^{1-1/k}\leq |T'|$. We let $T\subseteq T'$ be an arbitrary subset of $N^k$ vertices, that we refer to as \emph{terminals} for the current phase.

Next, we set up an instance of the \embedors problem on graph $G'$, the set $T$ of its vertices that we defined before, parameter $d$ replaced by $d'$, and the remaining parameters $N,k,\eta$ and $\delta$ unchanged. We first verify that this is indeed a valid instance of the problem. 

From the definition of the \embedorscatter problem, $\Delta\geq 16$ and $k,\eta\geq 1$ hold. Additionally, $d'\geq d\geq 1$ must hold. From our construction, $|T|= N^{k}$, that every vertex in $T$ has degree at least $\Delta$ in $G$. Since $d'=8dk$, $N\leq n^{1/k}$, and, from the definition of the \embedorscatter problem, $\eta=n^{4/k}\cdot d\cdot k^{ck}\cdot 2^{c/\delta^6}$ for a large enough constant $c$, we get that $\eta>N^4\cdot k\cdot d'\cdot 2^{c/\delta^6}$ holds.
Since $N=\floor{n^{1/k-1/k^2}}$, while $4\leq k< (\log n)^{1/3}$ and $n$ is sufficiently large, we get that $N\geq 1$.
From the definition of the \embedorscatter problem, $\frac{2}{(\log n)^{1/24}}<\delta<\frac{1}{400}$, and $1/\delta$ is an integer. Since $|T|=N^k\geq \frac{n^{1-1/k}}{2^k}\geq \sqrt{n}$, we get that $\log (|T|)\geq \frac{\log n}{2}$, and so
$\frac{2}{(\log (|T|))^{1/12}}<\delta<\frac{1}{400}$ holds. 

Finally, we need to show that $(|T|\cdot \Delta)^{\delta^5/2}\geq 2\log(|T|\cdot \Delta)$ holds.
Recall that that $|T|=N^k=\floor{n^{1/k-1/k^2}}^k$. Therefore, $\frac{n^{1-1/k}}{2^k}\leq |T|\leq n^{1-1/k}$, and so:

\[
\begin{split}
 \left (|T|\cdot \Delta\right )^{\delta^5/2}&\geq \left(\frac{n^{1-1/k}\cdot \Delta}{2^k}\right )^{\delta^5/2} \\
 &\geq n^{\delta^5/4}\\
 &\geq n^{1/(\log n)^{1/4}}\\
 &=2^{(\log n)^{3/4}}\\
 &\geq 8\log n\\
 &
 \geq 4\log(n\cdot \Delta)\\
 &\geq 2\log(|T|\cdot \Delta). 
\end{split}
 \]

For the second inequality, we have used the fact that $k\leq (\log n)^{1/3}$, so $2^k\le n^{1/8}$. For the third inequality, we have used the fact that $\delta\geq \frac{2}{(\log n)^{1/{24}}}$. For the following inequality, we used the fact that $\sqrt{\log n}>100\log\log n$, and for the penultimate inequality we used the fact that $\Delta\leq n$.

We conclude that we obtain a valid instance $(G',T,\Delta,N,k,d',\eta,\delta)$ of the \embedors problem. Next, we apply the algorithm from \Cref{thm: embedors} to this problem instance. Recall that the running time of the algorithm is $O\left(m^{1+O(\delta)}\cdot\eta^2\right )$. We now consider two cases. The first case happens if the algorithm from \Cref{thm: embedors}  returned a collection $F\subseteq E(W_k^{N,\Delta})$ of at most $\frac{N^k\cdot \Delta}{(512k)^{4k}}$ edges of $W_k^{N,\Delta}$, together with an embedding of $W_k^{N,\Delta}\setminus F$ into $G'$ via paths of length at most $d'\cdot 2^{O(1/\delta^6)}\leq d\cdot k\cdot 2^{O(1/\delta^6)}$ that cause congestion at most $\eta$. In this case  
we terminate the algorithm, and return the set $F$ of fake edges, and the resulting embedding of $W^{N,\Delta}_k\setminus F$ into $G$.

Assume now that the second case happens, that is, the algorithm from \Cref{thm: embedors}  returned 
a subset $\tdel\subseteq T$ of at least $\frac{N^{k-1}}{k\cdot (512k)^{4k}}$ terminals, and a subset $\edel_i\subseteq E(G')$ of at most $N^{k+1}\cdot \frac{k\cdot \Delta\cdot d'\cdot 2^{O(1/\delta^6)}}{\eta}$ edges, such that, for all $t\in \tdel$, $B_{G'\setminus \edel}(t,d')$ contains at most $\frac{N^k}{2}$ vertices of $T$. In the latter case, for the sake of analysis, we define a set $\Pi_i$ of pairs of terminals $(t,t')\in T$, such that $\dist_{G'\setminus\edel}(t,t')>d'$. Clearly, $|\Pi_i|\geq |\tdel|\cdot \frac{N^k}{2}\geq \frac{N^{2k-1}}{2k\cdot (512k)^{4k}}$. Since $N=\floor{n^{1/k-1/k^2}}$, we get that $N^k\geq \frac{n^{1-1/k}}{2^k}$, and so $|\Pi_i|\geq \frac{n^{2-3/k}}{(512k)^{6k}}$. We add the edges of $\edel$ to $E'$ and delete them from $G'$. We say that the edges of $\edel$ are added to $E'$ \emph{indirectly}. We then continue to the following phase.

Notice that, for every pair $(t,t')\in \Pi_i$ of vertices, $\dist_{G'}(t,t')>d'$ holds from now on. Therefore, in subsequent phases, we may never include both $t$ and $t'$ in the same set $T$ of terminals, since every pair of terminals in $T$ must be at distance at most  $d'$ from each other in $G'$. It then follows that all sets $\Pi_1,\Pi_2,\ldots$ of pairs of vertices are disjoint, and so the number of phases in the algorithm is bounded by $n^{3/k}\cdot (512k)^{6k}$. The number of edges added to set $E'$ in every phase is bounded by:
$N^{k+1}\cdot \frac{k^2\cdot \Delta\cdot d\cdot 2^{O(1/\delta^6)}}{\eta}\leq  \frac{n\cdot k^2\cdot \Delta\cdot d\cdot 2^{O(1/\delta^6)}}{\eta}$.
Therefore, at the end of the algorithm, the number of edges that are added indirectly to $E'$ is bounded by:

\[
\begin{split}
|E'|&\leq \frac{n\cdot k^2\cdot \Delta\cdot d\cdot 2^{O(1/\delta^6)}}{\eta}\cdot n^{3/k}\cdot (512k)^{6k}    \\
&\leq \frac{n^{1+3/k}\cdot \Delta\cdot d\cdot (512k)^{7k}\cdot 2^{O(1/\delta^6)}}{\eta} \\
&<n\cdot \Delta,
\end{split}
\]

since  $\eta=n^{4/k}\cdot d\cdot k^{ck}\cdot 2^{c/\delta^6}$ for a large enough constant $c$.

Recall that the preprocessing step takes time $\tilde O(m)$, and the running time of the algorithm from \Cref{claim: scattered or large ball} is $O(m^{1+1/k})$. The running time of the algorithm from \Cref{thm: embedors} is:

\[
O\left(m^{1+O(\delta)}\cdot\eta^2\right )\leq O\left(m^{1+O(\delta+1/k)}\cdot d^2 \cdot k^{O(k)}\cdot 2^{O(1/\delta^6)} \right )\leq O\left(m^{1+O(\delta+1/k)}\cdot d^2 \cdot 2^{O(1/\delta^6)} \right ).
\]

 since $\eta\leq n^{4/k}\cdot d\cdot k^{O(k)}\cdot 2^{O(1/\delta^6)}$. Additionally, we have used the fact that $k<(\log n)^{1/3}$, so $k^k\leq n^{1/k}$.

We conclude that the running time of a single phase is bounded by $O\left(m^{1+O(\delta+1/k)}\cdot d^2 \cdot 2^{O(1/\delta^6)} \right )$. Since, as observed before, the number of phases is bounded by $n^{3/k}\cdot k^{O(k)}\leq n^{O(1/k)}$, we get that the total running time of the algorithm is bounded by $O\left(m^{1+O(\delta+1/k)}\cdot d^2 \cdot 2^{O(1/\delta^6)} \right )$.

\subsection{Algorithm for \embedors -- Proof of \Cref{thm: embedors}}
\label{subsec: algorithm for embedorseparate}

In this subsection, we provide an algorithm for the \embedors problem,  proving \Cref{thm: embedors}. Throughout the proof, we denote $m=|E(G)|$ and $n=|V(G)|$. For convenience, we will denote $W^{N,\Delta}_k$ by $W_k$ in the remainder of this subsection.
We start by mapping every vertex of $W_k$ to a distinct vertex of $T$.
 For convenience, we will not distinguish between vertices of $W_k$ and the terminals to which they are mapped.
We initialize $F=\emptyset$, and we initialize the embedding $\qset$ of $E(W_k)\setminus F$ into $G$ by setting $\qset=\emptyset$. Over the course of the algorithm, we will gradually add paths to $\qset$, where every path $Q(\hat e)$ added to $\qset$ is an embedding of a distinct edge $\hat e\in E(W_k)$, and has length at most $d\cdot 2^{O(1/\delta^6)}$. We will ensure that every edge of $G$ participates in at most $\eta$ paths in $\qset$. To this end, for every edge $e\in E(G)$, we maintain a counter $n_e$, that counts the number of paths in $\qset$ that use the edge $e$. Once the counter $n_e$ reaches $\eta$, we will ensure that the edge will never participate in any subsequent embedding paths.
We initialize all such counters $n_e$ to $0$.

Throughout the algorithm, we maintain a set $\edel$ of edges that have been deleted from $G$. We also maintain a set $\tdel$ of terminals, such that, for every terminal $t\in \tdel$, $B_{G\setminus \edel}(t,d)$ contains at most $\frac{N^k}{2}$ vertices of $T$. We will gradually add vertices to $\tdel$, and, whenever $|\tdel|$ reaches $\frac{N^{k-1}}{k\cdot (512k)^{4k}}$, the algorithm terminates with the output $(\edel,\tdel)$. Therefore, throughout the algorithm, $|\tdel|\leq \frac{N^{k-1}}{k\cdot (512k)^{4k}}$ holds (except possibly when the algorithm terminates).

 The set $\edel$ of edges that the algorithm maintains is partitioned into two subsets, that we denote by $\edel_1$ and $\edel_2$, respectively.
An edge $e\in E(G)$ may be added to  $\edel_1$ only if it participates in at least $\eta/2$ paths in the current set $\qset$. Since $|E(W_k)|\leq N^k\cdot k\cdot \Delta$, and since we ensure that the length of every embedding path in $\qset$ is bounded by 
$d\cdot 2^{O(1/\delta^6)}$, we get that, throughout the algorithm, the cardinality of $\edel_1$ is bounded as follows:

\begin{equation}\label{eq: card of edel1}
|\edel_1|\leq \frac{N^k\cdot k\cdot \Delta\cdot d\cdot 2^{O(1/\delta^6)}}{\eta}
\end{equation}

Over the course of the algorithm, we will add paths to $\qset$ one by one, and we will never delete paths from $\qset$, so that the set $\edel_1$ of edges is incremental.

The second set, $\edel_2$ of edges, may contain some additional edges, and we ensure that it is also incremental.
We will ensure that, throughout the algorithm: 

\begin{equation}\label{eq: bound on edel 2 prop}
|\edel_2|\leq |\tdel|\cdot N\cdot \Delta \cdot\frac{32d}{\eta}
\end{equation}

 holds. Since $|\tdel|\leq N^{k-1}$ over the course of the algorithm, we get that, throughout the algorithm:

 \begin{equation}\label{eq: bound on edel}
 |\edel|\leq |\edel_1|+|\edel_2|\leq N^k\cdot\frac{k\cdot \Delta\cdot d\cdot 2^{O(1/\delta^6)}}{\eta} +N^{k}\cdot \Delta\cdot\frac{32d}{\eta}\leq  N^{k}\cdot\frac{k\cdot \Delta\cdot d\cdot 2^{O(1/\delta^6)}}{\eta}.
 \end{equation}
 
 Since, at the end of the algorithm, $|\tdel|\leq N^k$, we get that, at the end of the algorithm:
 \begin{equation}\label{eq: bound on edel end}
|\edel|\leq |\edel_1|+|\edel_2|\leq N^k\cdot\frac{k\cdot \Delta\cdot d\cdot 2^{O(1/\delta^6)}}{\eta} +N^{k+1}\cdot \Delta\cdot\frac{32d}{\eta}\leq  N^{k+1}\cdot\frac{k\cdot \Delta\cdot d\cdot 2^{O(1/\delta^6)}}{\eta},
 \end{equation}

as required.
The algorithm also maintains a set $F\subseteq E(W_k)$ of fake edges. At the beginning of the algorithm, we initialize $F=\emptyset$, and then, over the course of the algorithm, we may add edges to $F$. However, an edge of $W_k$ may only be added to $F$ if at least one of its endpoints currently lies in $\tdel$. Since $|\tdel|\leq \frac{N^{k-1}}{k\cdot (512k)^{4k}}$ holds throughout the algorithm, and since the degree of every vertex in $W_k$ is at most $N\cdot k\cdot \Delta$, this guarantees that $|F|\leq   \frac{N^{k-1}}{k\cdot (512k)^{4k}}\cdot N\cdot k\cdot \Delta\leq  \frac{N^k\cdot \Delta}{ (512k)^{4k}}$ always holds, as required (if, at the end of the algorithm, $|\tdel|\geq \frac{N^{k-1}}{k\cdot (512k)^{4k}}$, then the algorithm's output is $(\tdel,\edel)$, so the cardinality of the set $F$ is unimportant in that case).

We use a parameter ${z=(2m)^{32\delta}\cdot k\cdot  d\cdot 2^{c/\delta^6}}$, where $c$ is a large enough constant, for example that used in the condition on $\eta$ in the definition of the problem. 
The algorithm consists of a number of phases. We now provide a high-level description of a single phase.

\subsubsection*{High-Level Description of a Single Phase} 
At the beginning of the $i$th phase, we select an arbitrary subset $\hat E_i\subseteq E(W_k)\setminus F$ of $\floor{\frac{|E(W^k)|}{z}}$ edges that were not embedded yet into $G$, and whose endpoints do not currently lie in $\tdel$ (if $E(W_k)\setminus F$ contains fewer such edges, then we let $\hat E_i$ denote the set of all such remaining edges). Then, over the course of the $i$th phase, we will embed a subset $\hat E_i'\subseteq \hat E_i$ of the edges into $G$. We denote by $A'_i\subseteq E(G)$ the set of edges $e\in E(G)$, such that, prior to the i$th$ phase, $e$ participated in fewer than $\eta/2$ embedding paths, but after the $i$th phase is completed, $e$ participates in at least $\eta/2$ embedding paths (we will ensure that no edge participates in more than $\eta$ of the embedding paths). 
We will also compute another subset $A''_i\subseteq E(G)$ of edges, and we will identify a subset $\tdel_i\subseteq T\setminus\tdel$ of terminals, such that, for all $t\in \tdel_i$, $B_{G\setminus(\edel\cup A'_i\cup A''_i)}(t,d)$ contains at most $N^k/2$ vertices of $T$. We will ensure that the following hold:

\begin{properties}{P}
	\item either $|\tdel_i|\geq \frac{N^{k-1}}{m^{O(\delta^3)}}$ and $|A''_i|\leq |\tdel_i|\cdot N\cdot \Delta\cdot \frac{32d}{\eta}$, in which case we say that the $i$th phase is \emph{bad}; or
	
	\item $A''_i=\emptyset$, and every edge in $\hat E_i\setminus \hat E'_i$ has an endpoint in $\tdel_i$, in which case we say that the $i$th phase is \emph{good}.
\end{properties}

In either case, we add the edges of $A'_i$ to $\edel_1$, and the edges of $A''_i$ to $\edel_2$. We then update $\edel=\edel_1\cup \edel_2$, and add the vertices of $\tdel_i$ to $\tdel$. 
Notice that if, before the current phase, Inequality \ref{eq: bound on edel 2 prop} held, then it continues to hold at the end of the current phase.
Lastly, we add the embeddings of the edges in $\hat E'_i$ that our algorithm computed to $\qset$. If the $i$th phase is good, then the edges of  $\hat E_i\setminus \hat E'_i$ are added to the set $F$ of fake edges. Additionally, every edge of $W_k$ with at least one endpoint in $\tdel$ is added to $F$. If $|\tdel|\geq \frac{N^k}{k\cdot(512k)^{4k}}$, then we terminate the algorithm and return $(\tdel,\edel)$. If all edges of $W_k\setminus F$ are currently embedded, then we terminate the algorithm, and return the current set $F\subseteq E(W^k)$, and the current embedding $\qset$ of $W^k\setminus F$, that is guaranteed to have all required properties. Otherwise, we continue to the next iteration.

This completes the high-level description of a single phase. Before we provide an algorithm to implement a single phase, we bound the number of phases in the following simple observation.

\begin{observation}\label{obs: number of phases}
	The number of phases in the algorithm is bounded by $O\left(m^{O(\delta)}\cdot k\cdot d\cdot 2^{O(1/\delta^6)}\right )$.
\end{observation}
\begin{proof}
	Since every bad phase adds at least $\frac{N^{k-1}}{m^{O(\delta^3)}}$ new terminals to set $\tdel$, the number of bad phases is bounded by $m^{O(\delta^3)}$. Notice that, if the $i$th phase is good, then, at the end of the phase, every edge of $\hat E_i$ is either embedded into $G$, or added to $F$. Since $|\hat E_i|=\floor{\frac{|E(W^k)|}{z}}$ (except when $i$ is the last phase), there are at most $2z\leq m^{O(\delta)}\cdot k\cdot d\cdot 2^{O(1/\delta^6)}$ good phases.

	 Therefore, the total number of phases is bounded by $O\left(m^{O(\delta)}\cdot k\cdot d\cdot 2^{O(1/\delta^6)}\right )$.
\end{proof}

We now turn to describe an algorithm for implementing the $i$th phase, for $i\geq 1$. We assume that we are given a set $\hat E_i\subseteq E(W_k)\setminus F$ of at most $\floor{\frac{|E(W^k)|}{z}}$ edges that are not embedded yet into $G$, and whose endpoints do not currently lie in $\tdel$. For convenience, we will denote $\hat E_i$ and $\hat E_i'$ by $\hat E$ and $\hat E'$, respectively. The algorithm consists of two steps. In the first step, we compute a large well-connected graph $X$, and we attempt to embed it into $G$, using \Cref{cor: HSS witness}. If we fail to do so, then we will compute a subset $A''_i\subseteq E(G)$ of edges, and a subset $\tdel_i\subseteq T\setminus\tdel$ of terminals, such that, for all $t\in \tdel_i$, $B_{G\setminus(\edel\cup A''_i)}(t,d)$ contains at most $|T|/2$ terminals. In this case, we terminate the algorithm for the current phase with $\hat E'=\emptyset$ and $A'_i=\emptyset$. If we manage to successfully compute a well-connected graph $X$ and embed it into $G$, then we continue to the second step, in which we will employ the algorithm for \APSP in well-connected graphs from \Cref{thm: APSP in HSS full}, together with the standard Even-Shiloach Tree data structure, in order to compute an embedding of a large subset of edges of $\hat E$. If we fail to do so, then we will again compute the set $\tdel_i$ of terminals and a set $A''_i$ of edges with the required properties.
We now formally describe each of the two steps in turn.

\subsubsection{Step 1: Embedidng a Well-Connected Graph into $G$}

We assume that $\edel$ contains every edge $e$ that participates in at least $\eta/2$ paths of $\qset$ at the beginning of the phase.
Let $G'=G\setminus \edel$, and we let  $T'\subseteq T\setminus \tdel$
contain all terminals $t\in T\setminus \tdel$, such that the degree of $t$ in $G'$ is at least $\Delta/2$.
Recall that from Inequality \ref{eq: bound on edel}, and since we have assumed that $\eta>N^4\cdot k\cdot d\cdot 2^{c/\delta^6}$ for a large enough constant $c$, we get that: 

\[
|\edel|\leq \frac{N^k\cdot k\cdot \Delta\cdot d\cdot 2^{O(1/\delta^6)}}{\eta}\leq \frac{N^k\cdot \Delta}{16}.
\]

Since the algorithm did not terminate yet, $|\tdel|\leq \frac{N^{k-1}}{k\cdot(512k)^{4k}}$, and  so $|T\setminus \tdel|\geq \frac{N^k}{2}$ holds. Since the degree of every terminal $t\in T$ in $G$ is at least $\Delta$, we then get that $|T'|\geq \frac{N^k}{4}$ must hold.

We start by slightly modifying the graph $G'$, as follows.  Consider an edge $e=(u,v)\in E(G')$. If both endpoints of $e$ lie in $T'$, then we replace $e$ with a path $(u,u',v',v)$, where $u'$ and $v'$ are new vertices that we call \emph{copies} of $u$ and $v$, respectively. We also refer to the edges $(u,u'),(u',v')$ and $(v',v)$ as \emph{artificial edges corresponding to edge $(u,v)$}. If exactly one endpoint of $e$ (say $u$) lies in $T'$, then we replace $e$ with a path $(u,u',v)$, and we call $u'$ a copy of $u$, and we call all edges on the new path artificial edges corresponding to $(u,v)$. Otherwise, we do not perform any transformation of $e$.

Let $G''$ denote the resulting graph, and let $\tilde T\subseteq V(G'')$ be the set of vertices obtained by adding, for every terminal $t\in T'$, exactly $\ceil{\Delta/2}$ copies of $t$ into $\tilde T$. We refer to the vertices in $\tilde T$ as \emph{pseudo-terminals}, and we denote $q=|\tilde T|$. Notice that $q=|T'|\cdot \ceil{\frac{\Delta}{2}}\geq\frac{N^k\cdot \Delta}{8}$. Observe also that $|E(G'')|\leq O(m)$.

Let $\hat N=\floor{q^{\delta}}$, so $ \frac{q^{\delta}}{2}\leq \hat N\leq q^{\delta}$. Notice that:

\[\hat N^{1/\delta}\geq \frac{q}{2^{1/\delta}}\geq \frac{|T|\cdot \Delta}{8\cdot 2^{1/\delta}}. 
\]

From the problem definition,
$\delta>\frac{2}{(\log (|T|))^{1/12}}>\frac{2}{(\log q)^{1/12}}$, and so $\frac{2}{(\log q)^{1/12}}< \delta<\frac{1}{400}$ holds.

We apply the algorithm from \Cref{cor: HSS witness} to graph $G''$, the set $\tilde T$ of its vertices, with parameter $d$ replaced by $\hat d=16d$, parameter $\eta$ remaining unchanged, and parameter $\eps$ replaced by $\delta$.

We now consider two cases.
In the first case, the algorithm returns a graph $X$ with $V(X)\subseteq \tilde T$, and $|V(X)|=\hat N^{1/\delta}\geq \frac{|T|\cdot \Delta}{8\cdot 2^{1/\delta}}$, whose maximum degree is at most   $q^{32\delta^3}$, together with an embedding $\pset$ of $X$ into $G''$ via paths of length at most $\hat d=16d$, that cause congestion at most $ \eta\cdot q^{32\delta^3}\leq \eta\cdot \left(N^k\cdot \Delta\right )^{32\delta^3}\leq 2\eta\cdot m^{32\delta^3}$, and a level-$(1/\delta)$ \HSS for $X$, such that $X$ is $(\td,\eta')$-well-connected with respect to the set $S(X)$ of vertices defined by the support structure, where $\eta'=\hat N^{6+256\delta}$, and $\td=2^{\tilde c/ \delta^5}$, and $\tilde c $ is the constant used in the definition of the \HSS. In this case, we say that the first step has successfully embedded a well-connected graph into $G'$, and continue to the second step.

In the second case, the algorithm from \Cref{cor: HSS witness} returns a pair $\tilde T_1,\tilde T_2\subseteq \tilde T$ of disjoint subsets of pseudo-terminals of equal cardinality, and a set $E'$ of edges of $G''$ of cardinality at most:

\[|E'|\leq |\tilde T_1|\cdot \frac{16 d}{\eta}\leq \frac{|\tilde T_1|}{2},\]

since, from the problem definition, $\eta>N^4\cdot k\cdot d\cdot 2^{c/\delta^6}$ holds, for a large enough constant $c$.

The algorithm also ensures that  for every pair $t\in \tilde T_1,t'\in \tilde T_2$ of pseudo-terminals, $\dist_{G''\setminus E'}(t,t')>\hat d=16d$.
Finally, the algorithm ensures that:

\[|\tilde T_1|\geq \frac{q^{1-4\delta^3}}{4}\geq\frac{(|T'|\cdot \Delta/2)^{1-4\delta^3}}{4}\geq\frac{N^k\cdot \Delta}{32\cdot (N^k\cdot \Delta)^{4\delta^3}}\geq \frac{N^k\cdot \Delta}{64\cdot m^{4\delta^3}}.\]

We construct a set $E''$ of edges in $G$ as follows: for every edge $e\in E'$, if $e\in E(G)$, then we add $e$ to $E'$. Otherwise, $e$ is an artificial edge, corresponding to some edge $e'\in E(G)$. We then add $e'$ to $E''$. It is easy to verify that $|E''|\leq |E'|$.

Next, we construct two sets $T_1,T_2$ of terminals, as follows. 
First, we let $\tilde T_1'\subseteq \tilde T_1$ be a set of pseudo-terminals obtained from $\tilde T_1$ by deleting all pseudo-terminals that are incident to the edges of $E'$. 
We similarly define a set $\tilde T_2'\subseteq \tilde T_2$ of pseudo-terminals.
Since $|E'|<\frac{|\tilde T_1|}{2}$ and $|\tilde T_1|=|\tilde T_2|$, we get that $|\tilde T_1'|,|\tilde T_2'|\geq \frac{|\tilde T_1|}2$.
For every pseudo-terminal $\tilde t\in \tilde T_1'$, we add the corresponding terminal $t$ to $T_1$. Similarly, for every pseudo-terminal $\tilde t'\in \tilde T_2'$, we add the corresponding terminal $t'$ to $T_2$. Clearly:

\[|T_1|\geq \frac{|\tilde T'_1|}{\Delta}\geq \frac{|\tilde T_1|}{2\Delta} \geq \frac{N^{k}}{128m^{4\delta^3}}.  \]

Using the same reasoning, $|T_2|\geq  \frac{N^{k}}{128m^{4\delta^3}}$.  Assume w.l.o.g. that $|T_2|\geq |T_1|$.
We delete vertices from $T_2$, until the cardinalities of both sets become equal.
We also get that:

\[|E''|\le |E'|\leq   |\tilde T_1|\cdot \frac{16 d}{\eta}\leq |\tilde T'|\cdot \frac{32d}{\eta}\leq |T_1|\cdot \Delta\cdot \frac{32d}{\eta}.\]

We need the following simple observation.

\begin{observation}\label{obs: get distancing}
	$\dist_{G'\setminus E''}(T_1,T_2)>2d$.
\end{observation}
\begin{proof}
	Assume for contradiction that the claim is false. Then there are terminals $t\in T_1$, $t'\in T_2$, and a path $P$ in $G'\setminus E''$, connecting $t$ to $t'$, such that the length of $P$ is at most $d$.
	
	Let $P'$ be the path corresponding to $P$ in graph $G''$. Since graph $G''$ is obtained from $G'$ by subdividing each of its edges by at most $2$ new vertices, the length of $P$ in $G''$ is at most $8d$, and $P$ connects $t$ to $t'$ in $G''$. We claim that $P'$ must avoid the edges of $E'$. Indeed, consider any edge $e\in E'$, and assume for contradiction that $P'$ contains $e$. If $e\in E(G')$, then $e$ is an edge of $E''$ that lies on $P'$ and hence on $P$. Since $P$ is disjoint from $E''$, this is impossible. Therefore, $e$ is an artificial edge that corresponds to some edge $e'\in E(G')$. But then $e'\in E''$, and $e'$ does not lie on $P$. It is then impossible that $e'$ lies on $P'$. We conclude that $P'$ is a path of length at most $8d$ in $G'\setminus E'$, connecting $t$ to $t'$.
	
	Lastly, let $\tilde t$ be a pseudo-terminal in $\tilde T_1'$ that corresponds to $t$, and let $\tilde t'$ be a pseudo-terminal in $\tilde T_2'$ that corresponds to $t'$. From the definition of the sets $\tilde T_1',\tilde T_2'$ of pseudo-terminals, edges $(\tilde t,t)$ and $(t',\tilde t')$ are disjoint from set $E'$. Therefore, if we let $P''$ be the path obtained from $P'$ by appending $(\tilde t,t)$ at the beginning and $(t',\tilde t')$ at the end of the path, then $P''$ is a path of length $8d+2\leq 10d<\hat d$ in $G''\setminus E'$, connecting a pair $\tilde t\in \tilde T_1$, $\tilde t'\in \tilde T_2$ of terminals, a contradiction.
\end{proof}


Let $S\subseteq T$ be the set of all terminals lying in $B_{G'\setminus E''}(T_1,d)$, and let $S'\subseteq T$ be the set of all terminals lying in 
$B_{G'\setminus E''}(T_2,d)$. Since $\dist_{G'\setminus E''}(T_1,T_2)>2d$, it must be the case that either $|S|\leq \frac{N^k}{2}$, or $|S'|\leq \frac{N^k}{2}$. Assume w.l.o.g. that it is the former. Then we let $\tdel_i=T_1$, and $A''_i=E''$. We also let $\hat E'_i=\emptyset$, and $A'_i=\emptyset$. We return $\tdel_i,A'_i,A''_i$ and $\hat E'_i$ as the outcome of the phase, and terminate the phase as a bad phase. Since $|S|\leq \frac{N^k}{2}$, it is easy to verify that, for every vertex $t\in \tdel_i$, 
$B_{G\setminus(\edel\cup A''_i)}(t,d)=B_{G'\setminus A''_i}(t,d)$ contains at most $N^k/2$ vertices of $T$. As we have shown, $|A''_i|=|E''|\leq |T_1|\cdot \Delta\cdot\frac{32d}{\eta}=|\tdel_i|\cdot \Delta\cdot\frac{32d}{\eta}$, and $|\tdel_i|=|T_1|\geq \frac{N^k}{m^{O(\delta^3)}}$.


We now bound the running time of the first step. The running time of the algorithm from \Cref{cor: HSS witness} is bounded by: 

\[
\begin{split}
O\left (q^{1+O(\delta)}+|E(G)|\cdot q^{O(\delta^3)}\cdot(\eta+d\log n)\right )&\leq O\left((N^k\cdot \Delta)^{1+O(\delta)}+|E(G)|\cdot (N^k\cdot \Delta)^{O(\delta^3)}\cdot(\eta+d\log n)\right )\\
&\leq O\left(m^{1+O(\delta)}\cdot \eta\right ).
\end{split}
\]

The remaining time required to compute the set $E''$ of edges and the sets $T_1,T_2$ of terminals, together with the time required to compute the graph $G''$ and the sets $S$ and $S'$ of terminals, is asymptotically bounded by the above running time. Therefore, the total running time of the first step of the algorithm is at most: $O\left(m^{1+O(\delta)}\cdot \eta\right )$.

\subsubsection{Step 2: Computing the Embedding of the Edges of $\hat E$}

We assume from now on that the algorithm from Step 1
computed a graph $X$ with $V(X)\subseteq \tilde T$, and $|V(X)|=\hat N^{1/\delta}\geq \frac{|T|\cdot \Delta}{8\cdot 2^{1/\delta}}$, whose maximum degree is at most   $q^{32\delta^3}$, together with an embedding $\pset$ of $X$ into $G'$ via paths of length at most $16d$ that cause congestion at most $ 2\eta\cdot m^{32\delta^3}$, and a level-$(1/\delta)$ \HSS for $X$, such that $X$ is $(\td,\eta')$-well-connected with respect to the set $S(X)$ of vertices defined by the support structure, where $\eta'=\hat N^{6+256\delta}$, and $\td=2^{\tilde c/ \delta^5}$, and $\tilde c $ is the constant used in the definition of the \HSS. 
Our algorithm maintains two main data structures.

\paragraph{First Data Structure: \APSP in $X$.}
For our fist data structure, we use the algorithm for \APSP in well-connected graphs from \Cref{thm: APSP in HSS full}, 
that is applied to graph $X$, with parameter $N$ replaced by $\hat N$ and parameted $\eps$ replaced by $\delta$. We exploit the level-$(1/\delta)$ \HSS for  $X$ that was computed in Step 1 of the algorithm. Recall that we are guaranteed that $X$ is $(\td,\eta')$-well-connected with respect to the set $S(X)$ of vertices defined by the \HSS. In order to be able to use the algorithm from \Cref{thm: APSP in HSS full}, we need to verify that $\frac{\hat N^{\delta^4}}{\log \hat N}\geq 2^{128/\delta^6}$ holds.
Indeed, recall that  $ \frac{q^{\delta}}{2}\leq \hat N\leq q^{\delta}$, and that $q=|T'|\cdot\ceil{\Delta/2}\geq \frac{|T|\cdot \Delta}{8}$. Therefore:

\begin{equation}\label{eq: bound}
\frac{\hat N^{\delta^4}}{\log \hat N}\geq \frac{(|T|\cdot \Delta)^{\delta^5}}{8\delta\log(|T|\cdot\Delta)}\geq (|T|\cdot \Delta)^{\delta^5/2}\geq  2^{128/\delta^6},
\end{equation}

since
$(|T|\cdot \Delta)^{\delta^5/2}\geq 2\log(|T|\cdot \Delta)$
from the problem definition. We have also used the fact that $\frac{2}{\delta}\leq (\log (|T|))^{1/12}$, so, in particular, $\log |T|\geq \frac{2^{12}}{\delta^{12}}$, and $|T|\geq 2^{2^{12}/\delta^{12}}$. 

Let $\Lambda=|V(X)|^{1-10\delta}$, and recall that $|V(X)|\geq\frac{|T|\cdot \Delta}{8\cdot 2^{1/\delta}}$, so that: 

\begin{equation}\label{eq: bound on Lambda}
\Lambda\geq \frac{(|T|\cdot \Delta)^{1-10\delta}}{8\cdot 2^{1/\delta}}\geq (|T|\cdot \Delta)^{1-11\delta},
\end{equation}

since $8\cdot 2^{1/\delta}\leq 2^{256/\delta^{10}}\leq (|T|\cdot \Delta)^{\delta}$, where the last inequality follows from Equation \ref{eq: bound}.

Recall that the algorithm from \Cref{thm: APSP in HSS full} maintains a set $S'(X)$ of vertices of $X$, as graph $X$ undergoes a sequence of at most $\Lambda$ edge deletions, such that set $S'(X)$ is decremental: vertices may leave it but not join it. Moreover, the algorithm ensures that, at all times:

\begin{equation}\label{eq: bound on core}
|S'(X)|\geq \frac{|V(X)|}{2^{4/\delta}}\geq\frac{|T|\cdot |\Delta|}{2^{6/\delta}}
\end{equation}
 holds (we have used the fact that
 $|V(X)|\geq \frac{|T|\cdot \Delta}{8\cdot 2^{1/\delta}}$).
We let $S''(X)\subseteq T$ be a set that contains every terminal $t\in T$, such that at least one pseudo-terminal representing $t$ lies in $S'(X)$. It is easy to verify that $S''(X)$ is a decremental set, and, since every terminal has at most $\Delta$ corresponding pseudo-terminals, throughout the algorithm:

\begin{equation}\label{eq: bound on core2}
|S''(X)|\geq \frac{|T|}{2^{6/\delta}}
\end{equation} 

holds.
We may sometimes refer to the vertices of $S''(X)$ as the \emph{core}.
The algorithm from \Cref{thm: APSP in HSS full} supports short-path queries between vertices of $S'(X)$: given a pair $x,y\in S'(X)$ of vertices, it returns a path $P$ connecting $x$ to $y$ in the current graph $X$, whose length is at most $2^{O(1/\delta^6)}$, in time $O(|E(P)|)$. 
We denote the data structure maintain by the algorithm from \Cref{thm: APSP in HSS full} by $\DS_1$. 
The total update time of the algorithm from \Cref{thm: APSP in HSS full} is bounded by:

\[ O( |E(X)|^{1+O(\delta)})\leq O\left(\left (|E(G)|^{1+O(\delta^3)} \right )^{1+O(\delta)}\right )\leq O\left(|E(G)|^{1+O(\delta)}\right ),\]

since maximum degree in $X$ is at most $q^{32\delta^3}\leq (|T|\cdot \Delta)^{O(\delta^3)}\leq |E(G)|^{O(\delta^3)}$, and $|V(X)|\leq |T|\cdot\Delta\leq |E(G)|$.

\paragraph{Second Data Structure: \EST.}
For our second data structure, we construct a dynamic graph $H$, as follows. We start with the graph $G'$, and we add a source vertex $s$ to it, that connects with an edge to every vertex in set $S''(X)$ -- the core set maintained by $\DS_1$. As the algorithm progresses, whenever a vertex $v$ is deleted from set $S''(X)$, we delete the edge $(s,v)$ from $H$. Additionally, we may delete some edges from $G'$ and from $H$ directly -- these are edges that participate in too many of the embedding paths that we construct. Our second data structure, that we denote by $DS_2$, maintains an \EST in graph $H$, rooted at the source vertex $s$, up to depth $2d+1$. Recall that the total update time of this data structure is $O(|E(H)|\cdot d)=O(|E(G)|\cdot d)$. The data structure also maintains a set $U$ of vertices of $G$, that contains every vertex $u$ with $\dist_H(s,u)>2d+1$.

We are now ready to describe our algorithm for the second step. We start by initializing data structures $\DS_1$ and $\DS_2$. For every edge $e\in E(G')$, we initialize a list $L(e)$ of all edges $e'\in E(X)$, whose embedding path $P(e')\in \pset$ contains either $e$ or one of the artificial edges that subdivide $e$. The time that is required to initialize all these lists is asymptotically bounded by the running time of Step 1, that computed the embedding $\pset$ of $X$ into $G''$.

Recall that $G'=G\setminus \edel$.
For every edge $e\in E(G')$, we let $n_e$ the number of paths in the current set $\qset$ in which $e$ participates. We initialize $\edel_i=\emptyset$. For every edge $e\in E(G')$, once the number of embedding paths in $\qset$ that use $e$ reaches $\eta$, we add $e$ to $\edel_i$. We also initialize the set $F_i$ of new fake edges to $\emptyset$.

Next, we process the edges $\hat e\in \hat E$ one by one. 
Consider any such edge $\hat e=(u,v)$. If either of the vertices $u$ or $v$ lies in $U$, then we add $e$ to the set $F_i$ of the fake edges. Otherwise, we use the \EST data structure to compute a path $P_1$ of length at most $2d$ connecting $u$ to some vertex $u'\in S''(X)$ (by computing a path connecting $u$ to $s$, whose penultimate vertex must lie in $S''(X)$), and another path $P_2$ of length at most $2d$ connecting $v$ to some vertex $v'\in S''(X)$. Both paths can be computed in time $O(d)$. 
Recall that some pseudoterminal $\tilde u'$ correpsonding to $u'$, and some pseudoterminal $\tilde v'$ correpsonding to $v'$ must lie in $S'(X)$.
Next, we use the algorithm from \Cref{thm: APSP in HSS full} to compute a path $\hat P$ in graph $X$ connecting $\tilde v'$ to $\tilde u'$, such that the length of $\hat P$ is bounded by $2^{O(1/\delta^6)}$, in time $O(|E(\hat P)|)$. We then use the embedding of $X$ into $G''$ to transform path $\hat P$ into a path $P'_3$, connecting $\tilde u'$ to $\tilde v'$ in $G''$. Since the embedding of $X$ into $G''$ is via paths of length at most $O(d)$, the length of $P'_3$ is bounded by $d\cdot 2^{O(1/\delta^6)}$. 
Since graph $G''$ is obtained from $G'$ by subdividing some of its edges, it is immediate to convert $P'_3$ into a path $P_3$ in $G$ connecting $v'$ to $u'$, whose length is at most $d\cdot 2^{O(1/\delta^6)}$.  
By concatenating $P_1$, $P_3$, and the reversed path $P_2$, we obtain a path connecting $u$ to $v$ in $G$, of length at most $d\cdot 2^{O(1/\delta^6)}$. The time required to compute this path is $O\left (d\cdot 2^{O(1/\delta^6)}\right )$, and we can turn the resulting path into a simple one, in time $O\left (d\cdot 2^{O(1/\delta^6)}\right )$. Let $Q(\hat e)$ denote the resulting $u$-$v$ path in $G$, whose length is bounded by $d\cdot 2^{O(1/\delta^6)}$. We add $Q(\hat e)$ to set $\qset$ as the embedding of the edge $\hat e$. For every edge $e\in Q(\hat e)$, we increase the counter $n_e$, and, if it reaches $\eta$, we delete $e$ from graph $H$, and add it to $\edel_i$. In the latter case, for every edge $e'\in L(e)$ (that is, every edge  $e'\in E(X)$ whose embedding path $P(e')$ uses $e$ or one of the artificial edges subdividing $e$), we delete $e'$ from $X$, and we update data structure $\DS_1$ with this deletion.

Before we provide the remainder of the algorithm, we bound the total number of edges that are deleted from $H$, and the total number of edges that are deleted from $X$ over the course of the second step of the phase, in the following observation and its corollary.

\begin{observation}\label{obs: deleted from H}
	Let $E'$ be the set of all edges that are deleted from $H$ over the course of Step 2 of the current phase, excluding the edges incident to $s$. Then $|E'|\leq  \frac{|T|\cdot \Delta}{8\cdot (2m)^{32\delta}\cdot \eta}$.
\end{observation}
\begin{proof}
Recall that:

\[|\hat E|\leq \frac{|E(W^k)|}{z}\leq \frac{|T|\cdot \Delta}{(2m)^{32\delta}\cdot d\cdot 2^{c/\delta^6}}\]
	
since $|E(W_k)|=N^k\cdot k\cdot\Delta=|T|\cdot k\cdot \Delta$, and $z=(2m)^{32\delta}\cdot k\cdot  d\cdot 2^{c/\delta^6}$, for a large enough constant $c$. The length of the embedding path $Q(\hat e)$ of each edge $e\in \hat E\setminus F_i$ is bounded by $d\cdot 2^{O(1/\delta^6)}$. An edge $e'$ lies in $E'$ iff it participates in at least $\eta/2$ embedding paths that have been added to $\qset$ over the course of Step 2 of the current phase, since, at the beginning of the phase, for every edge $e'\in E(G')$, $n_{e'}\leq \eta/2$ held. Therefore:
	
	\[|E'|\leq \frac{|T|\cdot \Delta}{(2m)^{32\delta}\cdot  d\cdot 2^{c/\delta^6}}\cdot \frac{d\cdot 2^{O(1/\delta^6)}}{\eta}\leq \frac{|T|\cdot \Delta}{8\cdot (2m)^{32\delta}\cdot \eta}.\]
\end{proof}

\begin{corollary}\label{cor few deleted from X}
	The total number of edges ever deleted from $X$ over the course of the algorithm is at most $\Lambda$.
\end{corollary}
\begin{proof}
	Let $E^*$ be the set of all edges deleted from $X$ over the course of the algorithm.
	From Equation \ref{eq: bound on Lambda}, it is enough to show that $|E^*|\leq (|T|\cdot \Delta)^{1-11\delta}$.
Recall that the embedding of $X$ into $G$ causes congestion at most  $2\eta\cdot m^{32\delta^3}$ in $G''$. Since every edge of $G'$ has at most $3$ edges of $G''$ subdividing it, every edge deleted from $H$ (except for those incident to $s$) over the course of the current step may lead to the deletion of up to $6\eta\cdot m^{32\delta^3}$ edges of $X$.
Altogether:

\[\begin{split}
|E^*|&\leq 6\eta\cdot m^{32\delta^3}\cdot |E'|\\
&\leq 6 m^{32\delta^3}\cdot \frac{|T|\cdot \Delta}{(2m)^{32\delta}}\\
&\leq \frac{6|T|\cdot \Delta}{(|T|\cdot \Delta)^{16\delta}}\\
&\leq  (|T|\cdot \Delta)^{1-11\delta}.
\end{split}\]
\end{proof}

We are now ready to complete the algorithm.  Let $A'_i\subseteq E(G)$ be the set of edges $e\in E(G)$, such that, prior to the current phase, $e$ participated in fewer than $\eta/2$ embedding paths, but at the end of the current phase, $e$ participates in at least $\eta/2$ embedding paths.
Notice that, in particular $E'\subseteq A'_i$. We also let $A''_i=\emptyset$, and $\hat E'_i\subseteq \hat E$ the set of edges $\hat e\in \hat E$ for which an embedding path $Q(\hat e)$ was computed in the current phase; in other words, $\hat E'_i=\hat E\setminus F_i$.

Consider the set  $F_i\subseteq \hat E$ of the fake edges that the algorithm constructed.
Then, at the end of the algorithm, for every edge $\hat e\in F_i$, at least one endpoint of $e$ lies in $U$. Recall that, from the definition of the set $U$, $\dist_{G'\setminus E'}(U,S''(X))>2d$ holds at the end of the algorithm. Since $G'=G\setminus \edel$, and since $E'\subseteq A'_i$, we get that $\dist_{G\setminus(\edel\cup A'_i\cup A''_i)}(U,S''(X))>2d$.

Let $S_1$ be the set of all terminals lying in $B_{G'\setminus E'}(U,d)$, and let $S_2$ be the set of all terminals lying in $B_{G'\setminus E'}(S''(X),d)$ at the end of the algorithm. Clearly, either $|S_1|\leq |T|/2$, or $|S_2|\leq |T|/2$ must hold. Assume first that $|S_2|\leq |T|/2$. Recall that, from Inequality \ref{eq: bound on core2}, $|S''(X)|\geq \frac{|T|}{2^{6/\delta}}$. We then define $\tdel_i=S''(X)$.
 Clearly, for every vertex $t\in \tdel_i$, 
$B_{G\setminus(\edel\cup A'_i\cup A''_i)}(t,d)$ contains at most $|T|/2=N^k/2$ vertices of $T$. 
 We then output $\tdel_i$, $A'_i$, $A''_i$, $\hat E_i'$, and the embedding of the edges in $\hat E'_i$, as the outcome of the current phase, and the current phase is considered a bad phase. Notice that $|\tdel_i|=|S''(X)|\geq  \frac{|T|}{2^{6/\delta}}\geq \frac{N^k}{m^{O(\delta^3)}}$ as required.

Lastly, we assume that $|S_1|\leq |T|/2$. In this case, we let $\tdel_i=U$. As before, we are guaranteed that, for every vertex $t\in \tdel_i$, 
$B_{G\setminus(\edel\cup A'_i\cup A''_i)}(t,d)$ contains at most $|T|/2=N^k/2$ vertices of $T$. Moreover, every edge in $F_i$ now has an endpoint in $\tdel_i$. We then output $\tdel_i,A'_i,A''_i$, and $\hat E'_i$, together with the embedding of the edges in $\hat E'_i$, and the current phase is considered good.

We now bound the running time of Step 2. First, the total update time of data structure $\DS_1$, from  \Cref{thm: APSP in HSS full} is bounded by $O\left(|E(G)|^{1+O(\delta)}\right )$, as discussed above. The time required to maintain the \EST data structure is bounded by $O(md)$. For every edge $e\in \hat E_i'$, we may spend time $O(d\cdot 2^{O(1/\delta^6)})$ in order to compute its embedding. The time spent on inspecting the edges of $\hat E$ that are eventually added to $F_i$, and the time spent on computing the sets $S_1,S_2$ of vertices is bounded by $O(m)$. Therefore, the total running time of Step 1 is bounded by:

\[ O\left(m^{1+O(\delta)}\right )+O\left (|\hat E_i'|\cdot d\cdot 2^{O(1/\delta^6)}\right). \]

Since the running time of Step 1 is bounded by 
$O\left(m^{1+O(\delta)}\cdot \eta\right )$, we get that the total running time of the $i$th phase is bounded by:

\[ O\left(m^{1+O(\delta)}\cdot \eta\right )+O\left (|\hat E_i'|\cdot d\cdot 2^{O(1/\delta^6)}\right). \]

Lastly, since, from \Cref{obs: number of phases}, the number of phases is bounded by  $O\left(m^{O(\delta)}\cdot k\cdot d\cdot 2^{O(1/\delta^6)}\right )$, and since $\sum_i|\hat E'_i|\leq |E(W_k)|\leq N^k\cdot \Delta \cdot k\leq O(mk)$, we get that the total running time of the algorithm is bounded by:

\[O\left(m^{1+O(\delta)}\cdot k\cdot d\cdot 2^{O(1/\delta^6)}\cdot \eta\right )+O\left (mk\cdot d\cdot 2^{O(1/\delta^6)}\right)\leq O\left(m^{1+O(\delta)}\cdot\eta^2\right ).\]

since, from the problem definition, $\eta>k\cdot d\cdot 2^{c/\delta^6}$, for a large enough constant $c$.

\section{Third Tool: Router Witness}
\label{sec: router witness}

In this section we show that an embedding of a nice router $W^{N,\Delta}_k$ into a graph $C$ can be used to certify that the graph $C$ is a strong router, and also provide an algorithm to compute a sparse subgraph of $C$ that is a strong enough router. We start with the following lemma.

\begin{lemma}\label{lem: nice router to uniform degree router}
	Let $C$ be a graph, and let $\alpha\geq 1,\Delta\geq 1$ be parameters such that, for every vertex $v\in V(C)$, $\deg_C(v)\leq \alpha\cdot \Delta$. Assume that we are given a graph $W$, that is a properly pruned subgraph of $W^{N,\Delta}_k$, for some parameters $N$ and $k$, and an embedding $\pset$ of $W$ into $C$ via paths of length at most $d^*$, that cause congestion at most $\eta^*$. Assume further that every vertex $v\in V(C)$ lies on at least $\frac{\Delta}{\beta}$ paths in $\pset$. Then $C$ is a $(\vdeg_C,\tilde d,\tilde \eta)$-router with respect to the set $V(C)$ of its vertices, where $\tilde d=22d^*\cdot k^2$, and $\tilde \eta=2\alpha\cdot \beta\cdot k^{4k+1}\cdot d^*\cdot \eta^*$.
	\end{lemma}
\begin{proof}
	Consider any $\vdeg_C$-restricted demand $\dset=\left (\Pi,\set{D(a,b)}_{(a,b)\in \Pi}\right )$  on $V(C)$. It is enough to show that this demand can be routed in $C$ via paths of length at most $\tilde d$, with congestion at most $\tilde \eta$.

	Our algorithm uses a parameter  $q=\ceil{\frac{\Delta}{\beta}}$.
	For brevity, we will denote the graph $W^{N,\Delta}_k$ by $W_k$ in this proof.
	We will define a new demand $\dset'=\left (\Pi',\set{D(a',b')}_{(a',b')\in \Pi'}\right )$ over the vertices of $W$.
	In order to do so, we will define, for every pair $(a,b)\in \Pi$, a collection $M(a,b)$ of $q$ pairs of vertices in $V(W)$, called \emph{proxy pairs for $(a,b)$}. For every pair $(a',b')\in M(a,b)$, we will set its demand $D'(a',b')=\frac{D(a,b)}{q}$. We will also compute a partial routing $F(a,b)$ of the demand $(a,b)$, where, for every pair $(a',b')\in M(a,b)$, vertex $a$ sends $\frac{D(a,b)}{q}$ flow units to $a'$ and vertex $b$ sends  $\frac{D(a,b)}{q}$ flow units to $b'$. Demand $\dset'$ is then obtained by setting $\Pi'=\bigcup_{(a,b)\in \Pi}M(a,b)$.
	We will show that demand $\dset'$ is properly restricted, and then use \Cref{thm: pruning to routing} to compute its routing $f'$ in $W$. By using the embedding of $W$ into $C$, we can then obtain a routing $f''$ of $\dset'$ in $C$. Lastly, by combining $f''$ with the partial routings $F(a,b)$ that we computed for all pairs $(a,b)\in \Pi$, we will obtain the final routing of $\dset$. We now define the demand $\dset'$, the proxy pairs, and the partial routings $F(a,b)$ for pairs $(a,b)\in \Pi$ formally.

	Consider a vertex $v\in V(C)$, and let $\pset(v)\subseteq \pset$ be the collection of paths that contain $v$, so $|\pset(v)|\geq \frac{\Delta}{\beta}$. We discard paths from $\pset(v)$ until $|\pset(v)|=\ceil{\frac{\Delta}{\beta}}$ holds. 
	Next, we construct a multiset $S(v)$ of $q$ vertices of $V(W)$, and, for every vertex $u\in S(v)$, a path $R_v(u)$, as follows. Consider some path $P\in \pset(v)$, and let $e\in E(W)$ be the edge that is embedded into $P$. Recall that exactly one endpoint of $e$ is a leaf vertex of $W_k$; denote this vertex by $u$. We add $u$ to the multiset $S(v)$, and we set the corresponding path $R_v(u)$ to be the subpath of $P$ between $v$ and $u$. Note that $u$ may serve as an endpoint of several paths in $\pset(v)$. In such cases, we add several copies of $u$ to the set $S(v)$, and for each copy we include a corresponding path $R_v(u)$ that is a subpath of a distinct path in $\pset(v)$. To avoid clutter, abusing the notation, we denote each such path by $R_v(u)$, and we assume that the corresponding copy of $u$ in $S(v)$ is implicit in this definition. We denote $\rset(v)=\set{R_v(u)\mid u\in S(v)}$. Observe that $|\rset(v)|=q$, and every path in $\rset(v)$ is a subpath of a distinct path in $\pset(v)$. Let $\rset=\bigcup_{v\in V(C)}\rset(v)$.
	
	 Since the set $\pset$ of paths causes congestion at most $\eta^*$ in $C$, and the length of every path in $\pset$ is at most $d^*$, we get that the total congestion caused by the multiset of paths $\pset'=\bigcup_{v\in V(C)}\pset(v)$ in $C$ is at most $d^*\cdot \eta^*$. Therefore, the congestion caused by $\rset$ in $C$ is also at most $d^*\cdot \eta^*$.
	
	Next, we define a new demand  $\dset'=\left (\Pi',\set{D'(a,b)}_{(a,b)\in \Pi'}\right )$ over the vertices of $V(W)$, as follows.
	We start with $\Pi'=\emptyset$, so $\dset'$ is empty, and then process the pairs $(a,b)\in \Pi$ one by one. When a pair $(a,b)$ is processed, we define an arbitrary perfect matching $M(a,b)$ between vertices of $S(a)$ and vertices of $S(b)$; recall that $|S(a)|=|S(b)|=q$. We then add every pair $(a',b')\in M(a,b)$ to $\Pi'$, and set its demand  $D'(a',b')=\frac{D(a,b)}{q}$. Additionally, we define a \emph{partial routing} $F(a,b)$ of the demand $(a,b)$, in which $a$ sends $\frac{D(a,b)}{q}$ flow units to every vertex $a'\in S(a)$, and $b$ sends $\frac{D(a,b)}{q}$ flow units to every vertex $b'\in S(b)$, as follows: for every vertex $a'\in S(a)$, we send  $\frac{D(a,b)}{q}$ flow units from $a$ via the path $R_a(a')\in \rset(a)$, and for every vertex $b'\in S(b)$, we sent   $\frac{D(a,b)}{q}$ flow units from $b$ via the path $R_b(b')\in \rset(b)$. This completes the description of the demand $\dset'$, and of the partial routing $F=\bigcup_{(a,b)\in \Pi}F(a,b)$.

	Consider any vertex $v\in V(C)$, and let $\Pi(v)\subseteq \Pi$ be the collection of all demand pairs in wich $v$ participates in $\dset$. Since demand $\dset$ is $\vdeg_C$-restricted, $\sum_{(v,u)\in \Pi(v)}D(v,u)\leq \deg_C(v)\leq \alpha\cdot \Delta$. For every demand pair $(v,u)\in \Pi(v)$, the partial routing $F(v,u)$ sends $\frac{D(v,u)}{q}$ flow units on every path in $\rset(v)$. Therefore, overall, the partial flow $F$ sends at most $\frac{\alpha \Delta}{q}$ flow units on every path in $\rset(v)$. Since the paths in $\rset=\bigcup_{v\in V(C)}\rset(v)$ cause congestion at most $d^*\cdot \eta^*$, the congestion of the partial flow $F$ in $C$ is bounded by:
	
	\[ d^*\cdot \eta^*\cdot \frac{\alpha \Delta}{q}\leq d^*\cdot\eta^*\cdot\alpha\cdot \beta. \]

since $q=\ceil{\frac{\Delta}{\beta}}$.	

Consider now some leaf vertex $u\in V(W)$. From the definition of graph $W_k$, the degree of $u$ in $W$ is at most $\Delta\cdot k$. Therefore, $u$ serves as an endpoint of at most $\Delta\cdot k$ paths in $\pset$. Since the length of every path $P\in \pset$ is at most $d^*$, we get that $u$ serves as an endpoint of at most $\Delta\cdot k\cdot d^*$ paths in the multiset $\pset'=\bigcup_{v\in V(C)}\pset(v)$, and hence $u$ serves as an endpoint of at most $\Delta \cdot k\cdot d^*$ paths in $\rset$. Consider now the total demand $D^*(u)=\sum_{(u,x)\in \Pi'}D'(u,x)$ on the vertex $u$ in $\dset'$. For every pair $(u,x)\in \Pi'$, there is some path $R\in \rset$, such that the partial flow $F$ sends $D'(u,x)$ flow units along $R$ to $u$. Since $u$ serves as endpont of at most $\Delta\cdot k\cdot d^*$ paths in $\rset$, and since the total flow along each such path, as we have shown above, is bounded by $\frac{\alpha\Delta}{q}\leq \alpha\beta$, we get that $D^*(u)\leq \alpha\cdot \beta\cdot \Delta\cdot k\cdot d^*$. We conclude that demand $\dset'$ is $(\alpha\cdot \beta\cdot \Delta\cdot k\cdot d^*)$-restricted.

Let $\dset''$ be the demand obtained from $\dset'$, after we scale each demand $D'(a,b)$ down by factor $\alpha\cdot \beta\cdot k^{4k+1}\cdot d^*$. Then it is immediate to verify that $\dset''$ is  a  $\frac{\Delta}{k^{4k}}$-restricted demand  for $W$. From \Cref{thm: pruning to routing}, there is a routing $f''$ of $\dset''$ in $W$ via paths of length at most $20k^2$, with no edge-congestion. By scaling the flow $f''$ up by factor $\alpha\cdot \beta\cdot k^{4k+1}\cdot d^*$, we obtain a routing $f'$ of the demand $\dset'$ in $W$ via paths of length at most $20k^2$, with congestion at most $\alpha\cdot \beta\cdot k^{4k+1}\cdot d^*$.

Consider now any flow-path $Q$ that is used in the routing $f'$. Let $Q'$ be a path in graph $C$, that is obtained from $Q$, by replacing every edge $e\in E(Q)$ by its corresponding embedding path $P(e)\in \pset$. It is then easy to see that the length of $Q'$  is bounded by $20k^2\cdot d^*$. By setting the flow $f(Q')$ on each such path $Q'$ to be $f'(Q)$, we obtain a routing $f$ of $\dset'$ in graph $C$, via paths of length at most $20k^2\cdot d^*$. Since the congestion of the routing $f'$ in $W$ is at most  $\alpha\cdot \beta\cdot k^{4k+1}\cdot d^*$, and the congestion caused by the set $\pset$ of embedding paths in $C$ is at most $\eta^*$, it is easy to verify that the congestion of $f$ is at most  $\alpha\cdot \beta\cdot k^{4k+1}\cdot d^*\cdot \eta^*$.

Lastly, by combining the partial routing $F$ with the routing $f$, we obtain a final routing $f^*$ of the demand $\dset$ in $C$, via paths of length at most $22k^2\cdot d^*$, with congestion at most:

\[\alpha\cdot \beta\cdot k^{4k+1}\cdot d^*\cdot \eta^*+\alpha\cdot\beta\cdot d^*\cdot \eta^*\leq 2\alpha\cdot \beta\cdot k^{4k+1}\cdot d^*\cdot \eta^*.\]
\end{proof}

Recall that a Router Decomposition requires that, for every cluster $C\in \cset$, we provide a subgraph $C'\subseteq C$ with $V(C')=V(C)$ and $|E(C')|\leq \Delta^*\cdot |V(C)|$, such that $C'$ is a $(\vDelta^*,\tilde d,\tilde \eta)$-router for the set $V(C)$ of supported vertices.
We next describe how $C'$ is constructed, and how we certify that it is indeed a router with required properties. We will call $C'$ a \emph{sparsified router}.

\subsection{A Sparsified Router}
\label{subsec: sparsified router}


In this subsection we assume that we are given a simple graph $C$, together with parameters $1\leq \Delta^*\leq \Delta$.
Assume further that we are given another graph $W$, that is a properly pruned subgraph of $W^{N,\Delta}_k$, for some parameters $N$ and $k$, together with an embedding $\pset$ of $W$ into $C$ via paths of length at most $d^*$, that cause congestion at most $\eta^*$, such that every vertex $v\in V(C)$ lies on at least $\floor{\frac{\Delta^*}{2kd^*}}$ paths in $\pset$. Finally, assume that $\Delta^*\geq 2kd^*$.

We use a parameter $\Delta'=\floor{\frac{\Delta^*}{2kd^*}}$. In order to construct a sparsified router $C'$ for $C$, we select, for every superedge $(u,v)$ of $W$, an arbitrary collection $E'(u,v)\subseteq E(W)$ of $\Delta'$ parallel edges (recall that, from \Cref{def: properly pruned}, graph $W$ must contain at least $\frac{\Delta}{2}\geq \frac{\Delta^*}{2}\geq \Delta'$ such parallel edges). Let $W'\subseteq W$ be the graph obtained from $W$ by including, for every superedge $(u,v)$ of $W$, only the edges of $E'(u,v)$. Additionally, for every vertex $v\in V(C)$, we select an arbitrary subset $\pset(v)\subseteq \pset$ of $\Delta'$ paths that contain $v$, such that every leaf vertex $u\in V(W)$ serves as an endpoint in at most $\gamma\cdot \Delta'$ paths in the multiset $\bigcup_{v\in V(C)}\pset(v)$, for some parameter $\gamma$.  (Whenever using this tool, we will provide an explicit algorithm for constructing and maintaining the sets $\pset(v)$ of paths for vertices $v\in V(C)$).

We let $C'\subseteq C$ be the simple graph obtained by taking the union of all paths in $\bigcup_{v\in V(C)}\pset(v)$, and all paths of $\pset$ that serve as embedding paths for the  edges in $E(W')$. 
The following observation summarizes the properties of $C'$.

\begin{observation}\label{obs: sparsified router}
	Graph $C'$ is a $(\vDelta^*,\tilde d,\tilde \eta')$-router with respect to the set $V(C')$ of supported vertices, where  $\tilde d=22d^*\cdot k^2$, and $\tilde \eta'=8\gamma (d^*)^2\cdot \eta^*\cdot k^{4k+1}$. Moreover, $|E(C')|\leq |V(C)|\cdot \Delta^*$.
\end{observation}
\begin{proof}
	Observe first that, since graph $W'$ is embedded into $C$, $|V(W')|\leq |V(C)|$. Therefore, $|E(W')|\leq |V(W')|\cdot k\cdot \Delta'\leq |V(C)|\cdot k\cdot \Delta'$. Consider the collection $\qset$ of paths that contains the embedding paths $P(e)\in \pset$ for every edge $e\in E(W')$, and, for every vertex $v\in V(C)$, all paths in set $\pset(v)$. Then $|\qset|\leq |E(W')|+|V(C)|\cdot \Delta'\leq 2|V(C)|\cdot k\cdot \Delta'$, and the length of every path in $\qset$ is bounded by $d^*$. Therefore, $|E(C')|\leq \sum_{P\in \qset}|E(P)|\leq 2|V(C)|\cdot k\cdot \Delta'\cdot d^*\leq |V(C)|\cdot \Delta^*$, since $\Delta'=\floor{\frac{\Delta^*}{2k\Delta d^*}}$.
	
	We now turn to prove that graph $C'$ is a $(\vDelta^*,\tilde d,\tilde \eta')$-router with respect to the set $V(C')$ of supported vertices. 
	Consider any $\vDelta^*$-restricted demand	 $\dset=\left (\Pi,\set{D(a,b)}_{(a,b)\in \Pi}\right )$  on $V(C')$. We now show that this demand can be routed in $C'$ via paths of length at most $\tilde d$, with congestion at most $\tilde \eta'$. Note that, from the definition of a properly pruned subgraph of $W_k^{N,\Delta}$ (see \Cref{def: properly pruned}), graph $W'$ is a properly pruned subgraph of $W^{N,\Delta'}_k$. The remainder of the proof is almost identical to the proof of \Cref{lem: nice router to uniform degree router}. 
	
	For every vertex $v\in V(C)$, we define a multiset $S(v)$ of $\Delta'$ leaf vertices of $W'$ as follows. Consider some path $P\in \pset(v)$, and let $e\in E(W')$ be the edge that is embedded into $P$. Recall that exactly one endpoint of $e$ is a leaf vertex of $W^{N,\Delta'}_k$; denote this vertex by $u$. We add $u$ to the multiset $S(v)$, and we set the corresponding path $R_v(u)$ to be the subpath of $P$ between $v$ and $u$. As before, $u$ may serve as an endpoint of several paths in $\pset(v)$. In such cases, we add several copies of $u$ to the set $S(v)$, and for each copy we include a corresponding path $R_v(u)$ that is a subpath of a distinct path in $\pset(v)$. 
	We denote $\rset(v)=\set{R_v(u)\mid u\in S(v)}$. 
	Observe that $|\rset(v)|=\Delta'$, and every path in $\rset(v)$ is a subpath of a distinct path in $\pset(v)$. Let $\rset=\bigcup_{v\in V(C)}\rset(v)$.
	
	Since the set $\pset$ of paths causes congestion at most $\eta^*$ in $C$, and the length of every path in $\pset$ is at most $d^*$, we get that the total congestion caused by the multiset of paths $\pset'=\bigcup_{v\in V(C)}\pset(v)$ in $C$ is at most $d^*\cdot \eta^*$. Therefore, the congestion caused by $\rset$ in $C$ is also at most $d^*\cdot \eta^*$.
		
	Next, we define a new demand  $\dset'=\left (\Pi',\set{D'(a,b)}_{(a,b)\in \Pi'}\right )$ over the vertices of $V(W')$, as follows.
	We start with $\Pi'=\emptyset$, so $\dset'$ is empty, and then process the pairs $(a,b)\in \Pi$ one by one. When a pair $(a,b)$ is processed, we define an arbitrary perfect matching $M(a,b)$ between vertices of $S(a)$ and vertices of $S(b)$; recall that $|S(a)|=|S(b)|=\Delta'$. We then add every pair $(a',b')\in M(a,b)$ to $\Pi'$, and set its demand  $D'(a',b')=\frac{D(a,b)}{\Delta'}$. Additionally, we define a \emph{partial routing} $F(a,b)$ of the demand $(a,b)$, in which $a$ sends $\frac{D(a,b)}{\Delta'}$ flow units to every vertex $a'\in S(a)$, and $b$ sends $\frac{D(a,b)}{\Delta'}$ flow units to every vertex $b'\in S(b)$, as follows: for every vertex $a'\in S(a)$, we send  $\frac{D(a,b)}{\Delta'}$ flow units from $a$ via the path $R_a(a')\in \rset(a)$, and for every vertex $b'\in S(b)$, we sent   $\frac{D(a,b)}{\Delta'}$ flow units from $b$ via the path $R_b(b')\in \rset(b)$. This completes the description of the demand $\dset'$, and of the partial routing $F=\bigcup_{(a,b)\in \Pi}F(a,b)$.

	Consider any vertex $v\in V(C)$, and let $\Pi(v)\subseteq \Pi$ be the collection of all demand pairs in which $v$ participates in $\dset$. Since demand $\dset$ is $\vDelta^*$-restricted, $\sum_{(v,u)\in \Pi(v)}D(v,u)\leq  \Delta^*$. For every demand pair $(v,u)\in \Pi(v)$, the partial routing $F(v,u)$ sends $\frac{D(v,u)}{\Delta'}$ flow units on every path in $\rset(v)$. Therefore, overall, the partial flow $F$ sends at most $\frac{\Delta^*}{\Delta'}\leq 4kd^*$ flow units on every path in $\rset(v)$ (we have used the fact that  $\Delta'=\floor{\frac{\Delta^*}{2kd^*}}$). Since the paths in $\rset=\bigcup_{v\in V(C)}\rset(v)$ cause congestion at most $d^*\cdot \eta^*$, the congestion of the partial flow $F$ in $C'$ is bounded by $4k(d^*)^2\cdot \eta^*$.	
	
	Consider now some leaf vertex $u\in V(W')$, and recall that it may serve as an endpoint of at most $\gamma\cdot \Delta'$ paths in the multiset $\bigcup_{v\in V(C)}\pset(v)$. Consider the total demand $D^*(u)=\sum_{(u,x)\in \Pi'}D'(u,x)$ on the vertex $u$ in $\dset'$. For every pair $(u,x)\in \Pi'$, there is some path $R\in \rset$, such that the partial flow $F$ sends $D'(u,x)$ flow units along $R$ to $u$. Since $u$ serves as endpont of at most $\gamma\cdot \Delta'$ paths in $\rset$, and since the total flow along each such path, as we have shown above, is bounded by $4kd^*$, we get that $D^*(u)\leq 4\gamma k\cdot d^*\cdot \Delta'$. We conclude that demand $\dset'$ is $(4\gamma k\cdot d^*\cdot \Delta')$-restricted.
	
	Let $\dset''$ be the demand obtained from $\dset'$, after we scale each demand $D'(a,b)$ down by factor $4\gamma d^*\cdot k^{4k+1}$. Then it is immediate to verify that $\dset''$ is  a  $\frac{\Delta'}{k^{4k}}$-restricted demand  for $W'$. From \Cref{thm: pruning to routing}, there is a routing $f''$ of $\dset''$ in $W$ via paths of length at most $20k^2$, with no edge-congestion. By scaling the flow $f''$ up by factor $4\gamma d^*\cdot k^{4k+1}$, we obtain a routing $f'$ of the demand $\dset'$ in $W'$ via paths of length at most $20k^2$, with congestion at most $4\gamma d^*\cdot k^{4k+1}$.
	
	Consider now any flow-path $Q$ that is used in the routing $f'$. Let $Q'$ be a path in graph $C'$, that is obtained from $Q$, by replacing every edge $e\in E(Q)$ by its corresponding embedding path $P(e)\in \pset$. It is then easy to see that the length of $Q'$  is bounded by $20k^2\cdot d^*$. By setting the flow $f(Q')$ on each such path $Q'$ to be $f'(Q)$, we obtain a routing $f$ of $\dset'$ in graph $C$, via paths of length at most $20k^2\cdot d^*$. Since the congestion of the routing $f'$ in $W'$ is at most  $4\gamma d^*\cdot k^{4k+1}$, and the congestion caused by the set $\pset$ of embedding paths in $C$ is at most $\eta^*$, it is easy to verify that the congestion of $f$ is at most  $4\gamma d^*\cdot \eta^*\cdot k^{4k+1}$.
	
	Lastly, by combining the partial routing $F$ with the routing $f$, we obtain a final routing $f^*$ of the demand $\dset$ in $C$, via paths of length at most $22k^2\cdot d^*$, with congestion at most:
	
	\[4\gamma d^*\cdot \eta^*\cdot k^{4k+1}+4k(d^*)^2\cdot \eta^*\leq 8\gamma (d^*)^2\cdot \eta^*\cdot k^{4k+1}.\]
\end{proof}

\section{Fourth Tool: Decremental Low-Diameter Clustering}
\label{sec: clustering}

\subsection{Problem Definition}
We assume that we are given an undirected $n$-vertex graph $G$ that undergoes an online adversarial sequence of edge deletions.
We are also given a parameter $1<k<\log n$, and we use another parameter $d=4k^3$.
We will sometimes refer to subgraphs of $G$ as \emph{clusters}. Next, we need the definition of a settled cluster, and of a valid clustering. 

\begin{definition}[Settled Cluster]
	We say that a cluster $C\subseteq G$ is \emph{settled}, if either $|V(C)|\leq n^{1/k}$; or there is a vertex $v\in V(C)$, such that $|B_C(v,d)|\geq |V(C)|^{1-1/k}$. A cluster that is not settled is called \emph{unsettled}.
\end{definition}

\begin{definition}[Valid Clustering]
	A valid clustering is a collection $\cset$ of clusters of $G$, such that:
	\begin{itemize}
		\item every cluster $C\in \cset$ is settled; and
		\item every edge in the current graph $G$ lies in exactly one of the sets $\set{E(C)}_{C\in \cset}$.
	\end{itemize}
\end{definition}

We are now ready to define the Decremental Low-Diameter Clustering problem.

\begin{definition}[Low-Diameter Clustering problem]
In the Decremental Low-Diameter Clustering problem, the input is a simple  $n$-vertex graph $G$, and a parameter $1<k<\log n$. Let $d=4k^3$. 
At the beginning of the algorithm, an initial valid clustering $\cset$ of the initial graph $G$ must be produced by the algorithm. Over the course of the algorithm, the adversary maintains a partition of the set $\cset$ of clusters into two subsets: set $\cset^I$ of inactive clusters, and set $\cset^A$ of active clusters. Once a cluster $C\in \cset$ is added to  $\cset^I$, it may no longer be modified, and it remains in $\cset^I$ until the end of the algorithm. Immediately after the initialization, $\cset^A=\cset$ and $\cset^I=\emptyset$ holds.

The algorithm proceeds over a number of phases. At the beginning of every phase, the adversary may delete some edges and vertices from some of the clusters in $\cset^A$, and move some clusters from $\cset^A$ to $\cset^I$, after which every cluster that remains in $\cset^A$ is guaranteed to become unsettled.  

The algorithm is then required to update the current clustering, via a sequence of \emph{cluster splitting operations}, defined as follows. In every cluster splitting operation, we start with a cluster $C\in \cset^A$, and a vertex-induced subgraph $C'\subseteq C$ with $|V(C')|\leq |V(C)|^{1-1/k}$. Cluster $C'$ is then added to $\cset$ and $\cset^A$ as a new cluster, and the edges of $E(C')$ are deleted from $C$. Additionally, some vertices that have become isolated in $C$ can be deleted from $C$.
At the end of the phase, all clusters in $\cset^A$ must be settled.

For every vertex $v$, let $n_v$ be the total number of clusters to which $v$ ever belonged over the course of the algorithm. Then we additionally require that $\sum_{v\in V(G)}n_v\leq 2n^{1+1/k}$ holds.
\end{definition}

The main result of the current section is an algorithm for the Low-Diameter Clustering problem, that is summarized in the following theorem; we note that the algorithm relies on the standard ball-growing technique, and a standard approach for computing neighborhood covers.

\begin{theorem}\label{thm: clustering algorithm}
There is a deterministic algorithm for the Decremental Low-Diameter Clustering problem, such that, if $m$  is the number of edges in the initial graph $G$, then  the running time of the initialization, and the running time of each of the phases, is bounded by $O(m^{1+1/k^3})$.
\end{theorem}

In the remainder of this section, we prove the above theorem.  
We start by describing our main subroutine that is based on the Ball Growing technique, whose purpose is to perform a single split of a cluster. We then provide our second main subroutine, that processes a single cluster, via  a sequence of such split operations. Finally, we provide the description of the entire algorithm.

\subsection{First Main Subroutine: Ball Growing Procedure}

Our first main subroutine is given as input a cluster $C\in \cset$ (which, presumably, is currently unsettled). 
The subroutine selects a vertex $v\in V(C)$, and computes an \emph{eligible index} $1\leq i\leq d$, that is defined below. If $|B_C(v,i)|\geq |V(C)|^{1-1/k}$, then the subroutine correctly determines that $C$ is settled and terminates. Otherwise, it creates a new cluster $C'$, that is a subgraph of $C$ induced by the set $B_C(v,i)$ of vertices. Our algorithm will then split $C'$ off from the cluster $C$. Eventually, the same subroutine will be repeatedly applied to the cluster $C$ until it becomes is settled.

From now on we fix a cluster $C\in \cset$, and a vertex $v\in V(C)$, and describe our first main subroutine, whose purpose is to compute an \emph{eligible index} $1\leq i\leq d$, that we now define.

For all $i> 0$, let $L_i$ be the $i$th layer of the Breadth First Search in $C$ starting from $v$; in other words, $L_i=B_C(v,i)\setminus B_C(v,i-1)$. Let $L_0=\set{v}$. For convenience, we denote by $\hat N=|V(C)|$, and, for $i\geq 0$, we denote by $N_i=|L_0|+|L_1|+\cdots+|L_i|$. We also denote by $\hat E_i$ the set of all edges of $C$ with both endpoints in $L_0\cup\cdots\cup L_i$. We are now ready to define an eligible index.

\begin{definition}[Eligible Index]
	For $i\geq 1$, we say that index $i$  is \emph{eligible}, if the following three conditions hold:
	
	\begin{itemize}
		\item $N_i\leq N_{i-1}\cdot \hat N^{1/k^3}$;
		\item $N_i< \hat N^{1-1/k}$; and
		\item $|\hat E_i|\leq |\hat E_{i-1}|\cdot \hat N^{1/k^3}$.
	\end{itemize} 
\end{definition}

The following claim summarizes our first main subroutine.

\begin{claim}\label{claim: eligible index alg}
There is an algorithm, that, given a cluster $C$ in the adjacency list representation, and a vertex $v\in V(C)$ that is not isolated in $C$, either computes an eligible index $1\leq i\leq d$ for $v$, or certifies that $|B_C(v,d)|\geq \hat N^{1-1/k}$, so the cluster is settled. The running time of the algorithm is $O(|\hat E_i|)\leq O\left(|\hat E_{i-1}|\cdot \hat N^{1/k^3}\right )$ if an eligible index $i$ is returned, and $O(|E(C)|)$ otherwise.
\end{claim}
\begin{proof}
	The proof of the claim easily follows from the following observation.
	
	\begin{observation}\label{obs: eligible index}
	If $|B_C(v,d)|< \hat N^{1-1/k}$, then some index $1\leq i\leq d$ is eligible.
\end{observation}
\begin{proof}
	Assume otherwise, that is, $|B_C(v,d)|< \hat N^{1-1/k}$, but no index $1\leq i\leq d$ is eligible. We say that index $i$ is \emph{type-1 ineligible} if $N_i>N_{i-1}\cdot \hat N^{1/k^3}$, and we say that it is \emph{type-2 ineligible} if $|\hat E_i|>|\hat E_{i-1}|\cdot \hat N^{1/k^3}$. Since $|V(C)|=\hat N$ and $N_1=1$, it is easy to verify that at most $k^3$ indices may be type-1 ineligible. Since $|E(C)|\leq \hat N^2$, and vertex $v$ is not isolated, it is easy to verify that at most $2k^3$ indices may be type-2 ineligible. But $d=4k^3$, and every index $1\leq i\leq d$ must be either type-1 or type-2 ineligible, a contradiction.
\end{proof}
	
We are now ready to describe our algorithm. The algorithm performs a BFS in $C$, starting from vertex $v$, until it either encounters an eligible index $1\leq i\leq d$, or establishes that no such index is eligible. In the former case, it terminates the search once it reaches the $i$th layer of the BFS, and returns the index $i$. 	In this case, the running time of the algorithm is bounded by $O(|\hat E_i|)\leq O\left(|\hat E_{i-1}|\cdot N^{1/k^3}\right )$. In the latter case, the running time of the algorithm is bounded by $O(|E(C)|)$, and we are guaranteed that $|B_C(v,d)|>\hat N^{1-1/k}$.
\end{proof}

\subsection{Second Main Subroutine: Processing a Cluster}

Our second main subroutine is responsible for processing a single cluster. The input to the subroutine is a single cluster $C\in \cset^A$. The subroutine must compute a collection $\hset=\set{C_0,C_1,\ldots,C_r}$ of clusters, where $r\geq 0$, that will replace $C$ in $\cset^A$. We require that each of the resulting clusters is a subgraph of $C$ and is a settled cluster. Additionally, we require that, for all $1\leq j\leq r$, $|V(C_j)|\leq |V(C)|^{1-1/k}$ hold, and that every edge of $C$ lies in exactly one of the clusters $C_0,\ldots,C_r$. 
Generally, we would also like to ensure that $\sum_{j=1}^r|V(C_j)|\le |V(C)|^{1+1/k^3}$ holds. The problem is that we may need to apply this subroutine repeatedly to cluster $C$ (where we will view the cluster $C_0$ as replacing $C$ after each such application, while clusters $C_1,\ldots,C_r$ are viewed as newly created clusters). Eventually, we will need to ensure that the total number of vertices in all resulting new clusters is bounded by $|V(C)|^{1+1/k^3}$. In order to overcome this difficulty, we  require that, for every cluster $1\le j\leq r$, the algorithm computes a ``core'' $U_j$, containing a large fraction of vertices of $C_j$. The cores $U_1,\ldots,U_r$ must be disjoint from each other, and disjoint from $V(C_0)$. We can then ``charge'' each such core $U_j$ for the vertices of $V(C_j)\setminus U_j$, which may be duplicated between different clusters.
We provide an algorithm for computing such a decomposition of $C$ in the following lemma, which is a straightforward application of \Cref{claim: eligible index alg}.

\begin{claim}\label{claim: process single cluster}
	There is a deterministic algorithm, that is given as an input a graph $C$ in the  adjacency list representation with $|V(C)|\geq n^{1/k}$. The algorithm produces a collection $\hset=\set{C_0,C_1,\ldots,C_r}$ of subgraphs of $C$, 
	and, for all $1\leq j\leq r$, a subset $U_j\subseteq V(C_j)$ of vertices, such that the following hold:
	
	\begin{itemize}
		\item every edge of $E(C)$ lies in exactly one cluster of $\hset$;
		\item for all $1\leq j\leq r$, $|V(C_j)|\leq |V(C)|^{1-1/k}$; 
		\item every vertex $v\in V(C)$ lies in at most one of the sets $U_1,\ldots,U_r,V(C_0)$; 
		\item for all $1\leq j\leq r$, $|U_j|\geq  \frac{|V(C_j)|}{|V(C)|^{1/k^3}}$; and
		\item Every cluster $C_i \in  \hset$ is settled. 
	\end{itemize}
The running time of the algorithm is $O\left (|E(C)|\cdot |V(C)|^{1/k^3}\right )$.
\end{claim}

\begin{proof}
	At the beginning of the algorithm, we set $C_0=C$ and $\hset=\set{C_0}$, and then perform iterations. The $j$th iteration, for $j\geq 1$, is executed as follows. 
	Let $v\in V(C_0)$ be an arbitrary vertex. If vertex $v$ is isolated in $C_0$, then we set $C_j=\set{v}$ and $U_j=\set{v}$, delete $v$ from $C_0$, and continue to the next iteration. From now on we assume that $v$ is not isolated in $C_0$. 
	
	We apply the algorithm from \Cref{claim: eligible index alg} to cluster $C_0$, and vertex $v$. If the algorithm establishes that $|B_{C_0}(v,d)|\geq |V(C_0)|^{1-1/k}$, then we terminate the algorithm, and return the current set $\hset$. In this case, we say that the current iteration is good. The running time of the iteration in this case is bounded by $O(|E(C)|)$. Assume now that the algorithm from \Cref{claim: eligible index alg} computes an eligible index $1\leq i\leq d$. In this case, we let $C_j$ be the subgraph of $C_0$ induced by the vertices of $B_{C_0}(v,i)$. From the definition of the eligible index, $|V(C_j)|\leq |V(C_0)|^{1-1/k}\leq |V(C)|^{1-1/k}$. We also set $U_j=B_{C_0}(v,i-1)$. We then delete from $C_0$ all vertices of $U_j$, and all edges that lie in $E(C_j)$. If  the algorithm from \Cref{claim: eligible index alg} returned an eligible index $i$, we say that the $j$th iteration is bad. In this case, the running time of the iteration is guaranteed to be at most $O(|\hat E^j|)\cdot O(|V(C)|^{1/k^3})$, where $\hat E_j$ is the set of all edges that were deleted from $C_0$ in the current iteration. Observe also that, from the definition of the eligible layer, $|U_j|\geq \frac{|V(C_j)|}{|V(C_0)|^{1/k^3}}\geq \frac{|V(C_j)|}{|V(C)|^{1/k^3}}$. Lastly, since $i\leq d$, $B_{C_j}(v,d)=V(C_j)$, so the cluster $C_j$ is settled.
	
	It is easy to verify that the algorithm guarantees that every edge of $E(C)$ always lies in exactly one cluster of $\hset$, and that vertex sets $U_j$ for clusters $C_j\in \hset\setminus\set{C_0}$ are disjoint from each other and from $C_0$. It is also immediate to verify from our discussion that, for all $1\leq j\leq r$, $|C_j|\leq |V(C)|^{1-1/k}$, and $|U_j|\geq \frac{|V(C_j)|}{|V(C)|^{1/k^3}}$. We have shown already that all clusters in $\hset\setminus\set{C_0}$ are settled. It is easy to verify that cluster $C_0$ is also settled at the end of the algorithm, since the algorithm may only terminate following a good iteration, that establishes that $C_0$ is a settled cluster. 
	
The algorithm has at most one good iteration, whose running time is at most $O(|E(C)|)$. If the $j$th iteration is bad, then its running time is at most $O(|\hat E^j|)\cdot O(|V(C)|^{1/k^3})$, where $\hat E^j$ is the set of all edges that were deleted from $C_0$ in iteration $j$. Therefore, the total running time of the algorithm is $O(|E(C)|\cdot |V(C)|^{1/k^3}$, as required.
\end{proof}

\subsection{The Remainder of the Algorithm and its Analysis}

We are now ready to complete the description of our algorithm for the Low-Diameter Clustering problem. At the beginning of the algorithm, we let the set $\cset^A$ of active clusters contain a single cluster -- the entire graph $G$. For simplicity, we consider the process of constructing the initial clustering $\cset$ as the $0$th phase.

In every phase, we consider every cluster $C\in \cset^A$ that contains more than $n^{1/k}$ vertices one by one.
For each cluster $C\in \cset^A$, we apply the algorithm from \Cref{claim: process single cluster} to it. Let $\hset=\set{C_0,C_1,\ldots,C_r}$ be the outcome of the algorithm. We then add $C_1,\ldots,C_r$ to $\cset$ and to $\cset^A$ as new clusters, and we replace $C$ with $C_0$ in both sets, by deleting edges and vertices from $C$ that do not lie in $C_0$. We notice that this update can be implemented via a sequence of $r$ cluster splitting operations, where, for $1\leq j\leq r$, the $j$th such operation splits cluster $C_j$ from $C$.
This completes the description of the algorithm.

In order to analyze this algorithm, it will be convenient to define a \emph{partitioning tree}. For every cluster $C$ that ever lied in $\cset$, let $\hat C$ denote the same cluster when it was first added to $\cset$, and let $\hat \cset$ denote the collection of all graphs $\hat C$, where $C$ is a cluster that ever lied in $\cset$. Let $T$ be a tree, whose vertex set is $\set{v_{\hat C}\mid \hat C\in \hat \cset}$. Consider now any graph $\hat C\in \hat \cset$, and any application of the algorithm from \Cref{claim: process single cluster} to the corresponding cluster $C$. Let $\hset=\set{C_0,C_1,\ldots,C_r}$ be the collection of clusters computed by \Cref{claim: process single cluster}. Then we say that $C_1,\ldots,C_r$ are \emph{child clusters} of $\hat C$, and, for all $1\le j\leq r$, we add an edge $(v_{\hat C},v_{\hat C_j})$ to the tree. We also denote the set $U_j$ of vertices corresponding to $\hat C_j$ by $U(\hat C_j)$. This completes the description of the partitioning tree. 
We say that a cluster $\hat C\in \hat\cset$ is \emph{small} if $|V(\hat C)|\leq n^{1/k}$, and we say that it is \emph{large} otherwise.

For every cluster $\hat C\in \hat \cset$, we denote by $\wset(\hat C)$ the collection of its child clusters. 
Note that, if $\hat C$ is a small cluster, then $\wset(\hat C)=\emptyset$. Otherwise,
from \Cref{claim: process single cluster}, we are guaranteed that, for each child cluster $\hat C'$ of $\hat C$, $|V(\hat C')|\le |V(C)|^{1-1/k}\leq \frac{|V(\hat C)|}{n^{1/k^2}}$. Additionally, we are guaranteed that, for each such cluster $\hat C'$, $|U(\hat C')|\geq \frac{|V(\hat C')|}{|V(\hat C)|^{1/k^3}}$. Moreover, the sets $\set{U(\hat C')\mid \hat C'\in \wset(\hat C)}$ of vertices are mutually disjoint. It is then easy to verify that:

\begin{equation}\label{eq: vertex growth}
\sum_{\hat C'\in \wset(C)}|V(\hat C')|\leq \sum_{\hat C'\in \wset(C)}|U(\hat C')|\cdot |V(\hat C)|^{1/k^3}\leq |V(\hat C)|^{1+1/k^3} \end{equation}

For all $i\geq 0$, let $S_i$ denote the set of vertices of the tree $T$ that lie at distance exactly $i$ from the root, and let $\cset^i=\set{\hat C\in \hat \cset\mid v_{\hat C}\in S_i}$ be the corresponding collection of clusters. Then, from the above discussion, for every index $i\geq 0$, every cluster $\hat C\in \cset^i$ contains at most $n^{1-i/k^2}$ vertices, and so the height of the partitioning tree is at most $k^2$.

From Inequality \ref{eq: vertex growth}, it is easy to verify that, for all $1\le i\leq k^2$:

\[\sum_{\hat C\in \cset^i}|V(\hat C)|\leq \sum_{\hat C\in \cset^{i-1}}|V(\hat C)|^{1+1/k^3}\leq \left (\sum_{\hat C\in \cset^{i-1}}|V(\hat C)|\right)\cdot n^{1/k^3}.\]

Overall, we get that $\sum_{v\in V(G)}n_v=\sum_{\hat C\in \hat \cset}|V(\hat C)|\leq 2n^{1+1/k}$, as required.

It now remains to bound the running time of each phase. Recall that the time required to process a single cluster $C$ by the algorithm from 
\Cref{claim: process single cluster} is bounded by  $O(|E(C)|\cdot |V(C)|^{1/k^3}$. At the beginning of every phase, all clusters of $\cset^A$ that need to be processed are disjoint in their edges, and these are the only clusters processed in that phase. Therefore, the running time of each phase (including the initialization phase) is bounded by $O(\hat m^{1+1/k^3})$.


\section{Combining All Tools: the Proof of \Cref{thm: one level router decomposition}}
	\label{sec: tieup}

In this section we complete the prooof of \Cref{thm: one level router decomposition} by combining the tools presented in Sections \ref{sec: nice router}--\ref{sec: clustering}.

In order to simplify the notation in this subsection, we call a graph $G$ and parameters $n,k,\delta, \Delta$ and $\Delta^*$ that satisfy the conditions of  \Cref{thm: one level router decomposition} (except for the constraint on vertex degrees in $G$) as ``valid input graph and parameters''. For completeness we define them below.

\begin{definition}[Valid input graph and parameters]\label{def: input graph and parameters}
	Assume that we are given a parameter $n$, that is greater than a large enough constant, and additional parameters 
	 $512\leq k\leq (\log n)^{1/49}$,
	$\frac{1}k\leq \delta\leq \frac{1}{400}$,  $\Delta^*$, and $\Delta$, such that $n^{18/k}\leq \Delta^*\leq \frac{\Delta}{n^{8/k^2}}\leq n$ holds, and $k,\Delta,\Delta^*$ and $1/\delta$ are integers.
	Lastly, assume that we are given a simple graph $G$ with $|V(G)|\leq n$. We refer to a graph $G$ and parameters $n,k,\delta,\Delta,\Delta^*$ that satisfy these constraints as a \emph{valid input graph $G$ with valid input parameters $n,k,\delta,\Delta$ and $\Delta^*$}.
\end{definition}

Throughout, we use a large enough constant $c$ from from the definition of the \embedorscatter problem (see \Cref{def: embedorscatter}).

Given a valid input graph $G$ with valid input parameters $n,k,\delta,\Delta$ and $\Delta^*$, we define additional parameters
 $d^*=k^{10}\cdot 2^{c/\delta^6}$, 
$\eta^*=2^{c/\delta^6}\cdot k^{4ck^2}\cdot n^{4/k^2}$, 
and $\hk=k^2$. 
Note that, if  $n,k,\delta,\Delta$ and $\Delta^*$ are valid input parameters, then, since $k\leq (\log n)^{1/49}$ holds, we get that $k^{29}\leq \frac{\log n}{k^{20}}$, and so:

\begin{equation}\label{eq: k and n}
 2^{k^{29}}\leq n^{1/k^{20}}.
 \end{equation}

Additionally, since $n$ is sufficiently large, and $k^{49}\leq \log n$ holds, we can assume that $ck^{40}\leq c(\log n)^{40/49}\leq \log n$ holds, and so $2^{ck^{40}}\leq n$. Since $\delta\geq 1/k$, we get that:

\begin{equation}\label{eq: k and n2}
2^{c/\delta^6}\leq 2^{ck^{10}}\leq n^{1/k^{10}}.
\end{equation}

From the definition of $d^*$ and inequality \ref{eq: k and n2}, we get that:

\begin{equation}\label{eq: bound on d*}
 d^*=2^{c/\delta^6}\cdot k^{10k} \leq 2^{ck^6+11k}\leq n^{1/k^6}.
\end{equation}

Lastly, since $\eta^*=2^{c/\delta^6}\cdot k^{4ck^2}\cdot n^{4/k^2}$, we get that:

\begin{equation}\label{eq: bound d eta}
\eta^*\cdot d^*=2^{2c/\delta^6}\cdot k^{8ck^2}\cdot n^{4/k^2}\leq 2^{4ck^6}\cdot n^{4/k^2}\leq n^{5/k^2}.
\end{equation}

The proof of \Cref{thm: one level router decomposition} consits of two parts. In the first part, we provide an algorithm that maintains the collection $\cset$ of edge-disjoint clusters of $G$, such that every cluster $C\in \cset$ is a $(\vdeg_C,\tilde d,\tilde \eta)$-router with respect to the set $V(C)$ of supported vertices. The second part provides an algorithm that, for every cluster $C\in \cset$, maintains a subgraph $C'\subseteq C$ with $V(C')=V(C)$, such that $C'$ is a $(\vDelta^*,\tilde d,\tilde \eta)$-router for the set $V(C)$ of supported vertices, and $|E(C')|\leq \Delta^*\cdot |V(C)|$.

\subsection{Part 1: Maintaining the Clustering $\cset$}

In this subsection we provide an algorithm that  maintains a collection $\cset$ of edge-disjoint clusters of $G$, such that every cluster $C\in \cset$ is a $(\vdeg_C,d^*,\eta^*)$-router with respect to the set $V(C)$ of supported vertices. A central notion that we use in this part is a \emph{router certificate}, that we define next.

\begin{definition}[Router Certificate]\label{def: router certificate}
	Let $G$ be a valid input graph, and let $n,k,\delta,\Delta$ and $\Delta^*$ be the corresponding valid input parameters. Let $d^*=k^{10}\cdot 2^{c/\delta^6}$, $\eta^*=2^{c/\delta^6}\cdot k^{4ck^2}\cdot n^{4/k^2}$, and $\hk=k^2$. For a given subgraph (cluster) $C\subseteq G$, a \emph{router certificate} for $C$  consists of:
	\begin{itemize}
		\item a  graph $W_C$, that is a properly pruned subgraph of $W^{N_C,\Delta}_{\hk}$ for some parameter $\half\cdot |V(C)|^{1/\hk-1/\hk^2}\leq N_C\leq (2|V(C)|)^{1/\hk}$, and
		
		\item an embedding $\pset_C$ of $W_C$ into $C$ via paths of length at most $d^*$ that cause congestion at most $\eta^*$, such that every edge $e\in E(C)$ lies on some path in $\pset_C$, and every vertex $v\in V(C)$ participates in at least $\frac{\Delta}{n^{4/k^2}}$ paths in $\pset_C$. 
	\end{itemize}

	Moreover, graph $W_C$ must be obtained from $W^{N_C,\Delta}_{\hk}$ by applying the algorithm from \Cref{thm: pruning of Wk} for the $\pruning$ problem  to it, with a single phase $\Phi_0$ of updates, consisting of at most $\frac{N_C^{\hk}\cdot \Delta}{2\hk^{4\hk}}$ edge deletions. 
\end{definition}

Next, we describe the initialization step, in which we construct an initial clustering $\cset$, and, for every cluster $C\in \cset$, we construct a router certificate $\left (W^{N_C,\Delta}_{\hk},W_C,\pset_C\right )$. Intuitively, from \Cref{lem: nice router to uniform degree router}, the existence of the router certificate for each such cluster $C$ proves that $C$ is a $(\vdeg_C,\tilde d,\tilde \eta)$-router, for appropriately chosen parameters $\tilde d$ and $\tilde \eta$. Later, we describe an algorithm that maintains the clusters in $\cset$ as the graph $G$ undergoes $k$ batches of edge deletions. For each cluster $C\in \cset$, the algorithm will also maintain a subgraph $\hat W_C\subseteq W_C$, such that, for every vertex $v$ that remains in $C$, the number of the embedding paths paths in $\set{P(e)\in \pset_C\mid e\in E(\hat W_C)}$ that contain $v$ is sufficiently large. Again, from \Cref{lem: nice router to uniform degree router}, this is sufficient in order to ensure that each cluster $C\in \cset$ remains a $(\vdeg_C,\tilde d,\tilde \eta)$-router, for appropriately chosen parameters $\tilde d$ and $\tilde \eta$. We now provide the algorithm to initialize the clustering $\cset$, and to compute an initial router certificate for every cluster $C\in \cset$.

\subsubsection{Initialization: Constructing the Clustering and Router Certificates}

The algorithm for the first part of the initialization is summarized in the following theorem.

\begin{theorem}\label{thm: initialization router witness}
	There is a deterministic algorithm, that receives as input a valid input graph $G$ together with valid input parameters $n,k,\delta,\Delta$ and $\Delta^*$, such that,  for every vertex $v\in V(G)$, $\deg_G(v)\leq \Delta\cdot n^{16/k}$.
	The algorithm computes a collection $\cset$ of subgraphs of $G$ called \emph{clusters}  that are disjoint in their edges but may share vertices, with $\sum_{C\in \cset}|V(C)|\leq n^{1+20/k}$. For every cluster $C\in \cset$, it also computes a router certificate $\left (W^{N_C,\Delta}_{k^2},W_C,\pset_C\right )$.
	If we denote by  $\edel=E(G)\setminus \left(\bigcup_{C\in \cset}E(C)\right )$, then the algorithm guarantees that $|\edel|\leq |V(G)|\cdot \Delta\cdot n^{2/k^2}$.
	The running time of the algorithm is $O\left((n\Delta)^{1+O(\delta)} \right )$.
\end{theorem}

Recall that, from the definition of a router certificate, for every cluster $C\in \cset$, our algorithm initializes the algorithm from \Cref{thm: pruning of Wk} for the $\pruning$ problem on graph $W_C=W^{N_C,\Delta}_{\hk}$, that we denote by $\aset_C$. During the phase $\Phi_0$, Algorithm $\aset_C$ performs an initial pruning of $W_C$, and the resulting graph serves as the router certificate for $C$. After the initialization step, as our algorithm receives batches $\pi_1,\ldots,\pi_k$ of edge deletions for graph $G$, we may delete additional edges from $W_C$, over the course of $k$ phases $\Phi_1,\ldots,\Phi_k$, that correspond to the above batches of updates to $G$. We will employ the same algorithm $\aset_C$, that was initialized as part of the algorithm in \Cref{thm: initialization router witness}, in order to implement the additional pruning of graph $W_C$ over the course of the phases $\Phi_1,\ldots,\Phi_k$.

The proof of \Cref{thm: initialization router witness} easily follows from the following lemma, that is almost identical to \Cref{thm: initialization router witness}. The main difference is that the lemma only ensures that 
$\sum_{C\in \cset}|E(C)|\geq \frac{|E(G)|}{2 n^{18/k}}$,  while \Cref{thm: initialization router witness} requires that all but at most $|V(G)|\cdot \Delta\cdot n^{2/k^2}$ edges of $G$ lie in $\bigcup_{C\in \cset}E(C)$. The lemma also requires that all vertex degrees in $G$ are at least $4\Delta\cdot n^{1/k^2}$.

\begin{lemma}\label{lem: initialization router witness}
	There is a deterministic algorithm, that receives as input a valid input graph $G$ together with valid input parameters $n,k,\delta,\Delta$ and $\Delta^*$, such that,  for every vertex $v\in V(G)$,  $4\Delta\cdot n^{1/k^2}\leq \deg_G(v)\leq \Delta\cdot n^{16/k}$.
The algorithm computes a collection $\cset$ of subgraphs of $G$ called \emph{clusters}  that are disjoint in their edges but may share vertices, with $\sum_{C\in \cset}|V(C)|\leq  2n^{1+1/(4k^2)}$. For every cluster $C\in \cset$, it also computes a router certificate $\left (W^{N_C,\Delta}_{k^2},W_C,\pset_C\right )$.
The algorithm ensures that $\sum_{C\in \cset}|E(C)|\geq \frac{|E(G)|}{4n^{18/k}}$. 
	The running time of the algorithm is $O\left((n\Delta)^{1+O(\delta)}  \right )$.
\end{lemma}

We prove the lemma below, after we complete the proof of \Cref{thm: initialization router witness} using it.

\begin{proofof}{\Cref{thm: initialization router witness}}
Our algorithm consists of at most $n^{19/k}$ iterations.  We start with $G'=G$, $\cset=\emptyset$, and $\edel=\emptyset$. 

We now describe the $i$th iteration, for $i\geq 1$. We start by iteratively removing from $G'$ all vertices whose current degree is below $4\Delta\cdot n^{1/k^2}$. All edges that were removed from $G'$ in this step are added to set $\edel$. If no vertices remain in $G'$, then we terminate the algorithm.
Otherwise, we apply the algorithm from \Cref{lem: initialization router witness} to the resulting graph 
$G'$, and we let $\cset_i$ be the collection of clusters that the algorithm returns. We add the clusters in $\cset_i$ to $\cset$, and we delete from $G'$ all edges that lie in $\bigcup_{C\in \cset_i}E(C)$. 
Let $E_i=\bigcup_{C\in \cset'}E(C)$, and recall that, from \Cref{lem: initialization router witness}, $|E_i|\geq \frac{|E(G')|}{4n^{18/k}}$.
We then continue to the next iteration.

We now bound the number of iterations. At the beginning of the algorithm, $|E(G)|\leq n^{1+16/k}\cdot \Delta$ holds, since all vertex degrees are bounded by $\Delta\cdot n^{16/k}$. As long as the algorithm continues, $G'$ contains at least one vertex of degree at least $4\Delta\cdot n^{1/k^2}$. From the above calculations, after the $i$th iteration:

\[|E(G')|\leq |E(G)|\cdot \left(1-\frac{1}{4 n^{18/k}}\right )^i\leq  n^{1+16/k}\cdot \Delta\cdot \left(1-\frac{1}{4 n^{18/k}}\right )^i \]

holds. Clearly, after at most $8\cdot n^{18/k}\cdot \log n$ iterations, the algorithm must terminate.
Since $k\leq (\log n)^{1/10}$, we get that $n^{1/k}\geq 2^{(\log n)^{9/10}}\geq 16\log n$ (since we have assumed that $n$ is sufficiently large). Therefore, altogether, the number of iterations is bounded by $n^{19/k}$.

It is also easy to verify that the total number of edges that were ever added to $\edel$ is bounded by $4|V(G)|\Delta\cdot n^{1/k^2}\leq |V(G)|\cdot \Delta\cdot n^{2/k^2}$, and every edge of $G$ that does not lie in $\edel$ must lie in one of the clusters of $\cset$; moreover, the clusters in $\cset$ are disjoint in their edges. Lastly recall that, from \Cref{lem: initialization router witness}, for all $i\geq 1$, $\sum_{C\in \cset_i}|V(C)|\le  2n^{1+1/(4k^2)}$. Since the number of iterations is bounded by $ n^{19/k}$, we get that:

\[\sum_{C\in \cset}|V(C)|\le  2n^{1+1/(4k^2)}\cdot n^{19/k}\leq n^{1+20/k}.\]


It now only remains to bound the running time of the algorithm. The running time of the algorithm from \Cref{lem: initialization router witness} is $O\left((n\Delta)^{1+O(\delta)}\right )$, and the number of iterations is at most $n^{19/k}$. In every iteration, we also need to spend $O(|E(G)|)\leq O(n^{1+16/k}\cdot \Delta)$ time to remove low-degree vertices from $G'$. The total running time of the algorithm is then bounded by $O\left((n\Delta)^{1+O(\delta+1/k)} \right )\leq O\left((n\Delta)^{1+O(\delta)} \right )$, since $\delta\geq 1/k$.
\end{proofof}

Next, we prove \Cref{lem: initialization router witness}.

\begin{proofof}{\Cref{lem: initialization router witness}}
The algorithm consists of at most $z=16k^3$ phases. Throughout the algorithm, we maintain a collection $\cset$ of clusters that are disjoint in their edges but may share vertices. The set $\cset$ of clusters is partitioned into two subsets: set $\cset^I$ of \emph{inactive} clusters, and set $\cset^A$ of \emph{active} clusters. 
For every cluster $C\in \cset$, if $|V(C)|\leq \Delta$, then we say that it is a \emph{small cluster}, and we say that it is a \emph{large cluster} otherwise.
Throughout the algorithm, we use a parameter $d=256k^6=4(4\hk)^3$.
 Our algorithm also maintains two sets $\edel_1,\edel_2$ of edges.
We ensure that the following invariants hold throughout the algorithm.

\begin{properties}{I}
	\item If $C\in \cset^I$ is an inactive cluster that is large, then the algorithm computes a router certificate $\left (W^{N_C,\Delta}_{\hk},W_C,\pset_C\right )$ for $C$, when $C$ is first added to $\cset^I$. After that time, cluster $C$ and its corresponding router certificate remain unchanged; \label{inv: inactive witness}

\item  every edge of $G$ lies in exactly one of the sets $\edel_1$, $\edel_2$, and $\set{E(C)}_{C\in \cset}$ at all times; \label{inv: partition of edges}

\item $\sum_{C\in \cset}|V(C)|\leq 2n^{1+1/(4k^2)}$ holds at all times;\label{inv: sum of vertices}

\item  for all $j>0$, after the $j$th phase, $|\edel_1|\leq 4j\cdot n^{1+1/(4k^2)}\cdot \Delta$ holds;\label{inv: few type 1 deleted}
 and  
 
 \item $|\edel_2|\leq n^{18/k}\cdot \sum_{C\in \hat \cset^I}|E(C)|$ hold at all times, where $\hat \cset^I$ is the set of all large inactive clusters.\label{inv: few type 2 deleted}

\end{properties}

At the beginning of the algorithm, we initialize the algorithm from \Cref{thm: clustering algorithm} for the Low-Diameter Clustering problem on graph $G$, with the parameter $k$ replaced by $4\hk=4k^2$, and obtain an initial clustering $\cset$, such that every cluster $C\in \cset$ is settled. For every cluster $C\in \cset$, if $|V(C)|\leq n^{1/(4k^2)}$, then $C$ is small, since $\Delta\geq n^{1/(4k^2)}$ holds from the definition of valid input parameters. We add each cluster $C$ with $|V(C)|\leq n^{1/(4k^2)}$ to $\cset^I$; and we add all other clusters to $\cset^A$. We also initialize $\edel_1=\emptyset$ and $\edel_2=\emptyset$. 
Notice that Invariants \ref{inv: inactive witness}--\ref{inv: few type 2 deleted} hold at the beginning of the algorithm.
We perform phases as long as $\cset^A\neq \emptyset$ holds; we will show later that the total number of phases is bounded by $z=16k^3$. We now describe the $j$th phase, for some $j\geq 1$. We assume that, at the beginning of the phase, $\cset^A\neq\emptyset$, and Invariants \ref{inv: inactive witness}--\ref{inv: few type 2 deleted} hold.

\paragraph{Execution of the $j$th phase.}
In order to execute the $j$th phase, we process every cluster $C\in \cset^A$ one by one. We now provide an algorithm for processing the cluster $C$. First, we delete all isolated vertices from $C$, and, if $C$ becomes a small cluster, then we move it to $\cset^I$. In this case, no further processing of $C$ is needed. From now on we assume that $C$ remains a large cluter, so $|V(C)|\geq \Delta$ holds.

Our next step is to apply the algorithm from \Cref{thm: embedorscatter} for the $\embedorscatter$ problem to cluster $C$, with parameters $\Delta$, $\delta$ and $d=256k^6$ remaining unchanged, and parameter $k$ replaced with $\hk$. For convenience, we denote $n'=|V(C)|$. In the following simple claim, whose proof is deferred to Section \ref{subsec: valid input to embed or scatter} of Appendix, we verify that this is indeed a valid input to the $\embedorscatter$ problem.

\begin{claim}\label{claim: valid input to embedorscatter}
	Graph $C$ together with parameters $\Delta,\delta,d$ and $\hk$ as defined above are a valid input to the \embedorscatter problem.
\end{claim}

We now consider two cases. The first case happens if the algorithm for the \embedorscatter problem, when appied to cluster $C$, 
computed a subset $E_C\subseteq E(C)$ of at most $2|V(C)|\cdot \Delta$ edges, such that $C\setminus E_C$ is $(d,1/\hk)$-scattered. In this case, we delete the edges of $E_C$ from cluster $C$, and then delete isolated vertices from $C$. The edges of $E_C$ are then added to set $\edel_1$. If cluster $C$ becomes small, then we move it from $\cset^A$ to $\cset^I$. Otherwise, it remains in $\cset_A$. Note that, from the definition of a scattered graph, we are now guaranteed that, for every vertex $v\in V(C)$, $|B_C(v,d)|\leq |V(C)|^{1-1/k^2}\leq |V(C)|^{1-1/(4k^2)}$ holds. In particular, cluster $C$ is now unsettled.

Consider now the second case, when the algorithm for the \embedorscatter problem computes, for a parameter $N_C=\floor{|V(C)|^{1/\hk-1/\hk^2}}$, a collection $F\subseteq E(W_{\hk}^{N_C,\Delta})$ of at most $\frac{N_C^{\hk}\cdot \Delta}{(512\hk)^{4\hk}}$ fake edges of $W_{\hk}^{N_C,\Delta}$, together with an embedding $\pset_C$ of $W_{\hk}^{N_C,\Delta}\setminus F$ into $C$ via paths of length at most $d\cdot \hk\cdot 2^{O(1/\delta^6)}=256k^8\cdot 2^{O(1/\delta^6)}\leq d^*$ that cause congestion at most $n^{4/\hk}\cdot d\cdot \hk^{c\hk}\cdot 2^{c/\delta^6}\leq \eta^*$ (recall that $d=256k^6$, $\hk=k^2$, $d^*=k^{10}\cdot 2^{c/\delta^6}$ and 
$\eta^*=2^{c/\delta^6}\cdot k^{4ck^2}\cdot n^{4/k^2}$). Next, we will compute a router certificate for $C$, after possibly deleting some edges and vertices from it. 

In order to do so, for every vertex $v\in V(C)$, we initialize a counter $n_v$ to be the number of edges in graph $W_{\hk}^{N_C,\Delta}\setminus F$ whose embedding path in $\pset_C$ contains $v$, and we initialize a list $L(v)$ that contains all such edges. Next, we initialize the algorithm for the \pruning problem from \Cref{thm: pruning of Wk} on graph $W_{\hk}^{N_C,\Delta}$, with parameter $k$ replaced by $\hk$ and parameter $N$ replaced by $N_C$. 
Recall that $\hk=k^2\geq 256$ and $\Delta\geq n^{18/k}\geq 4k^4=4\hk^{2}$ holds from the definition of valid input parameters (see \Cref{def: input graph and parameters}) and Inequality  \ref{eq: k and n}. 
Additionally: 

\[N_C=\floor{|V(C)|^{1/\hk-1/\hk^2}}\geq (n')^{1/(2k^2)}\geq n^{9/k^3}\geq 2^{k^4}\geq k^{6k^2}=\hk^{3\hk},\]

since $C$ is a large cluster (so $n'=|V(C)| \geq \Delta\geq n^{18/k}$), and from Inequality \ref{eq: k and n}. Therefore, we obtain a valid input to the \pruning problem. For convenience, in the following discussion, we denote parameter $N_C$ by $N$.

We denote the algorithm for the \pruning problem from \Cref{thm: pruning of Wk} that we initialized on graph $C$ with the above parameters by $\aset_C$.
We now describe the online sequence of edge deletions that serves as input to $\aset_C$. All these edge deletions are executed as part of Phase $\Phi_0$. Initially, algorithm $\aset_C$ receives the set $F$ of fake edges as part of the deletion sequence. Recall that algorithm $\aset_C$ maintains a graph $W_C$, that is a properly pruned subgraph of $W_{\hk}^{N,\Delta}$, together with the corresponding sets $U_1,\ldots,U_{\hk}$ of its vertices. As the algorithm progresses,  some edges are deleted from $W_C$, either as part of the input sequence of edge deletions, or by algorithm $\aset_C$ itself. Whenever an edge $e$ is deleted from $W_C$, we consider its corresponding embedding path $P(e)\in \pset_C$. For every vertex $v\in P(e)$, we decrease the corresponding counter $n_v$ by $1$, and we delete $e$ from list $L(v)$. Whenever, for any vertex $v\in V(C)$, the counter $n_v$ falls below $\frac{\Delta}{n^{4/k^2}}$, all the edges of $L(v)$ are deleted from $W_C$, by providing them to Algorithm $\aset_C$ as part of its online sequence of edge deletions.

Recall that $|F|\leq \frac{N^{\hk}\cdot \Delta}{(512\hk)^{4\hk}}$, and, for every vertex $v\in V(C)$, when the counter $n_v$ falls below $\frac{\Delta}{n^{4/k^2}}$, we may delete at most $\frac{\Delta}{ n^{4/k^2}}$ edges from $W_C$. 

We denote by $E'_0$ the online sequence of edge deletions that serves as input to Algorithm $\aset_C$ so far. Then $|E'_0|$ is bounded by:

\[\begin{split}
|F|+|V(C)|\cdot \frac{\Delta}{n^{4/k^2}}&\leq  \frac{N^{\hk}\cdot \Delta}{(512\hk)^{4\hk}}+|V(C)|\cdot \frac{\Delta}{ n^{4/k^2}}  \\
&\leq  \frac{N^{\hk}\cdot \Delta}{(512\hk)^{4\hk}}+N^{\hk}\cdot n^{2/k^2}\cdot \frac{\Delta}{ n^{4/k^2}}\\
&\leq \frac{N^{\hk}\cdot \Delta}{2{\hk}^{4{\hk}}}.
\end{split}\]

We have used the fact that
 $N=\floor{|V(C)|^{1/\hk-1/\hk^2}}$, and so $N^{\hk}\geq \frac{|V(C)|}{2^{k^2}\cdot n^{1/k^2}}\geq \frac{|V(C)|}{ n^{2/k^2}}$. We have also used Inequality  \ref{eq: k and n}.

We update the set $\pset_C$ of embedding paths, so that it only contains paths corresponding to the edges that currently lie in $W_C$.
Let $V'\subseteq V(C)$ be the set of vertices $v\in V'$, such that $v$ participates in fewer than $\frac{\Delta}{n^{4/k^2}}$ of the paths in $\pset_C$. Note that our algorithm ensures that every vertex $v\in V'$ in fact does not participate in any paths in $\pset_C$. Consider now the graph $C'\subseteq C$ that contains all edges and vertices of $C$ that lie on the paths in $\pset_C$. In other words, graph $C'$ is obtained from $C\setminus V'$, by deleting from it all edges that do not lie on the paths in $\pset_C$.
In the next simple claim, whose proof is deferred to Section \ref{subsec: router certificate} of Appendix, we prove that $\left(W^{N,\Delta}_{\hk},W_C,\pset_C\right )$ is a valid router certificate for $C'$, and that $|E(C')|\geq \frac{|E(C)|}{n^{18/k}}$.

\begin{claim}\label{claim: router certificate}
	$\left(W^{N,\Delta}_{\hk},W_C,\pset_C\right )$ is a valid router certificate for $C'$, and  $|E(C')|\geq \frac{|E(C)|}{n^{18/k}}$.
\end{claim}

We add the edges of $E(C)\setminus E(C')$ to the set $\edel_2$, and we \emph{charge} these edges to the edges of $C'$, where the charge to every edge is at most $n^{18/k}$. Next, we delete from $C$ all edges and vertices, except those lying in $C'$, and we move the resulting cluster $C$ from $\cset^A$ to $\cset^I$. This completes the description of the procedure for processing a cluster $C\in \cset^A$. Next, we analyze its running time.

The running time of the algorithm from \Cref{thm: embedorscatter} for the \embedorscatter problem is bounded by:

\[O\left(|E(C)|^{1+O(\delta+1/\hk)}\cdot d^2 \cdot 2^{O(1/\delta^6)} \right ) \leq O\left(|E(C)|^{1+O(\delta+1/k^2)}\cdot k^6 \cdot 2^{O(1/\delta^6)} \right ).\]

The time required to initialize the counters $n_v$ and the lists $L_v$ for every vertex $v\in V(C)$, and to maintain them during the pruning procedure is subsumed by the above running time.
The running time of the algorithm from \Cref{thm: pruning of Wk} for the \pruning problem is bounded by:

\[O\left(\hat k^2\cdot N^{\hat k}+|E'_0|\cdot \hk^{O(\hk)}\right )\leq O\left(N^{\hk}\cdot \Delta\cdot k^{O(k^2)}\right )\leq O\left(|V(C)|\cdot \Delta\cdot   k^{O(k^2)}\right ).\]

Altogether, the time required to process a cluster $C$ is bounded by:

\[ O\left(|E(C)|^{1+O(\delta+1/k^2)}\cdot k^6 \cdot 2^{O(1/\delta^6)} \right )+  O\left(|V(C)|\cdot \Delta\cdot   k^{O(k^2)}\right ).\]

Recall that, from Invariant \ref{inv: partition of edges}, the clusters in $\cset$ do not share edges, and, from Invariant \ref{inv: sum of vertices}, $\sum_{C\in \cset}|V(C)|\leq 2n^{1+1/(4k^2)}$ holds. Therefore, the time required to process all clusters in a single phase is bounded by:

\[\begin{split}
&O\left(|E(G)|^{1+O(\delta+1/k^2)}\cdot k^6 \cdot 2^{O(1/\delta^6)} \right )+  O\left(n^{1+1/(4k^2)}\cdot \Delta\cdot   k^{O(k^2)}\right )\\
&\quad\quad\quad\quad\quad\quad\quad\quad\quad\leq O\left((n\Delta)^{1+O(\delta+1/k)}\cdot 2^{O(1/\delta^6)}\right )\\
&\quad\quad\quad\quad\quad\quad\quad\quad\quad\leq \left((n\Delta)^{1+O(\delta)}\right ),
\end{split}\]

since $|E(G)|\leq n\cdot \Delta\cdot n^{O(1/k)}$, that  $\delta\geq 1/k$, and Inequality \ref{eq: k and n2}.

Next, we verify that all invariants continue to hold after we finish processing all clusters in $\cset^A$. Invariants \ref{inv: inactive witness} and \ref{inv: partition of edges} are immediate from our algorithm, and Invariant \ref{inv: sum of vertices} continues to hold since the only changes to the clusters in $\cset$ are edge- and vertex-deletions. Invariant \ref{inv: few type 2 deleted} follows immediately from our charging scheme.
In order to show that Invariant \ref{inv: few type 1 deleted} continues to hold, it is enough to show that the total number of edges added to $\edel_1$ in the current iteration is bounded by $4n^{1+1/(4k^2)}\cdot \Delta$. Recall that, for every cluster $C\in \cset^A$ that the algorithm processes, at most $2|V(C)|\cdot \Delta$ edges are added to $\edel_1$. Since, from \ref{inv: sum of vertices}, 
$\sum_{C\in \cset}|V(C)|\leq 2n^{1+1/(4k^2)}$ holds, we get that the total number of edges added to $\edel_1$ in the current phase is bounded by:

\[2\Delta\cdot \sum_{C\in \cset}|V(C)|\leq 4n^{1+1/(4k^2)}\cdot \Delta. \]

Once all clusters in $\cset^A$ are processed, the algorithm for the Low-Diameter Clustering problem is notified of the changes to the clusters in $\cset^A$, and to the sets $\cset^I,\cset^A$ of clusters. 
Recall that the algorithm is then required to amend the current clustering in $\cset$, by performing cluster splitting operations to the clusters in $\cset^A$, to ensure that all clusters in $\cset^A$, including the newly added ones, are settled.
The algorithm does not add any clusters to $\cset^I$, and it does not add any edges to $\edel_1$ or $\edel_2$. Moreover, it ensures that 
$\sum_{C\in \cset}|V(C)|\leq 2n^{1+1/(4\hk)}=2n^{1+1/(4k^2)}$ continues to hold. Therefore, all invariants continue to hold at the end of the phase.
This completes the description of the algorithm. 
In the next claim we bound the number of phases in the algorithm.

\begin{claim}\label{claim: num of phases}
	After $16k^3$ phases, $\cset^A=\emptyset$ must hold.
\end{claim}
\begin{proof}
	We partition the algorithm's execution $\ceil{\log k}$ stages. For $1\leq j\leq \ceil{\log k}$, the $j$th stage terminates when, for every cluster $C\in \cset^A$, $|V(C)|\leq n^{1/2^j}$ holds. At the end of the last stage, for every cluster $C\in \cset^A$, $|V(C)|\leq n^{1/k}<\Delta$ must hold, so $\cset^A=\emptyset$.

	\begin{observation}\label{obs: cluster sizes go down}
		For all $1\leq j\leq \ceil{\log k}$, the number of phases in Stage $j$ of the algorithm is bounded by $16k^2$.
	\end{observation}

Note that the above observation implies that the number of phases in the algorithm is bounded by $16k^2\cdot \ceil{\log k}\leq 16k^3$.
It now remains to prove the observation.

\begin{proof}
	In order to prove the observation it is enough to show that, for all $1\leq i\leq 16k^2$, if $C$ is a cluster that lies in $\cset^A$ at the end of the $i$th phase of Stage $j$, then $|V(C)|\leq n^{\frac 2{2^j}-\frac{i-1}{2^{j+3}\cdot k^2}}$.

	The proof is by induction on $i$. For $i=1$, the claim clearly holds. Consider now some integer $i>1$, and assume that the claim holds at the end of phase $i-1$ of stage $j$. Let $C$ be any cluster that lies in $\cset^A$ at the beginning of phase $i$ of stage $j$, and assume that, once the cluster is processed by the algorithm for the \embedorscatter problem, it remains in $\cset^A$. In other words, the algorithm for the $\embedorscatter$ problem from \Cref{thm: embedorscatter}, when applied to cluster $C$, computed a subset $E_C\subseteq E(C)$ of edges, that were subsequently deleted from $C$, such that, after the deletion of these edges, for every vertex $v\in V(C)$, $|B_C(v,d)|\leq |V(C)|^{1-1/k^2}$ holds, and cluster $C$ becomes unsettled.
	
	The algorithm for the Decremental Low-Diameter Clustering problem then needs to ``fix'' this cluster, by splitting some clusters $C'\subseteq C$ from it; we call each such cluster $C'$ a \emph{child-cluster} of $C$. Recall that, from the definition of the Low-Diameter Clustering problem, for each child-cluster $C'$ of $C$:
	
	\[|V(C')|\leq |V(C)|^{1-\frac 1{4k^2}}.\]
	
	Recall that, from the induction hypothesis:
	
	\[|V(C)|\leq n^{\frac 2{2^j}-\frac{i-2}{2^{j+3}\cdot k^2}}.\]
	
	At the same time, since the algorithm is still at stage $j$, $|V(C)|\geq n^{\frac{1}{2^{j}}}$. Therefore:

\[	|V(C)|^{\frac 1 {4k^2}}\geq 
	\left(n^{\frac 1{2^{j}}}\right)^{\frac 1{4k^2}}\geq n^{\frac{1}{2^{j+3}\cdot k^2}}. \]
	
Altogether, we get that:

\begin{equation}\label{eq: bound2}
|V(C)|^{1-1/(4k^2)}\leq n^{\frac 2{2^j}-\frac{i-1}{2^{j+3}\cdot k^2}},
\end{equation}

and so $|V(C')|\leq n^{\frac 2{2^j}-\frac{i-1}{2^{j+3}\cdot k^2}}$, as required.
	
	Next, we denote  by $\hat C$ the cluster $C$ that is obtained at the end of the phase, that is, after all its child clusters were split off from it in the current phase. For convenience, we continue to use the notation $C$ for the cluster $C$ before the splitting operations. Then cluster $\hat C$ is settled, and so there is some vertex $v\in V(\hat C)$, with $|B_{\hat C}(v,d)|\geq |V(\hat C)|^{1-1/(4k^2)}$. Recall however that, after Algorithm $\embedorscatter$ was applied to the cluster, $|B_C(v,d)|\leq |V(C)|^{1-1/k^2}$ held. Since $|B_{\hat C}(v,d)|\leq |B_C(v,d)|$ must hold, we get that:
	
	\[|V(\hat C)|^{1-1/(4k^2)}\leq  |V(C)|^{1-1/k^2}, \]
	
	and so:
	
	\[
	|V(\hat C)|\leq |V(C)|^{1-1/k^2}\cdot |V(\hat C)|^{1/(4k^2)}
	\leq |V(C)|^{1-1/(4k^2)}\leq  n^{\frac 2{2^j}-\frac{i-1}{2^{j+3}\cdot k^2}},\]
	from Inequality \ref{eq: bound2}.
\end{proof}
\end{proof}

We are now ready to bound the running time of the algorithm so far. 
The running time of the algorithm from \Cref{thm: clustering algorithm} for the low-diameter clustering problem is bounded by $O(|E(G)|^{1+64/k^6})\leq O\left ((n\Delta)^{1+O(1/k)}\right)$.
The additional time required to execute each phase is bounded by: 
$O\left((n\Delta)^{1+O(\delta)}\right )$.

Therefore, the total running time of the algorithm so far is 
$O\left((n\Delta)^{1+O(\delta+1/k)}\cdot k^3\right )\leq O\left((n\Delta)^{1+O(\delta)}\right )$, from Inequality \ref{eq: k and n}, and since $\delta\geq 1/k$.

Next, we bound the cardinality of the set $\edel_1$ of edges.

\begin{observation}\label{obs: bound edel1}
	At the end of the algorithm, $|\edel_1|\leq \frac{|E(G)|}{8}$.
\end{observation}
\begin{proof}	
From \Cref{claim: num of phases}, the number of phases is bounded by $z=16k^3$. From Invariant \ref{inv: few type 1 deleted}, at the end of the last phase:

\[|\edel_1|\leq 4z\cdot  n^{1+1/(4k^2)}\cdot \Delta\leq 64k^3\cdot n^{1+1/(4k^2)}\cdot \Delta\leq \frac{\Delta\cdot n^{1+1/k^2}}{16}\leq \frac{|E(G)|}{8}\]

(we have used Inequality \ref{eq: k and n}, and the fact that all vertex degrees in $G$ are at least $4\Delta\cdot n^{1/k^2}$).
\end{proof}

Consider the set $\cset=\cset^I$ of clusters obtained at the end of the algorithm, and let $\hat \cset\subseteq \cset$ be the subset containing all large clusters. 
From Invariant \ref{inv: sum of vertices}, $\sum_{C\in \cset}|V(C)|\leq 2n^{1+1/(4k^2)}$ holds.
From Invariant \ref{inv: inactive witness}, for every cluster $C\in \hat \cset$, our algorithm has computed a router certificate $\left (W^{N_C,\Delta}_{\hk},W_C,\pset_C\right )$ for $C$. 
Additionally, from Invariant \ref{inv: few type 2 deleted}, 
$|\edel_2|\leq n^{18/k}\cdot \sum_{C\in \hat \cset}|E(C)|$ holds. Moreover, our charging scheme charges every edge of $\edel_2$ to some edge of $\bigcup_{C\in \hat \cset}E(C)$, such that the charge to every edge in $\bigcup_{C\in \hat \cset}E(C)$ is at most $n^{18/k}$. It now remains to process small clusters that remain in $\cset$. 
We use the following observation in order to bound the number of edges in the small clusters.

\begin{observation}\label{obs: few edges in small clusters}
	The total number of edges lying in the small clusters of $\cset^I$ is bounded by $\frac{|E(G)|}{4}$.
\end{observation}
\begin{proof}
	Recall that, if $C$ is a small cluster, then $|V(C)|\leq \Delta$. Since graph $G$ is simple, for every vertex $v\in V(C)$, $\deg_C(v)\leq \Delta$. Let $\cset'\subseteq \cset^I$ denote the collection of all small clusters. Then, from Invariant \ref{inv: sum of vertices}:
	
	\[\begin{split}
	\sum_{C\in \cset'}|E(C)|&\leq \Delta\cdot \sum_{C\in \cset}|V(C)|\\
	&\leq 2\Delta\cdot n^{1+1/(4k^2)}\\
	&\leq \frac{|E(G)|}{4},
		\end{split}\]
	
	since all vertex degrees in $G$ are at least $4\Delta\cdot n^{1/k^2}$.
\end{proof}

We simply delete all small clusters from $\cset^I$, and we add all their edges to set $\edel_1$. 
It is immediate to verify that $\sum_{C\in \cset}|V(C)|\leq 2n^{1+1/(4k^2)}$ continues to hold, and every edge of $G$ belongs to exactly one of the sets $\edel_1,\edel_2$, and $E(C)$ for $C\in \cset$. From our discussion, $|\edel_1|\leq \frac{|E(G)|}{2}$, and, from our charging scheme:

\[\sum_{C\in \cset}|E(C)|\geq \frac{|\edel_2|} {n^{18/k}}.\]

Therefore:

\[\sum_{C\in \cset}|E(C)|\geq \frac{\sum_{C\in \cset}|E(C)|}{2}+\frac{|\edel_2|} {2n^{18/k}}
\geq \frac{|E(G)\setminus\edel_1|}{2n^{18/k}}\geq  \frac{|E(G)|}{4n^{18/k}}.\]

This completes the proof of 
\Cref{lem: initialization router witness}.
\end{proofof}

\subsubsection{Updating the Clusters and the Router Certificates}

In this subsection we provide an algorithm for maintaining the clustering $\cset$. For every cluster $C\in \cset$, for all $1\leq i\leq k$, the algorithm updates the cluster $C$ and the corresponding graph $W_C$ following the batch $\pi_i^C=\pi_i\cap E(C)$ of edge deletions from $C$. The algorithm for updating a single cluster $C\in \cset$ is summarized in the following theorem. For simplicity, in the theorem, we denote $\pi_i^C$ by $\pi_i$. We note that the theorem statement requires that $|\pi_i^C|$ is sufficiently small; whenever this is not the case, we will destroy the cluster completely, and move all its edges to $\edel$. 

\begin{theorem}\label{thm: update one cluster}
	There is a deterministic algorithm, whose input consists of:
	
	\begin{itemize}
		\item a valid input graph $G$ together with valid input parameters $n,k,\delta,\Delta$ and $\Delta^*$, such that all vertex degrees in $G$ are at most $\Delta\cdot n^{16/k}$;
		
		\item a subgraph $C\subseteq G$ called a \emph{cluster}, together with a router certificate  $\left(W^{N_C,\Delta}_{\hk},W_C,\pset_C\right )$ for $C$; 
		
		\item for every edge $e\in E(C)$, a list $L(e)$ of all edges $\hat e\in E(W_C)$, whose embedding path $P(\hat e)\in \pset_C$ contains $e$; and
		\item for every vertex $v\in V(C)$, the number $n_v$ of paths in $\pset_C$ containing $v$.
	\end{itemize}
	
	Recall that, from the definition of the router certiciate,  graph $W_C$ must be obtained from $W^{N_C,\Delta}_{\hk}$ by applying the algorithm $\aset_C$ from \Cref{thm: pruning of Wk} for the $\pruning$ problem  to it, with a single phase $\Phi_0$ of updates, consisting of at most $\frac{N_C^{\hk}\cdot \Delta}{2\hk^{4\hk}}$ edge deletions. We assume that our algorithm is also given access to Algorithm $\aset_C$, including its inner state, at the end of Phase $\Phi_0$.
	The algorithm also receives as input $k$ online batches $\pi_1,\ldots,\pi_k$ of edge deletions from $C$, where, for all $1\leq i\leq k$, $|\pi_i|\leq \frac{|V(C)|\cdot \Delta}{n^{11/k^2}}$.
	
	The algorithm maintains, at all times, a graph $\hat C\subseteq C$ and a graph $\hat W_C\subseteq W_C$, that is a properly pruned subgraph of $W^{N_C,\Delta}_{\hk}$. Initially, $\hat C=C$ and $\hat W_C=W_C$ hold, and the only changes that graphs $\hat C$ and $\hat W_C$ undergo is the deletion of some of their edges and vertices over time. Throughout, we denote by $\pset_C=\set{P(e)\mid e\in E(\hat W_C)}$ the embedding of the current graph $\hat W_C$ into $C$, where the embedding paths $P(e)$ do not change over the course of the algorithm.  The algorithm guarantees that, at all times, for every edge $e\in E(\hat W_C)$, its embedding path $P(e)\in \pset_C$ is contained in the current graph $\hat C$; for every edge $e'\in E(\hat C)$, some path in $\pset_C$ contains $e$; and for every vertex $v\in V(\hat C)$, at least $\frac{\Delta}{n^{6/k^2}}$  paths in $\pset_C$ contain $v$. Lastly, the algorithm ensures that, for all $1\leq i\leq k$, at most $|\pi_i|\cdot  n^{5/k^2}$ edges are deleted from $\hat C$, and at most $|\pi_i|\cdot n^{3/k^2}$ edges are deleted from $\hat W_C$  while processing batch $\pi_i$.
	For all $1\leq i\leq k$, the time required to process batch $\pi_i$ is at most $O\left(|\pi_i|\cdot n^{O(1/k^2)}\right )$.
\end{theorem}

The remainder of this subsection is dedicated to the proof of \Cref{thm: update one cluster}. 
At the beginning of the algorithm, we let $\hat C=C$ and $\hat W_C=W_C$. Throughout the algorithm, we denote by $\pset_C=\set{P(e)\mid e\in E(\hat W_C)}$, where for each edge $e\in E(\hat W_C)$, its embedding path $P(e)$ remains unchanged throughout the algorithm.
We will ensure that, throughout the algorithm, the following three conditions hold:

\begin{properties}{C}
	\item  for all $1\leq i\leq k$, if an edge $e$ lies in $\hat W_C$ after batch $\pi_i$ is processed, then its embedding path $P(e)\in \pset_C$ is contained in the current graph $\hat C$; \label{cond: embedding contained}
	
	\item for all $1\leq i\leq k$, if an edge $e$ lies in $\hat C$ after batch $\pi_i$ is processed, then some path in $\pset_C$ contains $e$; and \label{cond: edge in emb path}

	\item for all $1\le i\leq k$, if a vertex $v$ lies in $\hat C$ after batch $\pi_i$ is processed, then $v$ belongs to at least $\frac{\Delta}{n^{4/k^2}\cdot \left(4k^{16k^2}\cdot  d^*\right )^i}$ paths in $\pset_C$. \label{cond: many paths contain a vertex}
\end{properties}

Clearly, Condition \ref{cond: embedding contained} holds at the beginning of the algorithm, and, from the definition of a Router Certificate (see \Cref{def: router certificate}), conditions \ref{cond: edge in emb path} and \ref{cond: many paths contain a vertex} hold as well. 
Recall that, over the course of the algorithm, some vertices and edges may be deleted from $\hat W_C$. Whenever an edge $e$ is deleted from $\hat W_C$, we delete its embedding path $P(e)$ from $\pset_C$. For every edge $e'\in E(P(e))$, we also update the list $L(e')$, by deleting edge $e$ from the list, and, if $L(e')$ becomes empty, we delete $e'$ from $C$. Additionally, for every vertex $v\in V(P(e))$, we decrease the counter $n_v$ by $1$. This ensures that, throughout the algorithm, for every edge $e'\in E(\hat C)$, list $L(e')$ contains all paths in the current set $\pset_C$ to which $e'$ belongs, and for every vertex $v\in V(\hat C)$, counter $n_v$ is equal to the number of paths in the current set $\pset_C$ that contain $v$. This also ensures that Condition \ref{cond: edge in emb path} always holds.

We now fix an integer $1\leq i\leq k$, and describe the processing of the $i$th batch $\pi_i$ of edge deletions from $\hat C$. Over the course of this algorithm, we will construct a sequence $E_i$ of edges to be deleted from graph $\hat W_C$; all these edges serve as an input to Algorithm $\aset_C$, as part of Phase $\Phi_i$ of the algorithm, and we call all such deletions from $\hat W_C$ \emph{direct edge deletions}.

Initially, for every edge $e\in \pi_i$, we add all edges in $L(e)$ to $E_i$. Since the paths in $\pset_C$ cause congestion at most $\eta^*$, this initial collection of edge deletions contains at most $|\pi_i|\cdot \eta^*$ edges. Recall that Algorithm $\aset_C$ for the \pruning problem may also delete some edges from $\hat W_C$, that do not lie in $E_i$. We call such edge deletions \emph{indirect}. It may also delete some vertices from $\hat W_C$. Whenever an edge $e\in E(\hat W_C)$ is deleted by Algorithm $\aset_C$ indirectly, we inspect the corresponding embedding path $P(e)\in \pset_C$. For every edge $e'\in E(P(e))$, we delete $e$ from the list $L(e')$, and, if the list becomes empty, we delete $e'$ from $\hat C$. Next, we consider every vertex $v\in V(P(e))$. If $e$ is the first edge  that is deleted from $\hat W_C$ in this phase, whose embedding path contains $v$, then we mark $v$ to indicate that some edge whose embedding path contains $v$ was deleted from $\hat W_C$ in this phase, and we also record the number $n_v$ of the embedding paths that contained $v$ at the beginning of the phase; this number is denoted by $\lambda_i(v)$. In any case, we decrease $n_v$ by $1$, and, if it falls below $\frac{\lambda_i(v)}{4k^{16k^2}\cdot d^*}$, then we delete $v$ and all its incident edges from $\hat C$. For every edge $\hat e$ that we just deleted from $\hat C$, we add all edges of $\hat W_C$ that lie in the list $L(\hat e)$ to set $E_i$, which are then deleted from $\hat W_C$ as direct deletions, and become part of the input sequence of edge deletions for $\aset_C$. This completes the description of the algorithm for processing the $i$th batch $\pi_i$ of edge deletions. It is immediate to verify that, if Conditions \ref{cond: embedding contained}--\ref{cond: many paths contain a vertex} held before batch $\pi_i$ was processed, then they continue to hold after it is processed. The key in the analysis of the algorithm is the following claim, the bounds the total number of edges that are deleted from $\hat W_C$ during the processing algorithm.

\begin{claim}\label{claim: upper bound number of deleted from witness graph}
	The total number of edges deleted from $\hat W_C$, either directly or indirectly, while processing batch $\pi_i$, is bounded by $|\pi_i|\cdot n^{9/k^2}$.
\end{claim}
\begin{proof}
	For simplicity, in this proof we denote $\hat W_C$ by $W$.
	Recall that an edge $e$ is deleted from $W$ \emph{directly}, if it lies in the input sequence $E_i$ of edge deletions, and it is deleted from $W$ \emph{indirectly} otherwise.
	We denote by $W\attime[0]$ the graph $W$ just before the batch $\pi_i$ of edge deletions from $C$ is processed.
	In order to analyze the algorithm, we will maintain, for every edge $e\in E(W\attime[0])$, a \emph{budget}, as follows. 
	Whenever an edge $e$ is deleted from $W$, for every vertex $v\in V(P(e))$, we initially set the value $\gamma(v,e)$ to be $1$, and, if vertex $v$ is ever deleted from $\hat C$, value $\gamma(v,e)$ is set to $0$.
	
	Whenever an edge $e$ is added to $E_i$, its budget $\beta(e)$ is set to be $2k^{16k^2}\cdot d^*$, and following that, it may only decrease, but it must always remain at least $1+\sum_{v\in P(e)}\gamma(v,e)$.  The budget $\beta(e)$ of an edge $e$ that was deleted from $W$ indirectly is always set to be $\beta(e)=1+\sum_{v\in P(e)}\gamma(v,e)$. So initially, immediately after $e$ is deleted from $W$, $\beta(e)$ is set to be $|V(P(e))|+1$, and afterwords it may only decrease, but it always remains at least $1$. The budgets of all other edges in $W\attime[0]$ are $0$. We denote by $\beta^*=\sum_{e\in E(W\attime[0])}\beta(e)$ the total budget of all edges of $W\attime[0]$.
	
	Recall that at the beginning of the algorithm, we add to $E_i$ all edges that lie in lists $L(e)$ for edges $e\in \pi_i$. Since each such list may contain at most $\eta^*$ edges, initially, $|E_i|\leq |\pi_i|\cdot \eta^*$ holds. Since the initial budget of every edge in $E_i$ is set to be $2k^{16k^2}\cdot d^*$, we get that initially: 
	
	\[\beta^*\leq 2|\pi_i|\cdot k^{16k^2}\cdot d^*\cdot \eta^*\leq |\pi_i|\cdot n^{9/k^2}.\]

	(We have used the fact that, from Inequality \ref{eq: k and n}, $2k^{16k^2}\leq
	2^{k^6}\leq n^{1/k^4}$ holds, and, From Inequality \ref{eq: bound d eta}, $\eta^*\cdot d^*\leq n^{5/k^2}$.)
	
	Observe that, throughout the algorithm, the budget of every edge in $E(W\attime[0])\setminus E(W)$ is at least $1$, and so the number of edges deleted from $W$ is bounded by the final budget $\beta^*$. Therefore, it is now enough to show that we can maintain the budgets of all edges of $W\attime[0]$, so that they obey the above rules, and the total budget $\beta^*$ never increases.
	
	Recall that, when an edge $e$ is added to set $E_i$, we set its initial budget to be $2k^{16k^2}\cdot d^*$. The budgets of the edges in $E_i$ change over the course of the algorithm as follows. Whenever the algorithm $\aset_C$ chooses to delete some edge $e'$ from $W$ indirectly, its initial budget is set to $1+|V(P(e'))|\leq 2+d^*$. In order to ensure that the total budget $\beta^*$ does not grow, we will \emph{move} the budget from some edges in $E_i$ to edge $e'$. In other words, we will reduce the total budget of the edges in the current set $E_i$ by (additive) $d^*+2$, while ensuring that the budget of every edge in $E_i$ remains at least $1+\sum_{v\in P(e)}\gamma(v,e)$. Recall that Algorithm $\aset_C$ for the \pruning problem must ensure that the total number of edges deleted from $W$ over the course of phase $\Phi_i$ is bounded by $|E_i|\cdot k^{8k^2}$ (recall that we use the algorithm with parameter $k$ replaced by $k^2$). Since the algorithm does not know when Phase $\Phi_i$ is going to end, it must guarantee that, at any time $\tau$ during the phase, the total number of edges deleted from $W$ so far is bounded by $|E\attime_i|\cdot  k^{8k^2}$, where $E\attime_i$ is the set of all edges that were added to $E_i$ by time $\tau$. Since, whenever an edge is added to $E_i$, its initial budget is $2k^{16k^2}\cdot d^*$, while the initial budget of an edge that is deleted from $W$ indirectly is set to be at most $2+d^*$, it is easy to verify that, whenever a new edge $e$ is deleted from $W$ indirectly, we can move $1+|V(P(e))|$ units of budget from the edges that are currently in $E_i$ to $e$, so that the budget of every edge in $E_i$ remains at least $1$, and the total budget $\beta^*$ does not increase.
	
	Consider now some time $\tau$ during the algorithm's execution, where a vertex $v$ is deleted from $C$. Recall that we have denoted by $\lambda_i(v)$ the initial number $n_v$ of paths $P(e)\in \pset_C$ that contained $v$ before batch $\pi_i$ was processed. Let $\hat E(v)\subseteq E(W\attime[0])$ be the set of all edges $e\in E(W\attime[0])$, whose embedding path $P(e)$ contained $v$, so that $|\hat E(v)|=\lambda_i(v)$. Let $\hat E'(v)\subseteq E(W)$ be the set of edges $e$ that remain in graph $W$ at time $\tau$, whose embedding path $P(e)$ contains $v$. Since the algorithm decided to delete vertex $v$ from $C$ at time $\tau$, we get that $|\hat E'(v)|\leq \frac{|\hat E(v)|}{4k^{16k^2}\cdot d^*}$.
	
	Note that, before vertex $v$ is deleted from $C$, for every edge $e\in \hat E(v)\setminus \hat E'(v)$, $\gamma(v,e)=1$ holds, and immediately after $v$ is deleted from $C$, $\gamma(v,e)=0$. Therefore, following the deletion of $v$ from $C$, the budget $\beta(e)$ of every edge $e\in \hat E(v)\setminus \hat E'(v)$ decreases by $1$, and the total decrease in the budgets of all such edges is at least $|\hat E(v)\setminus \hat E'(v)|\geq \frac{\lambda_i}{2}$. For every edge $e\in \hat E'(v)$, following the deletion of $v$ from $C$, its budget $\beta(e)$ grows from $0$ to $2k^{16k^2}\cdot d^*$. But since $|\hat E'(v)|\leq \frac{\lambda_i}{4k^{16k^2}\cdot d^*}$, the total budget $\beta^*$ does not grow.
\end{proof}

Recall that the algorithm $\aset_C$ for the \pruning problem requires that the total number of edges deleted from $\hat W_C$ directly during phase $\Phi_i$ is bounded by $\frac{N_C^{k^2}\cdot \Delta}{k^{16k^2}}$. Recall that, from the definition of the router certificate (see \Cref{def: router certificate}), $N_C\geq \frac{|V(C)|^{1/k^2}}{2|V(C)|^{1/k^4}}\geq \frac{|V(C)|^{1/k^2}}{2 n^{1/k^4}}$ holds.
Since  $|\pi_i|\leq \frac{|V(C)|\cdot \Delta}{n^{11/k^2}}$, from \Cref{claim: upper bound number of deleted from witness graph},
we get that:

\[\begin{split}
|E_i|&\leq |\pi_i|\cdot n^{9/k^2}\\
&\leq \frac{|V(C)|\cdot \Delta}{n^{2/k^2}}\\
&\leq \frac{|V(C)|}{2^{k^2}\cdot n^{1/k^2} }\cdot\frac{\Delta}{k^{16k^2}}\\
&\leq\frac{N_C^{k^2}\cdot \Delta}{k^{16k^2}},
\end{split}\]

as required (we have used the fact that, from Inequality \ref{eq: k and n}, $n^{1/k^2}\geq k^{16k^2}\cdot 2^{k^2}$).

In the next simple corollary of \Cref{claim: upper bound number of deleted from witness graph}, we bound the total number of edges deleted from $\hat C$ when processing the batch $\pi_i$, for all $1\leq i\leq k$. 

\begin{corollary}\label{cor: upper bound number of deleted from cluster}
	The total number of edges deleted from $\hat C$ when processing batch $\pi_i$ of edge deletions is bounded by $|\pi_i|\cdot  n^{10/k^2}$. 
\end{corollary}
\begin{proof}
		Let $\hat E$ be the set of all edges deleted from $\hat W_C$ while processing batch $\pi_i$. From \Cref{claim: upper bound number of deleted from witness graph}, $|\hat E|\leq |\pi_i|\cdot n^{9/k^2}$.
	
		Let $E'$ be the set of all edges that were deleted from $\hat C$ while processing batch $\pi_i$. 
		Recall that, from Condition \ref{cond: edge in emb path}, before the algorithm started processing batch $\pi_i$, for every edge $e\in E(\hat C)$, some path in $\pset_C$ contained $e$. Since Condition \ref{cond: embedding contained} holds when the algorithm finishes processing $\pi_i$, we get that, for every edge $e\in E'$, there is an edge $e'\in \hat E$ with $e\in P(e')$.  Since the length of each embedding path in $\pset_C$ is at most $d^*\leq n^{1/k^6}$ from Inequality \ref{eq: bound on d*}, it is easy to see that:
	
	\[|E'|\leq d^*\cdot |\hat E|\leq |\pi_i|\cdot n^{10/k^2}.\]
\end{proof}

Recall that, from Property \ref{cond: many paths contain a vertex}, if a vertex $v$ lies in $\hat C$ at any time $\tau$, then, at time $\tau$, the number of paths in $\pset_C$ that contain $v$ is at least:

\[\begin{split}
\frac{\Delta}{n^{4/k^2}\cdot \left(4k^{16k^2}\cdot  d^*\right )^k}&\geq \frac{\Delta}{n^{4/k^2}\cdot 2^{2ck^7}}\\
&\geq \frac{\Delta}{n^{6/k^2}}.
\end{split}
\]

since $d^*= k^{10}\cdot 2^{c/\delta^6}\leq k^{10}\cdot 2^{ck^6}$, as $\delta\geq 1/k$; we also used Inequality \ref{eq: k and n}.

It now remains to bound the time required to process a batch $\pi_i$. The running time of the algorithm $\aset_C$ from \Cref{thm: pruning of Wk} for the \pruning problem during phase $\Phi_i$ is bounded by:

\[O\left(|E_i|\cdot k^{O(k)}\right )\leq O\left(|\pi_i|\cdot n^{O(1/k^2)}\right ),\]

from  \Cref{claim: upper bound number of deleted from witness graph} and Inequality \ref{eq: k and n}. 
Additionally, for every edge $e$ that is deleted from $\hat W_C$, we need to spend time $O(|E(P(e))|)\leq O(d^*)$ in order to process all edges and vertices on path $P(e)$. Since, from  \Cref{claim: upper bound number of deleted from witness graph}, the total number of edges deleted from $\hat W_C$ while processing batch $\pi_i$ is bounded by $|\pi_i|\cdot n^{9/k^2}$, and since, from Inequality \ref{eq: bound on d*}, $d^*\leq n^{1/k^6}$, the total time required to process all such paths is bounded by $O\left(|\pi_i|\cdot n^{O(1/k^2)}\right )$. Therefore, the total time the algorithm spends to process batch $\pi_i$ is bounded by $O\left(|\pi_i|\cdot n^{O(1/k^2)}\right )$.

\subsubsection{Part 1: Summary}

The following corollary easily follows from \Cref{thm: initialization router witness} and \Cref{thm: update one cluster}, and it summarizes our algorithm for maintaining the clustering $\cset$, and, for every cluster $C\in \cset$, the corresponding graph $W_C$, along with its embedding into $C$.

\begin{corollary}\label{cor: part 1}
	There is a deterministic algorithm, that receives as input a valid input graph $G$ together with valid input parameters $n,k,\delta,\Delta$ and $\Delta^*$, such that,  for every vertex $v\in V(G)$, $\deg_G(v)\leq \Delta\cdot n^{16/k}$.
	The algorithm also receives $k$ online batches $\pi_1,\ldots,\pi_k$ of updates to graph $G$, where for $1\leq i\leq k$, the $i$th batch $\pi_i$ is a collection of at most $\Delta\cdot n$ edges that are deleted from $G$. 

The algorithm maintains a collection $\cset$ of subgraphs of $G$ called \emph{clusters}, with $\sum_{C\in \cset}|V(C)|\leq n^{1+20/k}$, such that every cluster $C\in \cset$ is a 
$\left(\vdeg_C,\tilde d,\tilde \eta\right)$-router  with respect to the set $V(C)$ of supported vertices, for
$\tilde d\leq 2k^{18}\cdot 2^{c/\delta^6}$ and
$\tilde \eta\leq n^{17/k}$, and every edge of $G$ lies in at most one cluster $C\in \cset$. 
The algorithm also maintains, for every cluster $C\in \cset$, a graph $W_C$, that is a properly pruned subgraph of $W^{N_C,\Delta}_{k^2}$, for some parameter $N_C$, and an embedding $\pset_C$ of $W_C$ into $C$ via paths of length at most $d^*$ that cause congestion at most $\eta^*$, such that every vertex of $C$ belongs to at least $\frac{\Delta}{n^{6/k^2}}$ of the paths in $\pset_C$ at all times. 
After the clustering $\cset$ is initialized, for every cluster $C\in \cset$, the only changes that $C$ and $W_C$ may undergo are the deletions of some of their edges and vertices; for every edge $e$ that remains in $W_C$, its embedding path $P(e)\in \pset_C$ cannot change after the initialization. Moreover, for all $1\leq i\leq k$, if  we denote by $\pi_i^C=\pi_i\cap E(C)$, then the number of edges deleted from $C$ while processing the batch $\pi_i$ is at most $|\pi^C_i|\cdot n^{13/k^2}$. The only additional change that the clustering $\cset$ may undergo after its initializaition is the deletion of empty clusters from $\cset$.

Throughout, we denote by  $\edel=E(G)\setminus \left(\bigcup_{C\in \cset}E(C)\right )$. 
At the beginning of the algorithm, after the initialiation, $|\edel|\leq |V(G)|\cdot \Delta\cdot n^{2/k^2}$  must hold, and after that edges may only be added to $\edel$. For all $1\leq i\leq k$, the number of edges added to $\edel$ while processing batch $\pi_i$ is bounded by $|\pi_i|\cdot n^{13/k^2}$.

The initialization time of the algorithm is $O\left((n\Delta)^{1+O(\delta)} \right )$, and for all $1\leq i\leq k$, the time required to process the $i$th update batch $\pi_i$ is bounded by  $O\left(|\pi_i|\cdot n^{O(1/k^2)}\right )$.
\end{corollary}

\begin{proof}
We start by employing the algorithm from \Cref{thm: initialization router witness} to compute an initial collection $\cset$ of clusters of $G$, that are disjoint in their edges, such that $\sum_{C\in \cset}|V(C)|\leq n^{1+20/k}$.
Recall that, for every cluster $C\in \cset$, the algorithm also computes a router certificate $\left (W^{N_C,\Delta}_{k^2},W_C,\pset_C\right )$, where graph $W_C$ is obtained from $W^{N_C,\Delta}_{k^2}$ by applying the algorithm from \Cref{thm: pruning of Wk} for the $\pruning$ problem  to it, with a single phase $\Phi_0$ of updates, consisting of at most $\frac{N_C^{\hk}\cdot \Delta}{2\hk^{4\hk}}$ edge deletions; we denote this algorithm for $\aset_C$. The algorithm from \Cref{thm: initialization router witness} also ensures that the cardinality of the set $\edel=E(G)\setminus \left(\bigcup_{C\in \cset}E(C)\right )$ of edges is at most  $|V(G)|\cdot \Delta\cdot n^{2/k^2}$, and its running time is $O\left((n\Delta)^{1+O(\delta)} \right )$.
We also need the following simple observation.

\begin{observation}\label{obs: num of edges in cluster}
	For all  $C\in \cset$,  $|E(C)|\leq |V(C)|\cdot \Delta\cdot n^{2/k^2}$ holds.
	\end{observation}
\begin{proof}
Consider the router certificate $\left (W^{N_C,\Delta}_{k^2},W_C,\pset_C\right )$ for $C$. Clearly, $|V(C)|\geq |V(W_C)|$ must hold. From the definition of the graph $W^{N_C,\Delta}_{k^2}$, for every leaf vertex $v\in W_C$, the degree of $v$ in $W_C$ is at most $k^2\cdot \Delta$, and every edge of $W_C$ is incident to at least one leaf vertex. Therefore, $|E(W_C)|\leq |V(W_C)|\cdot k^2\cdot \Delta\leq |V(C)|\cdot k^2\cdot\Delta$. 

Recall that, from the definition of a router certificate, every edge of $C$ must lie on some path in $\pset_C$, and moreover, the length of every path in $\pset_C$ is at most $d^*$. Therefore:

\[|E(C)|\leq d^*\cdot |\pset_C|=d^*\cdot |E(W_C)|\leq |V(C)|\cdot d^*\cdot k^2\cdot \Delta \leq |V(C)|\cdot \Delta\cdot n^{2/k^2} \]

from Inequalies \ref{eq: bound on d*} and \ref{eq: k and n2}.
\end{proof}

Consider now some  cluster $C\in \cset$, and  the router certificate $\left (W^{N_C,\Delta}_{k^2},W_C,\pset_C\right )$ that we have computed for $C$.
For every edge $e\in E(C)$, we compute a list $L(e)$ of all edges $\hat e\in E(W_C)$, whose embedding path $P(\hat e)\in \pset_C$ contains $e$. Additionally,  for every vertex $v\in V(C)$, we the number $n_v$ of paths in $\pset_C$ containing $v$. The time required to compute all such lists $L(e)$ for $e\in E(C)$ and integers $n_v$ for $v\in V(C)$, for all clusters $C\in \cset$, is asymptotically bounded by the time required to compute the router certificates $\left (W^{N_C,\Delta}_{k^2},W_C,\pset_C\right )$ for all clusters, which is in turn asymptotically bounded by running time of the algorithm from \Cref{thm: initialization router witness}.

For every cluster $C\in \cset$,
we initialize the algorithm from \Cref{thm: update one cluster} for $C$, that we denote by $\aset'_C$. The algorithm receives as input the router certificate  $\left (W^{N_C,\Delta}_{k^2},W_C,\pset_C\right )$, the lists $\set{L(e)}_{e\in E(C)}$, and the values $\set{n_v\mid v\in V(C)}$ that we have computed.
This finishes the description of the initialization algorithm. We will later prove that the every cluster $C\in \cset$ is a 
$\left(\vdeg_C,\tilde d,\tilde \eta\right)$-router. Next, we provide an algorithm for processing the update batches $\pi_1,\ldots,\pi_k$. 

Consider some index $1\leq i\leq k$, and the batch $\pi_i$ of updates. For every cluster $C\in \cset$, let $\pi_i^C=\pi_i\cap E(C)$. Assume first that $|\pi_i^C|> \frac{|V(C)|\cdot \Delta}{n^{11/k^2}}$. Using the same reasoning as in \Cref{obs: num of edges in cluster}, it is easy to see that $|E(C)|\leq |V(C)|\cdot \Delta\cdot n^{2/k^2}\leq |\pi_i^C|\cdot n^{13/k^2}$. We then delete all edges and all vertices from $C$, and delete $C$ from $\cset$. Assume now that $|\pi_i^C|\leq \frac{|V(C)|\cdot \Delta}{n^{11/k^2}}$.
We then provide the batch $\pi_i^C$, as an input batch of edge deletions, to Algorithm $\aset'_C$. Recall that the algorithm deletes at most $|\pi^C_i|\cdot  n^{10/k^2}$ edges, and possibly some vertices, from cluster $C$. It may also delete some edges and vertices from graph $W_C$. The algorithm ensures that $W_C$ remains 
a properly pruned subgraph of $W^{N_C,\Delta}_{\hk}$; that every edge $e\in E(C)$ lies on some embedding path in $\set{P(e')\mid e'\in E(W_C)}$, and every vertex $v\in V(C)$ lies on at least  $\frac{\Delta}{n^{6/k^2}}$ such paths. The algorithm also ensures that every embedding path in  $\set{P(e')\mid e'\in E(W_C)}$ is contained in the current cluster $C$. The running time required to process batch $\pi_i^C$ is bounded by  $O\left(|\pi^C_i|\cdot n^{O(1/k^2)}\right )$.

This completes the description of the algorithm for processing the batch $\pi_i$ of edge deletions from $G$. All edges that have been deleted from the clusters in $\cset$ while processing $\pi_i$ are added to set $\edel$. From our discussion, 
for every cluster $C\in \cset$, the number of edges deleted from $C$ while processing $\pi_i$ is bounded by $|\pi^C_i|\cdot n^{13/k^2}$, and so
the total number of edges added to $\edel$ is bounded by $\sum_{C\in \cset}|\pi_i^C|\cdot n^{13/k^2}\leq |\pi_i|\cdot n^{13/k^2}$. The time that is required to process the batch $\pi_i$ is bounded by:

\[\sum_{C\in \cset}O\left (|\pi^C_i|\cdot n^{O(1/k^2)}\right)\leq O\left(|\pi_i|\cdot n^{O(1/k^2)}\right ).\]

Finally, we show that, throughout the algorithm, every cluster $C\in \cset$ remains a $\left(\vdeg_C,\tilde d,\tilde \eta\right)$-router  with respect to the set $V(C)$ of supported vertices, for 
$\tilde d\leq 2k^{18}\cdot 2^{c/\delta^6}$ and
$\tilde \eta\leq n^{17/k}$.

  Consider some cluster $C$, and any time $\tau$ at which $C$ lied in $\cset$, so $C\neq\emptyset$ holds. Then our algorithm maintains a graph $W_C$, that is a properly pruned subgraph of $W^{N_C,\Delta}_{\hk}$, together with an embedding $\pset_C$ of $W_C$ into $C$ via paths of length at most $d^*$ that cause congestion at most $\eta^*$, such that every vertex $v\in V(C)$ lies on at least  $\frac{\Delta}{n^{6/k^2}}$ paths in $\pset_C$. Recall that the maximum vertex degree in $C$ is bounded by $\Delta\cdot n^{16/k}$. By using \Cref{lem: nice router to uniform degree router} with parameters $\alpha=n^{16/k}$, $\beta=n^{6/k^2}$, and parameter $k$ replaced with $k^2$, we get that $C$ is a $\left(\vdeg_C,\tilde d,\tilde \eta\right)$-router with respect to the set $V(C)$ of supported vertices, for $\tilde d=22\cdot d^*\cdot k^8\leq 2k^{18}\cdot 2^{c/\delta^6}$, and $\tilde \eta=2n^{16/k}\cdot n^{6/k^2}\cdot k^{8k^2+1}\cdot d^*\cdot \eta^*\leq n^{17/k}$,
since $d^*=k^{10}\cdot 2^{c/\delta^6}$; we have also used Inequalities \ref{eq: k and n} and \ref{eq: bound d eta}.
\end{proof}

\subsection{Part 2 of the Algorithm: Maintaining the Subgraphs $C'\subseteq C$}

In order to complete the proof of \Cref{thm: one level router decomposition}, it is enough to provide an algorithm that maintains, for every cluster $C\in \cset$,  a subgraph $C'\subseteq C$ with $V(C')=V(C)$ and $|E(C')|\leq \Delta^*\cdot |V(C)|$, such that $C'$ is a $\left(\vDelta^*,\tilde d,\tilde \eta\right )$-router for the set $V(C)$ of supported edges.

Throughout, we use the parameter $\Delta'=\floor{\frac{\Delta^*}{2k^2d^*}}$. Note that, from the definition of valid input parameters, 
and Inequalities \ref{eq: k and n} and \ref{eq: bound on d*},
$\Delta^*\geq n^{18/k}>4k^2d^*$ holds.
Recall that the algorithm from \Cref{cor: part 1} maintains, for every cluster $C\in \cset$, a graph $W_C$, that is a properly pruned subgraph of $W^{N_C,\Delta}_{k^2}$, for some parameter $N_C$, and an embedding $\pset_C$ of $W_C$ into $C$ via paths of length at most $d^*$ that cause congestion at most $\eta^*$, such that every vertex of $C$ belongs to at least $\frac{\Delta}{n^{6/k^2}}$ of the paths in $\pset_C$ at all times. 
Notice that:

\[\Delta'=\floor{\frac{\Delta^*}{2k^2d^*}}\leq \frac{\Delta}{ n^{8/k^2}}\leq \frac{\Delta}{2^k\cdot n^{6/k^2}}, \]

from the definition of valid input parameters, and since $2^k\leq n^{1/k^2}$ from Inequality \ref{eq: k and n}.

Consider now some cluster $C\in \cset$. For every vertex  $v\in V(C)$, let $\pset_C(v)\subseteq \pset_C$ denote the set of all embedding paths that contain $v$. Consider now some path $P\in \pset_C(v)$, and recall that $P$ must be an embedding path of some edge $e\in E(W_C)$, so $P=P(e)$. Recall that one of the endpoints of $e$ must be a leaf vertex of $W_C$;  we call it the \emph{distinguished endpoint of $P$}.
We denote by $U_C(v)$ the multiset of leaf vertices of $W_C$ that serve as distiguished endpoints of the paths in $\pset_C(v)$. Lastly, consider the bipartite graph $R_C=(A,B,\tilde E)$, where $A=V(C)$, and $B$ contains all leaf vertices of $W_C$ (so a vertex $v\in V(C)$ may lie in both $A$ and $B$; we will refer to these vertices as ``the copy of $v$ in $A$'' and ``the copy of $v$ in $B$''). For every vertex $v\in V(C)$, for each vertex $u$ multiset $U_C(v)$, we include an edge connecting a copy of $v$ in $A$ to a copy of $u$ in $B$ in $\tilde E$. Recall that, for every vertex $u\in U_C(v)$, there is a distinct path $P\in \pset_C(v)$, such that $u$ is the distinguished endpoint of $P$. We will say that the new edge \emph{represents} the path $P$.

\paragraph{Additional Data Structures.}

We fix again a cluster $C\in \cset$. Throughout the algorithm, for every vertex $v\in V(C)$, we maintain the set $U_C(v)$ of leaf vertices of $W_C$ explicitly. Additionally, we maintain the set $\pset_C(v)$ of paths implicitly, as follows: for every vertex $u\in U_C(v)$, we maintain a pointer from the copy of $u$ in $U_C(v)$ to the edge  $e\in E(W_C)$, such that $u$ serves as the distinguished endpoint of the path $P(e)\in \pset_C(v)$ (if several copies of $u$ lie in $U_C(v)$, then we associate each copy with a distinct path in $\pset_C(v)$, and maintain a pointer from each copy of $u$ in $U_C(v)$ to the corresponding edge of $W_C$).
We also explicitly maintain the bipartite graph $R_C$.

Recall that our algorithm from Part 1 maintains, for every large cluster $C\in \cset$ and vertex $v\in V(C)$, the counter $n_v=|\pset_C(v)|$. The counter is maintained in a straightforward manner: during the initialization step, for every edge $e\in E(W_C)$, we inspect all vertices lying on path $P(e)$ and update their counters $n_v$. 
At this time, we can also construct the sets $U_C(v)$ of vertices for all $v\in V(C)$, together with the pointers from every vertex $u\in U_C(v)$ to the corresponding edge of $W_C$, without increasing the asymptotic running time of the initialization algorithm. We can also initialize the graph $R_C$ within this running time.

During the update steps, whenever an edge $e$ is deleted from graph $W_C$, we inspect every vertex  $v\in V(P(e))$, and we decrease the corresponding counter $n_v$.
During this time, we can also update the list $U_C(v)$ for each such vertex $v$ with the deletion of this edge, and delete the edge connecting $v$ to the distinguished endpoint of $P(e)$ from graph $R_C$. All this can be done without increasing the asymptotic running time of the algorithm for processing each batch $\pi_i^C$ of updates to cluster $C$.

Next, we summarize some properties of the graph $R_C$, in the following simple observation.

\begin{observation}\label{obs: properties of RC}
Upon the initialization of graph $R_C=(A,B,\tilde E)$, every vertex $x\in B$ has degree at most $d^*\cdot k^2\cdot \Delta\leq  n^{2/k^2}\cdot \Delta$ in $R_C$.
For all $1\leq i\leq k$, after batch $\pi_i$ of edge deletions is processed by the algorithm from Part 1, at most $|\pi^C_i|\cdot n^{10/k^2}$ edges are deleted from $R_C$, and all vertices that become isolated are deleted from $R_C$ as well. These are the only updates to graph $R_C$. Moreover, for every vertex $v\in A$, the degree of $v$ in $R_C$ remains at least $\frac{\Delta}{n^{6/k^2}}$ at all times.
\end{observation}
\begin{proof}
	Recall that, from the definition of graph $W^{N_C,\Delta}_{k^2}$, every leaf vertex of the graph has degree at most $k^2\cdot \Delta$. Consider now any leaf vertex $u\in B$, and let $E_u\subseteq E(W_C)$ be the set of edges that are incident to $u$ in $W_C$ after the initialiation step, so $|E_u|\leq k^2\cdot \Delta$. For every edge $e\in E_u$, path $P(e)\in \pset_C$ contains at most $d^*$ vertices, and vertex $x$ lies in the set $U_C(v)$ of each such vertex $v$. Therefore, the degree of $x$ in $R_C$ is bounded by $d^*\cdot k^2\cdot \Delta\leq  n^{2/k^2}\cdot \Delta$ (from Inequality \ref{eq: bound on d*}).
	
	Consider now some integer $1\leq i\leq k$. From \Cref{thm: update one cluster}, while processing the batch $\pi_i$ of updates, the algorithm from Part 1 may delete at most $|\pi^C_i|\cdot n^{9/k^2}$ edges from $W_C$. Since the embedding path of each such edge has length at most $d^*$, this can lead to the deletion of at most $|\pi^C_i|\cdot n^{9/k^2}\cdot d^*\leq |\pi^C_i|\cdot n^{9/k^2}$ edges from $R_C$ (from Inequality  \ref{eq: bound on d*}). Since every vertex that remains in $C$ must participate in at least $\frac{\Delta}{n^{10/k^2}}$  paths in $\pset_C$, we get that the degree of every vertex in $A$ remains at least $\frac{\Delta}{n^{6/k^2}}$ throughout the algorithm.
\end{proof}

The main result of the current subsection is summarized in the following claim.

\begin{claim}\label{claim: maintain the paths}
	There is a deterministic algorithm, that, given an access to the adjacency-list representation of the graph $R_C$, maintains, for every vertex $v\in V(C)$, a subset $U'_C(v)\subseteq U_C(v)$ of vertices, whose cardinality is between $\Delta'$ and $\Delta'\cdot n^{11/k^2}$ at all times, such that, for every leaf vertex $x$ of $W_C$, the total number of times that $x$ appears in the sets $\set{U'_C(v)\mid v\in V(C)}$ is at most $\gamma\cdot \Delta'$, where $\gamma=n^{16/k}$. After processing all updates to graph $R_C$ following each batch $\pi^C_i$, the algorithm outputs all vertices that were deleted from or inserted into the sets $\set{U'_C(v)}_{v\in V(C)}$. The initialization time of the algorithm is $O(|V(C)|\cdot \Delta\cdot n^{O(1/k)})$, and, for all $1\leq i\leq k$, the time required for updates following the processing of batch $\pi_i^C$ of edge deletions to graph $C$ is bounded by $O\left(|\pi_i^C|\cdot n^{O(1/k^2)}\right )$.
\end{claim}

We provide the proof of \Cref{claim: maintain the paths} below, after 
we comlete the proof of \Cref{thm: one level router decomposition} using it.

We now fix a cluster $C\in \cset$, and provide an algorithm for maintaining the corresponding graph $C'\subseteq C$. We first define the graph $C'$ and prove that it must be a $\left(\vDelta^*,\tilde d,\tilde \eta\right )$-router for the set $V(C)$ of supported edges, and that $|E(C')|\leq \Delta^*\cdot |V(C)|$. Later, we provide an algorithm for maintaining $C'$.

\paragraph{Definition of Graph $C'$.} Recall that the algorithm from \Cref{claim: maintain the paths} maintains, for every vertex $v\in V(C)$, a subset $U'_C(v)\subseteq U_C(v)$ of vertices, whose cardinality is between $\Delta'$ and $\Delta'\cdot n^{11/k^2}$ at all times. We let $U''_C(v)\subseteq U'_C(v)$ be an arbitray subset of $\Delta'$ such vertices, and we let $\qset_C(v)\subseteq \pset_C(v)$ be the set of paths that correspond to these vertices, so that $|\qset_C(v)|=\Delta'$. Let $\qset_C=\bigcup_{v\in V(C)}\qset_C(v)$. Notice that, from \Cref{claim: maintain the paths}, every leaf vertex of $W_C$ serves as an endpoint in at most $\gamma\cdot \Delta'$ paths in $\qset_C$.
 Additionally, for every superedge $e=(a,b)$ of $W_C$, we select an arbitrary collection $E'(a,b)$ of $\Delta'$ parallel edges $(a,b)$ (since $W_C$ remains a properly pruned subgraph of $W_{k^2}^{N_C,\Delta}$, the number of parallel edges $(a,b)$ must be at least $\ceil{\frac{\Delta}{2}}\geq \Delta'$).
 We let $\qset'_C$ be the set of paths that contains, for every superedge $(a,b)$ of $W_C$, for every edge $e\in E'(a,b)$, the embedding path $P(e)\in \pset_C$.
 We let $C''$ be the multigraph, obtained by taking the union of all paths in set $\qset_C\cup \qset'_C$, and we let $C'\subseteq C$ be the corresponding simple graph, obtained from $C''$ by deleting parallel edges.
 Recall that, from the definition of valid input parameters,  $\Delta^*\geq n^{18/k}\geq 2k^2d^*$ (from Inequalities \ref{eq: k and n2} and \ref{eq: bound on d*}) must hold.
 From \Cref{obs: sparsified router}, we conclude that $|E(C')|\leq |V(C)|\cdot \Delta^*$, and that graph 
	 $C'$ is a $(\vDelta^*,\tilde d,\tilde \eta')$-router with respect to the set $V(C')$ of supported vertices, where:  
	 
\[\tilde d=22d^*\cdot k^8 \leq k^{19}\cdot 2^{c/\delta^6},  \]

since $d^*=k^{10}\cdot 2^{c/\delta^6}$; and
	 
\[\tilde \eta=8\gamma (d^*)^2\cdot \eta^*\cdot k^{4k+1}\leq n^{17/k},\]

since $\gamma=n^{16/k}$, $\eta^*\cdot d^*\leq n^{5/k^2}$ from Inequality \ref{eq: bound d eta}; and $k^{6k}\leq n^{1/k^2}$ from Inequality \ref{eq: k and n}. Since $\sum_{C\in \cset}|V(C)|\leq n^{1+20/k}$, we get that $\sum_{C\in \cset}|E(C')|\leq \sum_{C\in \cset}|V(C)|\cdot \Delta^*\leq n^{1+20/k}\cdot \Delta^*$, as required.
It is now enough to provide an algorithm for maintaining the graph $C'$.

\paragraph{Maintaining the Graph $C'$.}
Recall that the algorithm from \Cref{claim: maintain the paths} maintains, for every vertex $v\in V(C)$, a subset $U'_C(v)\subseteq U_C(v)$ of vertices, whose cardinality is between $\Delta'$ and $\Delta'\cdot n^{11/k^2}$ at all times, such that, for every leaf vertex $x$ of $W_C$, the total number of times that $x$ appears in the sets $\set{U'_C(v)\mid v\in V(C)}$ is at most $\gamma\cdot \Delta'$. Once the subset $U'_C(v)$ is initialized, the algorithm may modify it arbitrarily, that is, vertices may be both added to, and deleted from the set $U'_C(v)$. We assume w.l.o.g. that the algorithm maintains an ordered list $L_C(v)$ of all vertices in $U'_C(v)$, so that, when a new vertex is added to $U'_C(v)$, it is added to the end of the list.
Throughout the algorithm, we then let $U''_C(v)$ be the collection of the first $\Delta'$ vertices that lie in list $L_C(v)$. As before, we denote by $\qset_C(v)\subseteq \pset_C(v)$ all paths corresponding to the vertices of $U''_C(v)$, and we denote by $\qset_C=\bigcup_{v\in V(C)}\qset_C(v)$.
For every superedge $(u,v)$ of $W_C$, we also assume that the parallel edges of $W_C$ corresponding to $(u,v)$ are stored in an ordered list $L'_C(u,v)$. Recall that, as the algorithm progresses, edges may only be deleted from $L'_C(u,v)$. Throughout the algorithm, we will let $E'(u,v)$ be the set of the first $\Delta'$ edges in list $L'_C(u,v)$. 
 As before, we denote by $\qset'_C$ the set of paths that contains, for each superedge $(a,b)$ of $W_C$, for every edge $e\in E'(a,b)$, the embedding path $\pset(e)\in \pset_C$. As before, multigraph graph $C''$ is obtained by taking the union of all paths in $\qset_C\cup \qset'_C$ while keeping parallel edges, and graph $C'$ is the simple graph obtained from $C''$ by removing parallel edges.
 We define the multiset $U^*_C=\bigcup_{v\in V(C)}U''_C(v)$, and by $E^*_C$ the union of the sets $E'(a,b)$ of edges of $W_C$, for all superedge $(a,b)$ of $W_C$.

At the beginning of the algorithm, once the data structures from Part 1 and from \Cref{claim: maintain the paths} are initialized, we can compute, for every cluster $C\in \cset$, the initial sets $U''_C(v)$ of vertices for all $v\in V(C)$, and sets $E'(a,b)$ of edges for each superedge $(a,b)$ of the initial graph $W_C$, in time that is asymptotically bounded by the time required to initialize the data structures from Part 1 (that include graph $W_C$ and its embedding), and the data structures from \Cref{claim: maintain the paths} (that include sets $U'_C(v)$ of vertices for all $v\in V(C)$). 
The time required for this computation is then bounded by:

\[T=O\left((n\Delta)^{1+O(\delta)} +\sum_{C\in \cset}|V(C)|\cdot \Delta\cdot n^{O(1/k)}\right )\leq O\left((n\Delta)^{1+O(\delta+1/k)} \right )\leq O\left((n\Delta)^{1+O(\delta)} \right ),\]

since $\sum_{C\in \cset}|V(C)|\leq |V(G)|^{1+20/k}$ must hold, and $\delta\geq 1/k$.
Since, for every cluster $C\in \cset$, the length of every path in $\pset_C$ is bounded by $d^*\leq n^{1/k^6}$ (from Inequality \ref{eq: bound on d*}), the time required to compute the graphs $C''$ and $C'$, for all $C\in \cset$ is then bounded by:

\[O\left (d^*\cdot (|U^*(C)|+|E^*_C|)\right )\leq O\left (T\cdot d^*\right )\leq O\left((n\Delta)^{1+O(\delta+1/k)} \right ) \leq O\left((n\Delta)^{1+O(\delta)}\right ).\]

Therefore, the initialiation time of both Part 1 and Part 2 of our algorithm is bounded by $O\left((n\Delta)^{1+O(\delta)} \right )$.

Consider now some integer $1\leq i\leq k$, and the batch $\pi_i$ of edge deletions. Once the algorithm from Part 1 finishes processing the batch $\pi_i$, for some clusters $C\in \cset$, some of the edges may be deleted from graph $W_C$. If $(a,b)$ is a superedge of graph $W_C$, and if the number of parallel edges $(a,b)$ deleted from $W_C$ that lie in set $E'(a,b)$ is $m_{a,b}$, then we add $m_{a,b}$ new edges from the list $L'_C(a,b)$, to set $E'(a,b)$, so that $E'(a,b)$ contains the first $\Delta'$ edges in list $L'_C(a,b)$. For each edge $e$ of the original set $E'(a,b)$ that was deleted from $W_C$, for every edge $e'=(x,y)\in E(P(e))$, we delete $e'$ from $C''$, and, if no edges $(x,y)$ remain in $C''$, then we delete edge $(x,y)$ from $C'$ as well. 
Next, for every edge $e$ that was just added to $E'(a,b)$, we consider every edge $e'=(x,y)\in P(e)$. We add $e'$ to graph $C''$, and, if no other edges $(x,y)$ currently lie in $C''$, then we add edge $(x,y)$ to $C'$ as well.
It is easy to verify that, if $T_i$ is the time that the algorithm from Part 1 spends on processing batch $\pi_i$, then all the above updates, for all clusters $C\in\cset$, can be done in time $O(T_i\cdot d^*)\le O(T_i\cdot n^{2/k^2})$.

Consider now a cluster $C\in \cset$, and denote by $\pi_i^C=E(C)\cap \pi_i$. Once the algorithm from  \Cref{claim: maintain the paths} processes the batch $\pi_i^C$, we consider every vertex $v\in V(C)$, for which at least one vertex that lies in $U''_C(v)$ was deleted from $U'_C(v)$. Let $n_C(v)$ be the number of such vertices. First, for every vertex $u\in U''_C(v)$ that was deleted from $U'_C(v)$, we delete $u$ from $U''_C(v)$ as well. 
We then consider the path $Q$ of $\qset_C(v)$ that corresponds to vertex $u$. For every edge $e'=(x,y)$ of $Q$, we delete $e'$ from $C''$, and, if no other edges $(x,y)$ remain in $C''$, then we delete edge $(x,y)$ from $C'$ as well.
Next, we add $n_C(v)$ vertices from list $L_C(v)$ to $U''_C(v)$, so that $U''_C(v)$ contains the first $\Delta'$ vertices from list $L_C(v)$. For each such newly added vertex $u$, we consider the path $Q$ of $\qset_C(v)$ corresponding to $u$. For every edge $e'=(x,y)$ of $Q$, we add $e'$ to $C''$, and, if no other edges $(x,y)$ lie in $C''$, then we add edge $(x,y)$ to $C'$ as well. Note that the time required to perform  these updates a cluster $C\in \cset$ is bounded by:

\[O\left(\sum_{v\in V(C)}n_C(v)\cdot d^*\right ).\]

For every cluster $C\in \cset$, the algoritm from \Cref{claim: maintain the paths} must spend at least $\Omega\left (\sum_{v\in V(C)}n_C(v)\right )$ time on updating the sets $U'_C(v)$ of vertices for all $v\in V(C)$. Therefore, if $T_i^C$ denotes the time that the algorithm from \Cref{claim: maintain the paths} spends on processing the batch $\pi_i^C$ of updates, then the time required to update the sets $U''_C(v)$ of vertices for all $v\in V(C)$, and to update the graph $C'$ accordingly, is bounded by $O\left(T_i^C\cdot d^*\right )\leq O\left(T_i^C\cdot n^{2/k^2}\right )$, from Inequality \ref{eq: bound on d*}. Since, from \Cref{claim: maintain the paths}, for every cluster $C\in \cset$, $T_i^C\leq O\left(|\pi_i^C|\cdot n^{O(1/k^2)}\right )$, and since the time required  for the algorithm from Part 1 to process the batch $\pi_i$ of updates is   $O\left(|\pi_i|\cdot n^{O(1/k^2)}\right )$, we get that the total time required for processing the batch $\pi_i$ of updates, for both parts of our algorithm, is bounded by:

\[O\left(|\pi_i|\cdot n^{O(1/k^2)}\right )+\sum_{C\in \cset}O\left(|\pi_i^C|\cdot n^{O(1/k^2)}\right )\leq O\left(|\pi_i|\cdot n^{O(1/k^2)}\right ).\]

It is easy to verify that the total number of all edges that were deleted from or inserted into the graphs in $\set{C'\mid C\in \cset}$ is also bounded by $O\left(|\pi_i|\cdot n^{O(1/k^2)}\right )$.

In order to complete the proof of \Cref{thm: one level router decomposition}, it now remains to prove \Cref{claim: maintain the paths}, which we do next.

\begin{proofof}{\Cref{claim: maintain the paths}}
For simplicity, in this proof, we denote $R_C$ by $R$, and, for every vertex $v\in V(C)$, we denote $U_C(v)$ and $U'_C(v)$ by $U(v)$ and $U'(v)$, respectively. 

Let $\beta$ be the largest integer, such that $2^k\cdot \Delta' \cdot n^{10\beta/k^2}\leq \frac{\Delta}{n^{6/k^2}}$. Since, from the definition of valid input parameters, $\Delta\leq n^2$ holds, we get that $\beta\leq k^2$.
For convenience, we denote $\Delta_0=\frac{\Delta}{n^{6/k^2}\cdot n^{10\beta/k^2}}$.
Notice that:

\begin{equation}\label{eq: delta0}
2^k\cdot \Delta'\leq \Delta_0\leq 2^k\cdot \Delta'\cdot n^{10/k^2}.
\end{equation}

For $1\leq j\leq \beta$, we denote by $\Delta_j=\Delta_0\cdot n^{10j/k^2}$, so that $\Delta_{\beta}=\frac{\Delta}{n^{6/k^2}}$. We also denote by $\tilde r=n^{10/k^2}$. Notice that, from \Cref{obs: properties of RC}, we are guaranteed that, throughout the algorithm, the degree of every vertex $v\in A$ in $R$ is at least $\Delta_{\beta}$, while the degree of every vertex $u\in B$ is at most $n^{2/k^2}\cdot \Delta<\Delta_{\beta}\cdot n^{8/k^2}<\Delta_{\beta}\cdot \tilde r$.
Throughout the algorithm, for all $0\leq j\leq \beta$, we will maintain a subgraph $R_{j}\subseteq R$, such that, for all $1\leq i\leq k$, just before update batch $\pi_i$  of updates arrives, the following properties hold:

\begin{properties}{P}
	\item $V(R_{j})=V(R)$; \label{prop: vertex set}
	\item for every vertex $v\in A$, the degree of $v$ in $R_{j}$ is at least $\frac{\Delta_{j}}{2^i}$ and at most $\Delta_{j}$; and \label{prop: degree 1}
	\item for every vertex $u\in B$, the degree of $u$ in $R_{j}$ is at most $\Delta_{j}\cdot \tilde r\cdot k^{4(i-1)}\cdot (16\log n)^{\beta-j}$.\label{prop: degree2}
\end{properties} 

In order to initialize and to maintain the  graphs $R_0,\ldots,R_{\beta}$, we use  the following simple claim, whose proof is deferred to Section \ref{subsec: proof of degree lowering claim} of Appendix.
\begin{claim}\label{claim: constructing the subgraph}
	There is a deterministic algorithm, whose input is a connected bipartite multi-graph $H=(X,Y,E)$, and parameters $z\geq 2$, $\hat \Delta,\gamma'\ge 1$, and $\hat r\geq 8\gamma'\cdot \log(|X|)$, such that, for every vertex $x\in X$, $\deg_H(x)\geq z\cdot \hat \Delta$, and, for every vertex $y\in Y$, $\deg_H(y)\leq \gamma'\cdot z\cdot  \hat \Delta$.
	The algorithm computes, for every vertex $x\in X$, a collection $E'(x)\subseteq \delta_H(x)$ of its incident edges with $|E'(x)|=\ceil{\hat \Delta}$, such that every vertex $y\in Y$ serves as an endpoint in at most $\hat r\cdot \hat \Delta$ edges in  multiset $\bigcup_{x\in X}E'(x)$.
	The running time of the algorithm is $O(|E|\cdot \log (|X|))$.
\end{claim}

We are now ready to describe our algorithms for initializing the graphs $R_0,\ldots,R_{\beta}$, and for maintaining them as graph $R$ undergoes $k$ batches of updates.

\paragraph{Initialization.}
Recall that, from \Cref{obs: properties of RC}, the degree of every vertex $v\in A$ in $R$ is at least $\Delta_{\beta}$, while the degree of every vertex $u\in B$ is at most $n^{2/k^2}\cdot \Delta<\Delta_{\beta}\cdot n^{8/k^2}<\Delta_{\beta}\cdot \tilde r$. In order to construct the graph $R_{\beta}$, we let $V(R_{\beta})=V(R)$. For every vertex $v\in A$, we then include in $R_{\beta}$ an arbitrary subset of $\floor{\Delta_{\beta}}$ edges incdient to $v$ in $R$. The resulting graph $R_{\beta}$ is now guaranteed to have Properties 
\ref{prop: vertex set}--\ref{prop: degree2}. Consider now some integer $0\leq j<\beta$, and assume that we have computed the graph $R_{j+1}$ for which Properties \ref{prop: vertex set}--\ref{prop: degree2}. We now provide an algorithm for computing graph $R_j$.
Let $z=n^{10/k^2}$, $\hat r=2\tilde r\cdot (16\log n)^{\beta-j}$, and $\gamma'=2\tilde r\cdot (16\log n)^{\beta-j-1}$. Notice that $\hat r\geq 8\gamma'\cdot \log n$. Moreover, for every vertex $v\in A$, the degree of $v\in A$ in $R_{j+1}$ is at least $\frac{\Delta_{j+1}}{2}=\frac{\Delta_{j}}{2}\cdot n^{10/k^2}=\frac{\Delta_j}{2}\cdot z$, 
while the degree of every vertex $u\in B$ in $R_{j+1}$ is at most $\Delta_{j+1}\cdot \tilde r\cdot (16\log n)^{\beta-j-1}=\frac{\Delta_j}{2}\cdot z\cdot \gamma'$.
We apply the algorithm from \Cref{claim: constructing the subgraph} to graph $R_{j+1}$, with $\hat \Delta=\frac{\Delta_j}{2}$, and the parameters $z,\gamma'$ and $\hat r$ as defined above. We construct the graph $R_j$ by first setting $V(R_j)=V(R)$, and then adding, for every vertex $v\in A$, the collection $E'(v)$ of $\ceil{\frac{\Delta_j}{2}}$ edges incident to $v$ in $R_{j+1}$ computed by the algorithm. The algorithm guarantees that every vertex $u\in B$ is an endpoint of at most $\hat r\cdot \frac{\Delta_j}{2}=\tilde r\cdot (16\log n)^{\beta-j}\cdot \Delta_j$ edges in $R_j$, as required. This completes the initalization algorithm. Notice that  the algorithm from \Cref{claim: constructing the subgraph} is  executed $\beta\leq k^2$ times, and the time required for each such execution is $O(|E(R)|\cdot \log n)$. Therefore, the total time required to initialize the graphs $R_1,\ldots,R_{\beta}$ is at most $O(|E(R)|\cdot k^2\cdot \log n)\leq O(|V(C)|\cdot \Delta\cdot n^{O(1/k)})$ (since $|E(R)|\leq O(|V(C)|\cdot \Delta\cdot n^{O(1/k^2)})$ from \Cref{obs: properties of RC}, and $k^2\log n\leq n^{O(1/k)}$, as $k\leq (\log n)^{1/49}$ and $n$ is large enough.)

\paragraph{Updates.}
We now fix an index $1\leq i\leq k$, and consider the $i$th batch $\pi_i$ of updates. Recall that, from \Cref{obs: properties of RC}, following the batch $\pi_i$ of updates, at most $|\pi_i^C|\cdot n^{10/k^2}$ edges are deleted from $R_C$. We denote by $E_i'$ this set of edges. We assume that Properties \ref{prop: vertex set} -- \ref{prop: degree2} hold before $\pi_i$ is processed.

Next, we partition the vertices of $A=V(C)$ into groups $Z_0,\ldots,Z_{\beta}$, as follows. Group $Z_0$ contains all vertices $v\in A$, such that the number of edges incident to $v$ that lie in $E'_i$ is at most $\frac{\Delta_0}{2^{i+1}}$. For $1\leq j<\beta$, group $Z_j$ contains all vertices $v\in A$, such that $\frac{\Delta_{j-1}}{2^{i+1}}< |\delta_R(v)\cap E'_i|\leq \frac{\Delta_j}{2^{i+1}}$. Lastly, set $Z_{\beta}$ contains all vertices $v\in A$, such that more than $\frac{\Delta_{\beta-1}}{2^{i+1}}$ edges incident to $v$ lie in $E'_i$.
Our algorithm only explicitly computes the groups $Z_1,\ldots,Z_{\beta}$; notice that this can be done in time $O(|E'_i|)\leq O(|\pi_i^C|\cdot n^{10/k^2})$.

For all $0\leq j\leq \beta$, we start by deleting from $R_j$ all vertices and edges that were deleted from $R$ when processing $\pi_i$.

Consider first the group $Z_0$, and let $v\in Z_0$ be any vertex in this group. Then at most $\frac{\Delta_0}{2^{i+1}}$ edges that are incident to $v$ in $R$ have been deleted from $R$ when processing batch $\pi_i$. Therefore, for all $0\leq j\leq \beta$, graph $R_j$ still contains at least $\frac{\Delta_j}{2^i}-\frac{\Delta_0}{2^{i+1}}\geq \frac{\Delta_j}{2^{i+1}}$ edges that are incident to $v$, and so we do not need to process the vertices of $Z_0$ further.

Consider now some group $Z_j$, for $1\leq j\leq \beta$. Since $|E'_i|\leq |\pi_i^C|\cdot n^{10/k^2}$, and every vertex in $Z_j$ is incident to at least $\frac{\Delta_{j-1}}{2^{i+1}}$ edges of $E'_i$, we get that: 

\begin{equation}
|Z_j|\leq \frac{|\pi_i^C|\cdot n^{10/k^2}\cdot 2^{i+1}}{\Delta_{j-1}}.\label{eq: bound zj}
\end{equation}

Consider now any vertex $v\in Z_j$, and recall that at most $\frac{\Delta_j}{2^{i+1}}$ edges incident to $v$ lie in $E'_i$.
If $j<\beta$, then, for every index $j'>j$, the number of edges incident to $v$ in $R_{j'}$ that were deleted from $R_{j'}$ is bounded by $\frac{\Delta_{j}}{2^{i+1}}$, and so the degree of $v$ in $R_{j'}$ remains at least $\frac{\Delta_{j'}}{2^{i}}-\frac{\Delta_{j}}{2^{i+1}}\geq \frac{\Delta_{j'}}{2^{i+1}}$, as required. Therefore, for indices $j'>j$, the vertices of $Z_{j}$ satisfy Condition \ref{prop: degree 1} in graph $R_{j'}$, and they require no further processing in $R_{j'}$. For indices $0\leq j'\leq j$, however, it is possible that too many edges incident to some vertices of $Z_j$ have been deleted from $R_{j'}$. Intuitively, for all $0\leq j'\leq j$, our algorithm will compute a new collection $\tilde E^j_{j'}$ of edges, such that, on the one hand, every vertex of $Z_j$ is incident to at least $\frac{\Delta_{j'}}{2^{i+1}}$ such edges, while, on the other hand, every vertex of $B$ is incident to at most $\Delta_{j'}\cdot \tilde r\cdot k^{4(i-1)}\cdot (16\log n)^{\beta-j'}$ such edges. 
For all $0\leq j'\leq j$, we will then delete all edges incident to the vertices of $Z_j$ from graph $R_{j'}$, and insert the edges of $\tilde E^j_{j'}$ instead. The sets $\set{\tilde E^j_{j'}}_{j'=0}^j$ of edges are computed by repeatedly applying the algorithm from 	\Cref{claim: constructing the subgraph}, from higher to lower values of $j'$, similarly to the algorithm that we used in order to initialize the data structures. We summarize the algorithm for computing these edge sets in the following observation.

\begin{observation}\label{obs: compute updated edge sets}
	There is a deterministic algorithm that, given an index $1\leq j\leq \beta$, computes, for all $0\leq j'\leq j$, a collection $\tilde E_{j'}^j\subseteq E(R)$ of edges, such that:
	\begin{itemize}
		\item every edge in $\tilde E_{j'}^j$ is incident to a vertex of $Z_j$;
		\item every vertex $v\in Z_j$ is incident to at least $\frac{\Delta_{j'}}{2^{i+1}}$ and at most $\Delta_{j'}$ edges of $\tilde E_{j'}^j$; and
		\item every vertex $u\in B$ is incident to at most $\Delta_{j'}\cdot \tilde r\cdot k^{4(i-1)}\cdot (16\log n)^{\beta-j'}$ edges of $\tilde E_{j'}^j$.
	\end{itemize}
The running time of the algorithm is $O(|Z_j|\cdot \Delta_j\cdot \log n)$.
\end{observation}
\begin{proof}
We start by describing the algorithm for computing the set $\tilde E_j^j$ of edges. Assume first that $j=\beta$. In order to construct set $\tilde E_{\beta}^{\beta}$ of edges, for every vertex $v\in Z_{\beta}$, we include in $\tilde E_{\beta}^{\beta}$ arbitrary $\floor{\Delta_{\beta}}$ edges that are incident to $v$ in $R$; recall that, from \Cref{obs: properties of RC}, the degree of $v$ in $R$ must be at least $\Delta_{\beta}$. Moreover, the observation guarantees that the degree of every vertex $u\in B$ is at most $n^{2/k^2}\cdot \Delta<\Delta_{\beta}\cdot n^{8/k^2}<\Delta_{\beta}\cdot \tilde r$. Assume now that $j<\beta$. Recall that, from Property \ref{prop: degree 1}, every vertex $v\in Z_j$ had degree at least $\frac{\Delta_j}{2^i}$ and at most $\Delta_j$ in graph $R_j$ prior to the arrival of batch $\pi_i^C$. From the definition of set $Z_j$ of vertices, for every vertex $v\in Z_j$, at most $\frac{\Delta_j}{2^{i+1}}$ edges incident to $v$ were deleted from $R$. Therefore, the degree of every vertex $v\in Z_j$ in graph $R_j$ remains at least $\frac{\Delta_j}{2^i}-\frac{\Delta_j}{2^{i+1}}\geq \frac{\Delta_j}{2^{i+1}}$, as required. We let set $\tilde E_j^j$ contain, for every vertex $v\in Z_j$, all edges incident to $v$ in graph $R_j$. Clearly, the time required to compute the set $\tilde E_j^j$ of edges is bounded by $|Z_j|\cdot \Delta_j$.

Consider now some index $0\leq j'<j$, and assume that we have computed the set $\tilde E_{j'+1}^j$ of edges of $R$, such that every edge in $\tilde E_{j'+1}^j$ is incident to a vertex of $Z_j$; every vertex of $Z_j$ is incident to at least $\frac{\Delta_{j'+1}}{2^{i+1}}$ and at most $\Delta_{j'}$ edges of $\tilde E_{j'+1}^j$; and
every vertex $u\in B$ is incident to at most $\Delta_{j'+1}\cdot \tilde r\cdot k^{4(i-1)}\cdot (16\log n)^{\beta-j'-1}$ edges of $\tilde E_{j'+1}^j$.

We construct a graph $H_{j'}$, that contains all edges in $\tilde E_{j'+1}^j$, and all their endpoints. For convenience, we denote by $B_{j'}=V(H_{j
})\cap B$. We will compute the set $\tilde E_{j'}^j$ of edges by applying the algorithm from  \Cref{claim: constructing the subgraph} to graph $H_{j'}$. In order to do so, we set the parameters as follows. First, we let $\hat \Delta=\frac{\Delta_{j'}}{2^{i+1}}$, and $z=n^{10/k^2}$. Recall that every vertex $v\in Z_j$ is incident to at least $\frac{\Delta_{j'+1}}{2^{i+1}}=\frac{\Delta_{j'}\cdot n^{10/k^2}}{2^{i+1}}=\hat \Delta\cdot z$ edges in $H_{j'}$.
We set $\gamma'=\tilde r\cdot k^{4(i-1)}\cdot 2^{i+1}\cdot (16\log n)^{\beta-j'-1}$; recall that every vertex $u\in B_j$ is incident to at most $\Delta_{j'+1}\cdot \tilde r\cdot k^{4(i-1)}\cdot (16\log n)^{\beta-j'-1}=\Delta_{j'}\cdot n^{10/k^2}\cdot \tilde r\cdot k^{4(i-1)}\cdot (16\log n)^{\beta-j'-1}\leq \hat \Delta\cdot z\cdot \gamma'$ edges of $\tilde E_{j'+1}^j$. Lastly, we set $\hat r=8\gamma'\log n$.
We can now use the algorithm from 
\Cref{claim: constructing the subgraph} to compute, for every vertex $v\in Z_j$,  a collection $E'(v)\subseteq \delta_{H_j}(v)$ of its incident edges.
We then set $\tilde E_{j'}^j=\bigcup_{v\in Z_j}E'(v)$. Recall that \Cref{claim: constructing the subgraph} guarantees that every vertex of $Z_j$ is incident to  $\ceil{\hat \Delta}=\ceil{\frac{\Delta_{j'}}{2^{i+1}}}$ edges of $\tilde E_{j'}^j$, and every vertex of $B_j$ is incident to at most $\hat r\cdot \hat \Delta= \frac{\Delta_{j'}}{2^{i+1}} \cdot 8\gamma'\log n\leq \Delta_{j'}\cdot \tilde r\cdot k^{4(i-1)}\cdot (16\log n)^{\beta-j'}$ such edges, as required. The running time of the algorithm from \Cref{claim: constructing the subgraph} is bounded by $O(|E(H_j)|\cdot \log (n))\leq O(|Z_j|\cdot \Delta_{j'}\cdot \log n)$.

The total running time of our algorithm to compute all sets $\tilde E_{j'}^j$ of edges, for all $0\leq j'\leq j$ is bounded by:

\[O\left(|Z_j|\cdot \sum_{j'=0}^{j}\Delta_{j'}\log n\right )\leq O(|Z_j|\Delta_j\log n).\]
\end{proof}

We are now ready to complete our algorithm for updating the graphs $R_0,\ldots,R_{\beta}$. Consider an integer $0\leq \ell\leq \beta$, and the corresponding graph $R_{\ell}$. For all $\ell\leq j\leq \beta$, we delete from $R_{\ell}$ all edges that are incident to the vertices of $Z_j$, and insert the edges of $\tilde E_{\ell}^j$ instead (if $\ell=0$, then we only do so for indices $1\leq j\leq \beta$). Recall that, before the current update, from Property \ref{prop: degree2}, every vertex of $B$ was incident to at most $\Delta_{\ell}\cdot \tilde r\cdot k^{4(i-1)}\cdot (16\log n)^{\beta-\ell}$ edges in graph $R_{\ell}$. For all $\ell\leq j\leq \beta$, we are guaranteed that every vertex of $B$ is incident to at most $\Delta_{\ell}\cdot \tilde r\cdot k^{4(i-1)}\cdot (16\log n)^{\beta-\ell}$ edges of $\tilde E_{\ell}^j$. Since $\beta\leq k^2$, we get that, in the updated graph $R_{\ell}$, the degree of every vertex $v\in B$ is bounded by $(\beta+1)\cdot \Delta_{\ell}\cdot \tilde r\cdot k^{4(i-1)}\cdot (16\log n)^{\beta-\ell}\leq  \Delta_{\ell}\cdot \tilde r\cdot k^{4i}\cdot (16\log n)^{\beta-\ell}$. From our construction and the discussion above, every vertex of $A$ has degree at least $\frac{\Delta_{\ell}}{2^{i+1}}$ and at most $\Delta_{\ell}$ in $R_{\ell}$. Therefore Properties \ref{prop: vertex set} -- \ref{prop: degree2} continue to hold after the update.
From our discussion, the time required for processing the set $Z_j$ of vertices, for $0\leq j\leq \beta$, is bounded by:

\[O(|Z_j|\Delta_j\log n)\leq O\left(\frac{|\pi_i^C|\cdot n^{10/k^2}\cdot 2^{i+1}\cdot \Delta_j\cdot \log n}{\Delta_{j-1}}\right )\leq O\left(|\pi^c|\cdot n^{O(1/k^2)}\right ) \]

(we have used the fact that $i\leq k$, inequalities \ref{eq: k and n} and \ref{eq: bound zj}, together with the facts that $k\leq (\log n)^{1/49}$ and $n$ is large enough).
Since $\beta\leq k^2\leq n^{O(1/k^2)}$,	the total time required to process batch $\pi_i$ is bounded by $O\left(|\pi^c|\cdot n^{O(1/k^2)}\right )$.

We are now ready to complete the proof of \Cref{claim: maintain the paths}. For every vertex $v\in V(C)$, we simply let $U'_C(v)$ be the set of all vertices that serve as endpoints to the edges of $R_0$ that are incident to $v$. From Property \ref{prop: degree 1}, for every vertex $v\in V(C)$, $\frac{\Delta_{0}}{2^k}\leq |U'_C(v)|\leq \Delta_0$ holds at all times. Since, from Inequality \ref{eq: delta0}, 
$2^k\cdot \Delta'\leq \Delta_0\leq 2^k\cdot \Delta'\cdot n^{10/k^2}$, we get that, for every vertex $v\in V(C)$, $\Delta'\leq |U'_C(v)|\leq 2^k\cdot \Delta'\cdot n^{10/k^2}\leq \Delta'\cdot n^{11/k^2}$ always holds.

From Property \ref{prop: degree2}, for every vertex $x\in B$, the degree of $x$ in $R_0$ is bounded by:

\[\Delta_{0}\cdot \tilde r\cdot k^{4k}\cdot (16\log n)^{\beta}\leq  \Delta'\cdot n^{20/k^2}\cdot 2^{8k^2}\cdot (\log n)^{k^2}\leq \Delta'\cdot n^{16/k}. \]

since $\tilde r=n^{10/k^2}$, $\beta\leq k^2$, and $\Delta_0\leq 2^k\cdot \Delta'\cdot n^{10/k^2}$ as observed above.
We have also used the fact that, from the definition of valid input parameters, $k\leq (\log n)^{1/49}$, and so $n^{1/k^4}\geq 2^{(\log n)^{4/5}}\geq \log n$, since $n$ is large enough. Lastly, we have used Inequality \ref{eq: k and n}.
As we have shown, the initialization time of our algorithm is $O(|V(C)|\cdot \Delta\cdot n^{O(1/k)})$, and, for all $1\leq i\leq k$, the time required for updates following the processing of batch $\pi_i^C$ of edge deletions to graph $C$ is bounded by $O\left(|\pi_i^C|\cdot n^{O(1/k^2)}\right )$.
This completes the proof of \Cref{claim: maintain the paths}.
\end{proofof}

\section{Applications to (Fault-Tolerant) Graph Sparsification}\label{sec:bd-fault}

Consider a simple graph $G$, and a router decomosition $\left (\cset,\set{C'}_{C\in \cset}\right )$ of $G$ with parameters $\Delta^*,\tilde d,\tilde \eta$ and $\rho$ (see \Cref{def: router decomposition}). Recall that the decomposition ensures that, for every cluster $C\in \cset$, the corresponding subgraph $C'\subseteq C$ is  a $(\vDelta^*, \tilde d,\tilde \eta)$-router, for the set $V(C)$ of supported vertices. Also recall that we have denoted by $\edel=E(G)\setminus \left(\bigcup_{C\in \cset}E(C)\right )$ -- the set of edges that do not lie in any cluster of the decomposition.
Consider the graph $H'$, whose vertex set is $V(G)$, and edge set is $\edel\cup \left(\bigcup_{C\in \cset}E(C')\right )$, and recall that the algorithm from \Cref{thm: router decomposition main} maintains a graph $H$ with $H'\subseteq H$. In this subsection, we explore various properties of the graph $H'$, that imply that the  graph $H$ itself has many useful properties as well. This, in turn, implies our results for constructing and maintaining spanners, fault-tolerant spanners with respect to bounded-degree faults, low-congestion spanners, and connectivity certificates. We start with discussing general dynamic spanners and low-congestion spanners. 

\subsection{Dynamic Spanners and Low-Congestion Spanners}\label{sec:appnonFT-sparse}

\textbf{Dynamic Spanners.} Let $G$ be a simple $n$-vertex graph, and let $\mathcal{R}=(\mathcal{C}, \{C'\}_{C \in \mathcal{C}})$ be a router decomposition of $G$ with parameters $\Delta^*, \tilde d,\tilde \eta$ and $\rho$. Consider the corresponding graph $H'$ whose   vertex set is $V(G)$, and edge set is $\edel\cup \left(\bigcup_{C\in \cset}E(C')\right )$, where $\edel=E(G)\setminus \left(\bigcup_{C\in \cset}E(C)\right )$. In the following observation, we show that $H'$ is a $\tilde d$-spanner.

\begin{observation}\label{obs: spanner H'}
	Graph $H'$ is  $\tilde d$-spanner. 
\end{observation}
\begin{proof}
	Let $e=(u,v)$ be any edge of $G$. If $e\in E(H')$, then $\dist_{H'}(u,v)=1$. Otherwise, there must be some cluster $C\in \cset$, with $e\in E(C)$. Consider the subgraph $C'\subseteq C$ that the decomposition provides. Since $V(C)=V(C')$, and $C'$ is a $(\Delta^*, \tilde d,\tilde \eta)$ router for the set $V(C)$, it must be the case that $\dist_{C'}(u,v)\leq \tilde d$. The claim follows as $C' \subseteq H'$. 
\end{proof}
	

By combining \Cref{thm: router decomposition main} with \Cref{obs: spanner H'}, we obtain the following immediate corollary.

\begin{corollary}\label{cor: spanner}
There is a deterministic algorithm, whose input is a graph $G$ with $n$ vertices and no edges, that undergoes an online sequence of at most $n^2$ edge deletions and insertions, such that $G$ remains a simple graph throughout the update sequence. Each edge inserted into $G$ has an integral length $\ell(e)\in \set{1,\ldots,L}$. Additionally, the algorithm is given parameters $512\leq k\leq (\log n)^{1/49}$ and $\frac{1}{k}\leq \delta\leq \frac{1}{400}$, such that $k$ and $\frac 1{\delta}$ are integers. The algorithm maintains a subgraph $H\subseteq G$ with $V(H)=V(G)$ and $|E(H)|\leq O\left(n^{1+O(1/k)}\log L\right )$, such that $H$ is a $k'$-spanner for $G$, for $k'=k^{19}\cdot 2^{O(1/\delta^6)}$. The worst-case update time of the algorithm is $n^{O(\delta)}$ per operation, and the number of edge insertions and deletions that graph $H$ undergoes after each update to $G$ is bounded by $n^{O(1/k)}$.
\end{corollary}
\begin{proof}
	We partition all edges that ever belong to $G$ over the course of the algorithm into sets $E_0,\ldots,E_{\ceil{\log L}}$, where, for $0\leq i\leq \ceil{\log L}$, set $E_i$ contains all edges $e$ with
	$2^{i}\leq  \ell(e)< 2^{i+1}$. For all $0\leq i\leq \ceil{\log L}$, we let $G_i=G[E_i]$. For all $0\leq i\leq \ceil{\log L}$, we apply the algorithm from \Cref{thm: router decomposition main} to graph $G_i$, with parameters $k$ and $\delta$ that remain unchanged, and parameter $\Delta=\ceil{n^{20/k}}$. We denote the graph $H$ that the algorithm maintains by $H_i$. Recall that $|E(H_i)|\leq n^{1+O(1/k)}\cdot \Delta\leq n^{1+O(1/k)}$. From \Cref{obs: spanner H'}, it is immediate to verify that $H_i$ is a $k'$-spanner for $G_i$, for $k'=k^{19}\cdot 2^{O(1/\delta^6)}$. Finally, we let $H=\bigcup_{i}H_i$. Notice that $|E(H)|\leq O\left (n^{1+O(1/k)}\log L\right )$ must hold, and it is easy to verify that the algorithm from \Cref{thm: router decomposition main} can be extended to maintain the graph $H$ with  worst-case update time  $n^{O(\delta)}$, such that the number of edge updates that $H$ undegroes following each update to $H$ is at most $n^{O(1/k)}$.
\end{proof}

\paragraph{Low-Congestion Spanners.} We formalize the notion of \emph{low-congestion spanners}, that was implicitly introduced by Chen et al. \cite{ChenKLPGS22}. The new notion views a spanner $H$ of $G$  as a sparse subgraph of $G$, with the additional property that  $G$ can be embedded into $H$ with low congestion and path length bounds. 
Notice that, if we use the standard notion of a $t$-spanner $H$ of $G$, then $G$ can be embedded into $H$ via paths of length at most $t$, but the congestion of the embedding may be as large as $|E(G)|$. Low-congestion spanners aim at  optimizing the length and the congestion parameters of this embedding simultaneously. We now provide a formal definition of low-congestion spanners.

\begin{definition}[Low-Congestion Spanners]
Let $G$ be a graph, let $H \subseteq G$ be a subgraph of $G$ with $V(H)=V(G)$, and let $d,\eta>0$ be parameters. We say that $H$ is a $(d,\eta)$-low congestion spanner for $G$, if there is an embedding $\qset=\set{Q(e)\mid e\in E(G)}$ of $G$ into $H$, such that the length of every path in $\qset$ is a most $d$, and the paths in $\qset$ cause congestion at most $\eta$. 
\end{definition}


We note that low-congestion spanners can also be defined with respect to vertex congestion, and it is the spanners of the latter kind that were considered by  \cite{brand2023deterministic}. Their algorithm maintains such a spanner $H$ for a dynamic $n$-vertex graph $G$, with $|E(H)|\leq O(n^{1+o(1)})$, together with the embedding $\qset$ of $G$ into $H$, such that the lengths of the embedding paths are bounded by $d=n^{o(1)}$, and vertex-congestion of the embedding $\qset$ is bounded by $\eta=n^{o(1)}\cdot \max_{v\in V(G)}\set{\deg_G(v)}$. The amortized update time and recourse bounds are $n^{o(1)}$. 

We use the following simple lemma.

\begin{lemma}\label{lem:router-embed-subgraph}
Let $G$ be an $n$-vertex graph, and let $\mathcal{R}=\left (\mathcal{C}, \{C'\}_{C \in \mathcal{C}}\right )$ be a router decomposition of $G$ with parameters $\Delta^*, \tilde d,\tilde \eta$ and $\rho$. Then for every cluster $C\in \cset$, there is an embedding $\qset_C$ of $C$ into $C'$ via paths of length at most $\tilde d$, that cause congestion at most $\eta=\max\set{\frac{16\tilde \eta \cdot \Delta_{\max}(C)}{\Delta^*},16\log n}$, where $\Delta_{\max}(C)$ is the maximum vertex degree in $C$. 
\end{lemma}
\begin{proof}
Consider a cluster $C \in \mathcal{C}$, and recall that $C'\subseteq C$ is a subgraph of $C$ with $V(C')=V(C)$, that is  a $(\vDelta^*,\tilde d,\tilde \eta)$-router for the set $V(C)$ of supported vertices. 

We define a demand $\dset=\left(\Pi,\set{D(a,b)}_{(a,b)\in \Pi}\right)$ as follows: set $\Pi$ contains all pairs $(u,v)$ of vertices, for which edge $(u,v)$ lies in $C$, and the corresponding demand value $D(u,v)$ is $1$. 
Clearly, demand $\dset$ is $\Delta_{\max}(C)$-restricted in $C'$. It is now enough to show that demand $\dset$ can be routed integrally in $C'$ via paths of length at most $\tilde d$, that cause congestion at most $\eta$.

Consider the demand $\dset'$ that is obtained from $\dset$ by scaling every demand $D(u,v)$ down by factor $\frac{\Delta_{\max}(C)}{\Delta^*}$. Then it is immediate to see that $\dset'$ is a $\Delta^*$-restricted demand, and so there is a routing $f$ of this demand in $C'$ via paths of length at most $\tilde d$, that cause congestion at most $\tilde \eta$. By scaling the flow on each flow-path up by factor $\frac{\Delta_{\max}(C)}{\Delta^*}$, we obtain a routing $f$ of the demand $\dset$ in $C$ via paths of length at most $\tilde d$, that cause congestion at most $\frac{\tilde \eta \cdot \Delta_{\max}(C)}{\Delta^*}$. Our last step is to turn the resulting routing into an integral one, using the standard Randomized Rounding procedure of Raghavan and Thompson \cite{RaghavanTx87}. 

Consider any edge $e=(u,v)\in E(C)$, and let $\pset_{u,v}$ be the set of all $u$-$v$ paths in $C'$ of length at most $\tilde d$. Since flow $f$ sends $1$ flow unit from $u$ to $v$ over paths of length at most $\tilde d$, it defines a probability distribution $D_{u,v}$ over the paths of $\pset_{u,v}$, where every path $P\in \pset_{u,v}$ is assigned probability $f(P)$. For each such edge $e=(u,v)\in E(C)$, we choose a single path $P(e)\in \pset_{u,v}$ from this distribution, where the choices between the different edges are independent. Let $\qset_C=\set{P(e)\mid e\in E(C)}$ be the resulting collection of paths, that defines an embedding of $C$ into $C'$. Clearly, the paths in $\qset$ have length at most $\tilde d$ each. Next, we argue that with high probability, the paths in $\qset$ cause congestion  at most $\eta$ in $C'$. Indeed, the expected congestion for an edge $e\in E(C')$ is at most $\frac{\tilde \eta \cdot \Delta_{\max}(C)}{\Delta^*}  $. Using the Chernoff bound from \Cref{lem: Chernoff} with $t=\eta=\max\set{\frac{16\tilde \eta \cdot \Delta_{\max}(C)}{\Delta^*},16\log n}$,
	 we get that the probability that the congestion on $e$ is greater than $\eta$ is bounded by $2^{-\eta}<1/n^3$, since $\eta\geq 16\log n$. Using the union bound, with high probability, the congestion on every edge of $C'$ is at most $\eta$.
\end{proof}

We  obtain the following immediate corollary of \Cref{lem:router-embed-subgraph}.

\begin{corollary}\label{cor:router-decomp-to-LCS}
Let $G$ be an $n$-vertex graph, and let $\mathcal{R}=(\mathcal{C}, \{C'\}_{C \in \mathcal{C}})$ be a router decomposition of $G$ with parameters $\Delta^*, \tilde d,\tilde \eta$ and $\rho$. Consider the corresponding graph $H'$ whose   vertex set is $V(G)$, and edge set is $\edel\cup \left(\bigcup_{C\in \cset}E(C')\right )$, where $\edel=E(G)\setminus \left(\bigcup_{C\in \cset}E(C)\right )$. 
Then graph $H'$ is an $(\tilde d,\eta)$-low congestion spanner for $G$, where $\eta=\max\set{\frac{16\tilde \eta \cdot \Delta_{\max}(G)}{\Delta^*},16\log n}$,  and  $\Delta_{\max}(G)$ is the maximum vertex degree in $G$.
\end{corollary}
\begin{proof}
	We obtain an embedding $\qset$ of $E(G)$ into $H$, as follows. First, since $\edel\subseteq E(H')$, we embed every edge $e\in \edel$ into itself. Each remaining edge $e\in E(G)$ must lie in some cluster $C\in \cset$. For each such cluster $C$, we use the embedding $\qset_C$ of $C$ into $C'$ that is given by \Cref{lem:router-embed-subgraph}. Since, for every cluster $C\in \cset$, the corresponding graph $C'$ is contained in $H'$, and since the clusters in $\cset$ are edge-disjoint, we obtain an embedding of $G$ into $H'$ via paths of length at most $\tilde d$, that cause congestion at most $\eta$.
\end{proof}

Finally, by combining \Cref{thm: router decomposition main} with \Cref{cor:router-decomp-to-LCS}, we obtain the following immediate corollary; the proof is essentially identical to the proof of \Cref{cor: spanner} and is omitted here.

\begin{corollary}\label{cor: spanner2}
	There is an algorithm, whose input is a graph $G$ with $n$ vertices and no edges, that undergoes an online sequence of at most $n^2$ edge deletions and insertions, such that $G$ remains a simple graph throughout the update sequence. Each edge inserted into $G$ has an integral length $\ell(e)\in \set{1,\ldots,L}$. Additionally, the algorithm is given parameters $512\leq k\leq (\log n)^{1/49}$, $\frac{1}{k}\leq \delta\leq \frac{1}{400}$ and $\Delta\geq n^{20/k}$, such that $k,\Delta$ and  $\frac 1{\delta}$ are integers. The algorithm maintains a subgraph $H\subseteq G$ with $V(H)=V(G)$ and $|E(H)|\leq O\left(n^{1+O(1/k)}\cdot \Delta\cdot \log L\right )$, such that $H$ is a $(k',\eta)$-low-congestion spanner for $G$, for $k'=k^{19}\cdot 2^{O(1/\delta^6)}$ and $\eta\leq n^{O(1/k)} \cdot\max\set{\frac{ \Delta_{\max}(G)}{\Delta},1}$,  where $\Delta_{\max}(G)$ is the maximum degree in $G$. The worst-case update time of the algorithm is $n^{O(\delta)}$ per operation, and the number of edge insertions and deletions that graph $H$ undergoes after each update to $G$ is bounded by $n^{O(1/k)}$.
\end{corollary}

\subsection{Dynamic Fault-Tolerant Spanners and Connectivity Certificates}\label{sec:appFT-sparse}

In this subsection we provide  a key observation that routers with a sufficiently large minimum degree are inherently resilient to edge failures. We then show that, by setting the parameter $\Delta$ in \Cref{thm: router decomposition main} appropriately, we can ensure that the resulting graph $H$ is resilient to bounded-degree faults. Lastly, we prove that $H$ can be used as a connectivity certificate.

\subsubsection{Resilience of Routers to Bounded-Degree Faults}\label{sec:res-routers}

Let $G$ be a graph, and let $F$ be a subset of edges of $G$ (edge faults). We denote by $\deg(F)$ the maximum, over all vertices of $v$, of the number of edges in $F$ that are incident to $v$. We say that a set $F$ of edges is an $f$-BD faulty set, if $\deg(F)\leq f$. Recall that a graph $G$ is a $(\vDelta,d,\eta)$-router for the set $V(G)$ of supported vertices, if every $\Delta$-restricted demand $\dset$ over $V(G)$ can be routed with congestion at most $\eta$ in $G$, via paths of length at most $d$. In this subsection we show that routers are resilient to bounded-degree faults, and the sense that  $G \setminus F$ remains a router, albeit with slightly weaker parameters. The main result of this subsection is summarized in the following theorem, whose proof is provided below.

\begin{theorem}\label{thm:resilient-router}
Let $G$ be an $n$-vertex graph, and let $f,k,d>0$,  $1\leq \eta<n^2$ and $\Delta \geq 32f\cdot  n^{1/k} \cdot \eta$ be parameters. Assume further that $G$ is a $(\vDelta,d,\eta)$-router with respect to the set $V(G)$ of supported vertices. Then for every $f$-BD faulty set $F \subseteq E(G)$ of edges, graph $G \setminus F$ is a $(\vDelta, O(k) \cdot d,O(k) \cdot \eta)$-router for the set $V(G)$ of supported vertices.
\end{theorem}

The above theorem should be compared to the resilience of length-constrained expanders against $f$-BD faults, that was established in \cite{BodwinDR22}. Specifically, they show in Theorem 1.2 of  \cite{BodwinDR22} that, if $G$ is an $n$-vertex $(h, s)$-length $\phi$-expander with minimum degree $\Omega(fn^{\epsilon}/\phi)$, then, for any $f$-BD faulty set $F\subseteq E(G)$ of edges,  the graph $G \setminus F$ is an $(hs)^{1/\epsilon}$-spanner of $G$; in other words, the distances in $G \setminus F$ may grow by at most factor $(hs)^{1/\epsilon}$.
It then follows that the $f$-FD spanners obtained by  \cite{BodwinDR22} via LC-expanders  have size $\widetilde{O}(f \cdot n^{1+1/k})$ and stretch $t=k^{O(k)}$.
The question of improving the exponential dependence of the stretch on $k$, while preserving the size of the spanner was left open in \cite{BodwinDR22} (see Theorem 1.14 therein), and this improvement would immediately imply stronger bounds for $f$-FD spanners. We resolve this open problem in \Cref{thm:resilient-router}, for the case where the LC-expander has a bounded diameter (that is, it is a router).

\begin{proofof}{\Cref{thm:resilient-router}} 
Consider any demand $\hat \dset=\left(\hat \Pi,\set{\hat D(a,b)}_{(a,b)\in \hat \Pi}\right)$. We say that demand $\hat \dset$ is \emph{integral}, if,  for every pair $(a,b)\in \hat \Pi$, the demand value $\hat D(a,b)$ is an integer. It will be convenient for us to represent such an integral demand $\hat \dset$ by a multiset $I=\set{(a_1,b_1),\ldots,(a_{\chi},b_{\chi})}$ of pairs of vertices, with $\chi=\sum_{(a,b)\in \hat \Pi} \hat D(a,b)$, where each pair $(a,b)\in \hat \Pi$ appears in $I$ with multiplicity $\hat D(a,b)$. 
Throughout, we identify each pair (and its two vertices) by its index in the multiset, $(a_{\ell},b_{\ell})$. Conversely, any multiset $I$ of pairs of vertices in $G$ naturally defines the corresponding demand $\hat \dset(I)$, where the demand $\hat D(a,b)$ for every pair $(a,b)$ is the number of times it appears in $I$. The resulting demand is $\hat \Delta$-restricted, for a parameter $\hat \Delta$, if every vertex of $G$ appears in at most $\hat \Delta$ pairs in $I$. For convenience, we will not distinguish between the integral demand $\hat \dset$, and its corresponding multiset $I$ of vertex pairs. Consider now an integral demand $I=\set{(a_1,b_1),\ldots,(a_{\chi},b_{\chi})}$. A routing $f$ of $I$ is \emph{integral} if, for every pair $(a_i,b_i)$, there is exactly one flow-path $P$ connecting $a_i$ to $b_i$ with $f(P)>0$, and moreover, $f(P)=1$. An integral routing of an integral demand $I=\set{(a_1,b_1),\ldots,(a_{\chi},b_{\chi})}$ can equivalently be defined as a collection $\pset=\set{P_1,\ldots,P_{\chi}}$ of paths, where, for all $1\leq i\leq \chi$, path $P_i$ connects $a_i$ to $b_i$. 
We will use the following simple observation.

\begin{observation}\label{obs: integral routing of integral demand}
	Let $\hat \dset=\left(\hat \Pi,\set{\hat D(a,b)}_{(a,b)\in \hat \Pi}\right)$ be an integral demand in $G$, and assume that $\hat \dset$ is an $(\alpha\cdot \Delta)$-restricted demand, for some parameter $\alpha\geq 8\log n$. Then there is an integral routing $\pset$ of $\hat \dset$, where each path in $\pset$ has length at most $d$, and the paths cause congestion at most $8\alpha \eta$.
\end{observation}
\begin{proof}
Let $\hat \dset'$ be a demand that is obtained from $\hat \dset$,  by scaling all demands in $\hat \dset$ down by factor $\alpha$. It is easy to verify that demand $\hat \dset'$ is $\Delta$-restricted. Since $G$ is a $(\vDelta,d,\eta)$-router with respect to the set $V(G)$ of supported vertices, there is a routing $f'$ of $\hat\dset'$ in $G$ via paths of length at most $d$, that causes congestion at most $\eta$. By scaling the flow value $f'(P)$ on every flow-path up by factor $\alpha$, we obtain a routing $f$ of $\hat \dset$ via paths of length at most $d$, that cause congestion at most $\alpha\eta$. Lastly, in order to convert this routing into an integral one, we use the standard Randomized Rounding procedure of Raghavan and Thompson \cite{RaghavanTx87}.

Consider the multiset $I=\set{(a_1,b_1),\ldots,(a_{\chi},b_{\chi})}$ of vertex pairs that corresponds to the demand $\hat \dset$. We can naturally view the flow $f$ as routing the pairs in $I$ via paths of length at most $d$, with congestion at most $\alpha\eta$. Consider now
any demand pair $(a_i,b_i)\in I$, and let $\pset_{(a_i,b_i)}$ be the collection of all $a_i$-$b_i$ paths of length at most $d$ in $G$. Since flow $f$ sends $1$ flow unit from $a_i$ to $b_i$, it defines a probability distribution $D_i$ over the paths of $\pset_{(a_i,b_i)}$, where every path $P\in \pset_{(a_i,b_i)}$ is assigned probability $f(P)$. For all $1\le i\leq \chi$, we choose a single path $P_i\in \pset_{(a_i,b_i)}$ from this distribution, where the choices between the different pairs are independent. Let $\pset=\set{P_1,\ldots,P_{\chi}}$ be the resulting routing. Clearly, the paths in $\pset$ have length at most $d$ each. Next, we argue that with high probability, the congestion of the routing is at most $8\alpha \eta$. Indeed, the expected congestion for an edge $e\in E(G)$ is at most $\alpha\cdot \eta$. Using the Chernoff bound from \Cref{lem: Chernoff} with $t=8\alpha\eta$, we get that the probability that the congestion on $e$ is greater than $8\alpha\eta$ is bounded by $2^{-8\alpha\eta}<1/n^3$, since $\alpha\geq 8\log n$. Using the union bound, with high probability, the congestion on every edge is at most $8\alpha\eta$.
\end{proof}

Let  $\dset=\left(\Pi,\set{D(a,b)}_{(a,b)\in \Pi}\right)$ be a $\Delta$-restricted demand for $G$, and let $F\subseteq E(G)$ be an $f$-BD faulty edge set. We will denote by $V(F)$ the set of all vertices that serve as endpoints of the edges in $F$. Our goal is to show that $\dset$ can be routed in $G \setminus F$ via paths of length at most $O(k)\cdot d$, with congestion at most $O(k) \cdot \eta$. Our first step is to convert $\dset$ into an integral demand.

Let $\dset'=\left(\Pi,\set{D'(a,b)}_{(a,b)\in \Pi}\right)$ be the demand obtained from $\dset$, by setting $D'(a,b)=\ceil{D(a,b)\cdot n}$ for every pair $(a,b)\in \Pi$ of vertices.  It is easy to verify, that, since every vertex of $G$ may participate in at most $n$ pairs in $\Pi$, 
the demand $\dset'$ is $\Delta'$-restricted, for $\Delta'=2n\Delta$. Let $\eta'=16\eta\cdot n$. Assume that there is a routing $f'$ of $\dset'$ in $G\setminus F$ via paths of length $O(k)\cdot d$, with congestion at most $O(k)\cdot \eta'$. Then we can obtain a routing $f$ of $\dset$ in $G$ via paths of length $O(k)\cdot d$ and congestion at most $O(k)\cdot \eta$, by scaling the flow $f'$ down by factor $n$, and then reducing it as needed. Therefore, from now on, it is enough to show that there is a routing $f'$ of $\dset'$ in $G\setminus F$ via paths of length $O(k)\cdot d$, with congestion at most $O(k)\cdot \eta'$. 
Notice that the resulting demand $\dset'=\left(\Pi,\set{D'(a,b)}_{(a,b)\in \Pi}\right)$ is integral. Throughout, we denote by $I=\set{(a_1,b_1),\ldots,(a_{\chi},b_{\chi})}$ the corresponding multiset of vertex pairs. Our goal is to send $1$ flow unit between every pair $(a_{\ell},b_{\ell})$, such that the resulting flow only uses flow-paths of length at most $O(k)\cdot d$, and causes congestion at most $O(k\cdot \eta')$.
We will repeatedly use the following immediate corollary of \Cref{obs: integral routing of integral demand}.

\begin{corollary}\label{obs: integral routing of integral demand2}
	If $I'$ is any integral $\Delta'$-restricted demand in $G$, then there is an integral routing $\pset$ of $I'$, where each path in $\pset$ has length at most $d$, and the paths cause congestion at most $16\eta n=\eta'$.
\end{corollary}

Our algorithm consists of $z\leq O(k)$ iterations. In each iteration $i$, we define a $\Delta'$-restricted demand $\dset_i$ for $V(G)$, and then consider its routing $f_i$ in $G$. For each flow-path $P$ that carries non-zero flow, we discard all edges of $F\cap E(P)$, partitioning $P$ into segments, some of which continue to carry $f_i(P)$ flow units; we denote the resulting flow, that does not use the edges of $F$, by $f'_i$. The final routing of $\dset$ is obtained by combining the resulting flows $f'_1,\ldots,f'_z$.

Consider now the multiset $I$ of vertex pairs, corresponding to the demand $\dset'$. From \Cref{obs: integral routing of integral demand2}, there is an integral routing $\pset=\set{P_1,\ldots,P_{\chi}}$ of $I$ via paths of length at most $d$, with congestion at most $\eta'$, where, for all $1\leq \ell\leq \chi$, path $P_{\ell}$ connects $a_{\ell}$ to $b_{\ell}$. We say that a path $P\in \pset$ is \emph{safe}, if $E(P) \cap F=\emptyset$, and we say that it is  \emph{unsafe} otherwise. We say that a pair $(a_\ell, b_\ell)\in I$ is \emph{safe} if $P_{\ell}$ is a safe path, and we say that it is \emph{unsafe} otherwise. From now on, it is sufficient to compute a routing for unsafe pairs in $I$. 

Consider an unsafe pair $(a_\ell, b_\ell)\in I$. Let $x_{\ell}$ 
be the vertex of $V(F)\cap V(P_{\ell})$ lying closest to $a_{\ell}$ on the path, and let $P_{\ell}^1$ be the subpath of $P_{\ell}$ from $a_{\ell}$ to $x_{\ell}$. Similarly, let $y_{\ell}$ 
be the vertex of $V(F)\cap V(P_{\ell})$ lying closest to $b_{\ell}$ on the path, and let $P_{\ell}^2$ be the subpath of $P_{\ell}$ from $y_{\ell}$ to $b_{\ell}$. Notice that both $P_{\ell}^1$ and $P_{\ell}^2$ are safe paths.
Consider the following multiset of vertex pairs:

\[I'=\set{(x_{\ell},y_{\ell})\mid P_{\ell}\in\pset, P_{\ell}\cap F\neq \emptyset}\]

In the remainder of the proof, we will compute an integral routing of the pairs in $I'$  in $G\setminus F$ via paths of length at most $O(k)\cdot d$, that cause congestion at most $O(k)\cdot \eta'$. From the following observation, such a routing will complete the proof of \Cref{thm:resilient-router}.

\begin{observation}\label{obs:basic-FTrouting}
Suppose there is an integral routing $\qset$ of the demand $I'$ in $G\setminus F$ via paths of length at most $O(k)\cdot d$ that cause congestion at most $O(k)\cdot \eta'$. Then there is a routing $\pset'$ of the demand $I$ in $G\setminus F$ via paths of length at most $O(k)\cdot d$, that cause congestion at most $O(k)\cdot\eta'$.
\end{observation}
\begin{proof}
For every unsafe pair $(a_{\ell},b_{\ell})\in I$, let $Q_{\ell}\in \qset$ be the path connecting $x_{\ell}$ to $y_{\ell}$. Let $P'_{\ell}$ be the path  obtained by concatenating the paths $P^1_{\ell},Q_{\ell}$, and $P^2_{\ell}$. Then path $P'_{\ell}$ is contained in $G\setminus F$, it connects $a_{\ell}$ to $b_{\ell}$, and its length is at most $O(k)\cdot d$.
For every safe pair $(a_{\ell},b_{\ell})\in I$, we simply set $P'_{\ell}=P_{\ell}$. It is also immediate to verify that the resulting set $\pset'=\set{P'_1,\ldots,P'_{\chi}}$ of paths causes congestion at most $O(k)\cdot \eta'$, and it routes the pairs in $I$ in graph $G\setminus F$.
\end{proof}

In the remainder of the proof, we show that there exists an integral routing of the demands in $I'$ via paths of length at most $O(k)\cdot d$, that cause congestion at most $O(k)\cdot\eta'$. We construct such a routing over the course of $z=10k$ iterations. 
For all $1\leq i\leq z$, at the beginning of iteration $i$, we are given a partition of the multiset $I'$ into two subsets: set $\sset_i$ of pairs that we refer to as \emph{$i$-safe}, and set $\uset_i$ of pairs that we refer to as \emph{$i$-unsafe}. The set $\sset_i$ of $i$-safe pairs is guaranteed to satisfy the following property:

\begin{properties}{Q}
\item there is an integral routing $\pset_i$ of the pairs in $\mathcal{S}_i$  in graph $G\setminus F$ via paths of length at most $3i\cdot d$, that cause congestion at most $2i\cdot \eta'$. \label{prop: routing of safe}
\end{properties}

For all $1\leq i\leq z$, we denote by  $V'_i=\{x_{\ell},y_{\ell}\}_{ (x_{\ell},y_{\ell})\in \mathcal{U}_i}$ the multiset of all vertices that appear in the $i$-unsafe pairs. For every vertex $v\in V'_i$, we will construct a tree-like structure (that, for brevity, we refer to as a tree), denoted by $T_i(v)$, whose root vertex is $v$, and all other vertices lie in $V(F)$. Graph $T_i(v)$ is not a tree strictly speaking, in the sense that a vertex of $V(F)$ may appear multiple times in the tree. But if each such appearence is replaced with a new copy of the vertex, then we naturally obtain a tree graph, that is denoted by $T_i(v)$. We note that, if a vertex $v$ appears in several pairs in $\uset_i$, then we construct a separate tree for each such appearance of $v$, so $v$ will serve as a root for several such trees.  
Throughout, we use a parameter $\lambda=\floor{\frac{\Delta'}{f\cdot \eta'}}$.
We ensure that, for all $1\leq i\leq z$, at the beginning of iteration $i$, the following properties hold for every vertex $v\in V'_i$:

\begin{properties}[1]{Q}
\item $T_{i}(v)$ is a $\lambda$-regular tree of depth $i-1$; and \label{prop: tree depth reg}
\item the multiset $L_i(v)$ of the leaf vertices of $T_i(v)$ has cardinality $\lambda^{i-1}$. \label{prop: number of leaves}
\end{properties}

Additionally, the collection $\tset_i=\{T_i(w)\}_{w \in V'_i}$ of trees satisfies the following properties:

\begin{properties}[3]{Q}
\item every vertex of $V(F)$  appears with multiplicity at most $f \eta'$ in the union of multisets $\{L_i(v)\}_{v\in V'_i}$; and \label{prop: tree multiplicity}

\item For every tree $T\in \tset_i$, for every edge $e=(u,w)\in E(T)$, there is a path $R(e)$  connecting $u$ to $w$ in $G\setminus F$ of length at most $d$, such that the paths in set $\rset_i=\set{R(e)\mid e\in E(T),T\in  \tset_i}$ cause congestion at most $i\eta'$.\label{prop: embedding}
\end{properties}

We now show that, if the above invariants are always obeyed, then, for some $j\leq 10k$, $\uset_j=\emptyset$ and $\sset_j=I'$ holds. 

\begin{claim}\label{cl:FTrouter-stop}
Assume that Invariants \ref{prop: routing of safe}--\ref{prop: embedding} are always obeyed. Then, for some $1\leq j\leq 10k$, $\mathcal{U}_j=\emptyset$ and $\mathcal{S}_j=I'$ must hold.
\end{claim}
\begin{proof}
Assume for contradiction that the claim is false, and so $\mathcal{U}_{10k}\neq \emptyset$  must hold. Consider any pair $(x_{\ell}, y_{\ell})\in \mathcal{U}_{10k}$.  
Recall that $\lambda=\floor{\frac{\Delta'}{f\cdot \eta'}}$, while 
$\Delta'=2n\Delta\geq  64fn^{1+1/k}\cdot \eta\geq 2fn^{1/k}\eta'$, since
$\eta'=16\eta\cdot n$. Therefore, $\lambda=\floor{\frac{\Delta'}{f\cdot \eta'}}\geq 2n^{1/k}-1$. From Invariant \ref{prop: number of leaves}, the multiset $L_{10k}(x_{\ell})$  of the leaf vertices of $T_i(x_{\ell})$ has cardinality at least $\lambda^{10k-1}> 2n^5$. However, from  Invariant \ref{prop: tree multiplicity}, every vertex of $V(F)$ appears with multiplicity at most $f \eta' \leq 2n^4$ in $L_{10k}(x_{\ell})$. Therefore, the number of distinct vertices in 
$L_{10k}(x_{\ell})$ is strictly greater than $n$, leading to a contradiction. Since multisets $\mathcal{U}_j$ and $\mathcal{S}_j$ partition the multiset $I'$, the claim follows.
\end{proof}

At the beginning of the algorithm, we set $\sset_1=\emptyset$, $\pset_1=\emptyset$, and $\uset_1=I'$. As before, we let $V_1'$ be the multiset of all vertices that participate in the pairs in $\uset_1$. For every vertex $v\in V_1'$, we let $T_1(v)$ be a tree of depth $0$, that only consists of the root vertex $v$. It is immediate to verify that 
Invariants \ref{prop: routing of safe}--\ref{prop: number of leaves} and \ref{prop: embedding} hold. In order to see that Invariant \ref{prop: tree multiplicity} holds as well, recall that the collection $\pset$ of paths that defined the original routing of the demands in $I$ had congestion at most $\eta'$. Consider any vertex $u\in V(F)$. The number of times that $u$ appears in the trees of $\tset_1=\set{T_1(v)\mid v\in V_1'}$ is bounded by the number of paths $P\in \pset$, such that $P$ contains an edge of the faulty set $F$ that is incident to $u$. Since at most $f$ edges incident to $u$ may lie in $F$, and since each such edge participates in at most $\eta'$ paths, vertex $u$ may serve as the root of at most $f\cdot\eta'$ trees in $\tset_1$.

The algorithm is executed as long as $\uset_i\neq \emptyset$ holds. We now describe the $i$th iteration. We assume that we are given a partition $(\sset_i,\uset_i)$ of the set $I'$ of pairs, the corresponding multiset $V'_i$ of vertices, and the collection $\tset_i=\set{T_i(v)\mid v\in V'_i}$ of trees, for which invariants \ref{prop: routing of safe}--\ref{prop: embedding} hold. We now provide the description of the $i$th iteration.

\paragraph{Desciption of the $i$th Iteration.}

For every pair $(x_{\ell},y_{\ell})\in \uset_i$, we construct a new demand $\dset^i_{\ell}$, as follows. Consider the multisets $L_i(x_{\ell})$ and $L_i(y_{\ell})$ of vertices (the leaf vertices of the trees $T_i(x_{\ell})$ and $T_i(y_{\ell})$, respectively). From 
Invariant \ref{prop: number of leaves}, $|L_i(x_{\ell})|=|L_i(y_{\ell})|=\lambda^{i-1}$. Let $\Pi^i_{\ell}$ be an arbitrary complete matching between the vertices of $L_i(x_{\ell})$ and the vertices of $L_i(y_{\ell})$. For every pair $(u,v)\in \Pi^i_{\ell}$, we set the demand $D^i_{\ell}(u,v)=\lambda$. Since $\lambda$ is an integer, the resulting demand $\dset^i_{\ell}=\left(\Pi^i_{\ell},\set{D^i_{\ell}(u,v)}_{(u,v)\in \Pi^i_{\ell}}\right )$ is integral. We then let $\dset^i=\left(\Pi^i,\set{D^i(u,v)}_{(u,v)\in \Pi^i}\right )$ be the demand obtained by taking the union of the demands $\dset^i_{\ell}$ for all $(x_{\ell},y_{\ell})\in \uset_i$. Next, we prove that the demand $\dset^i$ is $\Delta'$-restricted.

\begin{observation}\label{obs: restricted demand}
	Demand $\dset^i$ is $\Delta'$-restricted.
\end{observation}
\begin{proof}	
	Notice that demand $\dset^i$ is only defined over the vertices of $V(F)$. Consider any such vertex $v$. From Invariant \ref{prop: tree multiplicity}, the number of copies of $v$ that appear in the sets $L_i(w)$ for $w\in V'_i$ is at most  $f \eta'$. Each such copy appears in a single demand pair with demand $\lambda$. Therefore, the total demand on vertex $v$ in $\dset^i$ is bounded by $\lambda\cdot f\cdot \eta'\leq \Delta'$, since $\lambda=\floor{\frac{\Delta'}{f\cdot \eta'}}$. We conclude that demand $\dset^i$ is $\Delta'$-restricted.
\end{proof}

Since demand $\dset^i$ is $\Delta'$-restricted, from \Cref{obs: integral routing of integral demand2}, there is an integral routing $\qset^i$ of $\dset^i$ in $G$ via paths of length at most $d$, that cause congestion at most $\eta'$. 

Consider now a pair $(x_{\ell},y_{\ell})\in \uset_i$, and a corresponding demand $\dset^i_{\ell}$. Let $\qset^i_{\ell}\subseteq \qset_i$ be the collection of $\lambda^i$ paths routing the demand $\dset^i_{\ell}$. We say that the pair $(x_{\ell},y_{\ell})$ is \emph{good}, if at least one path $P\in \qset^i_{\ell}$ is disjoint from $E(F)$, and we say that it is bad otherwise. 

Consider a single good pair $(x_{\ell},y_{\ell})\in \uset_i$, and let $P'_{\ell}\in \qset^i_{\ell}$ be any path that is disjoint from $E(F)$. Denote the endpoints of $P'_{\ell}$ by $w$ and $w'$, where $w\in L_i(x_{\ell})$, and $w'\in L_i(y_{\ell})$, respectively. Let $\hat P_1$ be the path in the tree $T_i(x_{\ell})$, connecting $x_{\ell}$ to $w$, so the number of edges on $\hat P_1$ is at most $i$. We obtain a path $P''_{\ell}$ in graph $G\setminus F$ connecting $x_{\ell}$ to $w$, by replacing every edge $e\in E(\hat P_1)$ with its embedding path $R(e)\in \rset_i$ given in Invariant \ref{prop: embedding}. Since the length of every path $R(e)$ is at most $d$, we get that the length of $P''_{\ell}$ is a tmost $id$. We use a similar procedure to compute a path $P'''_{\ell}$ in $G\setminus F$ connecting $w'$ to $y_{\ell}$. By concatenating the paths $P''_{\ell},P'_{\ell}$ and $P'''_{\ell}$, we obtain a path $P_{\ell}$ in $G\setminus F$ connecting $x_{\ell}$ to $y_{\ell}$, whose length is at most $3id$. Consider now the collection $\pset'_i$ of paths, that contains a path $P_{\ell}$ for every good pair $(x_{\ell},y_{\ell})\in \uset_i$. From our discussion, all such paths are contained in $G\setminus F$, and each such path has length at most $3id$. Since, from Invariant \ref{prop: embedding}, the paths in $\rset_i$ cause congestion at most $\eta'\cdot i$, and since the routing 
$\qset_i$ has congestion at most $\eta'$, it is immediate to verify that the paths in $\pset'_i$ cause congestion at most $(i+1)\eta'$. We let set $\sset_{i+1}$ contain all pairs of $\sset_i$, and all good pairs that lie in $\uset_i$. We also let $\pset_{i+1}=\pset_i\cup \pset'_i$. 
From Invariant \ref{prop: routing of safe} and our discussion above, the set $\pset_{i+1}$ of paths causes congstion at most $2(i+1)\cdot \eta'$, and so Invariant \ref{prop: routing of safe} continues to hold.

We let $\uset_{i+1}\subseteq \uset_i$ be the set of all bad pairs $(x_{\ell},y_{\ell})$, and we let $V_{i+1}'$ be the multisetset of all vertices that participate in the pairs in $U_{i+1}$. For every vertex $v\in V_{i+1}'$, we extend the tree $T_i(v)$, in order to obtain the tree $T_{i+1}(v)$ with the required properties. 

Consider a vertex $v\in V_{i+1}'$, and assume that it belongs to a pair $(x_{\ell},y_{\ell})\in U_{i+1}$. Let $L=L_i(v)$ be the set of all leaf vertices of the tree $T_i(v)$, and let $u\in L$ be any such vertex. Recall that there is some pair $(u,w)\in \Pi^i_{\ell}$ in which $u$ participates, and moreover, this pair is assigned demand $\lambda$ in $\dset^i_{\ell}$. Therefore, there is a collection $\qset(u)\subseteq \qset^i_{\ell}$ of $\lambda$ paths that originate at $u$. Consider any such path $Q\in \qset(u)$, and recall that it must contain an edge of $F$. Let $u'\in V(F)\cap V(Q)$ be the vertex lying closes to $u$ on path $Q$, and let $R(u,u')$ be the subpath of $Q$ from $u$ to $u'$. Then $R(u,u')$ is disjoint from $F$, and $u'\in V(F)$. We add vertex $u'$ as a child vertex of $u$ in the tree $T_i(v)$, and we let the embedding path of the new edge $(u,u')$ be the path $R(u,u')$ defined above; note that the length of the path is at most $d$. Once we process all leaf vertices of $T_i(v)$, we obtain a new tree $T_{i+1}(v)$, that remains a $\lambda$-regular tree, whose depth is $i+1$. From our construction, its set of leaf vertices $L_{i+1}(v)$ has cardinality $\lambda\cdot |L_i(v)|=\lambda^i$. For every edge $e$ of the tree, if it lied in $T_i(v)$, then its embedding path $R(e)\in \rset_i$ is as defined in Invariant \ref{prop: embedding}, and otherwise, its embedding path is a subpath of one of the routing paths in $\qset^i$ as defined above. Since the paths in $\rset_i$ caused congestion at most $i\eta'$, while the paths in $\qset$ cause congestion at most $\eta'$, it is easy to verify that the resulting set $\rset_{i+1}=\set{R(e)\mid e\in E(T_{i+1}(v)),v\in V_{i+1}'}$ cause congestion at most $(i+1)\cdot \eta'$.

So far we have shown that 
Invariants \ref{prop: routing of safe}--\ref{prop: number of leaves} and \ref{prop: embedding} hold. It now only remains to establish Invariant \ref{prop: tree multiplicity}.
Indeed, recall that the collection $\qset^i$ of paths causes congestion at most $\eta'$. Consider any vertex $u\in V(F)$. The number of times that $u$ appears in the sets $L_{i+1}(v)$ of vertices for $v\in V'_{i+1}$ is bounded by the number of paths $Q\in \qset$, such that $Q$ contains an edge of the faulty set $F$ that is incident to $u$. Since at most $f$ edges incident to $u$ may lie in $F$, and since each such edge participates in at most $\eta'$ paths, vertex $u$ may appear at  $f\cdot\eta'$ times in the sets $\set{L_{i+1}(v)}_{v\in V'_{i+1}}$.

Note that, when the algorithm terminates, which must happen after $i\leq 10k$ iterations, all pairs of $I'$ lie in  the set $\sset_i$ of $i$-safe pairs. From Invariant  \ref{prop: routing of safe}, we conclude that there is an integral routing of the pairs in $I'$  in graph $G\setminus F$ via paths of length at most $O(k)\cdot d$, that cause congestion at most $O(k)\cdot \eta'$, as required.
\end{proofof}

\begin{figure}[h!]
\begin{center}
\includegraphics[scale=0.35]{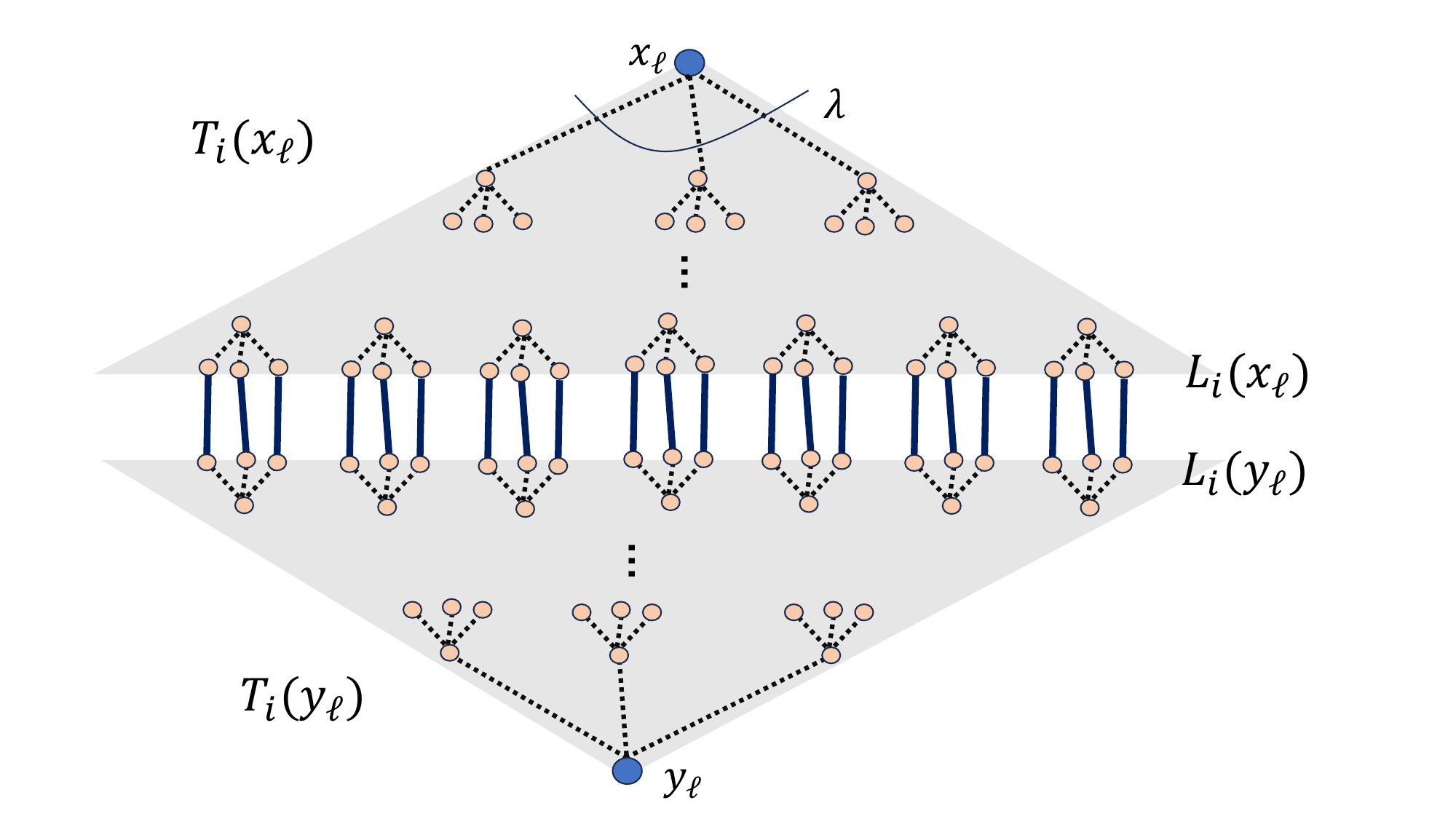}
\caption{\sf An illustration for the proof of Lemma \ref{thm:resilient-router}. Shown is an $i$-unsafe pair $(x_{\ell},y_{\ell})$ with their corresponding trees $T_{i}(x_{\ell})$ and $T_{i}(y_{\ell})$. The routing instance for iteration $i$ is defined by sending $\lambda$ flow units between matched leaf vertices in $L_{i}(x_{\ell})$ and $L_{i}(y_{\ell})$. The solid edges correspond to $\lambda$ units of demand between the matched pair.
}
\end{center}
\end{figure}

\subsubsection{From Router Decomposition to Fault Tolerant Spanners}\label{sec:graph-sparse}


\paragraph{Fault-Tolerant Spanners.} 
Recall that for a graph $G$, a stretch parameter $k \geq 1$, and an integer fault parameter $f \geq 1$, an $f$-FT $k$-spanner of $G$ is a subgraph $H \subseteq G$ satisfying that, for every subset $F \subseteq E(G)$ of at most $f$ edges, $H \setminus F$ is a $k$-spanner for $G \setminus F$.

Very recently, Bodwin, Haeupler and Parter \cite{BodwinHP24} introduced a considerably stronger notion of fault-tolerance, where only the degree of the faulty edge set, rather than its cardinality, is restricted.
For a subset $F \subseteq E(G)$ of faulty edges, its \emph{faulty-degree} $\deg(F)$ is the maximum, over all vertices $v$, of the number of edges of $F$ incident to $v$. For example, if $F$ is a matching, then $\deg(F)=1$, while $|F|$ may be as large as $n/2$. This naturally leads to the definition of faulty-degree spanners.

\begin{definition}[Faulty-Degree Spanners]\cite{BodwinHP24}
Let $G$ be a graph and let $H\subseteq G$ be a subgraph of $G$ with $V(H)=V(G)$. For parameters $f$ and $k>0$, we say that $H$ is an $f$-faulty-degree (FD) $k$-spanner for $G$ if, for every subset $F\subseteq E(G)$ of edges with $\deg(F)\leq f$, for every pair $u, v$ of vertices of $G$, $\dist_{H \setminus F}(u,v)\leq k \cdot \dist_{G \setminus F}(u,v)$ holds.
\end{definition}

In the following lemma, we show that, if $H'$ is a graph obtained from a router decomposition of the graph $G$ as before, then it is an $f$-FD $k'$-spanner, for appropriately chosen parameters $f$ and $k'$.

\begin{observation}[From Router Decomposition to $f$-FD Spanners]\label{obs:router-decomp-to-FDS}
	Let $G$ be an $n$-vertex graph, and let $k'>0$ and $f\in \set{1,\ldots,n}$ be parameters. Let $\mathcal{R}=\left (\mathcal{C}, \{C'\}_{C \in \mathcal{C}}\right )$ be a router decomposition of $G$ with parameters $\tilde d$, $1\leq\tilde \eta\leq n^2$, $\Delta^*\geq 32\cdot f\cdot \tilde \eta\cdot n^{1/k'}$,  and $\rho$. 
	 Consider the corresponding graph $H'$ whose   vertex set is $V(G)$, and edge set is $\edel\cup \left(\bigcup_{C\in \cset}E(C')\right )$, where $\edel=E(G)\setminus \left(\bigcup_{C\in \cset}E(C)\right )$. Then $H'$ is an $f$-FD $O(\tilde d\cdot k')$-spanner.
\end{observation}

\begin{proof}
	Consider an $f$-BD faulty set $F \subseteq E(G)$ of edges, and an edge $e=(u,v)\in E(G)\setminus F$. If $e\in \edel$, then $e$ is included in $H'$, and so $\dist_{H'\setminus F}(u,v)=1$. Otherwise, there must be some cluster $C\in \cset$, with $u,v\in C$. Consider the corresponding subgraph $C'\subseteq C$, and recall that $V(C)=V(C')$, and $C'$ is a $(\vDelta^*,\tilde d,\tilde \eta)$-router for the set $V(C)$ of supported vertices. Then from \Cref{thm:resilient-router}, graph $C'\setminus F$ remains a $(\vDelta^*, O(k \cdot \tilde d),O(k \cdot \tilde \eta))$-router. In particular, the diameter of $C' \setminus F$ is bounded by $O(k' \tilde d)$ which implies that 
$\dist_{C' \setminus F}(u,v)=O(k' \tilde d)$. Since $E(C') \subseteq H'$, the observation follows. 
\end{proof}

\paragraph{Fault-Tolerant Low-Congestion Spanners.} Finally, we show the strongest property of subgraph $H$ maintained by the algorithm from Theorem \ref{thm: router decomposition main}. We will use the fact that routers retain their routing properties (with a small loss), to show that $H$ is a fault-tolerant low-congestion spanner even against bounded degree faults. We start by formally defining the latter  notion.  

\begin{definition}[Fault-Tolerant Low-Congestion Spanners]
Let $G$ be a graph, let $H\subseteq G$ be a subgraph of $G$, and let $f,d$ and $\eta$ be parameters. We say that $H$ is an $f$-FT $(d,\eta)$-low congestion spanner if, for every subset $F \subseteq E(G)$ of edges with $|F|\leq f$, graph $G \setminus F$ embeds into $H \setminus F$ via paths of length at most $d$, with congestion at most $\eta$. Similarly, $H$ is an $f$-FD $(d,\eta)$-low congestion spanner if the above holds for every subset $F\subseteq E(G)$ of edges with $\deg(F)\leq f$. 
\end{definition}

In the following observation we show that the graph $H'$ obtained from a router decomposition of $G$ with an appropriately chosen parameter $f$ must be an $f$-FD low-congestion spanner.

\begin{observation}[From Router Decomposition to $f$-FD Low-Congestion Spanners]\label{obs:router-to-FDLCS}
Let $G$ be an $n$-vertex graph, and let $k'>0$ and $f\in \set{1,\ldots,n}$ be parameters. Let $\mathcal{R}=\left (\mathcal{C}, \{C'\}_{C \in \mathcal{C}}\right )$ be a router decomposition of $G$ with parameters $\tilde d$, $1\leq \tilde \eta\leq n^2$, $\Delta^*\geq 32\cdot f\cdot \tilde \eta\cdot n^{1/k'}$,  and $\rho$. 
Consider the corresponding graph $H'$ whose   vertex set is $V(G)$, and edge set is $\edel\cup \left(\bigcup_{C\in \cset}E(C')\right )$, where $\edel=E(G)\setminus \left(\bigcup_{C\in \cset}E(C)\right )$. Then $H'$ is an $f$-FD $(d,\eta)$-low congestion spanner for $d=O(k' \cdot \tilde d)$ and $\eta=\max\set{O\left(\frac{k'\cdot \tilde \eta \cdot \Delta_{\max}(G)}{\Delta^*}\right),16\log n}$, where $\Delta_{\max}(G)$ is maximum vertex degree in $G$. 
\end{observation}
\begin{proof}
Let $F \subseteq E(G)$ be a set of edges with $\deg(F)\leq f$. Our goal is to show that $H' \setminus F$ is a $(d,\eta)$-low congestion spanner for $G \setminus F$. In other words, we need to show that $G\setminus F$ can be embedded into $H' \setminus F$ via paths of length at most $d$ and congestion at most $\eta$. Since $\edel \subseteq E(H')$, and the clusters of $\mathcal{C}$ are edge-disjoint, it is sufficient to show that, for every cluster $C\in \cset$, there is an embedding $\qset_C$ of $C \setminus F$ into $C' \setminus F$ via paths of length at most $d$, that cause congestion at most $\eta$.

Since, for every cluster  $C \in \mathcal{C}$, the corresponding graph $C'\subseteq C$ is a $(\vDelta^*,\tilde \eta,\tilde d)$ router for the set $V(C)=V(C')$ of supported vertices, and $\Delta^*\geq 32f n^{1/k'} \cdot \tilde \eta$ holds, from \Cref{thm:resilient-router} we get that that $C' \setminus F$ is a  $(\Delta^*, O(k' \cdot \tilde d), O(k' \cdot \tilde \eta))$-router. In particular, $\left(\cset,\set{C'\setminus F}_{C\in \cset}\right )$ remains a router decomposition of $G$ with parameters $\Delta^*,O(k')\cdot \tilde d$, $O(k')\cdot \tilde \eta$, and $\rho$. From Lemma \ref{lem:router-embed-subgraph}, there is an embedding $\qset_C$ of $C$ into $C'$ via paths of length at most $O(k' \cdot \tilde d)$, that cause congetion at most 
$\max\set{O\left(\frac{k'\cdot \tilde \eta \cdot \Delta_{\max}(G)}{\Delta^*}\right),16\log n}$.
\end{proof}

Finally, we obtain the following analogue of Corollaries \ref{cor: spanner} and 
\ref{cor: spanner2} for $f$-FD low-congestion spanners.

\begin{corollary}\label{cor: spanner3}
	There is a deterministic algorithm, whose input is a graph $G$ with $n$ vertices and no edges, that undergoes an online sequence of at most $n^2$ edge deletions and insertions, such that $G$ remains a simple graph throughout the update sequence. Each edge inserted into $G$ has an integral length $\ell(e)\in \set{1,\ldots,L}$. Additionally, the algorithm is given parameters $512\leq k\leq (\log n)^{1/49}$, $\frac{1}{k}\leq \delta\leq \frac{1}{400}$, $f\in\set{1,\ldots,n}$, and $\Delta\geq f\cdot n^{c/k}$, for a large enough constant $c$, such that $k,\Delta$ and  $\frac 1{\delta}$ are integers. The algorithm maintains a subgraph $H\subseteq G$ with $V(H)=V(G)$ and $|E(H)|\leq O\left(n^{1+O(1/k)}\cdot \Delta\cdot \log L\right )$, such that $H$ is an a $f$-FD $(d,\eta)$-low congestion spanner for $G$, for $d\leq \poly(k)\cdot 2^{O(1/\delta^6)}$ and $\eta\leq n^{O(1/k)} \cdot\max\set{ \frac{ \Delta_{\max}(G)}{\Delta},1}$,  where $\Delta_{\max}(G)$ is the maximum degree in $G$. The worst-case update time of the algorithm is $n^{O(\delta)}$ per operation, and the number of edge insertions and deletions that graph $H$ undergoes after each update to $G$ is bounded by $n^{O(1/k)}$.
\end{corollary}

In order to prove the corollary, we use the algorithm from \Cref{cor: spanner2}. By using \Cref{obs:router-to-FDLCS} with parameter $k'$ replaced with $k$, it is easy to verify that the graph $H$ that the algorithm maintains has all required properties.

\paragraph{Dynamic Connectivity Certificates.}
The notion of connectivity certificates was introduced by  \cite{NagamochiI92}.
For a graph $G=(V,E)$ and integer $f \in \{1,\ldots, n\}$, an $f$-connectivity certificate is a subgraph $H \subseteq G$ with the following property: for every pair $u,v \in V$ of vertices, for every subset $F \subseteq E$ of at most $f$ edges, $u$ and $v$ are connected in $H \setminus F$ if and only if they are connected in $G \setminus F$. 
The recent work of \cite{BodwinHP24} introduced a stronger notion of $f$-FD connectivity certificate: a subgraph $H\subseteq G$ that satisfies the above requirement for any subset $F \subseteq E$ of edges with $\deg(F)\leq f$. It is well known that any $n$-vertex graph admits an $f$-connectivity certificate containing $O(fn)$ edges, and such certificates can be computed in linear time in the static setting. 
Using expander decompositions and expander routing, \cite{BodwinHP24} presented an algorithm that constructs an $f$-FD connectivity certificates with $\widetilde{O}(f n)$ edges, which is nearly tight, in polynomial time. 

Clearly, if a graph $H$ is an $f$-FD $(d,\eta)$-low congestion spanner for $G$, for any parameters $d$ and $\eta$, then $H$ is an $f$-FD connectivity certificate and an $f$-connectivity certificate for $G$ as well. Therefore, by using the algorithm from \Cref{cor: spanner3} with a parameter $\Delta=f\cdot n^{O(1/k)}$ and $\delta=\frac{1}{k}$, we get that the graph $H$ that the algorithm maintains has $|E(H)|\leq 
O\left(f\cdot n^{1+O(1/k)}\right )$, and it is an $f$-FD connectivity certificate for $G$. The algorithm has $n^{O(1/k)}$ worst-case update time and $n^{O(1/k)}$ recourse. Note that both of the latter quantities are independent of $f$.  To the best of our knowledge, all previous dynamic algorithms for (the classical notion of) $f$-connectivity certificates  had a worst-case update time with a polynomial dependency on $f$. Therefore, for high values of the parameter $f$, we obtain a faster algorithm.

\newpage
\appendix

\section{Proof of \Cref{claim: well connected to routers}}
\label{subsec: appx: proof well connected to router1}

Let $G$ be an $n$-vertex graph, and assume that it is  $(d,\eta)$-well-connected with respect to a set $S\subseteq V(G)$ of supported vertices. Let $1\leq k\leq \log n$ be a parameter. Our goal is to prove that $G$ is a $(1,d',\eta')$-router with respect to $S$.

Consider any demand $\dset=\left(\Pi,\set{D(a,b)}_{(a,b)\in \Pi}\right )$. We say that $\dset$ is a \emph{matching} demand, if $\Pi$ is a matching, and, for all $(a,b)\in \Pi$, $D(a,b)=1$. Notice that any matching demand can be specified by just the set $\Pi$ of pairs of its vertices, so for a matching demand $\dset$, abusing the notation, we may denote $\dset=\Pi$.
We show that it is enough to prove that any matching demand can be routed via paths of length at most $d'$ with congestion at most $\frac{\eta'}{16\log n}$, in the following simple claim.

\begin{claim}\label{claim: integral enough}
	Let $G$ be an $n$-vertex graph with a set $S\subseteq V(G)$ of supported vertices, and assume that, for every matching demand $\dset$ defined over $S$, there is a routing of $\dset$ in $G$ via paths of length at most $\hat d$ and congestion at most $\hat \eta$. Then $G$ is a $(\overline 1,\hat d,16\hat \eta\cdot \log n)$-router with respect to $S$.
\end{claim}
\begin{proof}
Consider any $1$-restricted demand $\dset=\left(\Pi,\set{D(a,b)}_{(a,b)\in \Pi}\right )$ over the vertices of $S$. Let $q=\ceil{\log n}+1$.
For all $1\leq i<q$, we define a demand $\dset_i=\left(\Pi_i,\set{D'(a,b)}_{(a,b)\in \Pi_i}\right )$ as follows: set $\Pi_i$ contains all pairs $(a,b)$ with $\frac{1}{2^i}\leq D(a,b)<\frac{1}{2^{i-1}}$, and $D'(a,b)=\frac{1}{2^i}$. Additionally, we define a demand $\dset_q=\left(\Pi_q,\set{D'(a,b)}_{(a,b)\in \Pi_q}\right )$, where $\Pi_q$ contains all pairs $(a,b)$ with $D(a,b)<\frac{1}{n}$, and we set $D'(a,b)=\frac{1}{n}$. It is easy to verify that, for all $1\leq i\leq q$, demand $\dset_i$ is $1$-restricted. Moreover, it is now enough to prove that each such demand $\dset_i$ can be routed in $G$ via paths of length at most $\hat d$ and congestion at most $4\hat \eta$.

From now on we focus on any such demand $\dset_i$, that we denote by $\dset'=\left(\Pi',\set{D'(a,b)}_{(a,b)\in \Pi'}\right )$ for convenience. Recall that there is an integer $z$, so that $D'(a,b)=\frac{1}{z}$ for all $(a,b)\in \Pi'$, and that demand $\dset'$ is $1$-restricted. Therefore, every vertex of $S$ participates in at most $z$ pairs in $\Pi'$. We can then compute a collection $\Pi'_1,\ldots,\Pi'_{2z+1}$ of pairs of vertices that partition $\Pi'$, such that, for all $1\leq j\leq 2z+1$, set $\Pi'_j$ is a matching, using a standard greedy algorithm.

From our assumption, for all $1\leq j\leq 2z+1$, there is a flow $f_j$ that routes the matching demand between the pairs of vertices in $\Pi'_j$, where, for each pair $(a,b)\in \Pi'_j$ of vertices, one flow unit is routed from $a$ to $b$ in $f_j$. Moreover, all flow-paths in $f_j$ have length at most $\hat d$, and the flow causes congestion at most $\hat \eta$. By scaling each such flow $f_j$ down by factor $z$ (so it causes congestion at most $\frac{\hat \eta}{z}$), and taking the union of all such flows, for all $1\leq j\leq 2z+1$, we obtain a flow $f$ that routes the demand $\dset'$ via paths of length at most $\hat d$, with congestion at most $4\hat \eta$.
\end{proof}

From now on, it is enough to prove that, for every matching demand $\dset$ defined over $S$, there is a routing of $\dset$ in $G$ via paths of length at most $d'$, with congestion at most $\frac{\eta'}{16\log n}$.

Consider any matching demand $\dset=\Pi$. We say that $\dset$ is \emph{half-routable}, if there is a subset $\Pi'\subseteq\Pi$ with $|\Pi'|\geq \frac{|\Pi|}{2}$, such that there is a routing of $\Pi'$ via paths of length at most $d'$, and congestion at most $\frac{\eta'}{32\log^2 n}$. It is easy to see that, if every matching demand $\dset$ defined over the set $S$ of vertices is half-routable in $G$, then $G$ is a $(\overline 1,d',\eta')$-router with respect to $S$, as we show in the next observation.

\begin{observation}\label{obs: half routable}
	Suppose that every matching demand $\dset$ defined over the vertices of $S$ is half-routable in $G$. Then $G$ is a $(\overline 1,d',\eta')$-router with respect to $S$.
\end{observation}
	\begin{proof}
		We assume that every matching demand defined over the vertices of $S$ is half-routable in $G$. From \Cref{claim: integral enough}, it is enough to prove that, for every matching demand $\dset$ defined over $S$, there is a routing of $\dset$ in $G$ via paths of length at most $d'$, that cause congestion at most $\frac{\eta'}{16\log n}$.
		
		Let $\dset=\Pi$ be any matching demand. We perform at most $\ceil{\log n}$ iterations, where in the $i$th iteration we route a subset $\Pi_i\subseteq \Pi$ of the demand pairs, which are then deleted from $\Pi$.
		
		Specifically, the input to the $i$th iteration is the current set $\Pi$ of pairs. Since any matching demand is half-routable, there is a subset $\Pi_i\subseteq \Pi$, containing at least $\frac{|\Pi|}{2}$ pairs, and a routing $f_i$ of $\Pi_i$ in $G$ via paths of length at most $d'$, that causes congestion at most $\frac{\eta'}{32\log^2n}$. We then delete the pairs of $\Pi_i$ from $\Pi$, and continue to the next iteration. It is immediate to verify that, after at most $\ceil{\log n}$ such iterations, set $\Pi$ becomes empty, and the union of the resulting flows $f_i$ provides a routing of $\Pi$, via paths of length at most $d'$, and congestion at most $\ceil{\log n}\cdot \frac{\eta'}{32\log^2n}\leq \frac{\eta'}{16\log n}$.
	\end{proof}

In order to complete the proof of \Cref{claim: well connected to routers}, it is now enough to prove that every matching demand $\dset$ defined over the set $S$ of supported vertices is half-routable. Assume otherwise, and let $\dset=\Pi$ be a matching demand that is not half-routable. Consider the following iterative process for constructing a collection $\pset$ of paths routing some pairs in $\Pi$. We start from $\pset=\emptyset$, and then iterate, as long as there is a pair $(a,b)\in \Pi$, for which there is a path of length at most $d'$ connecting $a$ to $b$ in $G$. In each iteration, we select any such pair $(a,b)\in \Pi$, and any path $P$ of length at most $d'$  connecting $a$ to $b$ in $G$, add $P$ to $\pset$, delete the pair $(a,b)$ from $\Pi$, and delete from $G$ every edge that has been used in at least $\frac{\eta'}{32\log^2 n}$ of the paths in $\pset$.

Let $\Pi'$ be the collection of paths that remain in $\Pi$ at the end of this process. Let $E'$ be the set of edges deleted from $G$, and let $G'=G\setminus E'$. Recall that an edge $e$ lies in $E'$ if and only if it has been used by  at least $\frac{\eta'}{32\log^2 n}$ of the paths in $\pset$. Therefore, $|E'|\leq \frac{|\Pi|\cdot d'\cdot 32\log^2 n}{\eta'}$. Note also that $\pset$ is a routing of the pairs in $\Pi\setminus \Pi'$ via paths of length at most $d'$, that cause congestion at most $\frac{\eta'}{32\log^2 n}$. Since demand $\dset$ is not half-routable, we get that $|\Pi'|\geq\frac{|\Pi|}{2}$, and, in particular, $|E'|\leq \frac{64|\Pi'|\cdot d'\cdot \log^2 n}{\eta'}$. Lastly, for every pair $(a,b)\in \Pi'$, $\dist_{G'}(a,b)\geq d'$.

We use the following claim that easily follows from Lemma 3.10 in \cite{chuzhoy2023distanced}, by setting the parameter $\alpha$ to $1/2$ and parameter $\Delta$ to $64k$:

\begin{lemma}[Lemma 3.10 from \cite{chuzhoy2023distanced}]\label{lem: find separation of terminals}
	There is an efficient algorithm, whose input is a graph $G$, a set $T\subseteq V(G)$ of $z$ vertices called terminals, a distance parameter $\tilde d$, and another parameter $k>0$. The algorithm either computes a terminal  $t\in T$ with $|B_G(t,64k \tilde d)\cap T|> \frac{z}{2}$, or it computes two subsets $T_1,T_2$ of terminals, with $|T_1|=|T_2|$, such that $|T_1|\geq \frac{z^{1-1/k}}{3}$, and for every pair $t\in T_1$, $t'\in T_2$ of terminals, $\dist_G(t,t')\geq \tilde d$. 
\end{lemma}

We apply the lemma to graph $G'$, with the set $T$ of terminals containing all vertices that participate in the pairs of $\Pi'$, so that $z=|T|=2|\Pi'|$, and parameter $\tilde d$ replaced by $2d$, and parameter $k$ that remains unchanged. Recall that $d'=512kd$. Note that it is impossible that the algorithm from \Cref{lem: find separation of terminals} returns a terminal $t\in T$ with $|B_{G'}(t,64k\tilde d)|>\frac{z}{2}$, since in this case it would mean that there is a pair $(a,b)\in \Pi'$ of vertices, with both $a,b\in B_{G'}(t,64k\tilde d)$; in other words, $\dist_{G'}(a,b)\leq 256kd<d'$, which is impossible. Therefore, the algorithm from \Cref{lem: find separation of terminals} must return two equal-cardinality sets $A,B$ of vertices, with $|A|,|B|\geq \frac{z^{1-1/k}}{3}\geq \frac{|\Pi'|}{3n^{1/k}}$, such that $\dist_{G'}(A,B)\geq 2d$. 

Since graph $G$ is $(d,\eta)$-well-connected with respect to $S$, there is a collection $\qset$ of $|A|$ paths in $G$, connecting vertices of $A$ to vertices of $B$, so that the length of every path is at most $d$, and the paths cause congestion at most $\eta$. Since $\dist_{G'}(A,B)\ge 2d$, every path in $\qset$ must use an edge of $E'$. Therefore, $|E'|\geq\frac{|\qset|}{\eta}\geq \frac{|A|}{\eta}\geq \frac{|\Pi'|}{3\eta n^{1/k}}$ must hold.
But, as we have established above,
$|E'|\leq \frac{64|\Pi'|\cdot d'\cdot \log^2 n}{\eta'}$. Since $\eta'=\left(2^{17}\cdot kdn^{1/k}\log^2n\right )\cdot \eta=\left(2^{8}\cdot d'n^{1/k}\log^2n\right )\cdot \eta$, we get that $|E'|\leq \frac{|\Pi'|}{4\eta n^{1/k}}$, a contradiction. 

Therefore, every matching demand defined over the set $S$ of supported vertices must be routable in $G$, and so $G$ is a $(\overline 1,d',\eta')$-router with respect to $S$.

\section{Proof of \Cref{cor: one level router decomposition}}
\label{sec:appx:split one level}

	Consider the input graph $G$. Since $|E(G)|\leq \Delta\cdot n^{1+15/k}$, the average vertex degree in $G$ is at most $2\Delta \cdot n^{15/k}$. 	
	Note that, since $k\leq (\log n)^{1/49}$ holds, we get that $k^{45}\leq \frac{\log n}{k^4}$, and so:
	
	\begin{equation}\label{eq: k and n3}
	2^{k^{45}}\leq n^{1/k^4}.
	\end{equation}

	We use the algorithm from \Cref{claim: uniformize degree} to compute, in time $O(|E(G)|)\leq O\left (n^{1+O(1/k)}\cdot \Delta \right )\leq O(n^{1+O(\delta)}\cdot \Delta)$, a vertex-split graph $G'$ for $G$, such that the degrees of all vertices in $G'$ are at most $4n^{15/k}\cdot \Delta$, and $|V(G')|\leq 2|V(G)|$. For convenience, we denote $\hn=|V(G')|$, so $n\leq \hn\leq 2n$.
 Recall that every edge $e=(u,v)\in E(G)$ corresponds to a unique edge $e'=(u',v')$ of $G'$, where $u'$ is a copy of $u$, and $v'$ is a copy of $v$; we do not distinguish between these edges. We apply the algorithm from \Cref{thm: one level router decomposition} to the resulting graph $G'$, with the same parameters $k,\delta,\Delta^*$, and parameter $\ceil{\Delta}$ replacing $\Delta$. Notice that $512\leq k\leq (\log n)^{1/49}\leq (\log \hn)^{1/49}$, $\frac{1}k\leq \delta<\frac{1}{400}$, and $k,\Delta^*, \ceil{\Delta}$ and $\frac 1{\delta}$ are integers, as required.
Moreover, $\hn^{18/k}\leq (2n)^{18/k}\leq n^{19/k}\leq \Delta^*$ must hold, and, since $\hn^{8/k^2}\leq (2n)^{8/k^2}\leq n^{9/k^2}$, we get that $\Delta^*\leq \frac{\Delta}{n^{9/k^2}}\leq \frac{\ceil{\Delta}}{\hat n^{8/k^2}}$ holds. Therefore, altogether, $\hn^{18/k}\leq \Delta^*\leq \frac{\ceil{\Delta}}{\hn^{8/k^2}}\leq \hn$ must hold. The maximum vertex degree in $G'$ is bounded by  $4n^{15/k}\cdot \Delta\leq 4\hn^{15/k}\cdot \Delta\leq \hn^{16/k}\cdot \ceil{\Delta}$.
Therefore, graph $G'$, with parameters $\hat n$,  $k,\delta,\Delta^*$ and $\ceil{\Delta}$ is a valid input to the algorithm from \Cref{thm: one level router decomposition}.
Consider the router decomposition $\left (\cset,\set{C'}_{C\in \cset}\right )$ of graph $G'$ with parameters $\Delta^*,\tilde d, \tilde \eta$ and $\rho$, that the algorithm maintains, where 
$\tilde d= k^{19}\cdot 2^{O(1/\delta^6)}$, 
$\tilde \eta= \hn^{17/k}\leq (2n)^{17/k}\leq n^{19/k}$, and $\rho=\hat n^{20/k}\leq (2n)^{20/k}\leq \frac{n^{21/k}}{2}$.

Consider a cluster $C\in \cset$, and recall that it must be $(\vdeg_C,\tilde d,\tilde \eta)$-router with respect to the set $V(C)$ of supported vertices. Let $\hat C\subseteq G$ be the graph obtained by including all the edges of $C$, and all vertices whose copies lie in $C$. It is then easy to verify that $C$ is a vertex-split graph of $\hat C$. From \Cref{obs: split graph router}, graph $\hat C$ is a $(\vdeg_{\hat C},\tilde d,\tilde \eta)$-router with respect to the set $V(\hat C)$ of supported vertices. Let $\hat \cset=\set{\hat C\mid C\in \cset}$. It is immediate to verify that all clusters in $\hat \cset$ are edge disjoint, and that every edge $e\in E(G)\setminus \edel$ lies in some cluster of $\hat \cset$. It is also immediate to verify that $\sum_{\hat C\in \hat \cset}|V(\hat C)|\leq \sum_{C\in \cset}|V(C)|\leq \rho\cdot \hat n\leq 2\rho n$.

Consider again a cluster $C\in \cset$, and the corresponding subgraph $C'\subseteq C$. We can again define a subgraph $\hat C'\subseteq \hat C$, by including in it all edges of $C'$, and all vertices of $G$ whose copies lie in $C'$. Since graph $C'$ is a $(\vDelta^*,\tilde d,\tilde \eta)$-router with respect to the set $V(C')$ of supported vertices, from \Cref{obs: split graph router fixed delta}, $\hat C'$ is a $(\vDelta^*,\tilde d,\tilde \eta)$-router with respect to the set $V(\hat C')$ of supported vertices. Therefore, throughout the algorithm $\left (\hat \cset,\set{\hat C'}_{\hat C\in \hat \cset}\right )$ is a router decomposition of $G$, for parameters $\Delta^*,\tilde d,\tilde \eta$ and $2\rho\leq n^{21/k}$. 

Once  the algorithm from \Cref{thm: one level router decomposition} initializes the router decomposition $(\cset,\set{C'}_{C\in \cset})$ of $G'$, we can initialize the router decomposition $\left (\hat \cset,\set{\hat C'}_{\hat C\in \hat \cset}\right )$ of $G$ within the same asymptotic running time. Recall that the algorithm guaranees that initially, $|\edel|\leq \hn\cdot \ceil{\Delta}\cdot \hn^{2/k^2}\leq n^{1+4/k^2}\cdot \Delta$. 

For all $1\leq i\le k$, once the algorithm from  \Cref{thm: one level router decomposition} processes the batch $\pi_i$ of edge deletions from $G'$ and updates the clustering $\cset$ and the subgraphs $C'\subseteq C$ for $C\in \cset$ accordingly, we update the clustering $\hat \cset$ and the graphs $\hat C'\subseteq \hat C$ for $\hat C\in \hat \cset$ similarly: whenever an edge is deleted from a cluster $C\in \cset$, we delete it from $\hat C$; whenever an edge is deleted from or inserted into a graph $C'$ for some cluster $C\in \cset$, we insert or delete the same edge from the corresponding graph $\hat C'$; and whenever a vertex $v$ is deleted from some cluster $C$, if $v$ is a copy of some vertex $v'\in V(G)$, and no other copies of $v'$ remain in $C$, then we delete $v'$ from $\hat C$ and $\hat C'$ as well. It is easy to verify that the time required to update the router decomposition  $\left (\hat \cset,\set{\hat C'}_{\hat C\in \hat \cset}\right )$ of $G$ is asymptotically bounded by the time that the algorithm from \Cref{thm: one level router decomposition} spends on processing the batch $\pi_i$ of updates. Lastly, recall that, following the processing of batch $\pi_i$, the number of edges that may be added to $\edel$ is bounded by  $|\pi_i|\cdot \hn^{13/k^2}\leq |\pi_i|\cdot (2n)^{13/k^2}\leq |\pi_i|\cdot n^{15/k^2}$, as required, and the number of edge insertions and deletions from the graphs in $\set{\hat C'\mid \hat C\in \hat \cset}$ while batch $\pi_i$ is processed is bounded by the number of edge insertions and deletions from the graphs in $\set{ C'\mid  C\in  \cset}$, which, in turn, is bounded by $|\pi_i|\cdot n^{O(1/k^2)}$.

\section{Proof of \Cref{claim: scattered or large ball}}
\label{sec: appx: scattered or large ball} 
	We say that a vertex $v\in V(G)$ is scattered if $|B_G(v,d)|<n^{1-\eps}$. We start with $G'=G$, and then delete all isolated vertices from $G'$. Over the course of the algorithm, we may delete additional vertices from $G'$.
	
	The algorithm consists of a number of iterations. In each iteration, we may delete some vertices from $G'$ (which must be scattered vertices), and we may mark some additional vertices of $G'$ as scattered. Throughout, we denote by $U\subseteq V(G')$ the set of vertices that were not marked as scattered yet. We maintain the following invariants:
	
	\begin{properties}{I}
		\item If a vertex $v\in V(G)$ is not scattered, then $v\in U$ holds at all times; and\label{inv: nonscattered in U}
		\item For every vertex $u\in U$, $B_G(u,d)\subseteq G'$. \label{inv: nonscattered ball kept}
	\end{properties}

The algorithm continues as long as $U\neq \emptyset$ and $|V(G')|\geq n^{1-\eps}$ hold. If either of these conditions does not hold, then we are guaranteed that $G$ is $(d,\eps)$-scattered. Indeed, if $G$ contains a vertex $v$ with $|B_G(v,d)|\geq n^{1-\eps}$, then, from Invariant \ref{inv: nonscattered in U}, $v\in U$, and from Invariant \ref{inv: nonscattered ball kept} $B_G(u,d)\subseteq G'$ must hold, so $U\neq \emptyset$ and $|V(G')|\geq n^{1-\eps}$ must hold. Therefore, if either $U\neq \emptyset$ or $|V(G')|\geq n^{1-\eps}$ hold, we terminate the algorithm and report that $G$ is $(d,\eps)$-scattered.

At the beginning of the algorithm, $U=V(G')$, and it is easy to see that both invariants hold. We now describe a single iteration. We assume that both invariants hold at the beginning of the iteration, and we will ensure that they hold at the end of the iteration.

At the beginning of the iteration, we delete from $G'$ all isolated vertices. It is easy to verify that both invariants continue to hold.
Next, we select an arbitrary vertex $u\in U$. We denote $L_0=B_{G'}(u,2d)$, and, for all $j> 0$, we denote by $L_j=B_{G'}(u,2(j+1)d)\setminus B_{G'}(u,2jd)$. We also denote by $E_j$ the set of all edges with both endpoints in $L_0\cup\cdots\cup L_j$.

We perform a BFS in $G'$, starting from vertex $u$, until we  encounter the first index $j$, for which $|E_{j+1}|\leq m^{\eps}\cdot |E_j|$. It is easy to verify that $1\leq j\leq 2\floor{1/\eps}$ must hold, since otherwise, for all $1\leq j\leq 2\floor{1/\eps}$, $|E_j|>m^{\eps}\cdot |E_{j-1}|\geq m^{(j-1)\cdot \eps}$, and so $|E_{2\floor{1/\eps}}|>m$, which is impossible.

Let $S=L_0\cup\cdots \cup L_{j}$, let $S''=L_0\cup\cdots\cup L_{j+1}$, and let $S'=B_{G'}(u,2(j+1)d+d)$. In other words, $S'$ contains all vertices of $S$, and the subsequent $d$ layers of the BFS. We now consider two cases. The first case happens if $|S''|\geq n^{1-\eps}$. In this case, we say the current iteration is good. Notice that $S''\subseteq B_G(u,4d/\eps)$. We then return vertex $u$ as the outcome of the algorithm. The running time of the current iteration is bounded by $O(m)$ in this case.

Consider now the second case, where $|S''|<n^{1-\eps}$. In this case, we say that the iteration is bad. Notice that, from our invariants, we are guaranteed that, for every vertex $v\in S'\cap U$, $B_G(u,d)\subseteq B_{G'}(u,d)$, and moreover, since $B_{G'}(u,d)\subseteq S''$, we get that $|B_G(u,d)|\leq n^{1-\eps}$. Therefore, every vertex in $S'$ is scattered. We delete the vertices of $S$ from $G$, and we mark the vertices of $S'\setminus S$ as scattered, so they are deleted from $U$. Notice that, for every vertex $x$ that remains in $U$, if $B_G(x,d)\subseteq B_{G'}(x,d)$ held at the beginning of the current iteration, then $B_G(x,d)\subseteq B_{G'}(x,d)$ continues to hold at the end of the current iteration, since $\dist_{G'}(x,S)>d$. The running time of the current iteration in the second case is bounded by $O(|E_{j+1}|)\leq O(\vol_G(S)\cdot m^{\eps})$.

It now remains to analyze the running time of the algorithm. There is at most one good iteration in the algorithm, and its running time is $O(m)$. The running time of a bad iteration is $O(\vol_G(S)\cdot m^{\eps})$, where $S$ is the set of vertices deleted from $G'$ in that iteration. Therefore, the total running time of the algorithm is $O(m^{1+\eps})$.

\section{Proofs Omitted from \Cref{sec: tieup}}
\subsection{Proof of \Cref{claim: valid input to embedorscatter}}
\label{subsec: valid input to embed or scatter}

Observe first that our algorithm ensures that the graph $C$ has no isolated vertices. It also ensures that $C$ is a large cluster, so 
$n'=|V(C)|\geq \Delta$ holds. From the definition of valid input parameters (see \Cref{def: input graph and parameters}), $n$ is greater than a large enough constant, and $k\leq (\log n)^{1/49}$, so $n'>\Delta\geq n^{18/k}\geq 2^{18(\log n)^{48/49}}$ can also be assumed to be sufficiently large. 

Since $n'\geq \Delta$, and since, from the definition of valid input graph and parameters, $\Delta\geq 16$, we get that $16\leq \Delta\leq n'$ holds as required.
From the definition of valid input parameters, we are also guaranteed that $512\leq k\leq (\log n)^{1/49}$ holds.
Since $n'>\Delta\geq n^{18/k}\geq n^{1/(4k^2)}$ (from the definition of valid input parameters), and $k\leq (\log n)^{1/49}$, we get that:

\begin{equation}\label{eq: log n'}
 (\log n')\geq \frac{1}{4k^2}\cdot \log n\geq \frac{\log n}{4(\log n)^{1/20}}\geq (\log n)^{3/5}.
 \end{equation}

Since $\hk=k^2$, we then get that:

\[\hk\leq (\log n)^{1/20}\leq (\log n')^{1/3}. \]

Lastly, from the definition of valid input parameters, $1/\delta$ must be an integer, with $\frac 1 k\leq \delta\leq \frac{1}{400}$. Since $k\leq (\log n)^{1/49}$, we get that $\frac{1}{(\log n)^{1/49}}\leq \frac 1 k\leq \delta$, and so $\frac{2}{(\log n')^{1/24}}\leq \delta\leq \frac{1}{400}$,
 as required.

\subsection{Proof of \Cref{claim: router certificate}}
\label{subsec: router certificate}

Recall that the algorithm for the \pruning problem ensures that the total number of vertices that initially lied in $W_{\hk}^{N,\Delta}$, but no longer lie in $U_{\hk}$ at the end of Phase $\Phi_0$ is bounded by:

\[\frac{|E'_0|}{\Delta}\cdot \hk^{4\hk}\leq  \frac{N^{\hk}}{2}. \]

Therefore, at the end of Phase $\Phi_0$, graph $W_C$ and set  $U_{\hk}$  contain at least $\frac{N^{\hk}}{2}$ vertices. From the definition of a properly pruned subgraph (see \Cref{def: properly pruned}), graph $W_C$ must contain at least $\frac{|U_k|\cdot \Delta}{4}\geq \frac{N^{\hk}\cdot \Delta}{8}$ edges. 

Since $\pset$ is an embedding of $W_C$ into $C'$, we get that:

\[
|V(C')|\geq |V(W_C)|\\
\geq  \frac{N^{\hk}}{2}
\]

holds. 
Therefore, $N\leq (2|V(C')|)^{1/\hk}$ holds. Additionally, since $N=\floor{|V(C)|^{1/\hk-1/\hk^2}}$, we get that $N\geq \half\cdot |V(C)|^{1/\hk-1/\hk^2}\geq \half \cdot |V(C')|^{1/\hk-1/\hk^2}$. Since every edge of $C'$ lies on some path of $\pset_C$, and every vertex of $C'$ participates in at least $\frac{\Delta}{n^{4/k^2}}$ such paths,
we conclude that $\left(W^{N,\Delta}_{\hk},W_C,\pset\right )$ is a valid router certificate for $C'$.

Since the edges of $W_C$ are embedded into graph $C'$ with congestion at most $\eta^*$, we get that:

\[|E(C')|\geq \frac{|E(W_C)|}{\eta^*}\geq \frac{N^{\hk}\cdot \Delta}{8\eta^*}.\]

Recall that all vertex degrees in graph $G$ are bounded by $\Delta\cdot n^{16/k}$, and that $N^{\hk}\geq \frac{|V(C)|}{2^{\hk}\cdot n^{1/\hk}}$. 
Therefore:

\[|E(C)|\leq |V(C)|\cdot \Delta\cdot n^{16/k}\leq N^{\hk}\cdot \Delta\cdot n^{17/k}\cdot 2^{k^2}. \]

Altogether, we get that:

\[|E(C')|\geq \frac{N^{\hk}\cdot \Delta}{8\eta^*}\geq \frac{|E(C)|}{8\eta^*\cdot n^{17/k}\cdot 2^{k^2}}\geq \frac{|E(C)|}{n^{18/k}},\]

since, from Inequalty \ref{eq: bound d eta}, $\eta^*\leq n^{5/k^2}$; we also used Inequality \ref{eq: k and n2}.

\subsection{Proof of \Cref{claim: constructing the subgraph}}
\label{subsec: proof of degree lowering claim}
The claim easily follows from the following observation.

\begin{observation}\label{obs: one step}	
	There is an algorithm, whose input is a connected bipartite graph $H=(X,Y,E)$, and parameters $z\geq 2$, $\hat \Delta,\gamma'\geq 1$ and $ r> 4\gamma'$, such that, for every vertex $x\in X$, $\deg_H(x)\geq z\cdot \hat \Delta$, and, for every vertex $y\in Y$, $\deg_H(y)\leq \gamma'\cdot z\cdot \hat \Delta$.
	The algorithm computes a collection $S\subseteq X$ of at least $\frac{|X|}{2}$ vertices, and, for every vertex $x\in S$, a collection $E'(x)\subseteq \delta_H(x)$ of its incident edges with $|E'(x)|=\ceil{\hat \Delta}$, such that every vertex $y\in Y$ serves as an endpoint in at most $r\cdot \ceil{\hat \Delta}$ edges in $\bigcup_{x\in S}E'(x)$.
	The running time of the algorithm is $O(|E|)$.
\end{observation}

We provide the proof of the observation below, after we complete the proof of  \Cref{claim: constructing the subgraph} using it.
We let $r=\frac{\hat r}{\ceil{\log |X|}}$; notice that
$r> 4\gamma'$ must hold.
We start with a graph $H'=H$, and perform at most $\ceil{\log |X|}$ iterations. In the $i$th iteration, we apply the algorithm from \Cref{obs: one step} to the current graph $H'$, with the parameters $\hat \Delta,z,\gamma'$ and $r$ that remain unchanged. Let $S_i\subseteq X$ be the collection of vertices computed by the algorithm, and, for every vertex $x\in S_i$, let $E'(x)$ be the correponding subset of its incident edges. We then delete from $H'$ all vertices of $S_i$, and any additional vertices of $Y$ that become isolated, and continue to the next iteration. It is immediate to verify that, after at most $\ceil{\log (|X|)}$ iteration, graph $H'$ becomes empty, and, for every vertex $x\in X$, we obtain a collection $E'(x)$ of its incident edges with $|E'(x)|=\ceil{\hat \Delta}$. Moreover, it is immediate to verify that every vertex $y\in Y$ serves as an endpoint in at most $r\cdot \ceil{\hat \Delta}\cdot  \ceil{\log (|X|)}\leq \hat r\cdot \hat \Delta$ edges in $\bigcup_{x\in S}E'(x)$. Since the number of iteration is bounded by $O(\log (|X|))$, we get that the running time of the algorithm is $O(|E|\cdot \log (|X|))$.

It now remains to prove \Cref{obs: one step}.

\begin{proofof}{\Cref{obs: one step}}
	Throughout the algorithm, we maintain a subset $S\subseteq X$ of vertices, starting with $S=\emptyset$. Whenever a vertex $x$ is added to set $S$, we also define a collection  $E'(x)\subseteq \delta_H(x)$ of its incident edges with $|E'(x)|=\ceil{\hat \Delta}$. Throughout the algorithm, we denote by $E'=\bigcup_{x\in S}E'(x)$. For every vertex $y\in Y$, we maintain a counter $n_y$, that counts the number of edges in $E'$ incident to $n_y$. We say that vertex $y$ is \emph{overloaded}, if $n_y\geq (r-1)\cdot \ceil{\hat \Delta}$ holds.
	At the beginning of the algorithm, we initalize the counter $n_y$ of every vertex $y\in Y$ to $0$. 
	We then process the vertices of $X$ one by one. We now describe the processing of a vertex $x\in X$. 
	
	We consider the set $\delta_H(x)$ of edges incident to $x$ in $H$, and we denote by $E''(x)\subseteq \delta_H(x)$ the subset of all such edges, whose endpont $y\in Y$ is not overloaded. If $|E''(x)|\geq \hat \Delta$, then we let $E'(x)\subseteq E''(x)$ be any subset containing $\ceil{\hat \Delta}$ edges. We add $x$ to $S$, and, for every vertex $y\in Y$, that serves as an enpoint of at least one edge in $E'(x)$, we increase the counter $n_y$ by the number of edges in $E'(x)$ for which $y$ is an endpoint; notice that, since $|E'(x)|=\ceil{\hat \Delta}$, $n_y\leq r\cdot \ceil{\hat \Delta}$ continues to hold.
	Otherwise, if $|E''(x)|<\hat \Delta$, then we say that $x$ is a \emph{discarded vertex}.
	Once all vertices of $X$ are processed, we consider the final set $S\subseteq X$, and, for every vertex $x\in S$, the subset $E'(x)\subseteq \delta_H(x)$ of its incident edges. Our algorithm ensures that $|E'(x)|=\ceil{\hat \Delta}$, and that every vertex $y\in Y$ serves as an endpoint of at most $r\cdot \ceil{\hat \Delta}$ edges in $E'=\bigcup_{x\in S}E'(x)$. It now only remains to prove that $|S'|\geq \frac{|X|}{2}$.
	
	Assume for contradiction that this is not the case, and let $S''=X\setminus S'$ be the set of all discarded vertices, so $|S''|\geq \frac{|X|}{2}$. Let $\Pi$ be the collection of pairs $(x,y)$, where $x\in S''$, and $y\in Y$ is an overloaded vertex, that is an endpoint of at least one edge in $\delta_H(x)$. From the definition of the discarded vertices, and since the degrees of all vertices in $X$ are at least $z\cdot \hat \Delta$, we get that $|\Pi|\geq (z-1)\cdot \hat \Delta\cdot |S''|\geq \frac{z\cdot \ceil{\hat \Delta}\cdot |X|}{4}$. On the other hand, since every vertex $y\in Y$ has degree at most $\hat \Delta\cdot z\cdot \gamma'$ in $H$, we get that every overloaded vertex may participate in at most $\hat \Delta\cdot z\cdot  \gamma'$ pairs in $\Pi$. Therefore, at least $\frac{|X|}{4\gamma'}$ vertices in $Y$ are overloaded. 
	
	Recall that we have assumed that $|S|\leq \frac{|X|}{2}$, and so $|E'|=|S|\cdot \ceil{\hat \Delta}\leq \frac{|X|\cdot \ceil{\hat \Delta}}{2}$. If a vertex $y\in Y$ is overloaded, then the number of edges in $E'$ that are incident to $y$ is at least $(r-1)\cdot \ceil{\hat \Delta}$. Therefore, the number of overloaded vertices in $Y$ is bounded by:
	
	\[\frac{|E'|}{(r-1)\cdot \ceil{\hat \Delta}}\leq \frac{2|E'|}{r\cdot \ceil{\hat \Delta}}\leq  \frac{|X|}{r}<\frac{|X|}{4\gamma'},  \]
	
	since we have assumed that $r> 4\gamma'$, a contradiction.
\end{proofof}

\newpage
\bibliographystyle{alpha}

\bibliography{fault-tolerant-spanners-main}

\end{document}